\documentclass[fleqn,usenatbib]{mnras}

\usepackage{newtxtext,newtxmath}

\usepackage[T1]{fontenc}

\DeclareRobustCommand{\VAN}[3]{#2}
\let\VANthebibliography\thebibliography
\def\thebibliography{\DeclareRobustCommand{\VAN}[3]{##3}\VANthebibliography}

\usepackage{graphicx}	
\usepackage{amsmath}	
\usepackage{orcidlink}  
\usepackage{multirow}   
\usepackage{textgreek}  
\usepackage[export]{adjustbox} 

\usepackage{color,soul} 
\usepackage{colortbl} 
\usepackage{threeparttable}
\usepackage{multicol}
\usepackage{multirow}
\usepackage{cancel}

\newcommand{\tbc}[1]{\textcolor[rgb]{1.0,0.0,0.0}{#1}}
\newcommand{\teffone}{$T_{\rm eff,1}$}
\newcommand{\tefftwo}{$T_{\rm eff,2}$}

\newcommand{\rev}[2]{\textcolor{black}{#2}}

\title[EBLM XX: Mass Comparisons]{The EBLM project -- XX. Absolute dynamical masses of five low-mass eclipsing binaries, and an evaluation of mass estimation methods}

\author[A. T. Stevenson et al.]{
Adam T. Stevenson,$^{1}$\thanks{E-mail: a.t.stevenson@bham.ac.uk}\orcidlink{0000-0003-2399-7619}
Yasmin T. Davis,$^{1}$\orcidlink{0009-0000-6625-137X}
Daniel Sebastian,$^{2}$\orcidlink{0000-0002-2214-9258}
Amaury H.M.J. Triaud,$^{1}$\orcidlink{0000-0002-5510-8751}
\newauthor
~Pierre F. L. Maxted,$^{3}$\orcidlink{0000-0003-3794-1317}
Alix V. Freckelton,$^{1}$\orcidlink{0009-0007-1053-0004}
Thomas A. Baycroft,$^{1,4}$\orcidlink{0000-0002-3300-3449}
Neda Heidari,$^{5}$ 
\newauthor
~David V. Martin,$^{6}$ 
Nicola J. Miller,$^{7}$\orcidlink{0000-0001-9550-1198} 
Alexandre Santerne$^{8,9}$\orcidlink{0000-0002-3586-1316} 
\\
\\
$^{1}$School of Physics \& Astronomy, University of Birmingham, Edgbaston, Birmingham B15 2TT, UK\\
$^{2}$Th\"uringer Landessternwarte Tautenburg, Sternwarte 5, 07778 Tautenburg, Germany\\
$^{3}$Astrophysics Group, Keele University, Staffordshire ST5 5BG, UK\\
$^{4}$Tsung-Dao Lee Institute, Shanghai Jiao Tong University, 1 Lisuo Road, Shanghai 201210, China\\
$^{5}$ Institut d’astrophysique de Paris, UMR7095 CNRS, Université
Pierre \& Marie Curie, 98bis boulevard Arago, 75014 Paris, France\\
$^{6}$ Department of Physics \& Astronomy, Tufts University, Medford, MA 02155, USA\\
$^{7}$ Department of Physics and Astronomy, Uppsala University, Box 516,
Uppsala, 75120, Sweden \\
$^{8}$ Université Grenoble Alpes, CNRS, IPAG, 38000 Grenoble, France \\
$^{9}$ Aix Marseille Univ, CNRS, CNES, LAM, Marseille, France
}
 
\date{Accepted 2026 September 22. Received 2026 August 31; in original form 2026 July 03}

\pubyear{\the\year{}}

\begin{document}
\label{firstpage}
\pagerange{\pageref{firstpage}--\pageref{lastpage}}
\maketitle

\begin{abstract}
Eclipsing binaries consisting of a solar-type primary and an M dwarf secondary are key laboratories to study objects at the bottom of the main sequence. Fundamental parameters of such stars have a direct impact upon exoplanetary studies, as searches for temperate Earth-sized planets focus on M dwarf hosts. Eclipsing binaries are essential to measure these parameters. However for stellar masses $< 0.35 \,\rm M_\odot$, few double-lined eclipsing systems are known. Thus, stellar evolution models and empirical relations have had to rely on single-lined eclipsing binaries, for instance those from the EBLM project (Eclipsing Binary -- Low Mass). Using high resolution spectroscopy to extract the weak signal of the secondary (flux ratio $<1$~per~cent in the optical), we transform five more systems from single to double-lined, deriving absolute masses for both stars. We then compare methods typically employed to estimate stellar masses for field stars or single-lined binaries with the absolute dynamical masses we obtain, to assess their accuracy. Our results for SB1 and SB2 analyses are mostly consistent within $1$-$\sigma$, though we identify a small systematic overestimation of SB1 masses derived from a combination of radial velocities and eclipse photometry. We show that the bias originates from primary radius overestimation during isochronal analysis, wherein the secondary light reduces photometric magnitudes and impacts the recovered radii by modifying the spectral energy distribution. With proper calibration, we can ensure the accuracy of using the full EBLM SB1 sample to constrain mass-radius-luminosity relationships for the smallest, yet most abundant, stars in the Galaxy.
\end{abstract}





\section{Introduction}

In exoplanetary studies, inferred planetary qualities like mass, radius, bulk density, and interior composition, are all highly dependent on the estimated stellar mass and radius. 
Lately, significant effort has been devoted to low-mass M dwarf star systems, as this opens up the possibility of detecting and characterising temperate $\sim$Earth-sized planets \citep[][]{Bonfils2013,Proxima-b,Dittmann2017,Trappist-1,Delrez2022-speculoos2,Peterson2023,SM2023-GJ1002,Dransfield2024,Brady2025,Turner2025,Scott2026}, and allows investigations into the atmospheres these small planets may or may not retain \citep[][]{Lustig-Yaeger2023,May2023,Banerjee2024,JWST-trappist-community-initiative,Taylor2025-GJ357b,Kreidberg-Stevenson-2025,Ducrot2025}. Such analyses require stellar parameters to be known to a high degree of accuracy and precision -- yet measuring parameters of field stars is an indirect process, where observed properties are typically compared to evolution models to provide an estimated radius, mass, and age of the host star \citep{SME1996,Mortier2013,tayar-2022,freckelton-2024-bebopv}. The ubiquitous adage ``know thy star, know thy planet'' is particularly relevant for Earth-like planets, to correctly interpret that a particular world's atmosphere is consistent with liquid water at its surface \citep{triaud24}, or to correctly infer the presence of biosignatures.

To create relationships linking mass, radius, luminosity, and metallicity for the smallest M dwarfs, we require a sample of accurately characterised fundamental parameters for these stars. Typically, it is only possible to measure stellar masses and radii accurate to below $2$~per~cent from observations of detached double-lined eclipsing binary star systems \citep{Andersen1991,Torres2010,Southworth-debcat}.

Herein lies the problem: current catalogues suffer from an underpopulation of M dwarf eclipsing binaries (EBs). The relative scarcity of such systems is exacerbated below the fully-convective transition at $\sim0.35$\,M$_{\odot}$ \citep{ChabrierBaraffe1997}
-- only 16 systems with a component below this mass have precisely-measured fundamental parameters and are included in the Catalogue of Detached Eclipsing Binary Stars \citep[DEBCat][]{Southworth-debcat}.
The low number of late-type double-lined (SB2) EBs currently known \citep{Schweitzer2019,Pass2023} can be explained by numerous factors, including
their intrinsic faintness, and their relatively low multiplicity fraction compared to earlier types \citep[e.g.][]{duchene-kraus-2013,Winters2019,Winters2026,Offner2023}. Despite the fact that M~dwarfs are the most common type of star, based on estimates of the stellar initial mass function \citep{Salpeter1955,Kroupa2002,Chabrier2003} and observations \citep{Henry2006,Pass2023}, accumulating a substantial sample of fully convective double-lined EBs is no mean feat. Furthermore, many of these EBs are very short-period systems \citep[e.g.][]{Martin2024-CMDra} where the spin-up of the star(s) by tidal forces will make them magnetically active \citep{Skumanich1972,Hurley2002}. These are typically not useful for calibrating mass-radius relations for slowly-rotating planet hosts, as the interactions may also affect the stellar radii.

One solution is to target single-lined eclipsing binaries, which are far easier to identify due to the increased luminosity of the $\sim$solar type primary star. The EBLM project (Eclipsing Binary -- Low Mass) studies such systems, initially flagged as false positives in transiting exoplanet searches, to measure masses and radii for a larger sample of M~dwarfs \citep{triaud-2013-eblm-i,Maxted-2023-eblm}. These can be employed to calibrate relationships for the lower main sequence \citep{triaud-2013-eblm-i}. 


For EBLM systems with a light curve and a single-lined (SB1) radial velocity (RV) orbit, one additional piece of information ($M_{1}$ or $R_{1}$) is needed to derive measurements of $M_{2}$ and $R_{2}$. Most often, these estimates depend on stellar evolution models. Consequently estimates of the low mass star's physical parameters may be impacted by this. Ideally, we require a way to benchmark the SB1 results from the EBLM project. Since double-lined detections provide the gold standard for accuracy and precision, 
efforts have gone into detecting the spectra of faint secondary stars, effectively converting SB1 systems into SB2s \citep{PojmanskiUdalski1997,Maxted2022,sebastian_2024_saltire,sebastian-2024-bebop1,sebastian_2025-kepler16}.

At short orbital separations, binary stars will be unresolved on the sky, and the light entering a spectrograph fibre will contain the spectral signature of both stars. For EBLM targets, the\rev{}{average} contrast ratio may be on the order of $\sim10^{-3}$\rev{,}{($10^{-2}$) across the visible (near-infrared) wavelength range,} yet it is not zero. As the weak signature from the secondary is hidden within the noise produced by the primary, we require bespoke techniques to extract these signals. 

This task is similar to detecting the thermal emission spectrum from the dayside of an exoplanet atmosphere, using ground-based high-resolution spectroscopy \citep[e.g.][]{Brogi2012-tauBoo,deKok2013,Birkby2017-51peg,BrogiLine2019}.
As a short period exoplanet's velocity significantly varies throughout an orbit, it is possible to isolate the weak but predictably moving spectrum of the planet from that of the host star, or from Earth's telluric lines. This field has had numerous successes over recent years \citep[e.g.][]{Snellen2008,Snellen2010,BrogiLine2019,Lafarga2023,Snellen2025}, by cross-correlating a model spectrum with the observed data to tease the planet's moving signature out from the noise. Techniques developed to facilitate the detection of such weak features can now be applied to the study of high contrast ratio binary stars, not just EBLMs \citep{Thomsen25}.

For EBLM systems, with a typical contrast ratio an order of magnitude less extreme than that of exoplanets, high resolution cross-correlation spectroscopy (HRCCS) has proved extremely successful \citep{sebastian-2024-bebop1,sebastian_2025-kepler16,Baycroft-2025-bebop3,triaud-bebop4}. 
\citet{sebastian_2024_saltire} recently highlighted that this method could measure a companion M dwarf's semi amplitude from optical spectra, and this allows us to effectively transform EBLM systems into double-lined binaries.
This opens the door to two distinct analysis options: traditionally, as an SB1, and with HRCCS as an SB2. Performing both methods in tandem can serve as a validation of the methods used for the remainder of the EBLM sample, testing the accuracy of the results and allowing for them to be calibrated.

In this paper, we (i) apply the HRCCS method to five EBLM targets to measure accurate and precise absolute dynamical masses for both stellar components in each system, and (ii) compare these masses to those produced by model-dependent methodologies. We outline the selection of our sample and the spectroscopic observations in Section~\ref{sec:sample}, and describe the mass estimation methods applied to these observations in Section~\ref{sec:method}. Our results are presented in Section~\ref{sec:results}, alongside a comparison between methods and investigation of any discrepancies. The implications of our results for the validity of modelling these systems as single-lined binaries are discussed in Section~\ref{sec:validity-of-sb1}. We compare to benchmark stars and catalogues of detached eclipsing binaries in Sections~\ref{sec:benchmarks} and~\ref{sec:debcat}, calculate fundamental effective temperatures of these EBLMs in Section~\ref{sec:teff}, and conclude our findings in Section~\ref{sec:conclusion}.

\section{Sample and Observations}\label{sec:sample}

A substantial collection of spectra have been obtained for EBLM targets as part of the BEBOP search for circumbinary planets \citep{martin-2019-bebopI,standing-2022-bebopii,triaud-2022-bebopiii}. The availability of high-resolution spectra makes a small number of these datasets amenable to secondary companion characterisation with HRCCS. For this work, we select a sample of these targets where we expect to be able to detect the secondary with a minimum significance of 3-$\sigma$. Northern targets have been observed with the SOPHIE spectrograph, mounted on the T193 telescope at the Observetoire de Haute-Provence \citep{perruchot-sophie}. Spectra for the southern BEBOP sample are obtained with HARPS, at the ESO 3.6-m telescope at La Silla \citep{mayor-harps}. Both instruments have their data automatically reduced by \rev{the Data Reduction Software (DRS)}{their respective data reduction softwares\footnote{The \href{https://ohp.osupytheas.fr/wp-content/uploads/2024/06/Doc_SOPHIE_v4.pdf}{SOPHIE DRS} \citep[e.g.][]{Bouchy2009} is adapted from the \href{https://www.eso.org/sci/facilities/lasilla/instruments/harps/doc/DRS.pdf}{HARPS DRS} \citep[e.g.][]{Pepe2004}} (DRS)}. The DRS debiases, flattens and deblazes the data (using calibration files), performs the wavelength solution, and extracts each individual \'echelle order into what are referred to as the ``2D spectrum''.
Individual orders are cross-correlated, and the cross-correlation is fit with a Gaussian to extract the radial velocity of the primary star. Individual orders are also spliced together to produce the final data product, the ``1D spectrum".
SOPHIE has a typical spectral resolution around 75,000 in high resolution (HR) mode, whereas HARPS achieves 115,000.

To investigate the difference between dynamical and assumption-dependent masses (for SB1s), we also strive to detect secondary companions with masses spanning as much of the M dwarf domain as possible. Our targets here range from $\lesssim0.35$\,M$_{\odot}$ fully convective stars (at the extreme of what is possible with sub--$4$-m telescopes), to the upper limit of the spectral type at $\sim0.65$\,M$_{\odot}$. Our sample is listed in Table~\ref{tab:sample}, alongside information relating to the spectroscopic observations.

Furthermore, we include three previously published SB2 analyses of BEBOP targets to give a final complement of eight systems where we can compare fundamental parameters. These are EBLM J0608-59 \citep[BEBOP-1][]{standing-2023-bebop1,sebastian-2024-bebop1}, Kepler-16 \citep{triaud-2022-bebopiii,sebastian_2025-kepler16}, and EBLM J0723-79 \citep[BEBOP-3][]{Baycroft-2025-bebop3}. 
To be consistent between methods and ensure our study is as homogeneous as possible, we limit our sample selection to those SB2 analyses carried out in the visible, and require that all systems have primary eclipses in \textit{TESS} photometry to perform the independent joint modelling analysis typical of SB1 systems in the EBLM project \citep{triaud-2013-eblm-i,davis-2024}. As such, NIR analyses of EBLM J0113+31 and J0040+01 are not included in the present work \citep{Maxted2022,2025CoSka..55c.412S}, nor is EBLM J1250+26 as it lacks a \textit{TESS} eclipse \citep{triaud-bebop4}. These targets will be included in future studies.

\renewcommand{\arraystretch}{1.4}
\begin{table*}
	\centering
	\caption{The targets studied in this work, and the details regarding the spectra. Exposure times are occasionally varied, thus the $t_{\rm exp}$ column lists the typical exposure times for these observations under optimal observing conditions. \rev{}{The median effective rank used in SVD cleaning is reported in the final column.}
    }
	\label{tab:sample}
	\begin{tabular}{lcccccccccc} 
		\hline
		EBLM \#  & TIC ID & $V$ & Prelim. $M_{2}$ [M$_{\odot}$] & Instrument & Mode & Fiber B & $t_{\rm exp}$ [s] & $N_{\rm spectra}$ & $\overline{S/N}$ (@550nm) & \rev{}{Med(Rank)} \\
		\hline
        J0310-31  & 89045042 & 9.33 & $\sim0.4$ & HARPS & HR& SKY & 1800  &  71 & 96.58 & \rev{}{2} \\
        J1540-09  & 32431480 & 10.87 & $\sim0.4$ & HARPS & HR & SKY & 1800  & 98  & 44.42 & \rev{}{4}  \\
        J1928-38  & 469755925 & 11.20 & $\sim0.25$ & HARPS & HR& SKY & 1800  &  73 & 37.27 & \rev{}{4} \\
        J1555+26  & 258108067 & 8.31 & $\sim0.65$ & SOPHIE & HR& SKY & 900  &  64 & 76.04 & \rev{}{3} \\
        J1840+34  & 68271160 & 10.33 & $\sim0.6$ & SOPHIE & HR& SKY & 1800 &  84 & 46.08 & \rev{}{6} \\
        \hline
	\end{tabular}
\end{table*}

\section{Mass measurement methods}\label{sec:method}
In this section, we describe the methods used to estimate $M_2$. Firstly, we describe the dynamical mass measurement made possible by effectively transforming these systems into double lined binaries with HRCCS, used to determine absolute $M_2$ as well as $M_1$. We then introduce \rev{other, non-absolute methods}{methods that are dependent on empirical or stellar evolutionary models} to measure masses of secondary M dwarfs in EBLM binaries, that we will employ as comparators. These include using a combination of RV and photometric data to solve the binary mass function after adopting an \rev{indirect}{isochrone} primary radius estimate (Section~\ref{sec:rv+phot}), or estimating the secondary mass  by combining RV observables with inclination when assuming a MIST (Section~\ref{sec:spectra}), empirical relationship (Section~\ref{sec:emprel}), or \textit{TESS} input catalogue (Section~\ref{sec:tic}) primary mass.

\subsection{Dynamical masses from double-lined detections}\label{sec:saltire}

To measure dynamical masses for our EBLM binaries, we use the same HRCCS methodology as in \citet{sebastian_2024_saltire,sebastian-2024-bebop1,sebastian_2025-kepler16}, and \citet{Baycroft-2025-bebop3}, which we \rev{rapidly}{briefly} describe in the following subsections.

\subsubsection{Preparation of the high-res spectra}
 
For HRCCS, we use the 1D spectra produced by the data reduction software.
We start by splitting these HARPS/SOPHIE pipeline-processed
one-dimensional spectra into chunks of $\sim50$\,\AA, to make the detrending processes computationally feasible.

Each chunk of 1D spectrum is renormalised using a polynomial of order 6, fitted to the continuum regions. To fit this continuum, 
outlying regions associated with stellar lines, emission features, and hot pixels are masked by 
two stages of median clipping applied to five subsets of the chunk. Remaining values that are $0.5\sigma$ greater than the median are subsequently used to fit the continuum. The 1D spectral chunks are divided by this continuum, producing a spectrum that is normalised to 1 \citep{sebastian-2024-bebop1}.

The next step involves removal of the primary star's absorption spectrum. Following \citet{sebastian_2024_saltire,sebastian-2024-bebop1,sebastian_2025-kepler16}, we apply singular value decomposition (SVD). This is applied to each chunk containing all spectra for an individual star, shifted into the primary's rest frame. As described in \citet{sebastian-2024-bebop1}, we benefit from the fact that the orbital parameters of the primary star are known to an extremely high precision, enabling a successful shift into the rest frame. The required RV analysis is performed as described in \citet{standing-2022-bebopii} and \citet{baycroft-2023-binaries}, with the \textsc{kima} analysis framework \citep{kima-joss,kima-ascl}. Orbital parameters for our target binaries are reported in Table~\ref{tab:kima-solutions}.

\begin{table*}
	\centering
	\caption{Orbital parameters from the \textsc{kima} analysis of the five targets studied with HRCCS in this work. All quoted values are posterior medians, and $1$-$\sigma$ errors are displayed as the last two significant figures in brackets. All orbits are characterised to an extremely high degree of precision. Similar solutions for J0310-31, J1540-09, and J1928-38 were published by \citet{standing-2022-bebopii}, though are now updated with additional radial velocity measurements. Period, eccentricity, argument of periastron, and $T_{\rm peri}$ allow Equation~\ref{eqn:rv2} to be solved for each trial value of $K_{2}$.
    }
	\label{tab:kima-solutions}
	\begin{tabular}{lllllllcc} 
		\hline
		  Parameter & J0310-31 & J1540-09 & J1928-38 & J1555+26 & J1840+34 \\
		\hline
        $P_{\rm bin}$ (d) & 12.64279255(45) & 26.3382840(34) & 23.3229514(41) & 12.9683317(55) & 19.228803(31) \\
        $K_{\rm 1}$ (km\,s$^{-1}$) & 27.87480(51) & 23.17846(63) & 17.26199(50) & 39.0040(44) & 35.3917(88) \\
        $e_{\rm bin}$ & 0.308670(16) & 0.120469(26) & 0.073102(29) & 0.345890(99) & 0.43084(23) \\
        $\omega_{\rm 1}$ (rad) & 3.243647(58) & 1.09820(24) & 2.39060(66) & 3.67786(34) & 0.46276(49) \\
        $T_{\rm peri}$ [BJD-2400000] & 58895.49837(12) & 59051.00922(96) & 59424.7348(24) & 59293.61832(52) & 59344.9113(13) \\
        $\varv_{\rm sys}$ (km\,s$^{-1}$) & 28.97562\rev{(0.35)}{(35)} & -55.41480(47) & 16.3896(76) & -32.2001(27) & -13.7247(52) \\
        \hline
	\end{tabular}
\end{table*}

\begin{figure}
    \centering
    \includegraphics[width=0.95\columnwidth]{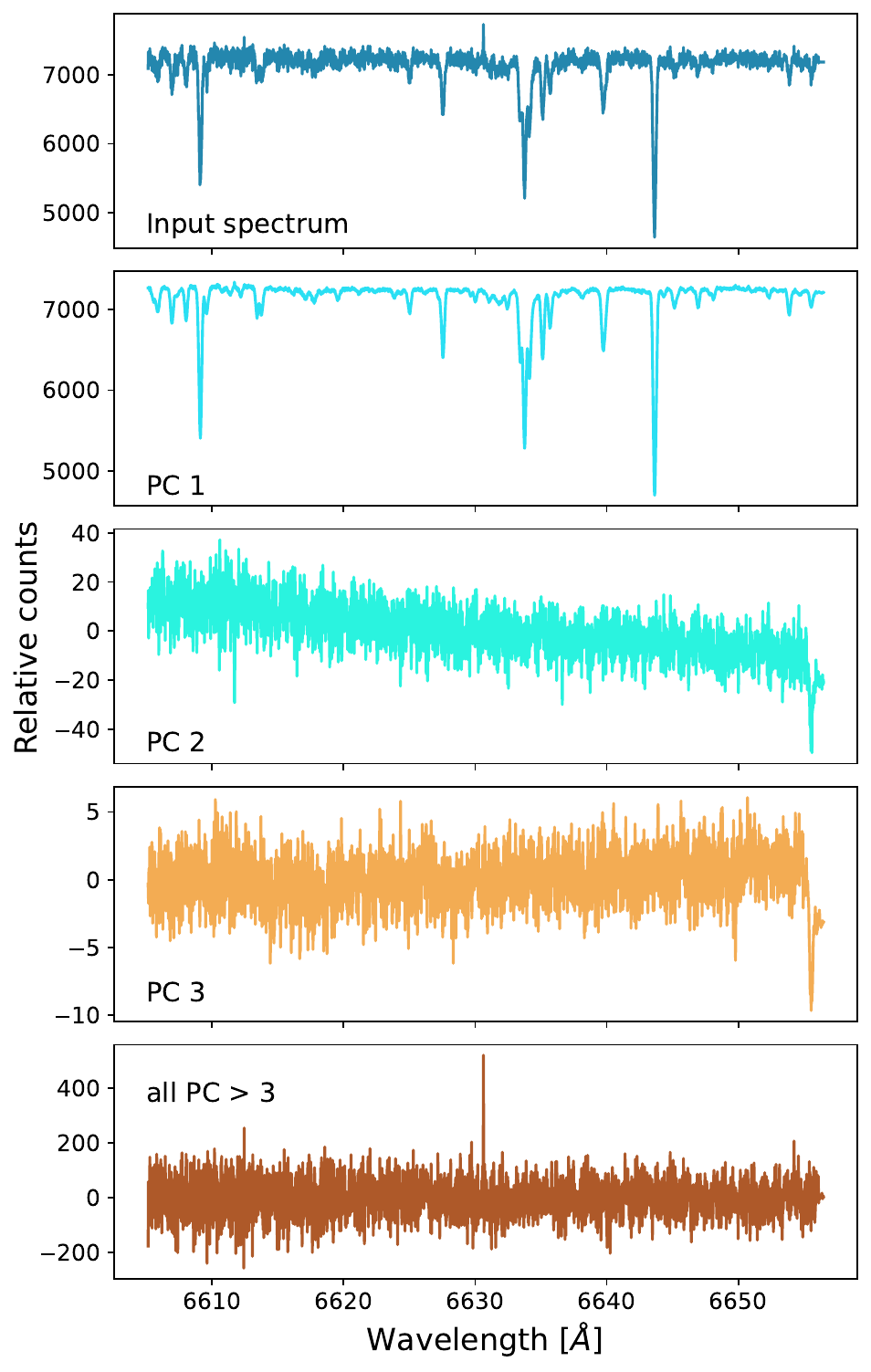}
    \caption{Singular value decomposition and reconstruction for a given spectral chunk. The principal components (PCs) are identified from all available spectra for a particular target. An example spectral chunk is shown in the top panel of this plot. The first few PCs of this spectrum are reconstructed from the identified common features of the primary star. Removing more than 2 or 3 PCs, the reconstructed `cleaned' spectrum no longer shows the absorption features present in the input spectrum. Each spectrum/component is shifted arbitrarily for visibility, and components beyond the first are scaled up to show the variation (or lack thereof).}
    \label{fig:svd_example}
\end{figure}

The SVD produces an eigenvalue decomposition of a matrix, here the spectral array for each chunk. The first few components of this decomposition display the highest correlation/level of similarity, indicating features that are common between the rows and columns of the spectral matrix -- the primary's spectrum. Beyond a certain number of components, the residual variation will be related to noise or any other weak signals in the data, as exemplified by Figure~\ref{fig:svd_example}. 

We use the `effective rank' \citep{roy-vetterli-2007} to quantify the number of components to be retained for each spectral chunk -- though impose that no more than 12 components should be removed, so that the secondary's signal is not excessively degraded when noise is significant. \rev{}{As each chunk is cleaned separately, the number of principal components varies across the wavelength coverage. The median number or removed components is reported in Table~\ref{tab:sample}.} This process produces `cleaned' spectra, where the absorption features from the primary star's spectrum have been suppressed. 

\subsubsection{Creation of CCF maps}
To extract the faint signal of the secondary star from the noise, we use the \textit{K}-focusing method \citep{sebastian_2024_saltire}. This involves enhancing the signal in each spectrum by cross correlating the spectrum with a template spectrum or line mask, and then combining each cross-correlation function (CCF) in the rest frame of the secondary star.

The secondary shares a number of parameters with the primary's reflex orbit, namely the systemic velocity, binary orbital period, time of periastron passage, eccentricity, and inclination ($\varv_{\rm sys}$, $P_{\rm bin}$, $T_{\rm peri}$, $e$, and $i$, respectively). The argument of periastron ($\omega$) can easily be calculated via $\omega_{2} = \omega_{1}-\pi$. The semi amplitude of the primary star (caused by the secondary) is $K_1$ (measured in the previous step). 
The only unknown is the secondary's semi amplitude, $K_{2}$. To derive a CCF in the secondary's rest frame, we sample over a grid of $K_{2}$ and $\varv_{\rm sys}$. We may not measure the same mean Doppler shift for the two stars \citep{sebastian-2024-bebop1}, because this depends on the assumed rest wavelengths of the measured features, and the different sets of lines used for each star. The lines can also be affected by differing convective blueshifts and gravitational redshifts \citep{LD2003}, so we leave $\varv_{\rm sys}$ as a free parameter in our search. If the CCF maximum is close to the expected systemic velocity calculated from analysis of the primary RVs, this helps inspire confidence in a detection. A similar approach has been used in the exoplanet literature \citep[e.g.][]{Snellen2010}.  

To derive the CCF, we cross-correlate the spectra with a stellar line mask moved into the secondary's rest frame, shifting by a radial velocity that is the combination of trial systemic velocity and epoch secondary RV, calculated via
\begin{equation}
    \varv_{\rm r,2} = K_{2}[\cos{(\nu+\omega_{2})+e\cos{(\omega_{2})}}] + \varv_{\rm sys}.
    \label{eqn:rv2}
\end{equation}
\rev{The argument of periastron from the secondary is derived from that of the primary ($\omega_{2}=\omega_1-\pi$), and the}{The} true anomaly $\nu$ is calculated from the period and the time of periastron \citep[e.g.][]{sebastian-2024-bebop1}. 
The Keplerian orbit described by Equation~\ref{eqn:rv2} is justified as none of these binaries show significant additional variability in the long-term monitoring of their RVs, due to, e.g. a tertiary star or other massive circumbinary object. For such systems, additional Keplerian signals first need removed before performing this HRCCS technique -- as was done for BEBOP-4b \citep{triaud-bebop4}.

The stellar line masks employed here are the HARPS data reduction pipeline\footnote{Publicly available from \url{https://ww.eso.org}} M2 or K5 masks. These are optimised for high-precision RV measurements, and provide the added benefit of excluding any spectral regions that suffer from telluric contamination. The K5 mask was trialled and used for the more massive companions at the transition between late-K and early-M type (J1555+26\,B, J1840+34\,B) as this resulted in a stronger CCF signal.

To measure semi-amplitude and systemic velocity, we sample over a $K_2$--$\varv_{\rm sys}$ grid, roughly centred on the semi-amplitude expected from combining the primary semi-amplitude and an estimate for the primary mass, and expected systemic velocity from the primary's RV solution. The grid has resolution of $1.5\times1.5~\rm{km\,s}^{-1}$, found by \citet{sebastian_2024_saltire} to provide the optimal balance: the CCF maps remain well sampled, yet are coarse enough to avoid erroneous correlations in the data and retain the computationally efficient nature of the algorithm.

For each point in the grid, we compute the CCF for each individual post-processed spectrum, creating a 2D CCF. These are normalised, then combined with an SNR-weighted mean. This creates a CCF map, where the CCF signal becomes strongest (largest contrast) at the location where $K_2$ and $\varv_{\rm sys}$ match the true parameters of the secondary. CCF maps for our five targets are shown in the left column of Figure~\ref{fig:saltire}.

\subsubsection{Applying the \texttt{Saltire} model}
To measure the position of the signal in the CCF maps, we use the \texttt{Saltire} model, described in detail in \citet{sebastian_2024_saltire}. The model uses a double Gaussian function to account for side-lobes commonly seen in M dwarf CCFs \citep[e.g.][]{Bourrier2018}, which also goes through the $K$-focusing process. We effectively create a model map to fit the measured map, and extract the parameter values. As this model is created with the same observational time-stamps as the data, it accurately recreates the signal and enables up to 10 times more accurate measurements of $K_{2}$ than the traditional approach of fitting a 1D Gaussian to a slice through the map at $\varv_{\rm sys}$ of maximum contrast \citep{sebastian_2024_saltire}. The model fits the whole shape of the signal, meaning the resulting parameters are less biased by local noise structures close to the location of maximum CCF amplitude \citep{sebastian-2024-bebop1}.

The double Gaussian model used in \texttt{Saltire} is described by five parameters: standard deviations ($\sigma_{1},\sigma_{2}$) and intensities ($A_{1},A_{2}$) of the two Gaussian functions, together with the height ($h$) of the CCF outside the signal. To fit the model, we use the sum ($\Sigma=A_1 +A_2$) and quotient ($\Delta = A_2 /A_1$) of the intensities and free parameters. This reparameterization allows us to fit symmetric side-lobes, by ensuring the two intensities have opposite signs \citep{sebastian_2024_saltire,sebastian-2024-bebop1}.

The \texttt{Saltire} model is initially fit with least-squares minimisation, and the parameter posterior distributions are sampled with Markov chain Monte Carlo (MCMC) exploration. The MCMC samples an additional `jitter' term, a weighting in the log-likelihood function that prevents over-fitting, by accounting for noise, spurious correlations, or artefacts in the data. This allows us to estimate detection significance, as the summed intensity of the two Gaussians divided by the jitter ($\Sigma/\sigma_{\rm jit}$).

\begin{figure*}
    \centering \includegraphics[trim={0.0cm 0 0.0cm 0},clip=true,width=0.92\linewidth]{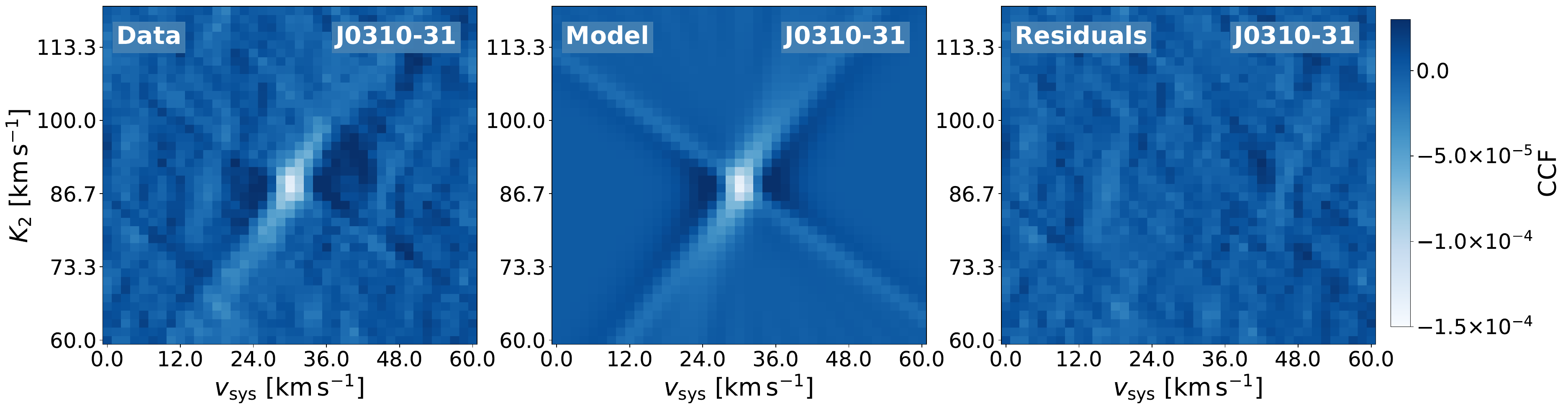}\\
    \hspace{0.15cm}\centering \includegraphics[trim={0.0cm 0 0.0cm 0},clip=true,width=0.915\linewidth]{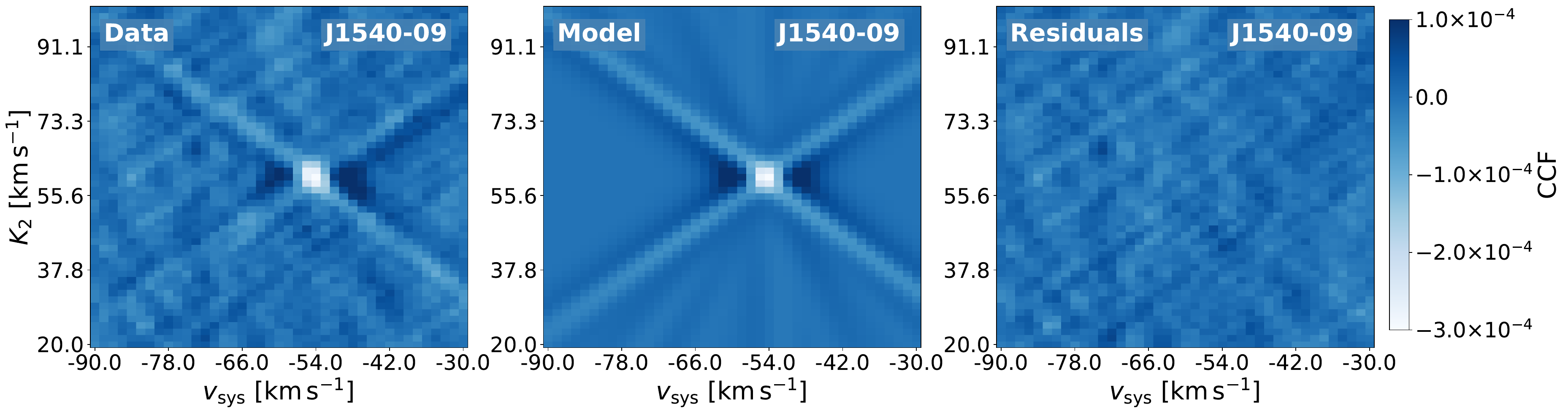}\\
    \hspace{0.15cm}\centering \includegraphics[trim={0.0cm 0 0.0cm 0},clip=true,width=0.915\linewidth]{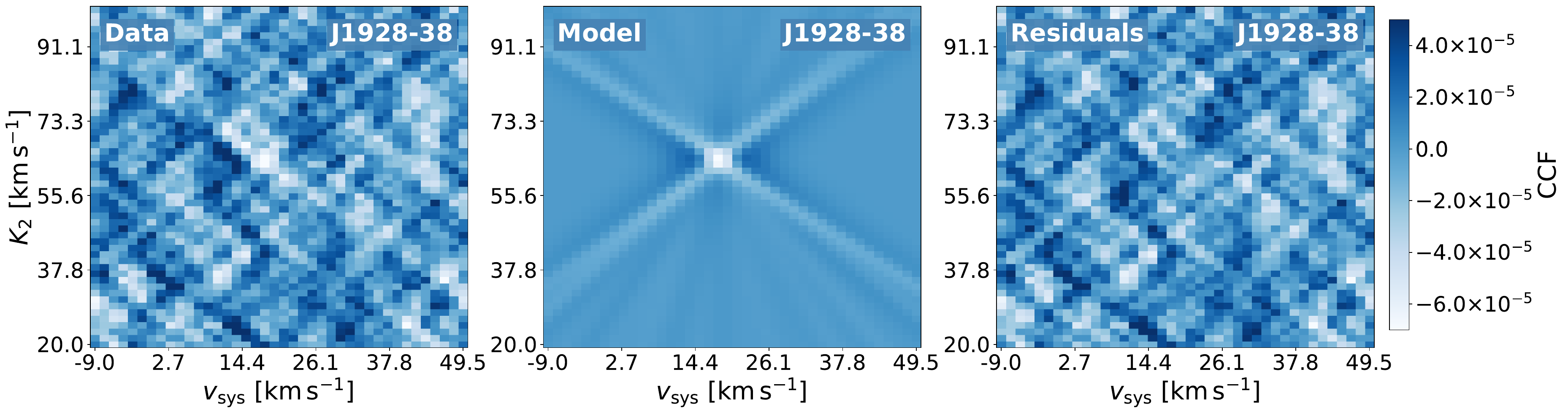}\\
    \centering \includegraphics[trim={0.0cm 0 0.0cm 0},clip=true,width=0.92\linewidth]{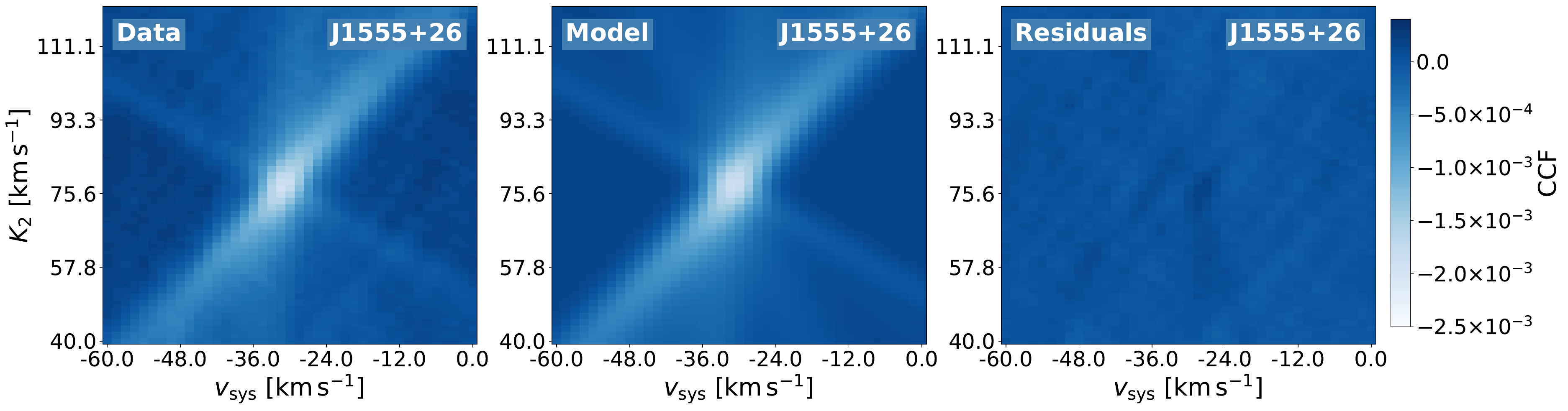}\\
    \centering \includegraphics[trim={0.0cm 0 0.0cm 0},clip=true,width=0.92\linewidth]{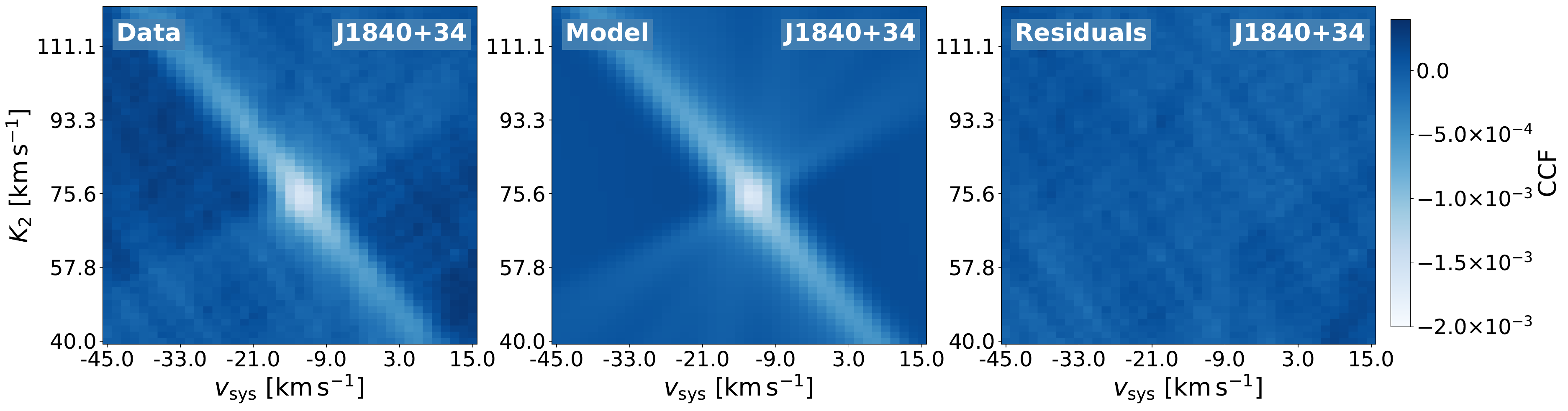}
    \caption{HRCCS detections of targets listed in Table~\ref{tab:sample}, with names indicated on the plot. \textbf{Left:} $K$-focused cross-correlation maps. \textbf{Center:} \texttt{Saltire} models for the SOPHIE or HARPS data, with parameters from MCMC. \textbf{Right:} residual CCF maps after model subtraction. Data and model have been median-subtracted to allow for a common scale to be shared between the three panels in each row.}
    \label{fig:saltire}
\end{figure*}

For our targets, we follow the prescription laid out by \citet{sebastian-2024-bebop1} to sample the parameter space. For all posterior distributions we report the parameters as the 50th percentile, and errors by averaging the 16/84 percentiles\rev{}{(as our posteriors are symmetric to first order, see Figure~\ref{fig:example_corner})}. Our \texttt{Saltire} results are presented in Table~\ref{tab:saltire-results}. The fitted models, with parameters from MCMC, are shown alongside CCF maps in Fig.~\ref{fig:saltire}. The residual maps are also shown, in the right column of the figure. The models successfully reproduce the observed structures seen in real data, accurately measuring the signal position in the $K_2$--$\varv_{\rm sys}$ space.

\begin{table*}
\caption{Measured fit parameters from \texttt{Saltire} MCMC analyses. We sample for 4000 calls on 42 parallel chains, reject 1500 burn-in samples for each walker, and thin the chains by a factor of 5. The posterior distributions each contain 21000 samples \citep{sebastian-2024-bebop1}. Uncertainties are averages of the 16/84 percentiles of the posterior, \rev{and are precision errors only}{except from those reported for $K_{\rm 2}$ and $\varv_{\rm sys,2}$: these errors are the combination of both fit and systematic errors, where contributions to the error budget are shown in the rows directly below these parameters}.}
\begin{adjustbox}{width=0.995\textwidth}
	\centering
	\label{tab:saltire-results}
	\begin{tabular}{lccccc} 
		\hline
		  Parameter & J0310-31 & J1540-09 & J1928-38 & J1555+26 & J1840+34 \\
		\hline
        \multirow{2}{*}{$K_{\rm 2}$ [km\,s$^{-1}$]} &  $88.2684 \pm 0.4552$  & $60.0282\pm 0.2620$  &  $64.3490\pm 2.8245$  & $77.8851\pm 0.3442$  & $75.0271\pm 0.2166$   \\
         &  \rev{}{(0.0666,0.4503)}  & \rev{}{(0.0474,0.2577)}  & \rev{}{(0.2340,2.8148)} & \rev{}{(0.0459,0.3412)} & \rev{}{(0.0589,0.2084)}   \\
        \multirow{2}{*}{$\varv_{\rm sys,2}$ [km\,s$^{-1}$]} & $30.1739 \pm 0.1038$  & $-54.5764 \pm 0.1862$ & $18.1638\pm 1.7584$ & $-30.9198\pm 0.2551$ & $-12.9830\pm 0.2306$  \\
        & \rev{}{(0.0400,0.0958)} & \rev{}{(0.0356,0.1828)} & \rev{}{(0.1700,1.7502)} &  \rev{}{(0.0244,0.2539)} & \rev{}{(0.0322,0.2283)} \\
        $\Sigma~(A_{1}+A_{2})$ & $-(1.4352\pm0.0374)\times10^{-4} $ & $-(3.2523\pm0.0752)\times10^{-4} $&  $-(8.3267\pm1.0025)\times10^{-5}$ & $-(2.0816\pm0.0102)\times10^{-3}$ &  $-(1.7995\pm0.0161)\times10^{-3}$ \\
        $\Delta~(A_{2}/A_{1})$ & $-0.3394\pm0.0336$ & $-0.4647\pm0.0218$ & $-0.3285\pm0.0815$  & $-0.0018\pm0.0024$ & $-0.0378\pm0.0058$ \\
        $\sigma_{1}$ [km\,s$^{-1}$] & $1.9243\pm0.0826$ & $2.1200\pm0.0579$ & $1.6426\pm0.2170$  & $3.5305\pm0.0218$ & $3.0155\pm0.0311$\\
        $\sigma_{2}$ [km\,s$^{-1}$] & $4.9733\pm0.2937$ & $4.9900\pm0.1765$ & $5.2758\pm1.1531$ & $30.8527\pm14.5272$ &  $19.6597\pm2.2088$ \\
        $\sigma_{\rm jit}$ & $(8.53\pm0.15)\times10^{-6}$ & $(2.00\pm0.03)\times10^{-5}$ & $(2.278\pm0.035)\times10^{-5}$ & $(4.872\pm0.074)\times10^{-5}$ &$(6.140\pm0.096)\times10^{-5}$  \\
        SNR~($\Sigma/\sigma_{\rm jit}$) & $16.80$ & $16.27$ & $3.65$ & $42.73$ & $29.31$  \\
        \hline
	\end{tabular}
\end{adjustbox}
\end{table*}

\subsubsection{Estimating Systematic Uncertainties}\label{sec:systematic}
As noted in previous works \citep[e.g.][]{sebastian-2024-bebop1}, the \texttt{Saltire} model fit to cross-correlation maps often overestimates the precision on measurements of $K_{2}$. The uncertainties returned by the least-squares minimisation and MCMC sampling are simply precision errors. Systematic uncertainty from correlated noise will be introduced during creation of the CCF map, and can account for spurious detections, particularly when working at low significance. To account for this, \citet{sebastian-2024-bebop1} performed the same analysis on partial samples, created by splitting the whole dataset into subsets. Each partial sample contains a different set of correlated noise, assuming white noise. By assuming the signal from the secondary component is constant, we can separate the noise from the signal. Uncertainties caused by correlated noise can be estimated by fitting CCF maps from partial samples.

We thus divide the datasets into four parts, usually corresponding to four observing seasons, and create and fit CCF maps, as done in \citet{sebastian-2024-bebop1}. Choosing to split by observing season ensures we have adequate orbital phase coverage in each part, and enough spectra to detect the signal, whilst remaining fully independent of the other parts. 
The systematic uncertainty is taken as the standard error of the best fitting $K_{2}$ parameters, from the MCMC sampler of \texttt{Saltire}. We apply Bessel's correction \citep{UaaoS} to account for the small sample size , and note that the uncertainty itself has fractional error of 41~per~cent \citep{Topping1972}. 

Unfortunately, for one target we could not employ time splitting analysis due to low detection significance. For J1928-38, the 73 spectra only facilitate a 3.7-$\sigma$ detection of the companion star. 
As a result, we employ the bootstrapping technique described by \citet{sebastian_2025-kepler16}. In this process, we select samples with a constant number of spectra (50~per~cent of total). These are randomly chosen, and we ensure that there is a minimal level of overlap between any two partial samples. We aim to keep the samples as independent as possible, to limit the similarity of the sets of correlated noise.
Drawing 200 partial samples, we create a CCF map for each sample and fit a \texttt{Saltire} model, deriving the systematic uncertainty on $K_{2}$ as the $1$-$\sigma$ 16/84 percentiles of the resulting $K_{2}$ values \citep{triaud-bebop4}.


\citet{sebastian_2025-kepler16} found that the bootstrapping uncertainty was consistent with that from the time-splitting, and thus could be successfully employed for faint signals that cannot be separated into several parts. We could have also derived bootstrapping errors for the other targets in this paper, though since these are amenable to time-splitting (and because the bootstrapping method is far more computationally demanding), we adopt the part-wise estimates for systematic error on $K_{2}$ where possible. Adding these in quadrature to the fit errors produces our final estimate of $\Delta K_{2}$, recorded in Table~\ref{tab:saltire-results}.

The sizes of our estimated systematic errors are consistent with previous works employing \texttt{Saltire}: our systematic error on $K_{2}$ is, on average, $\sim 8.3$ times larger than the fit (precision) error. \citet{sebastian-2024-bebop1} measured a systematic error $7\times$ larger than the fit error for ESPRESSO \citep{espresso} observations of J0608-59, and \citet{sebastian_2025-kepler16} measured systematics $3.6\times$ larger than the fit error for \texttt{Saltire} analysis of Kepler-16 SOPHIE spectra. This highlights the importance of estimating systematic errors for each system, as the fit error consistently overestimates the precision on $K_{2}$.

\subsubsection{Absolute Dynamical Masses}
 By combining high precision measurements of $K_{1}$ from SB1 RVs, and $K_{2}$ from the \texttt{Saltire} HRCCS detection, we measure dynamical masses for both stars in the binary. Using the inclination from a joint photometric and spectroscopic analysis (Section~\ref{sec:rv+phot}), we derive these masses following the IAU recommended equations for double-lined binaries \citep{prsa-2016-iau}. Dynamical masses from eclipsing SB2 systems are evolutionary model-independent, and thus provide ground-truth measurements to benchmark evolution models and other mass estimation methodologies.
 
 \rev{Here we note,}{It is important to consider} that the HRCCS method for measuring dynamical masses is different from previous SB2 approaches. \rev{(notably the SVD process can affect the signal from the secondary).}{Notably, the SVD process can in some ways reduce the signal from the secondary \citep{sebastian-2024-bebop1} at epochs where the secondary is stationary with respect to the primary.}
 This paper is about comparing this method to an SB1 approach, in an effort to provide a validation. We discuss more about this topic in Sec.~\ref{sec:results}.

\subsection{Masses using primary radial-velocities and photometry}\label{sec:rv+phot}

In this section, we use the same spectra but measure the secondary masses using a more traditional approach, where the secondary lines are not resolved (i.e. $K_2$ is not measured), and one must rely on an external assumption (here an estimate the radius of the primary star, $R_1$, as has been done for the EBLM project).
From here until the end of Section~\ref{sec:method}, we perform the analysis \emph{as if} $K_{2}$ was unavailable, to test the consistency between resulting masses from single-lined and double-lined methods, and assess the impact that stellar model dependencies impart on our conclusions.

For EBLM targets with sufficient photometric and spectroscopic measurements, we perform joint modelling of the data to extract system parameters, as described in \citet{davis-2024}. Primary and secondary masses can be calculated almost solely from physical observables, with the only required assumption being the radius of the primary star. Stellar radii are often estimated more accurately than masses when using empirical relations \citep[e.g.][]{Torres2010,Moya2018-empiricalrels}, and isochrone radii typically require smaller additive errors than masses do to account for any inaccuracies in the models \citep{tayar-2022}. For this method, we use the primary radii obtained through comparison with MESA Isochrones \& Stellar Tracks \citep[MIST; see Section~\ref{sec:spectra} and][]{freckelton-2024-bebopv}.

\renewcommand{\arraystretch}{1.4}
\begin{table*}
	\centering
	\caption{Results from the joint RV and photometry analysis\rev{,}{in Section~\ref{sec:rv+phot},} including total number of eclipses used during modelling alongside fitted and derived parameters.
    }
	\label{tab:transitparams}
    \begin{adjustbox}{width=0.999\textwidth}
	\begin{tabular}{lccccccc} 
		\hline
		Target  & $a/R_1$ & $R_2/R_1$ & $J$\tnote{*}* & $i\, [\rm deg]$ & $K_1\, [\rm km\,s^{-1}]$ & $P\, [\rm days]$ & $e$\\
		\hline 
        J0310-31  & $17.781_{-0.078}^{+0.095}$ & $0.272_{-0.016}^{+0.017}$ & $0.0650_{-0.0019}^{+0.0020}$ & $86.531\pm0.096$ & $27.86463\pm0.00061$ & $12.64279527\pm0.00000073$ & $0.308727\pm0.000019$ \\
        J1540-09  & $34.74_{-0.34}^{+0.35}$ & $0.344_{-0.051}^{+0.062}$ & - &  $87.92_{-0.13}^{+0.12}$ & $23.17269\pm0.00064$ & $26.3382853\pm0.0000030$ & $0.120479\pm0.000029$ \\
        J1928-38  & $24.87\pm0.23$ & $0.19328_{-0.00091}^{+0.00084}$ & $0.0809\pm0.0043$ & $89.133_{-0.062}^{+0.072}$ & $17.25821\pm0.00089$ & $23.3229526\pm0.0000048$ & $0.073001\pm0.000041$\\
        J1555+26  & $14.907\pm0.028$ & $0.2845_{-0.0021}^{+0.0023}$ & $0.2320_{-0.0027}^{+0.0025}$ & $86.295\pm0.016$ & $38.9883\pm0.0061$ & $12.9683360\pm0.0000035$ & $0.346237\pm0.000076$ \\
        J1840+34  & $27.681_{-0.032}^{+0.029}$ & $0.43040_{-0.00042}^{+0.00038}$ & $0.14080\pm0.00047$ &  $89.965_{-0.035}^{+0.025}$ & $35.3906\pm0.0085$ & $19.22888919\pm0.00000097$ & $0.431081\pm0.000093$ \\
        J0608-59  & $21.367_{-0.053}^{+0.056}$ & $0.23424\pm0.00039$ & $0.0730\pm0.0014$ & $89.639_{-0.075}^{+0.088}$ & $21.61200_{-0.00074}^{+0.00078}$ & $14.60855453\pm0.00000089$ & $0.155511\pm0.000031$ \\
        J0723+79  & $18.779\pm0.019$ & $0.19877_{-0.00021}^{+0.00019}$ & $0.06466\pm0.00042$ & $88.7488\pm0.0092$ & $19.35870\pm0.0010$ & $13.21766215_{-0.00000052}^{+0.00000055}$ & $0.063186\pm0.000048$ \\
        Kepler-16 & $74.373\pm0.091$ & $0.35035\pm0.00066$ & $0.2077\pm0.0018$ & $89.6556\pm0.0021$ & $13.6709\pm0.0022$ & $41.0775943\pm0.0000028$ & $0.15980\pm0.00016$\\
        \hline
	\end{tabular}
    \end{adjustbox}
    \begin{tablenotes}
  \item[*] *Average surface brightness ratio $= l_{T}/k^{2} = \frac{{\rm flux}_{2,TESS}}{{\rm flux}_{1,TESS}}/ \left (\frac{R_{2}}{R_{1}}\right ) ^{2}$.
  \end{tablenotes}
\end{table*}

\begin{table}
    \centering
    \caption{Target identifiers and total number of primary and secondary eclipses (120s cadence, SPOC pipeline) included in modelling.}
    \label{tab:transitinfo}
    \begin{tabular}{cccc}
    \hline
        Target & TIC ID  & N-Events (pri)  & N-Events (sec) \\
    \hline
        J0310-31  & 89045042 & 3 & 4 \\
        J1540-09  & 32431480 & 1 & 0 \\
        J1928-38  & 469755925 & 2\tnote{$\dagger$}$\dagger$ & 2$\dagger$ \\
        J1555+26  & 258108067 & 4 & 2 \\
        J1840+34  & 68271160 & 4 & 8 \\
        J0608-59  & 260128333  & 42 & 43 \\
        J0723+79  & 289949453  & 16 & 17 \\
        Kepler-16 & 299096355 & 9 & 7 \\
        \hline
    \end{tabular}
    \begin{tablenotes}
          \item[$\dagger$] $\dagger$ Total number of events includes both SPOC and TESS-SPOC (1800s cadence) pipelines.
    \end{tablenotes}
\end{table}

\begin{figure} 
    \centering
    \includegraphics[clip=true,trim={0pt 0pt 0pt 0pt},width=0.99\columnwidth]{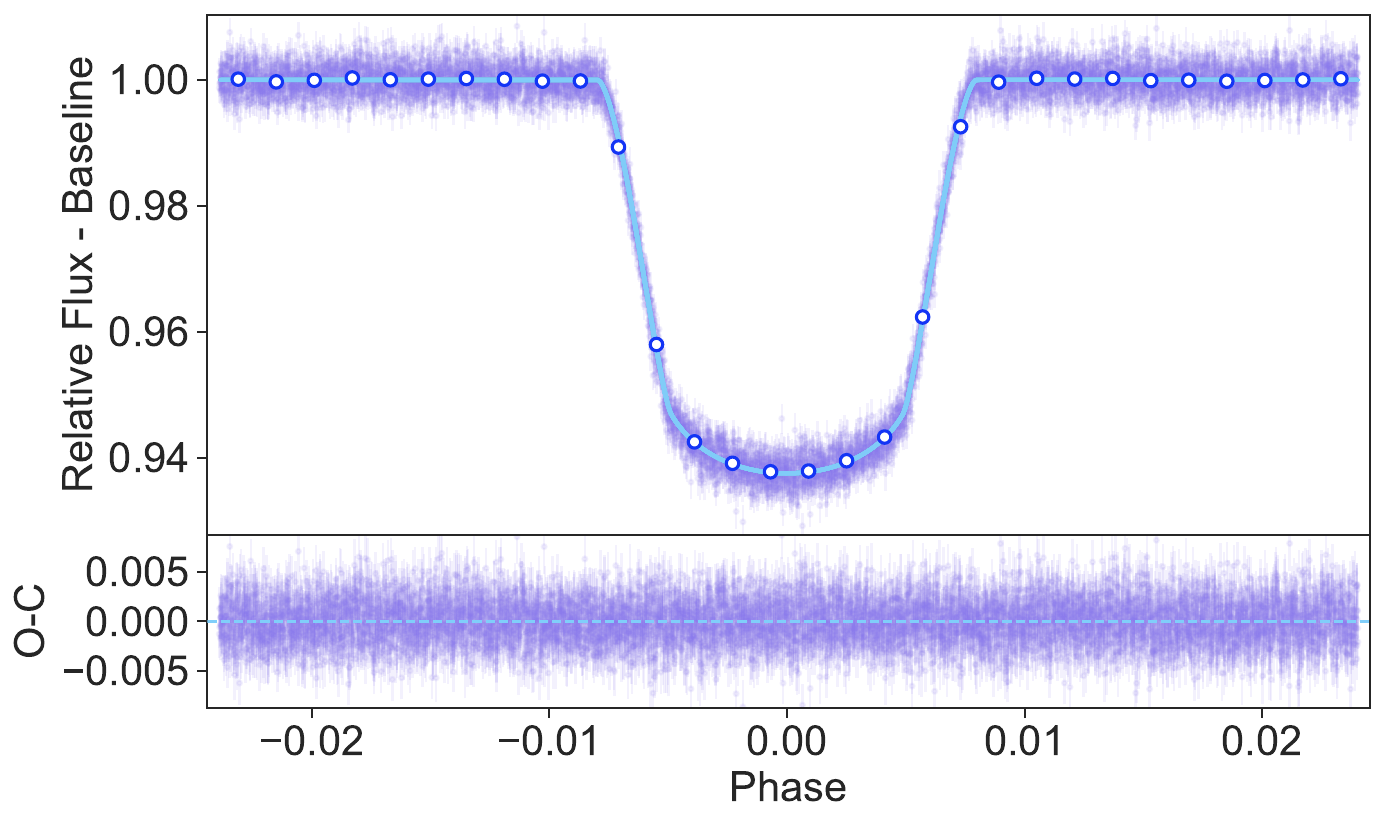}
    \includegraphics[clip=true,trim={0pt 0pt 0pt 0pt},width=0.99\columnwidth]{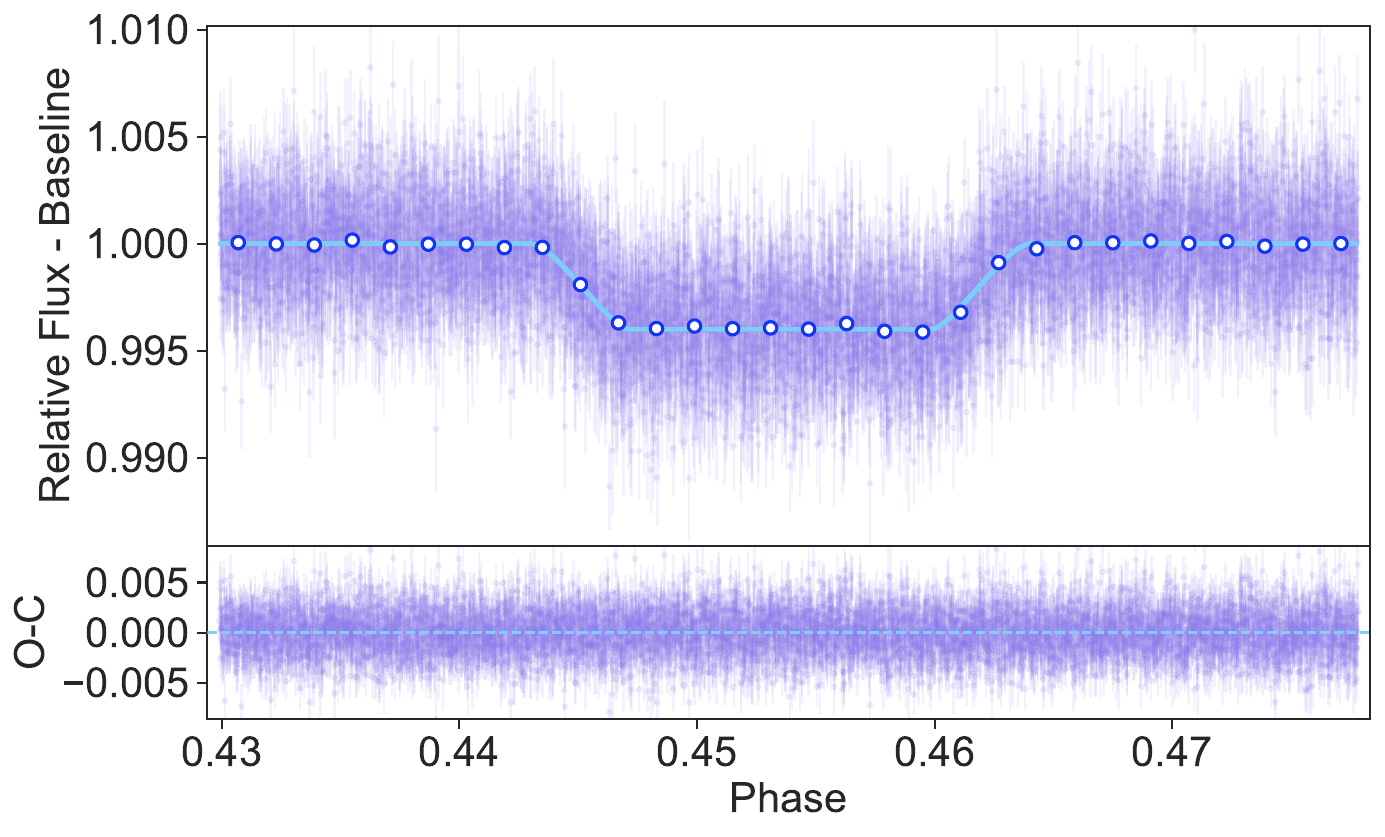}
    \caption{
    An example light curve fit, shown for EBLM J0608-59. The \textit{TESS} SPOC data are plotted as purple points, with empty circles representing binned data. \textbf{Top panels:} phase-folded primary eclipses and residuals for the best-fitting model. \textbf{Bottom panels:} phase-folded secondary eclipses and residials for the best-fitting model. In both plots, 50 random posterior samples are displayed (cyan curves).
    }
    \label{fig:example_lcfit}
\end{figure}

Our joint modelling is performed using \textsc{allesfitter} \citep{allesfitter-2019-ascl,allesfitter-2021-apjs}, where a short nested sampling (NS; \citealt{Skilling2006,dynesty}) run is initially employed on RV data alone to get the starting parameters, before including the wealth of photometric data from \textit{TESS}. An example fit to \textit{TESS} data is shown in Figure~\ref{fig:example_lcfit}. The \textit{TESS} data are processed by the SPOC pipeline \citep{spoc} and have 120\,s cadence\footnote{For one target (J1928-38) additional data authored by the \textit{TESS}-SPOC pipeline, with 1800s cadence, are also included in the modelling.}.

\rev{}{The \textsc{allesfitter} analysis incorporates \textsc{ellc} \citep{maxted-ellc} to accurately model the light curves and RV data of both detached eclipsing binary stars and transiting exoplanets.}
\rev{This combination of velocities and photometry allows us to measure $P,~K,~e,~i,~\textrm{and}~a/R_{1}$, and estimate the radius ratio and flux ratio in the \textit{TESS} band.}{This holistic approach allows us to measure $P,~K_1,~e,~i,~a/R_{1}$, and average surface brightness ratio in in the \textit{TESS} band ($J^{*}$). The parameters $P,~K_1,~\cos{i},~R_{2}/R_{1}~\textrm{and}~J^{*}$ are directly fed into the model. Orbital eccentricity $e$ is directly calculated from rearranging the reparameterized model inputs $\omega$, as $f_c = \sqrt{e}\cos{\omega}$ and $f_s = \sqrt{e}\sin{\omega}$ (where $\omega$ continues to denote the argument of periastron). Similarly, $a/R_{1}$ is calculated directly from the fitted parameters $R_2/R_1$ and $(R_1+R_2)/a$}

Parameters from the light curve fits are listed in Table~\ref{tab:transitparams}, with information relating to the photometry in Table~\ref{tab:transitinfo}. With these observables, one can estimate the total mass of the system ($M$) from a rearrangement of the stellar density derived from Kepler's law \citep[following][]{davis-2024}:


\begin{align}
    \frac{M_{1}}{R_{1}^{3}}&=\frac{4\pi^{2}}{G\,P^{2}}\bigg(\frac{a}{R_1}\bigg)^{3} - \frac{M_2}{R_{1}^{3}}, \\
    \implies M &= \frac{4\pi^{2}}{G\,P^{2}}\bigg(\frac{a}{R_1}\bigg)^{3} R_{1}^{3} = M_1 + M_2.   
    \label{eqn:stellar-density}
\end{align}

\noindent The mass function, also calculated from observed quantities, is given by \begin{equation}
    \label{eqn:massfunction}
    f = \frac{M_2 ^{3}}{(M_1 + M_2)^2} = \frac{P\,K^{3}}{2\pi G\,\sin^{3}{i}}(1-e^2)^{3/2} = C,
\end{equation}
\noindent which can be set equal to a constant, $C$, as both parts of the expression are fixed. The masses can therefore be calculated via

\begin{align}
    M_{2} &= \sqrt[3]{CM^2},~\textrm{and}\\
    M_{1} &= M - M_{2}.
    \label{eqn:M1M2}
\end{align}

This methodology has also been employed in the study of the eclipsing binary \textit{NGTS}-EB-7 \citep[][]{Rodel2025}, where the resulting masses were found to be
be consistent with those obtained from isochrone analysis, to $\sim1$-$\sigma$.
By performing a similar analysis for multiple targets considered herein, we can assess the consistency between techniques used to estimate the mass of primary stars in low-mass eclipsing binaries and, by extension, transiting exoplanetary systems.

\subsection{Spectral Analysis}\label{sec:spectra}
The atmospheric parameters for the majority of the primary stars discussed in the present paper were derived by \citet{freckelton-2024-bebopv}, using the \texttt{PAWS}\footnote{\url{https://github.com/alixviolet/PAWS}} pipeline. The available SOPHIE or HARPS spectra for each target were co-added, and spectral analysis performed on a high signal-to-noise ratio combined spectrum. For most of these targets, we do not repeat the analysis with any spectra collected since. However, Kepler-16 was not included in \citet{freckelton-2024-bebopv}, and previous analysis of J1928-38 was discovered to be impacted by instrumental effects disrupting the continuum of the spectra. We thus perform an identical analysis for these systems, to ensure a set of homogeneously-derived parameters for our primary stars. 

The \texttt{PAWS} pipeline implements both curve-of-growth equivalent widths (EW) and spectral synthesis, where the former is first employed to determine initial parameters for the computationally-demanding latter. The EW method uses \ion{Fe}{I} and \ion{Fe}{II} lines, abundant in the FGK primaries considered here, to return an initial estimate of stellar atmospheric parameters (aside from rotational and macroturbulent velocities, as EW is conserved under these two parameters). A Gaussian profile is fitted to each line separately using \texttt{ARES}, and the EW is measured as the width of the rectangle with height equal to that of the continuum and area equal to area under the continuum. The \texttt{WIDTH} radiative transfer code derives abundances for the Fe lines. 

These EW results are the starting point for spectral synthesis, where the \texttt{SPECTRUM} radiative transfer code is used to generate synthetic spectra for each input value set. The parameters are iterated in conjunction with a minimization algorithm until the optimal agreement is found between the synthesised and observed spectra. The resultant parameters are used to determine mass, radius, and age, through comparison with stellar evolution models. MIST isochrones \citep{dotter2016} are applied to interpolate the stellar parameters using the \texttt{isochrones} package \citep{morton2015}, sampled with \texttt{multinest} \citep{multinest}. Following \citet{mortier2020}, the spectroscopic $T_{\rm eff}$ and \text{[Fe/H]} are used with the Gaia DR3 parallax, 2MASS J, H, and K magnitudes \citep{skrutskie2006}, and AllWISE W1, W2, and W3 magnitudes \citep{cutri2013} to estimate the primary radii. 

Noise floors are added to the uncertainties on \texttt{PAWS} effective temperatures ($100$\,K), [Fe/H] (0.04 dex), $\log{g}$ (0.1 dex), and isochrone radii ($4.2$~per~cent), masses ($5$~per~cent), and ages ($20$~per~cent). This accounts for under-estimated uncertainty in the parameters, which is based solely on precision errors \citep[cf.][]{sousa-2011,tayar-2022}. 

With primary masses from atmospheric parameters and isochrones, we can solve the mass function (Equation~\ref{eqn:massfunction}) using a numerical method \citep[e.g.][]{Stevenson2023-BDD}, to derive an estimate for the mass of the secondary star (in combination with the inclinations from Section~\ref{sec:rv+phot}). A similar methodology provides the main avenue for characterising the fundamental parameters of exoplanet host stars, and as such it is key that we test the accuracy of our stellar models via comparison with absolute dynamical masses from HRCCS.

\subsection{Empirical Relations}\label{sec:emprel}

Relationships linking the fundamental parameters of stars and their stellar density, effective temperature, and metallicity can be derived by fitting polynomial functions to samples of stars with accurately-measured parameters \citep[e.g.][]{Torres2010,Enoch2010,Moya2018-empiricalrels}. Recently, \citet{Maxted2025} updated these relations using parameters from DEBCat, the catalogue of well-studied eclipsing binaries, where entries typically have mass and radius measurements accurate to within 2~per~cent \citep{Southworth-debcat}. Short-period ($<5$\,d, based on results from \citealt{Swayne2024-eblmXI}) or oblate stars were excluded, to make the sample as representative of single stars as possible. The relationships are valid for FGK dwarf stars with metallicities in the range $-0.4<$ [Fe/H] $<+0.3$, due to the demographics of the calibration sample selected from DEBCat. For convenience, we reproduce the 
scaling relation from \citet{Maxted2025} we have used in this work:

\begin{align}
    \log_{10}{(M/{\rm M_{\odot}})} &= 0.764084 + 3.014102~X + 2.272415~X^{2} \nonumber \\ 
    &- 0.042698~Y + 0.041479~Y^{2} - 0.019598~Y^{3} \nonumber \\
    &+ 0.124842~{\rm [Fe/H]},
    \label{eqn:emp-rel}
\end{align}
\noindent using the re-parametrised variables $X=\log_{10}{(T_{\rm eff}/{\rm K})}$ and $Y=\log_{10}{(\rho_{\star}/{\rm \rho_{\odot}})}$. We calculate the mean solar density using IAU/CODATA recommended values for the solar mass and radius \citep{IAU2015b3,prsa-2016-iau,CODATA2018}. Stellar density is calculated iteratively via Kepler's law, using the observables $P$ \rev{(obtained from the fit to RVs)}{}and $a/R_{1}$ (derived via photometric analysis in Section~\ref{sec:rv+phot}), where mass ratio ($q$) is updated for each iteration. 

It is important to note that due to the range of metallicities used in the calibration sample, we are unable to determine an empirical relation mass for J1540-09. Spectral analysis from \citet{freckelton-2024-bebopv} suggests the primary is a metal-poor star (with [Fe/H] $= -0.62\pm0.07$~dex), meaning that it is not advisable to use the \citet{Maxted2025} empirical relations. In the figures and tables throughout the following sections, this mass estimate has been omitted.


The use of empirical relations provides us with yet another means to explore mass estimation for EBLM targets, and an extra value to be compared to SB2 dynamical masses as we measure them in Sec~\tbc{\ref{sec:saltire}}. The relationships are not dependent on stellar evolution models, but they do rely on parameters estimated from analysis of the spectra. Where these are required, we adopt atmospheric parameters produced by \texttt{PAWS} \citep[Section~\ref{sec:spectra} and][]{freckelton-2024-bebopv}.

\subsection{\textit{TESS} Input Catalogue}\label{sec:tic}

As a final set of stellar mass to compare to the {\tt Saltire}-measured masses, we also include mass estimates from the \textit{TESS} input catalogue (TIC)\rev{}{version 8 \citep{stassun-2019-revisedTIC}}. A large, ideally homogeneous, science-grade catalogue can provide an additional consistency check, or a preliminary estimate to use before detailed study on a particular system is performed.

All primary stars studied in the present paper have been observed by \textit{TESS}, and thus have fundamental parameters reported in the TIC. For each target, the apparent magnitude in the \textit{TESS} bandpass (\textit{T}) is first calculated. The following stages require an estimate of the effective temperature of the star, either calculated using an empirical relation between \textit{Gaia} BP/RP colours and the spectroscopically determined $T_{\rm eff}$ for a subset of 20\,000 stars, or adopted from previous publications\footnote{Introducing effective temperatures from independent sources means the catalogue is not strictly homogeneous, though the majority of stars will not have readily available high-resolution spectra and any previously estimated $T_{\rm eff}$}.



Stellar mass is then inferred for all stars on the main sequence using a spline relationship between mass and $T_{\rm eff}$. For a complete description of the astrophysical parameter calculation involved in creating the TIC, we direct the reader to \citet{stassun-2019-revisedTIC}.

By design, the TIC primary masses are highly dependent on empirical relationships between (model dependent) parameters of the star and physical observables. Some of these observables that are based on photometry can be biased by the presence of a companion star. For instance, the stellar radii reported in the TIC will suffer for $\sim$equal mass binary stars. However, this is negligible for EBLM binaries with low contrast ratios \citep{Maxted2025}.
Nevertheless, TIC primary and secondary masses (after solving the binary mass function) provide a baseline estimate with which to compare our other methods and results.

\section{Resultant stellar masses}\label{sec:results}

The stellar masses calculated with the aforementioned methods (Section~\ref{sec:method}) are listed in Table~\ref{tab:all-masses}. \rev{}{We include the three previously published \texttt{Saltire} masses from \citet{sebastian-2024-bebop1,sebastian_2025-kepler16} and \citet{Baycroft-2025-bebop3}}. In this table, we highlight the results from \texttt{Saltire} analysis (model independent absolute masses) and those from the RV + Photometry method (interchangeably ``SB1'' throughout this work) that is typical of EBLM mass--radius studies \citep[][]{davis-2024,eblm-northern-sample}. The remaining methods only produce an estimate for the primary mass (TIC, MIST, empirical relation), so the secondary mass is derived by combining the primary mass with the parameters measured during RV fitting\footnote{We have two sets of parameters from RV fitting: either from \texttt{kima}, or from \texttt{allesfitter} when photometry is also included. These results are ever-so-slightly different, yet have a completely negligible impact on the derived SB2 masses -- the difference is below 10~per~cent of the uncertainty level. To be consistent with previously published dynamical masses, we use the results from the \texttt{kima} analysis, though note that the choice is arbitrary.}
(Table~\ref{tab:kima-solutions}). Masses for each system are plotted in Figures~\ref{fig:primaries} and~\ref{fig:secondaries}, and are seen to be mostly consistent with one another. The targets along the abscissa are ordered by the measured \texttt{Saltire} (SB2) mass.

The estimated masses are typically in good agreement with the absolute SB2 masses (within $1\sigma$ for the majority), thanks in part to the size of the uncertainty on these values. It is immediately apparent that the SB2 results can reach a significantly higher precision than other methods. Model dependencies introduce other errors based on the accuracy of our physical understanding, accounted for by inflating the estimated uncertainty. One target, J1928-38, has comparable uncertainties for all methods. This is a $\sim3\sigma$ detection with \texttt{Saltire} (see Table~\ref{tab:saltire-results}), representing an edge-case: this system pushes the boundaries of HRCCS detections with a $2$--$4$-m class telescope.
To raise the detection strength from $\sim3\sigma$ to beyond $10\sigma$ for J1928-38, and reach precision that is comparable to other targets in our sample, we would require $\mathcal{O}\sim100$ ESPRESSO observations, as demonstrated by \citet{sebastian-2024-bebop1} for J0608-59. This could also be achieved with fewer NIR observations on $2$--$4$-m telescopes, leveraging the improved contrast ratio in the infrared \citep[e.g.][]{Maxted2022}. 
In summary, HRCCS and ``SB2'' analyses for EBLMs provides both accurate and precise masses with which to benchmark our other methodologies.

\definecolor{Blue}{cmyk}{0.16,0.08,0,0.02}
\newcolumntype{b}{>{\columncolor{Blue}}c}
\definecolor{yazblue}{cmyk}{0.488,0.176,0,0.0196}
\newcolumntype{y}{>{\columncolor{yazblue}}c}
\definecolor{cyan}{cmyk}{0.15,0.0,0,0.0}
\newcolumntype{x}{>{\columncolor{cyan}}c}
\definecolor{Pink}{cmyk}{0.0,0.16,0.12,0.02}
\newcolumntype{p}{>{\columncolor{Pink}}c}
\begin{table*}
    \caption{
    Primary and secondary masses for the eight systems considered in this work, with sources described in Section~\ref{sec:method}. Secondary masses are estimated directly in the RV+photometry (SB1) and \texttt{Saltire} (SB2) methods, whereas the others are calculated from primary mass estimates in combination with inclination and primary RV semi-amplitude. 
    Results from bold methods are focused on throughout Sections~\ref{sec:results} and~\ref{sec:validity-of-sb1}. The targets are ordered by increasing \texttt{Saltire} primary mass, in accordance with Figure~\ref{fig:primaries}.
    }
    \label{tab:all-masses}
\begin{adjustbox}{width=0.99\textwidth}
\begin{tabular}{lcccccccccc}
\hline
& \multicolumn{2}{c}{\textbf{\texttt{Saltire} (SB2)}} & \multicolumn{2}{c}{TIC} & \multicolumn{2}{c}{MIST} & \multicolumn{2}{c}{\textbf{RV + Photometry (SB1)}} & \multicolumn{2}{c}{Maxted Empirical Relationship}  \\
Target & \textbf{$M_{1}$ [M$_{\odot}$]} & \textbf{$M_{2}$ [M$_{\odot}$]} & $M_{1}$ [M$_{\odot}$] & $M_{2}$ [M$_{\odot}$] & $M_{1}$ [M$_{\odot}$] & $M_{2}$ [M$_{\odot}$] & \textbf{$M_{1}$ [M$_{\odot}$]} & \textbf{$M_{2}$ [M$_{\odot}$]} & $M_{1}$ [M$_{\odot}$] & $M_{2}$ [M$_{\odot}$] \\
\hline
Kepler-16 & $\mathbf{0.704\pm 0.011}$ & $\mathbf{0.205\pm 0.002}$ & $0.630\pm 0.078$ & $0.192\pm 0.014$ & $0.701\pm 0.039$ & $0.205\pm 0.007$ & $\mathbf{0.704\pm 0.09}$ & $\mathbf{0.205\pm 0.016}$ & $0.704\pm 0.038$ & $0.205\pm 0.007$ \\
J1928-38 & $\mathbf{1.027\pm 0.117}$ & $\mathbf{0.276\pm 0.019}$ & $1.001\pm 0.128$ & $0.271\pm 0.021$ & $1.091\pm 0.070$ & $0.286\pm 0.011$ & $\mathbf{1.016\pm 0.092}$ & $\mathbf{0.274\pm 0.015}$ & $1.144\pm 0.074$ & $0.294\pm 0.012$ \\
J0723+79 & $\mathbf{1.084\pm 0.026}$ & $\mathbf{0.262\pm 0.004}$ & $1.127\pm 0.142$ & $0.268\pm 0.021$ & $1.157\pm 0.078$ & $0.272\pm 0.011$ & $\mathbf{1.165\pm 0.154}$ & $\mathbf{0.273\pm 0.023}$ & $1.136\pm 0.068$ & $0.269\pm 0.010$ \\
J0608-59 & $\mathbf{1.098\pm 0.017}$ & $\mathbf{0.308\pm 0.003}$ & $1.127\pm 0.068$ & $0.313\pm 0.012$ & $1.009\pm 0.073$ & $0.292\pm 0.013$ & $\mathbf{1.265\pm 0.179}$ & $\mathbf{0.335\pm 0.028}$ & $0.985\pm 0.066$ & $0.288\pm 0.012$ \\
J1540-09 & $\mathbf{1.112\pm 0.012}$ & $\mathbf{0.429\pm 0.003}$ & $1.049\pm 0.127$ & $0.415\pm 0.03$ & $0.853\pm 0.044$ & $0.368\pm 0.011$ & $\mathbf{1.252\pm 0.175}$ & $\mathbf{0.461\pm 0.039}$ & / & / \\
J1555+26 & $\mathbf{1.189\pm 0.012}$ & $\mathbf{0.595\pm 0.004}$ & $1.345\pm 0.212$ & $0.639\pm 0.058$ & $1.417\pm 0.106$ & $0.658\pm 0.029$ & $\mathbf{1.380\pm 0.196}$ & $\mathbf{0.648\pm 0.055}$ & $1.326\pm 0.115$ & $0.634\pm 0.032$ \\
J1840+34 & $\mathbf{1.339\pm 0.009}$ & $\mathbf{0.632\pm 0.003}$ & $1.257\pm 0.194$ & $0.609\pm 0.054$ & $1.297\pm 0.096$ & $0.62\pm 0.026$ & $\mathbf{1.594\pm 0.227}$ & $\mathbf{0.699\pm 0.06}$ & $1.203\pm 0.119$ & $0.594\pm 0.033$ \\
J0310-31 & $\mathbf{1.350\pm 0.017}$ & $\mathbf{0.426\pm 0.003}$ & $1.190\pm 0.173$ & $0.395\pm 0.035$ & $1.255\pm 0.082$ & $0.408\pm 0.016$ & $\mathbf{1.462\pm 0.198}$ & $\mathbf{0.447\pm 0.038}$ & $1.205\pm 0.089$ & $0.398\pm 0.018$ \\
\hline
\end{tabular}
\end{adjustbox}
\end{table*}

\begin{figure*}
    \centering
    \includegraphics[width=0.99\linewidth]{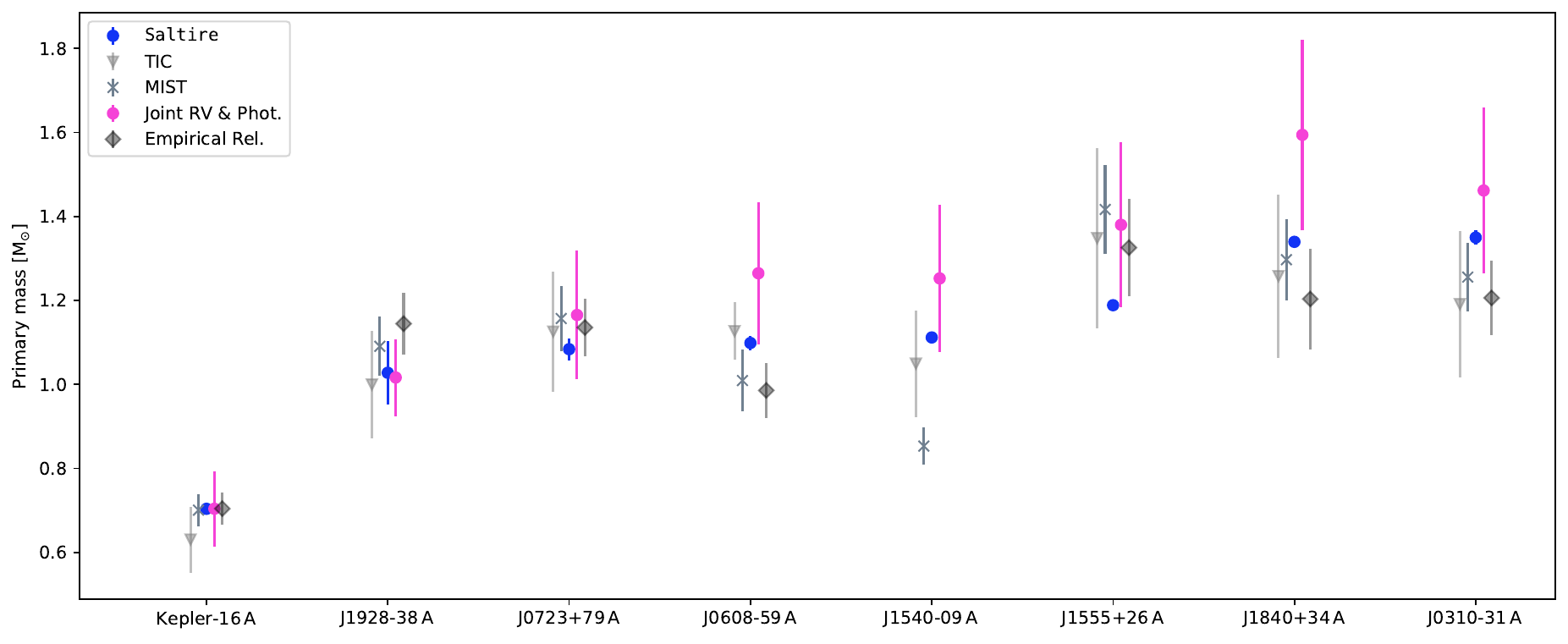}
    \includegraphics[width=0.99\linewidth]{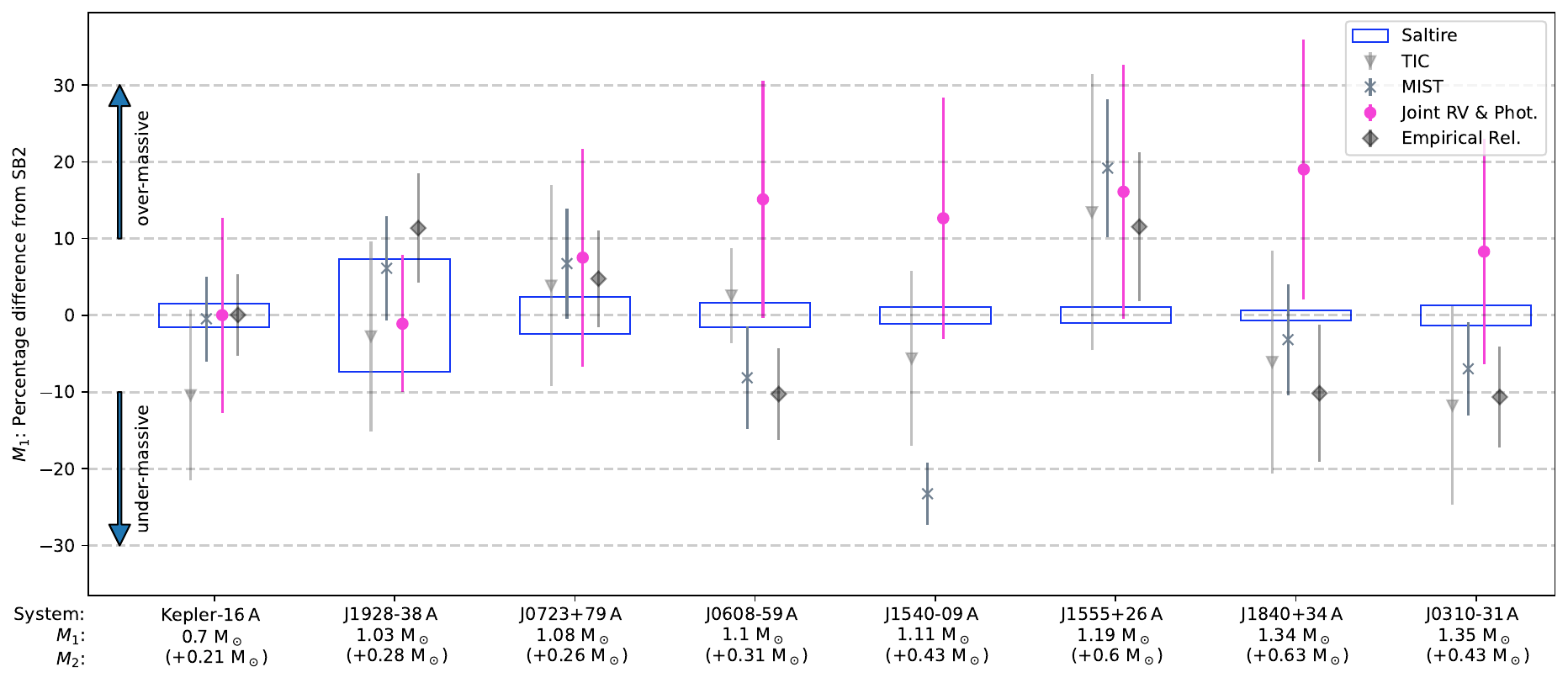}
    \caption{Figures comparing primary masses from different methods. \textbf{Top:} primary masses of the eight systems included in our sample (Table~\ref{tab:all-masses}). Marker colours and styles represent each method. \textbf{Bottom:} Masses plotted as percentage deviation from the \texttt{Saltire} SB2 results (blue boxes, including uncertainty). A systematic increase is seen for the RV+Photometry (SB1) method. The 0~per~cent difference is denoted with a horizontal dashed line. Results from other sources are scattered around the plots. Targets along the x-axes are ordered by primary mass, denoted below the name in the lower panel. The analogous plots for the secondary stars is shown in Figure~\ref{fig:secondaries}, which is very similar though scaled differently.
    }
    \label{fig:primaries}    
\end{figure*}

\subsection{A Systematic bias?}\label{sec:bias}

When interpreting the range of masses produced by different analysis methods, a curious feature becomes apparent. Considering the \texttt{Saltire} SB2 value as the ``absolute'' mass, we see the TIC, MIST, and empirical relation values are scattered around the SB2 result. However, the RV+Photometry results (Section~\ref{sec:rv+phot}) are often larger than the SB2 measurement -- especially for the more massive primaries and secondaries. This can be more easily demonstrated by inspecting the percentage `deviation-from-\texttt{Saltire}', shown in the bottom panel of Figure~\ref{fig:primaries}.

The boxes in Figure~\ref{fig:primaries} represent the SB2 value and error, and other methods are plotted around these for each star. For this plot of percentage deviations from a reference value, we have simply scaled the error bars by the value of the reference:
\begin{equation}
    \sigma_{f} = \frac{\sigma_{M}}{M_{\rm reference}}.
\end{equation}
\noindent Whilst the RV+Photometry results are typically consistent with SB2 values to within $\sim1\sigma$ (or just outside that for J1840+34), the median primary masses for six out of eight stars in the sample lie above the SB2 result. 

To calculate statistics on the percentage differences, we now compute the errors on the percentage under/over-estimation ($\Delta$) by error propagation (where the reference SB2 measurement is denoted as $M_{\rm ref}$):

\begin{align}
    \Delta &= \bigg(\frac{M}{M_{\rm ref}} -1\bigg), \\ \
    \implies \sigma_{\Delta} &= \sqrt{\bigg(\frac{\partial \Delta}{\partial M}\sigma_{M}\bigg)^{2} + \bigg(\frac{\partial \Delta}{\partial M_{\rm ref}}\sigma_{M_{\rm ref}}\bigg)^{2}}, \\
        & = \sqrt{\bigg(\frac{1}{M_{\rm ref}}\sigma_{M}\bigg)^{2} + \bigg(\frac{-M}{M_{\rm ref}^{2}}\sigma_{M_{\rm ref}}\bigg)^{2}}. 
\end{align}
\noindent Using \textsc{combine} from \textsc{pycheops} \citep{maxted-pycheops}, we calculate a weighted mean, including external errors for noise, of $\overline\Delta = 7.9\pm5.2$~per~cent, with $\chi^{2}=1.97$. Although the sample size is small, we find the mean is $>1$-$\sigma$ different from zero, indicating there could be some form of systematic bias. Although a larger sample size is required to fully explore the possibility of a systematic bias, we can begin to investigate possible causes for the discrepancy.

With such large radial velocity and transit/eclipse signals for our EBLM binaries, it is unlikely that any systematic bias stems from \rev{the measured or derived parameters returned by}{inaccuracies in observable parameters measured by} the RV+Photometry method. \rev{These}{The fitted parameters} are constrained by the data to an extremely high precision and accuracy, so the bias is likely to be introduced by the only quantity input to the method: the primary radius $R_{1}$ (Equation~\ref{eqn:stellar-density}). A biased isochrone radius would be a worrying result, as these are typically employed to estimate the mass/radius of single exoplanetary host stars. 

It may also be the case that the SB1 results are not in fact over-estimated, but that the SB2 masses are fairly systematically under-estimated. In the latter parts of this section, we also assess any possible sources of deviation that would mean the ``absolute'' masses we measure do not constitute the final word.

\subsection{Terminal age bias}\label{sec:tams}

Isochrone fitting can tend to produce unlikely results in cases where uncertainties on atmospheric parameters are large, or inaccurately determined. As stars are typically very stable over the main sequence (MS) lifetime, inaccurately determined/inconsistent parameters can place the star on an evolutionary track corresponding to rapid evolution away from the main sequence, where characteristics become more variable. 

This is known as the `terminal age bias' \citep{PontEyer04,PontEyer05,Maxted2015}, where best-fit methods produce too many stars in short-lived evolutionary stages. If not adequately modulated by a prior, there is a disproportionate chance that a star will be estimated to lie in the high-age region of the Hertzsprung-Russell diagram. As radii increase as stars evolve off the main sequence, perhaps our methods are affected by this bias and we are finding stars with overly inflated ages/evolutionary stage, and radii that are larger than they are in reality. 

We inspect the ages of our primary stars produced by MIST isochrone interpolation to see if this may be the case, and if the star ages are near the end of the predicted MS lifetime. To calculate the expected MS lifetime ($t_{\rm MS}$), we use the following equation:

\begin{equation}
    t_{\rm MS} \approx t_\odot \left(\frac{M}{\rm M_\odot}\right)^{-\nu},
\end{equation}
\noindent with $\nu$, the power law exponent, varying throughout literature, though typically $-3.5\leq\nu\leq-2$ \citep[e.g.][]{Davies2016,sse-lectures}. The solar lifetime $t_\odot$ is assumed to be $10$\,Gyr. 

Using the absolute masses calculated with \texttt{Saltire}, and $\nu=-2$ as a conservative estimate, we find that J0608-59 is predicted to be close to the $t_{\rm MS}$ age. With a MIST age of $\sim13$~Gyr, J1540-09\,A is well beyond the theoretical MS lifetime ($\sim8$~Gyr): this may be why the MIST isochrone analysis has attributed the primary star with a mass of $0.853$~M$_{\odot}$. The remainder of the stars appear to be comfortably within the main sequence, though using a more negative value for $\nu$ would bring J0310-31 closer to terminal age. The terminal age bias may affect some of our systems, yet is unlikely to be the dominant cause of uncertainty and biased radii for the primary stars studied in this work.

\subsection{Gaia Parallax offset}\label{sec:gaia-plx-zpt}

To derive the radii for our primary stars through isochrone interpolation, we require both \textit{Gaia} DR3 parallaxes and catalogued colours (to estimate the spectral energy distribution), which allow us to estimate the bolometric luminosity of the star from the apparent brightness.
With temperature in hand, we can also estimate $R_{1}$ \citep{freckelton-2024-bebopv}. 

So far, we have used parallaxes as initially published by \citet{GaiaEDR3}. 
Recent studies have shown that the \textit{Gaia} parallaxes may be systematically under-estimated, based on comparison with quasars \citep[e.g.][]{Lindegren2018,Groenewegen2021}. \citet{Lindegren2021} investigated the systematic parallax bias in \textit{Gaia} (E)DR3 by mapping the main dependencies of the bias, and provide a tentative recipe to correct for the offset. The provided analytical functions\footnote{\href{https://gitlab.com/icc-ub/public/gaiadr3_zeropoint}{Gitlab: \textit{Gaia} DR3 zero point}} compute an estimate for the parallax zero-point based on the ecliptic latitude, magnitude, and colour of any \textit{Gaia} source. 

Typically, works employing isochrone analysis to estimate masses and radii of field stars tend to not take the \textit{Gaia} parallax zero point offset into account. 
Presented with the possibility that our MIST radii may be systematically biased towards larger values, we investigate if the use of the uncorrected \textit{Gaia} parallax could affect the radii to such an extent that it has a significant impact on our derived stellar parameters.

\begin{figure}
    \centering
    \includegraphics[width=0.99\columnwidth]{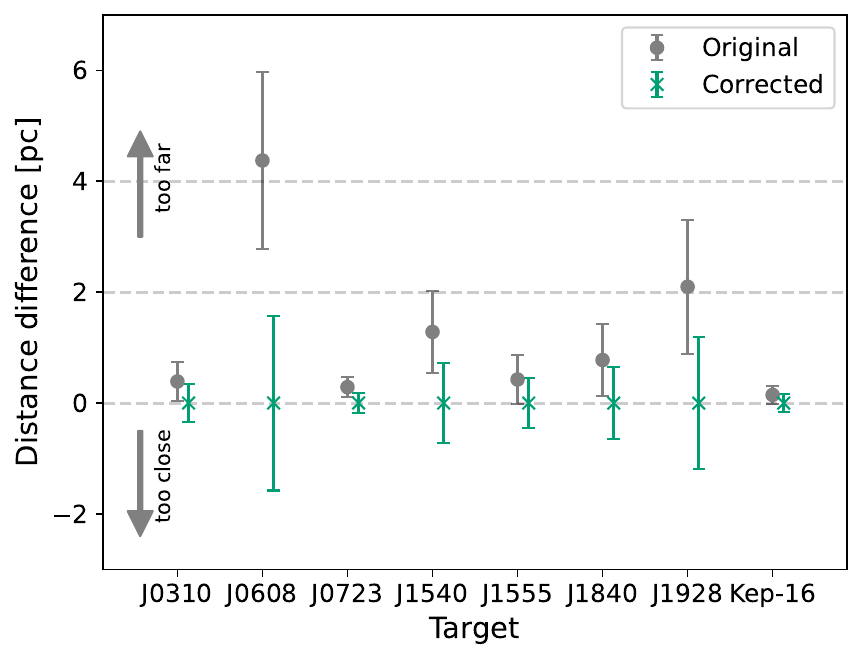}
    \includegraphics[width=0.99\columnwidth]{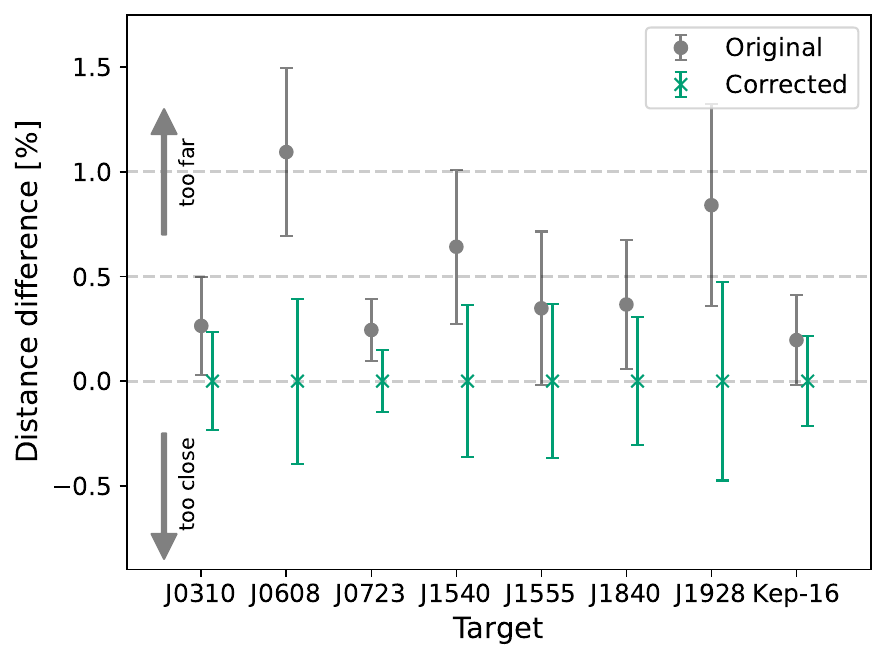}
    \caption{
    Plots showing the relative (\textbf{upper}) and fractional (\textbf{lower}) change in distance to each primary star, after applying the parallax zero point correction for \textit{Gaia} DR3 \citep{Lindegren2021}. The names for each system have been shortened for readability. For the most affected systems, the original distance can be up to $\sim1$~per~cent too large. The corrected and original distances are consistent to within $1\sigma$ uncertainties for all targets except J0608-59. In both panels, the corrected distances lie along $y=0$ by design.}
    \label{fig:distances}
\end{figure}

We have used the \citet{Lindegren2021} correction to estimate updated parallaxes for our current sample of eight targets. The distances, based on both corrected and uncorrected parallaxes, are plotted in Figure~\ref{fig:distances}. These are presented as absolute distance offsets (upper panel) and percentage offsets (lower panel) from the corrected parallax distance. For most systems, both distances are in agreement to within 1$\sigma$, with overlapping error bars. J0608-59 (BEBOP-1) is only consistent to slightly over 1$\sigma$. Nevertheless, the parallax zero point offset in \textit{Gaia} DR3 does represent a systematic bias towards larger distances for the targets in our small sample. The magnitude varies from star to star, though the sign of the correction required is always constant. 
 
\begin{figure}
    \centering
    \includegraphics[width=0.99\columnwidth]{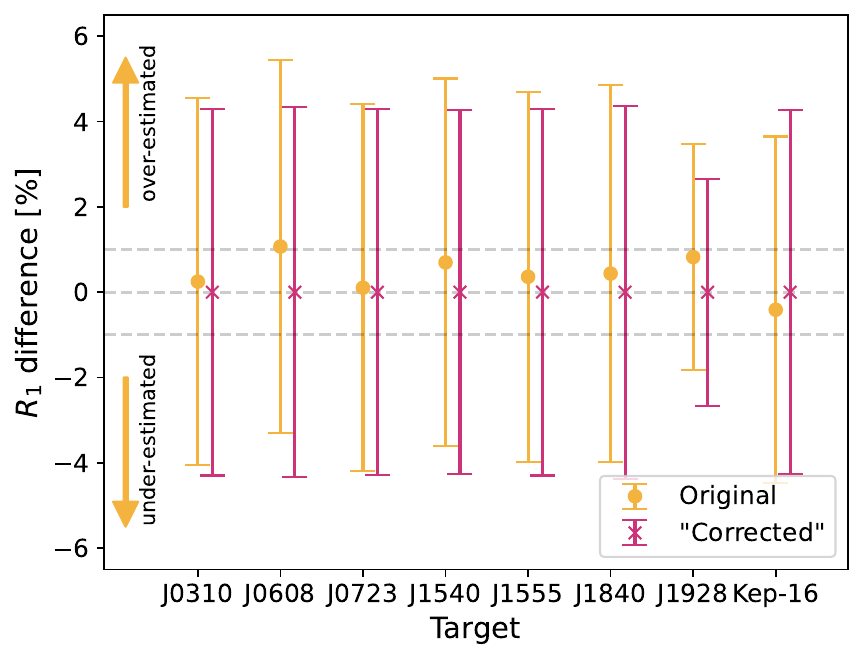}
    \caption{
    Radii calculated from isochrones using corrected \textit{Gaia} DR3 parallaxes are plotted along $y=0$. Originally determined radii (from e.g. \citealt{freckelton-2024-bebopv} and Section~\ref{sec:method}) are plotted as a percentage overestimation of the corrected value. The largest impact is observed for J0608-59, where the radius was $\sim1$~per~cent too large.}
    \label{fig:new_radii}
\end{figure}

The re-interpolated isochrones, using ``corrected'' \textit{Gaia} parallaxes, produce primary radii plotted in Figure~\ref{fig:new_radii} as a fractional difference. The updated radii lie along $y=0$. All systems, apart from Kepler-16, have a decreased $R_{1}$, and the scale of this effect is below $\sim2$\,per~cent. \rev{}{The radius increases for Kepler-16 as a result of degeneracies in the parameters involved in isochrone interpolation, allowing for a slightly larger radius at the expense of a somewhat older, cooler star, less affected by interstellar absorption.}

\begin{figure*}
    \centering
    \includegraphics[width=0.99\linewidth]{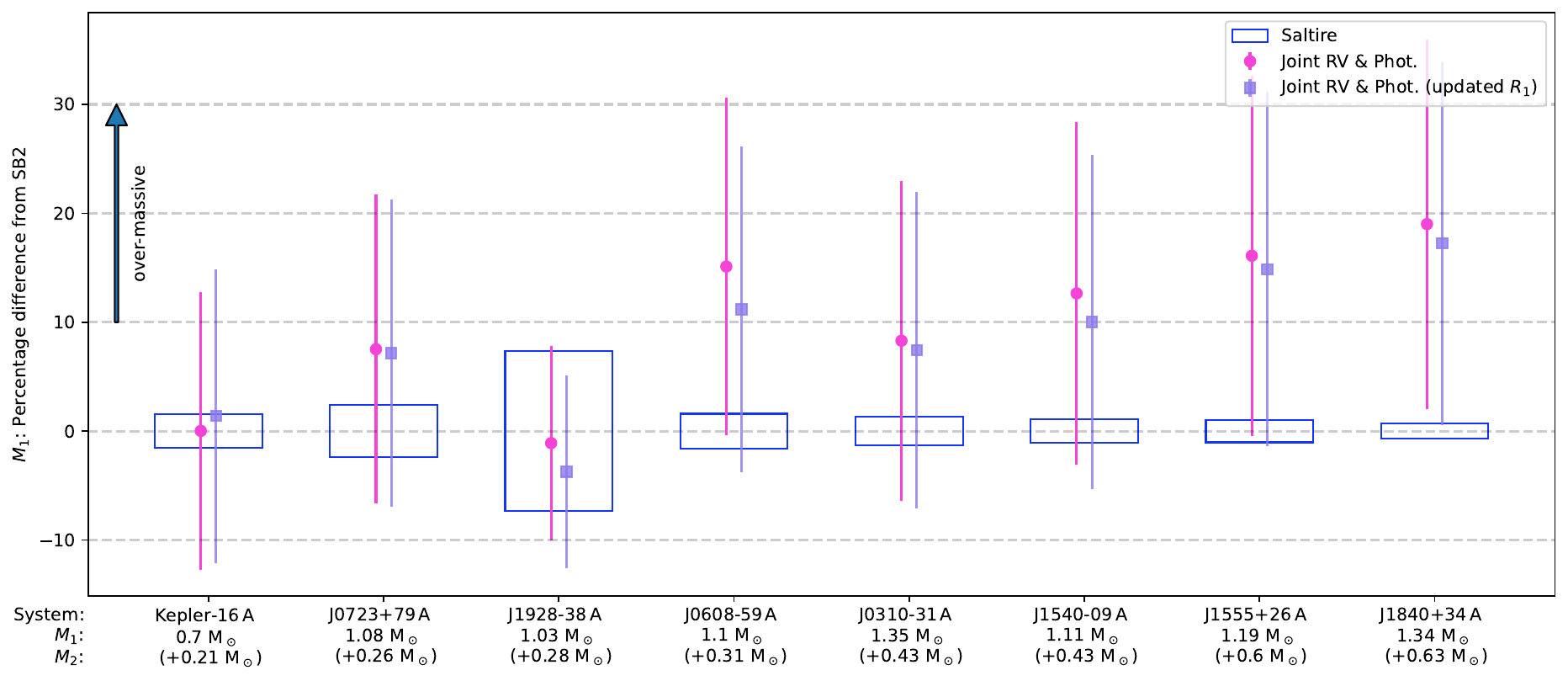}
    \caption{A comparison between SB1 results when using either original or parallax-zero-point-corrected primary radii. The values are plotted as percentage deviation from the \texttt{Saltire} SB2 results. The updated SB1 masses are in a different colour, and indicated with square markers. For most systems, the difference is almost negligible, and we see that this does not account for the identified systematic bias in results from this method. J1928-38 shows a larger modification (albeit fully within $1\sigma$ uncertainties) that brings the quoted median mass more into alignment with the SB2 measurement. The updated radii have reduced the mass for J1840+34, now enabling formal agreement (within $1\sigma$) between SB1 and SB2 methods for all eight targets considered herein. Targets along the x-axes are ordered by primary mass.}
    \label{fig:deviation-from-saltire-newR1}
\end{figure*}

We feed the modified radii through equations~\ref{eqn:stellar-density}--\ref{eqn:M1M2} to assess changes to primary and secondary masses. Updated primary mass comparisons are plotted in Figure~\ref{fig:deviation-from-saltire-newR1}, alongside the original RV+Photometry masses (using uncorrected \textit{Gaia} parallax in radii evaluation), as a difference from the \texttt{Saltire} SB2 masses. The equivalent plot for the secondary stars is included in an appendix (Fig.~\ref{fig:deviation-from-saltire-newR1-secondary}).
For most systems, the updated parallax has not had a major effect on the primary radius, and hence the derived masses. \rev{Apart from Kepler-16, all masses are decreased by some amount, and this also results}{The adjustment in masses caused by the parallax correction has, however, resulted} in all eight systems being in formal $1\sigma$ agreement between these methodologies (see J1840+34, before and after). Though remaining well within uncertainties of the masses derived with uncorrected radii, this shows the scale of the influence the zero point correction can have, and why it should be given consideration in works requiring fundamental stellar parameters known to a high degree of precision and accuracy.

Despite slightly altering the derived SB1 masses, the parallax zero point correction does not fully account for the possible bias between the two methods. The most massive SB1 secondary masses are still $\sim10$~per~cent \emph{larger} than those analysed as SB2s with \texttt{Saltire}. The discrepancy must originate from another source.

\subsection{Influence of binarity in `single-lined' systems}
A cornerstone assumption underlying the EBLM project is that, for this sample of low mass ratio binary systems, the single-lined nature of the system allows the target to be successfully treated as a single star \citep[e.g.][]{triaud-2013-eblm-i,triaud-eblm-iv,Maxted-2023-eblm}. In this way, they are considered to be not massively dissimilar to a star + hot Jupiter system. The EBLM secondary stars are typically 3--12 magnitudes dimmer than the primaries in the visible \citep{triaud-eblm-iv}, helping substantiate this assumption. Many of these binaries have also been studied as part of the BEBOP search for circumbinary planets, wherein a remarkable RV precision can be achieved on a number of these SB1 systems, indicating an almost negligible impact from the secondary \citep[cf.][]{standing-2022-bebopii,triaud-2022-bebopiii,standing-2023-bebop1,Baycroft-2025-bebop3}. 

However, this paper and previous works \citep{sebastian_2024_saltire,sebastian-2024-bebop1,sebastian_2025-kepler16} have shown that the secondary star in these systems is detectable in the optical spectra, albeit through bespoke post-processing and cross-correlation analysis. Where we have appreciable signals, we must consider the non-single nature of the star, and if this could impact other facets of our spectral or photometric analysis.

\subsubsection{Stellar parameters}
As the observed spectrum for an unresolved binary is the superposition of the two individual spectra, modelling the combined spectrum as a single star can introduce systematic biases in derived stellar parameters, especially for mass ratios approaching unity. \citet{ElBadry2018} investigated this by simulating combined binary spectra, mimicking those obtained through large scale surveys like APOGEE, GALAH, and LAMOST. They found that best-fitting effective temperatures can be systematically under-estimated in systems where the secondary contributes a non-negligible fraction of the total light, as the primary spectrum is contaminated by the cooler companion. The systematic errors in parameters calculated by \citet{ElBadry2018} were largest for $0.6\leq q \leq 0.9$, though there is a considerable effect noticed down to $q\sim0.4$ for near-infrared surveys/wavelengths, where the cooler secondaries provide a larger fraction of the total light. The systematic bias identified is also strongest for hotter primary stars, because $T_{\rm eff}$ varies more rapidly with mass as one moves higher up the main sequence.

Our sample of eight EBLM systems consists of binaries with $q\lesssim0.5$, and this is representative of the EBLM sample as a whole (by design, consisting of a solar-type primary and an M dwarf). Even for the most massive M dwarf secondary stars, our systems all lie within the region where the fraction of light in the optical spectra remains below the level where the stellar atmospheric parameters are likely to be significantly biased \citep[see, e.g., the rightmost two columns of fig.~2 in][]{ElBadry2018}. Furthermore, the spectral analysis by \citet{freckelton-2024-bebopv} involves co-adding numerous spectra to create a high S/N master, used to estimate atmospheric parameters. By virtue of this co-adding, the lines from the secondary will become smeared-out because they lie at a variety of relative velocities (with respect to the primary's lines), rather than creating one pronounced set of lines in the co-added spectrum. 

\subsubsection{Isochrone fitting}

To place a star on a given MIST isochrone, we require atmospheric parameters for the star, a distance to the star (from the \textit{Gaia} parallax), and photometric magnitudes in various passbands \citep[see Section~\ref{sec:spectra}, and e.g.][]{mortier2020,freckelton-2024-bebopv}. These bands cover the visible and near- and mid-infrared parts of the electromagnetic spectrum. As discussed above, unresolved stellar companions will contribute some small amount of flux to the observations. Although this is fairly negligible for EBLM systems in the optical, the fraction will become more pronounced as we move into the infrared and the contrast between the two stars ($\Delta$mag) lessens. 

\citet{Berger2020} investigated the effects of binarity on isochrone classification when including photometry for a variety of unresolved companions. 
\rev{The resultant isochrone ages are changed by the addition of a secondary companion, either because the binary system photometry stars to mimic that of an older, more evolved star, or because it resembles a higher-mass system that would be younger whilst remaining at a similar $T_{\rm eff}$.}{The isochrone results are changed by the addition of a secondary companion in one of two ways: either 1) the photometric magnitudes are increased at redder wavelengths and thus mimic an older, more-evolved star; or 2) for contrast ratios close to 1, the finite limit of MIST model ages means the only tracks consistent with $\sim\sqrt{2}$ higher luminosity correspond to both hotter \textit{and} larger stars of similar age.}
Regardless of the impact on age, the inset panel of figure~4 in \citet{Berger2020} shows that the inferred stellar radius\rev{}{continually} increases as one increases the mass of the unresolved companion, for all three primary star masses ($0.85,~1.15,~1.45$\,M$_{\odot}$).

This is especially relevant to our current study, as we have identified a possibly biased isochrone radius for those EBLM systems with the highest--mass companions. To quantify this effect, we perform a similar investigation. 

Using the \textsc{isochrones} package, we simulate mock MIST isochrone tracks \citep{morton2015,dotter2016} for a grid of binaries spanning the primary and secondary mass ranges typical for EBLM systems \citep[see fig.~5 in][]{triaud-eblm-iv}. We use inputs of $M_{1}\in[0.65,1.5]$ and $M_{2}\in[0.1,0.65]$, and vary both in steps of $0.05$. We assume a distance of 300\,pc, extinction $A_{V}=0.2$, solar-like metallicity, an age\footnote{Assuming an age of e.g. 3, 5, 10~Gyr results in the super-solar-mass primary stars evolving off the main sequence.} of $1$\,Gyr, and keep these parameters fixed. The generated tracks provide values for the simulated radii which we use as a ``ground truth'', and photometric magnitudes for each of the two stars ($BP,~G,~H,~J,~K_{s},~W_{1},~W_{2},~W_{3}$, similar to those used in Section~\ref{sec:spectra}). We combine the magnitudes and assign uncertainties representative of those measured for EBLM targets in the given surveys \citep{skrutskie2006,cutri2013,GaiaDR3}. 

We provide the combined magnitudes and parallax as input to the isochrone interpolator, along with the simulated $T_{\rm eff}$ and [Fe/H] for the primary star. We assign uncertainties on temperature and metallicity to be $150$~K and $0.1$~dex, respectively, to reflect typical measurement uncertainties for EBLM primaries, and to account for the systematic uncertainty that is added to these parameters output from spectral analysis \citep{tayar-2022}. To test the influence of secondary light, we interpolate the isochrones with a \texttt{SingleStarModel}, as done in e.g. \citet{freckelton-2024-bebopv}, which provides an inferred radius for the primary star (assuming it can be effectively treated as a \emph{single star}), to compare with the simulated input radius.

\begin{figure}
    \centering
    \includegraphics[width=0.99\columnwidth]{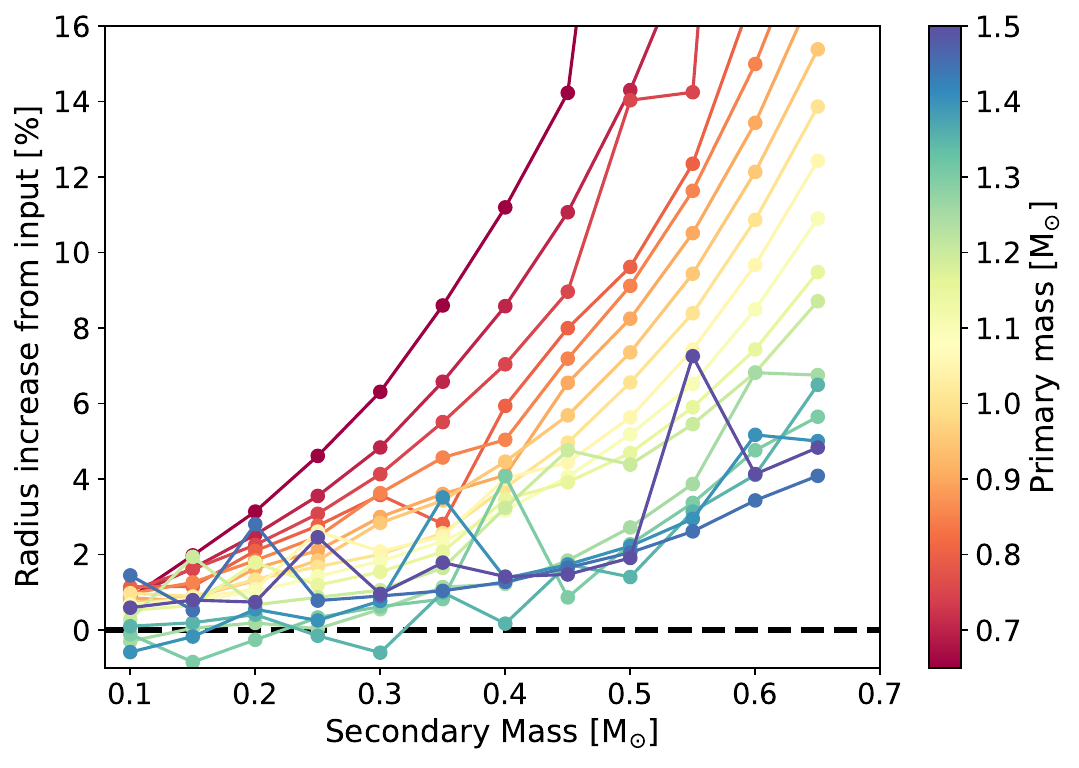}
    \includegraphics[width=0.99\columnwidth]{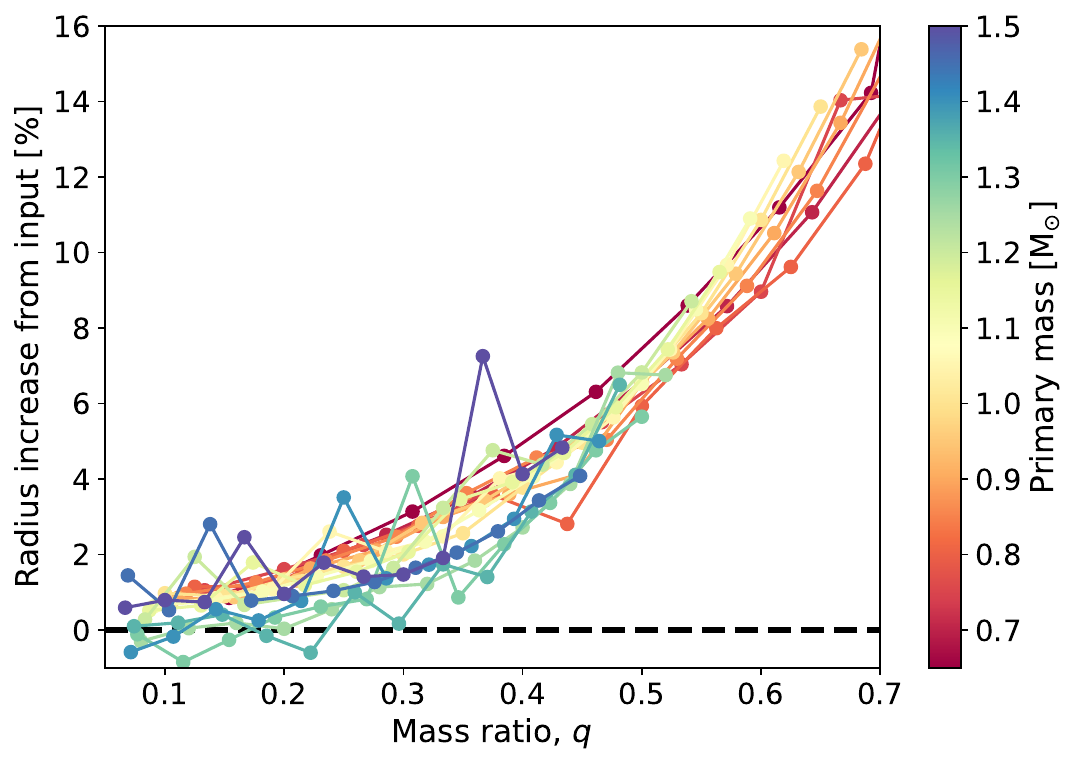}
    \caption{Figures showing the radius difference between the recovered \texttt{SingleStarModel} isochrone for our simulated binary systems, and the ground truth primary radius from the MIST track used to generate the system parameters. \textbf{Top panel:} as a function of secondary mass. \textbf{Bottom panel:} as a function of mass ratio. The amount of radius over-estimation increases in a consistent manner as a function of $q$. The over-estimation is similar no matter the mass of the simulated primary (indicated by line colours), though is slightly stronger for the least massive primaries -- a consequence of cooler stars being closer in temperature to their M dwarf companions.}
    \label{fig:isochrone_sim}
\end{figure}

Figure~\ref{fig:isochrone_sim} shows the results of this grid-based injection-recovery style simulation. For each primary mass, the inferred radius is typically over-estimated compared to the ground truth, and this over-estimation increases as the secondary mass increases. This goes hand-in-hand with the secondary star contributing more significantly to the total light of the system. The most dramatic radius overestimation appears to be for the lowest mass primaries, though the upper panel does not account for how massive/luminous each secondary is in relation to the primary. 

The lower panel of Fig.~\ref{fig:isochrone_sim} shows the same results for a given mass ratio, $q$. The radius over-estimation varies fairly consistently over $q$ for the range of simulated primary masses, with a possible slight increase for the lowest-mass primaries. The error envelopes typically extend $\pm2$~per~cent from the line, shown for each primary mass individually in Figure~\ref{fig:radius_isochrone_18panel}.
For the majority of the EBLM systems ($q\lesssim0.3$, $M_{1} \gtrsim 1$\,M$_{\odot}$, \citealt{triaud-eblm-iv}), the median radius over-estimation stemming from using a single-star model is minimal, at the 0--2~per~cent level.

The effect is most pronounced for targets with larger mass ratios, beyond $\sim0.3$, and this echoes what we see in Figure~\ref{fig:primaries}: those most discrepant or ``biased'' systems have secondaries which contribute a non-negligible fraction of the total light. For the purposes of this paper, we naturally selected a sample of systems where it would be possible to detect the secondary using HRCCS, so it is no surprise that these systems are also those that are least suitable for isochrone modelling as a single star. In works studying the EBLM sample as a whole \citep{eblm-northern-sample}, we posit that there will not be the same magnitude of apparent systematic overestimation in the derived radii or masses for primary or secondary star, and any effect caused by the inherent binarity will be far less noticeable. As we increase our exploitation of HRCCS to EBLM systems in the infrared \citep{2025CoSka..55c.412S}, lower mass ratio binaries will be probed as both SB1 and SB2s, allowing us to fully investigate this hypothesis.

\subsubsection{Correcting $R_1$}\label{sec:correction}
To perform a consistency check that the magnitude of this $R_1$ over-estimation could account for the increased masses measured with the SB1 method, we derive a basic correction. We begin by fitting an exponential function to the percentage radius increase against mass ratio (bottom panel, Fig.~\ref{fig:isochrone_sim}), simply combining the arrays for all primary masses. The fit follows the form 
\begin{align}
    f(q) &= A\exp{(-B\times q)} + C,
    \label{eqn:r-correction}
\end{align}
\noindent where the optimal parameter values are $A=1.2786\pm0.0773$, $B=-3.6498\pm0.0603$, and $C=-1.5167\pm0.1774$. The fit is valid for mass ratios $\sim0.05< q <1$, because of the input stellar masses. In reality, the majority of points influencing the fit are located at $q \lesssim0.7$, due to the imposed upper limit on $M_2$ lying at the edge of the M dwarf parameter space. For $q\gtrsim0.7$, the non-single nature of the star should be immediately apparent, and these systems are unlikely to have been analysed as a single star in the first instance. 

\begin{figure*}
    \centering
    \includegraphics[width=0.99\linewidth]{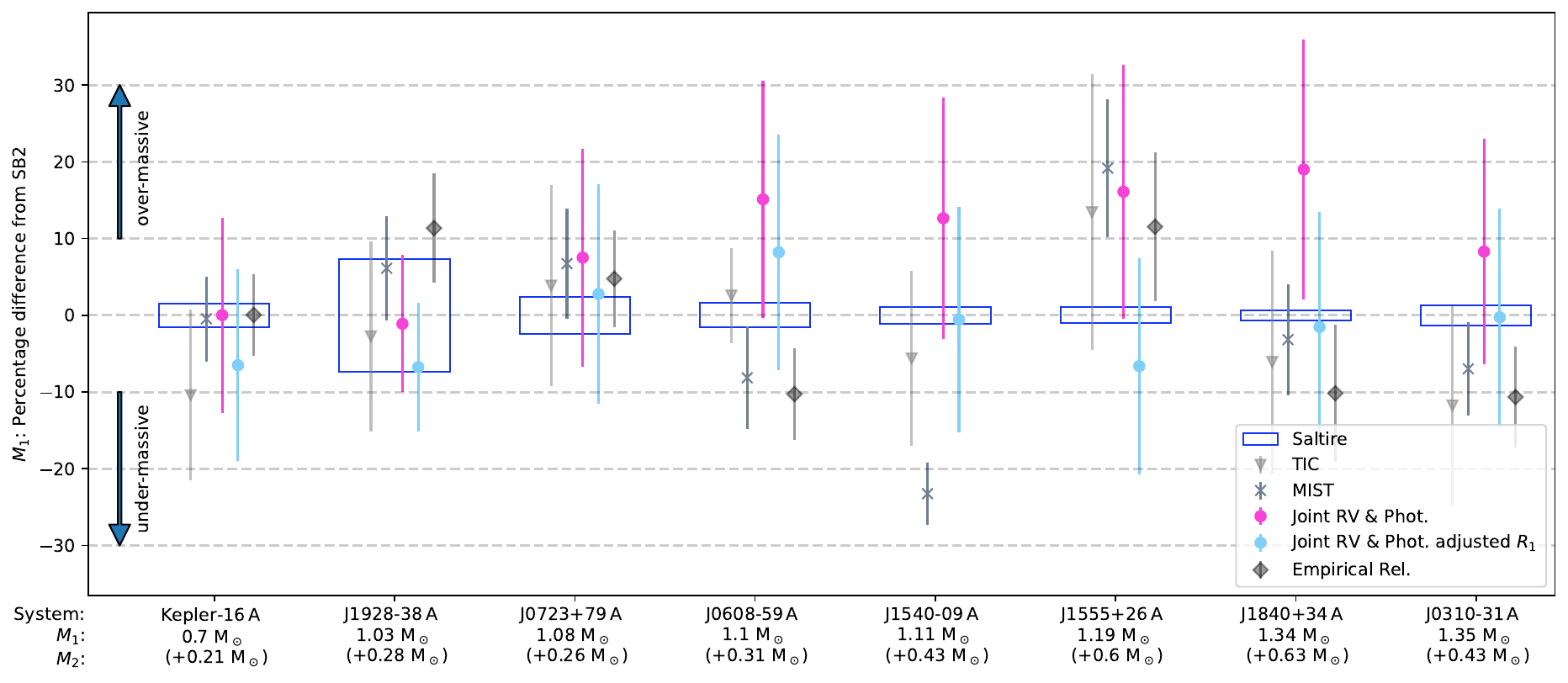}
    \includegraphics[width=0.99\linewidth]{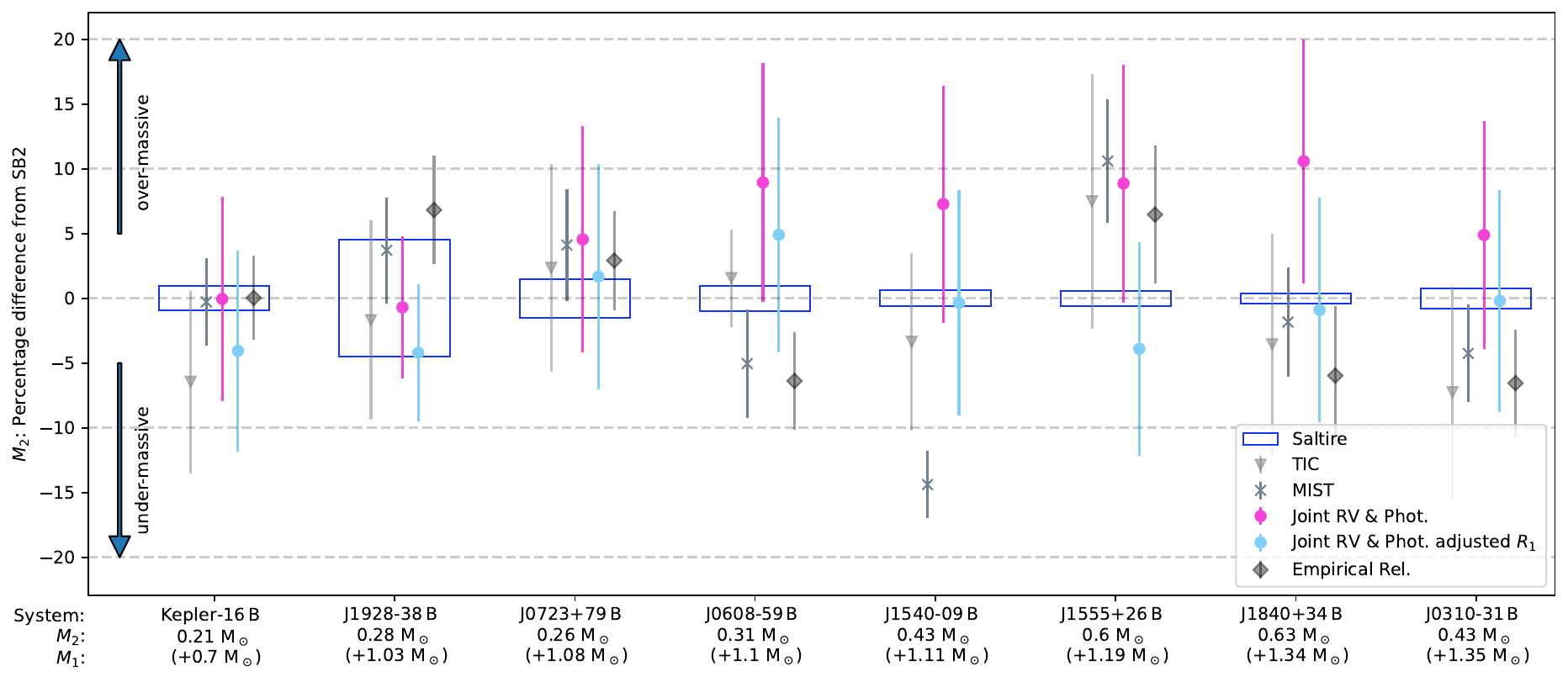}
    \caption{Similar to Figures~\ref{fig:primaries} and~\ref{fig:secondaries}, we now include the SB1 mass estimates when adjusting the input primary isochrone radius based on the over-estimation caused by using a single star model (light blue points). The adjusted values are far closer to the SB2 measurements, with typical scatter around zero difference.  \textbf{Top:} Primary Masses. \textbf{Bottom:} Secondary masses. Both panels are ordered from left to right by increasing primary mass.
    }
    \label{fig:r1-reestimated}    
\end{figure*}

From Equation~\ref{eqn:r-correction}, we predict the percentage radius over-estimation for our eight systems studied this work, given their mass ratios calculated using \texttt{Saltire} $M_1$ and $M_2$. We divide the input $R_1$ mock distribution by an over-estimation factor ($1+f(q)[\%]/100$), then recompute the SB1 masses following the methodology laid out in Section~\ref{sec:rv+phot}, and plot these new estimates atop previous percentage-mass-difference plots in Figure~\ref{fig:r1-reestimated}.

The new estimates lie far closer to the SB2 measurements, and seem to be scattered around zero in an expected manner. Figure~\ref{fig:r1-reestimated} shows that any discrepancy between our two approaches can be almost solely attributed to the single star isochrone modelling of these binary systems. 

It must be stressed that this correction is only in preliminary stages: numerous factors have not been taken into account, which will be addressed in a future work centred around deriving a complete correction. Firstly, the assumed exponential form is based primarily on visual inspection and the noticed behaviour, yet the correction may be better addressed through use of a different function. Secondly, one can see from the lower panel of Figure~\ref{fig:isochrone_sim} that there is a small amount of spread introduced by the differing primary masses. Even when expressed in terms of mass ratio $q$, the higher mass primaries are slightly less affected than those with lower masses. In reality, a better parameterisation to derive the correction would be related to the flux or the luminosity of the stars in the system. Involving parameters that are dependent on the measured radii inside a radius correction would introduce further difficulties that we have not tackled at this time. These inadequacies notwithstanding, the ``corrected'' MIST $R_{1}$ produce SB1 masses entirely consistent with the SB2 measurements (Fig.~\ref{fig:r1-reestimated}), providing evidence that there is merit in this simplistic correction.

In Figures~\ref{fig:isochrone_sim} and~\ref{fig:radius_isochrone_18panel}, there are a number of curves showing an uneven/irregular radius increase as $M_2$ increases. Most notably, the simulation for $1.50$\,M$_{\odot}$ displays considerably erratic behaviour. Repeating the experiment $10\times$ for just $1.5$\,M$_{\odot}$, we find that the ``spikes'' in the plot are distributed around different secondary masses, and correspond to outlandish ages where the isochrone interpolation has gone awry. Taking the median in each bin of $q$ displays a smooth increase in radius difference. This is an additional reason as to why we have combined the simulation results for all primary masses here, to average out any of this behaviour when performing the exponential fit. 

To further test our correction, we have investigated the effect of modifying parameters input to the isochrone generation. Changing the metallicity to $0.3$\,dex, the age to $5$\,Gyr, or the distance to 50\,pc, causes no appreciable difference to the fits (where lines are consistent to within their $1\sigma$ error envelopes). Increasing $A_{V}$ to $1.0$ generates a significant number of absurd isochrone results, such that fitting an exponential function to the combined simulation output is unsuccessful. However, the parameters from our original fit in Equation~\ref{eqn:r-correction} agreeably trace the lower envelope of the trend, indicating that the fit is actually fairly insensitive to this parameter. As a result of these tests, we consider the correction to be suitable for the purposes of the current work.

\subsection{SVD degradation}

To facilitate our HRCCS detections of the secondary stars in EBLM binaries, we use SVD to post-process the spectra (Section~\ref{sec:saltire}). This is performed in the primary's rest frame, to remove the stationary contribution of the primary. There is a chance, however, that the spectrum of the secondary may be degraded as a result. At phases where the secondary's lines are also aligned, the SVD can remove too much signal. Any phases with velocity separation of the secondary  below instrumental resolution will be most affected. Additionally, the number of principal components could also impact the degradation.

\citet{sebastian-2024-bebop1} performed a thorough investigation into the degradation caused by SVD cleaning the spectra. The authors split J0608-59 data into three sub-samples, containing: 1) spectra where the secondary would be `aligned' with two or more other spectra; 2) spectra where the secondary would be aligned with one other observation; and 3) unaligned spectra. The impact that the number of principal components retained would have on the degradation was investigated by repeating the cleaning whilst varying the reduced rank, $\overline{k}/R$ -- the average number of removed components divided by the maximal rank of the input spectra matrix. When fitting a \texttt{Saltire} model to the created CCF maps, \citet{sebastian-2024-bebop1} found that the whole data set showed consistent $K_{2}$ and $\varv_{\rm sys}$ independent of the reduced rank, and that the partial samples were all consistent for reduced rank $<0.3$. For our systems and the results listed in Table~\ref{tab:saltire-results}, we find a maximal reduced rank of $\overline{k}/R = 0.071$, far below the limit where poor phase coverage could impact the recovered CCF parameters. 

Furthermore, \citet{sebastian-2024-bebop1} performed a series of simulations to assess the SVD degradation across a range of spectral SNRs, which in turn affects the SNR of the secondary per resolution element. They found that as the SNR increases, and the secondary is no longer buried in the noise (SNR$_{\rm sec}>1$), phases containing aligned spectra are degraded more than the non-aligned phases. For the system with the least extreme brightness ratio (J1555+26), we derive that the average signal to noise of the secondary is $\sim0.76$, lying within the regime where the SVD cleaning does not disproportionally affect aligned phases more than the non-aligned phases. \rev{}{For targets with SNR$_{\rm sec}\gtrsim0.3$, we enter the high-SNR regime, where the noise is no longer just white noise, but also has contributions from spurious correlations between the secondary's spectrum and the line mask (known as wiggles; \citealt{DOLBY,sebastian_2024_saltire}). However, \citet{sebastian-2024-bebop1} showed that even in the presence of wiggles, the injected $K_{2}$ is accurately recovered, well within the measurement uncertaitny.}

By comparison with table~4 in \citet{sebastian-2024-bebop1}, we estimate that, in the worst-case scenario, the secondary's signal will be degraded by at most $15$\,per~cent for this target ($85$\,per~cent remaining). \rev{}{This pessimistic estimate for SVD degradation is at its maximum for the highest SNR secondaries, and diminishes when $\Delta {\rm mag}$ between the two stars increases. With an already strong signal, a small reduction will have little effect on the derived parameters and uncertainties.} Thanks to our excellent phase coverage of the binary orbits and low values of reduced rank used, we are confident that these results are unaffected by the SVD detrending. 

To verify that the results are consistent in spite of the cleaning used, we have also tried to detect the secondary in J1555+26 without removing the primary. This system has an average contrast ratio of $0.0245$ in the SOPHIE wavelength range, the largest in our small sample of targets. Inspecting the CCFs, we see that the secondary is not visible above the noise level in an individual CCF, in agreement with SNR$_{\rm sec}<1$. The system is therefore not suitable for a double-lined analysis \textit{without} HRCCS, using e.g. \texttt{DOLBY-CCF} \citep{DOLBY}. We can, however, plot the CCFs aligned in phase, and begin to identify the pattern showing the secondary's presence (shown in Figure~\ref{fig:ccftrails}). Comparing the predicted velocity of the secondary derived from our \texttt{Saltire} analysis to the trace seen in the CCF, we find good visual agreement, further reinforcing that the measured $K_{2}$ and $\varv_{\rm sys}$ are accurate.

\subsection{Stellar activity and SB2s}

Our \texttt{Saltire} SB2 results are assumed to be accurate, as we measure a semi-amplitude for both stars in the system and use these to derive absolute, model-independent masses. However, we assess the possibility of there being a systematic error in our fitted $K$ values, due to the impact of stellar activity. 

In exoplanetary investigations into stellar activity, the activity signature manifests at periodicities independent from the Keplerian signals of bona-fide orbiting bodies. Activity indicators, alongside techniques like Gaussian Process regression, can help to diagnose the origin of these signals \citep[e.g.][]{Desort2007,Haywood2014,Faria2016,Lafarga2021,Costes2021,Aigrain2023}. For instance, were an RV signal to be observed at the rotation period of the star, we would model out this contribution, removing the RV variation associated with this periodicity. In other words, a signal is almost always treated as single origin \citep[with examples of exceptions described in e.g.][]{Shkolnik2005}.

The presence of a companion star or a massive, close-orbiting planet can affect the activity of the host through tidal or magnetic interactions. Tidal forces and angular momentum transfer can lead to higher rotation rates for binary stars, and consequently, higher activity levels. A number of stars also display even greater activity than expected from their rotation rates, which may be caused by magnetic interaction. This is supported by observational evidence of coronal X-ray emission and hot-spots on facing hemispheres \citep{Ding2024}. Tidal or magnetic interaction can be seen to modulate activity measurements on timescales of either $P_{\rm orb}/2$, or $P_{\rm orb}$, respectively \citep{Shkolnik2005}. 

Usually, the rotation of a stellar surface strewn with active regions introduces additional scatter to a series of measured radial velocities. The scatter will typically be unevenly distributed across an orbital revolution of a planet or secondary star because of the differing orbital and rotational timescales \citep{Strassmeier2011}. In the case of planetary studies, one would remove/model out this activity variation, as it will often be of similar scale or larger than the RV signals of interest. For binary stars, the relevant RV amplitudes are typically $\sim$10s of km\,s$^{-1}$, and dwarf the activity-related jitter. If spread out over orbital phase, it may slightly increase the uncertainties in the recovered parameters, though not the accuracy of the recovered values \citep{Strassmeier2011}.

However, in those systems with significant star-star or star-planet interaction, stellar spins can eventually become synchronised with the orbital period of the companion, and both activity and RVs will be modulated on the exact same timescale. The measured RV `amplitude' will not be entirely dynamical, and will not represent the true system parameters. Although the majority of the RV variation is caused by the orbiting body, there remains the potential for a small systematic bias in the case of correlated RV and activity. 

By studying the rotational photometric modulation of primary stars in low mass eclipsing binaries, \citet{SethiMartin2024} found that $\sim50$\,per~cent of their sample showed variability consistent with overdensities of active regions at the sub- and anti-stellar points. In such scenarios where the primary rotation and secondary orbit are synchronised, stellar activity will also be modulated on half the orbital timescale if there are preferential locations for active longitudes. 


\citet{Wright2013} investigated the interactions between host and companion as a source of unusual RV and activity indicator measurements. A secondary star could heat the host's surface, altering the spectrum and the convective flows, cause prolate distortion, or excite pulsation modes within the primary star. These could create RV signals with correlated activity measurements, though would likely also be seen in photometric monitoring \citep[e.g.][]{SethiMartin2024}, thus may be easier to diagnose.

Contaminating spectra from background objects can also create spurious RV signatures. As the primary spectrum is repeatedly Doppler shifted by the companion, there could be periodic alignment with a stationary set of contaminating lines. The match would be weak, yet could confound RV measurements using CCF-based methodologies. The CCF peak would be ``pulled'' towards the contaminating velocity, altering the RV and activity indicator measurements relating to the width and skewness of the CCF. The time variable peak-pulling could cause systematics in the measurements \citep{Wright2013}, and could feasibly be modulated on the timescale of the binary period as this causes the large Doppler shift.

To rule out the possibility of systematically incorrect \textit{dynamical} semi-amplitudes, we assess the correlation between measured RVs and three commonly-used CCF-based activity indicators: the full width at half-maximum (FWHM), representing the width of the mean spectral line; the bisector inverse slope (BIS), the difference between velocities measured in the upper and lower regions of the CCF profile; and the contrast, the `strength' of the average absorption line profile \citep[e.g.][]{Queloz2001,Queloz2009,CollierCameron2019,Barnes2024}. The correlation plots are shown in Figures~\ref{fig:harps_correlations} and~\ref{fig:sophie_correlations}.

There is only one `significant' linear correlation, with $p$-value less than $0.05$ ($5$~per~cent). This is RV-BIS in HARPS data for J0608-59 ($r=0.35,p=0.0026$). As this is a weak correlation only observed in a single activity indicator, and because active regions are well-documented to introduce a \textit{negative} linear correlation between RV and BIS \citep[e.g.][]{Queloz2001,Barnes2024}, we are suspicious that this may not be a true correlation. Figure~\ref{fig:harps_correlations} does not show overly-convincing evidence that the bisector span is varying concurrently with the radial velocity measurements. Simply removing two outliers drastically reduces the correlation significance for RV-BIS in J0608-59 ($r=0.27,p=0.021$). Furthermore, to transform J0608-59 into an SB2, \citet{sebastian-2024-bebop1} favoured ESPRESSO spectra over those from HARPS, obtained as part of the discovery of BEBOP-1\,c \citep{standing-2023-bebop1}. In the ESPRESSO data, there is a complete absence of RV-BIS correlation ($r=0.005,p=0.940$), indicating that the correlation in the HARPS data can be safely disregarded.

It is therefore highly unlikely that the RVs are being significantly impacted by stellar activity induced by the secondary or modulated on the binary orbital period in any of these systems. There is no evidence thus far that the stellar rotation periods are synchronised with the orbital periods of the M dwarf companions in this sample.

\section{Validity of SB1 method}\label{sec:validity-of-sb1}

In this paper, we have expanded the sample of EBLM binaries with measured absolute dynamical masses. However, a large fraction of these systems will not be amenable to characterisation with HRCCS. For the lower-mass secondaries (towards the boundary between brown dwarf and late M-types), the contrast ratio is too low to allow for detection in the visible with 2--4-m class telescopes. These late-M stars are of great interest to the EBLM project, as they are the main focus of many an exoplanetary study concerning Earth-mass planets and habitable zones. Instead, we use the combined RV + Photometry (SB1) method \citep{davis-2024} to study low mass M~dwarfs when they are the secondary component in a binary, for investigations into their fundamental characteristics \citep{eblm-northern-sample}.

\subsection{Investigating M~dwarf radius inflation}

EBLM binaries analysed using the SB1 method above are able to provide insight towards the long-standing discussion topic of M~dwarf radius inflation \citep{Hoxie1970,Hoxie1973,Lacy1977,Popper1997,Clausen1999,TorresRibas2002,Swayne2021}. Theoretical models often predict radii for fully convective M~dwarfs that are significantly different to those we observe \citep[e.g.][]{Swayne2024-eblmXI}, which has far-reaching implications for those attempting to construct empirical mass-radius-temperature relationships. In turn, we require such highly-accurate relationships in order to infer physical characteristics of exoplanets hosted by fully-convective stars, and test theories of stellar evolution. 

The EBLM project provides an unprecedented opportunity to investigate the extent of radius inflation with a large, homogeneously analysed sample of fully-convective stars \citep{eblm-northern-sample}. Our comparison between the SB1 and SB2 methods in this work found resultant masses that are consistent to within $1\sigma$ uncertainties on the measured parameters, though with a possible bias for largest $q$ (Section~\ref{sec:bias}). Using a simplistic correction to $R_1$, we recalculated the SB1 masses and find that these are in extremely good agreement with the SB2 results. As such, the SB1 method is able to produce the accurate masses required to quantify the departure from evolutionary model tracks in mass-radius space. 

\begin{figure*}
    \centering
    \includegraphics[width=0.9\linewidth]{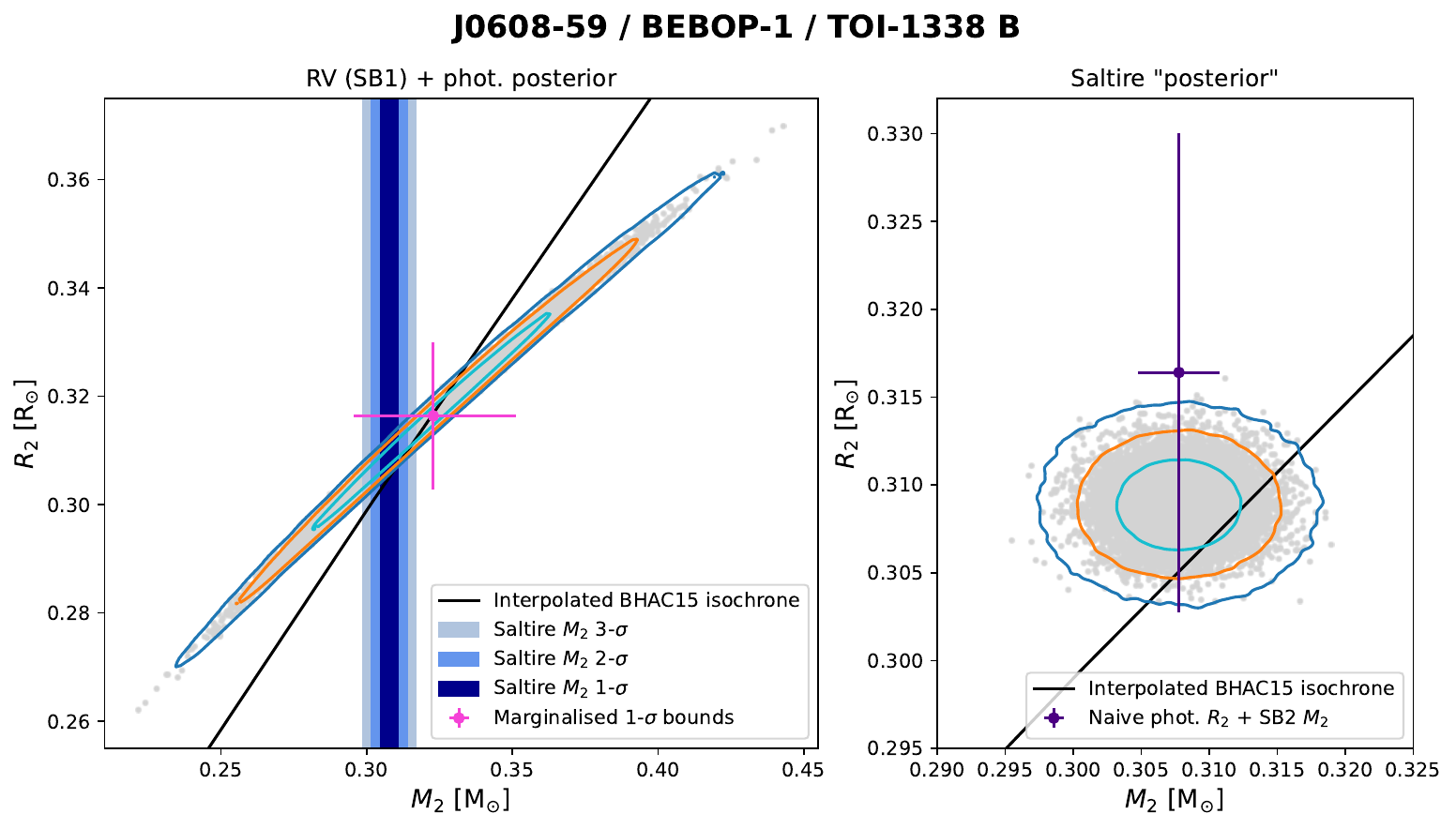}
    \caption{\textbf{Left:} The joint $M_2$, $R_2$ posterior from the SB1 method (adjusted for $R_1$ over-estimation), with a band over-plotted indicating the secondary mass from an SB2 analysis. Contours are drawn at the 1, 2, and 3$\sigma$ levels. One can clearly see that the marginalised $1\sigma$ medians and uncertainties do not accurately represent such a correlated posterior distribution.
    \textbf{Right:} A re-calculation of the secondary radius using observables from the RV + photometric modelling and the total system mass from \texttt{Saltire} as the one piece of assumed information. The resulting distribution lies ever-so-slightly above the BHAC15 5\,Gyr isochrone, though still passes through the 1$\sigma$ contour. The plotted point represents a na\"ive combination of the values from two different methods.}
    \label{fig:combining_posteriors}
\end{figure*}

To compare SB1 \& SB2 results for our systems with regards to radius inflation, we are also concerned with determining accurate radii, to place the stars on a mass--radius plane. In the left panel of Figure~\ref{fig:combining_posteriors} we plot an example SB1 posterior, marginalised $1\sigma$ uncertainties, and the $M_2$ estimate from \texttt{Saltire}. As a consequence of simultaneously solving Equations~\ref{eqn:stellar-density}--\ref{eqn:M1M2}, the posteriors for $R_{2}$, $M_{2}$ are naturally very correlated. Changing the inferred secondary mass therefore also changes the radius.

Inaccuracies in the SB1 results (e.g. Section~\ref{sec:bias}) do not translate to a drastic impact in terms of radius inflation, instead causing the system to traverse the space with a similar slope to that of a theoretical 5\,Gyr, ${\rm[Fe/H]}=0$ isochrone\footnote{The \citet{Baraffe2015} isochrone has been employed to study radius inflation in works such as \citet{davis-2024,eblm-northern-sample}, as it extends down to the lowest mass M dwarfs. For our purposes, the differences caused by changing the age (whilst remaining $>1$\,Gyr) are negligible.} \citep[][henceforth BHAC15]{Baraffe2015}.

Due to this correlation, it is overly na\"{i}ve to simply combine the median and uncertainty values for SB1 $R_2$ with those for SB2 $M_2$. The reported values and uncertainty bounds do not accurately reflect the shape of the posterior distributions constructed during the SB1 analysis. Conditioning a posterior predictive by applying knowledge of an SB2 mass measurement would also shift the inferred radius. The parameter space spanned by the reported $1\sigma$ uncertainties on the SB1 ($M_2$, $R_2$) measurement is clearly not reflective of the actual posterior distribution. Here, marginalisation does not give an accurate picture of the spread over which these parameters may/may not lie. This can also be seen for J0608-59/BEBOP-1 in the left panel of Figure~\ref{fig:combining_posteriors}.

Instead, the optimal approach would be to properly combine the posteriors. In Section~\ref{sec:rv+phot}, we have to assume one piece of knowledge to solve the system parameters via a combined analysis of the RVs and photometry. This has previously been $R_1$, though we can swap this for the total mass of the system that is inferred from the \texttt{Saltire} SB2 analysis. 

Starting from Equation~\ref{eqn:stellar-density}, we can rearrange for $R_{1}$:
\begin{align}
    M &= \frac{4\pi^{2}}{G\,P^{2}}\bigg(\frac{a}{R_1}\bigg)^{3} R_{1}^{3},\\ 
    \implies R_{1} &= \bigg(\frac{M\,G\,P^{2}}{4\pi^{2}\,\big(\frac{a}{R_1}\big)^{3}}\bigg)^{\frac{1}{3}}.
    \label{eqn:R1fromMtot}
\end{align}
\noindent We have a value for $M~(= M_1 + M_2)$ from the SB2 analysis, and an estimated period and $(a/R_1)$ from the joint photometric + radial velocity modelling (Section~\ref{sec:rv+phot}). The secondary radius is computed by combining $R_1$ with the radius ratio parameter fit for with \texttt{allesfitter}. We can therefore compute pseudo-posterior\footnote{As a result of the \texttt{Saltire} $K_{2}$ systematic uncertainty incorporation, we no longer have a posterior distribution for this parameter. We generate one assuming the mass samples are approximately Gaussian-distributed, with standard deviation equal to the uncertainty quoted in Table~\ref{tab:all-masses}.} distributions for the radii, to combine information effectively rather than na\"ively assuming $M_{2}$ from one work/method, and $R_{2}$ from another. 

This is shown in the right panel of Figure~\ref{fig:combining_posteriors}, where the posterior distribution and $1,2,3-\sigma$ contours demonstrate excellent agreement with the BHAC15 isochrone. The marker point represents what one would obtain by simply plotting independent $M_{2}$ and $R_{2}$ from two different methods/papers. Whilst technically in agreement, this does not accurately represent how the parameters are spread across parameter space, and may compel one to assume that a higher degree of radius inflation is observed compared to that estimated from a complete treatment.

\begin{figure*}
    \centering
    \includegraphics[width=0.95\linewidth]{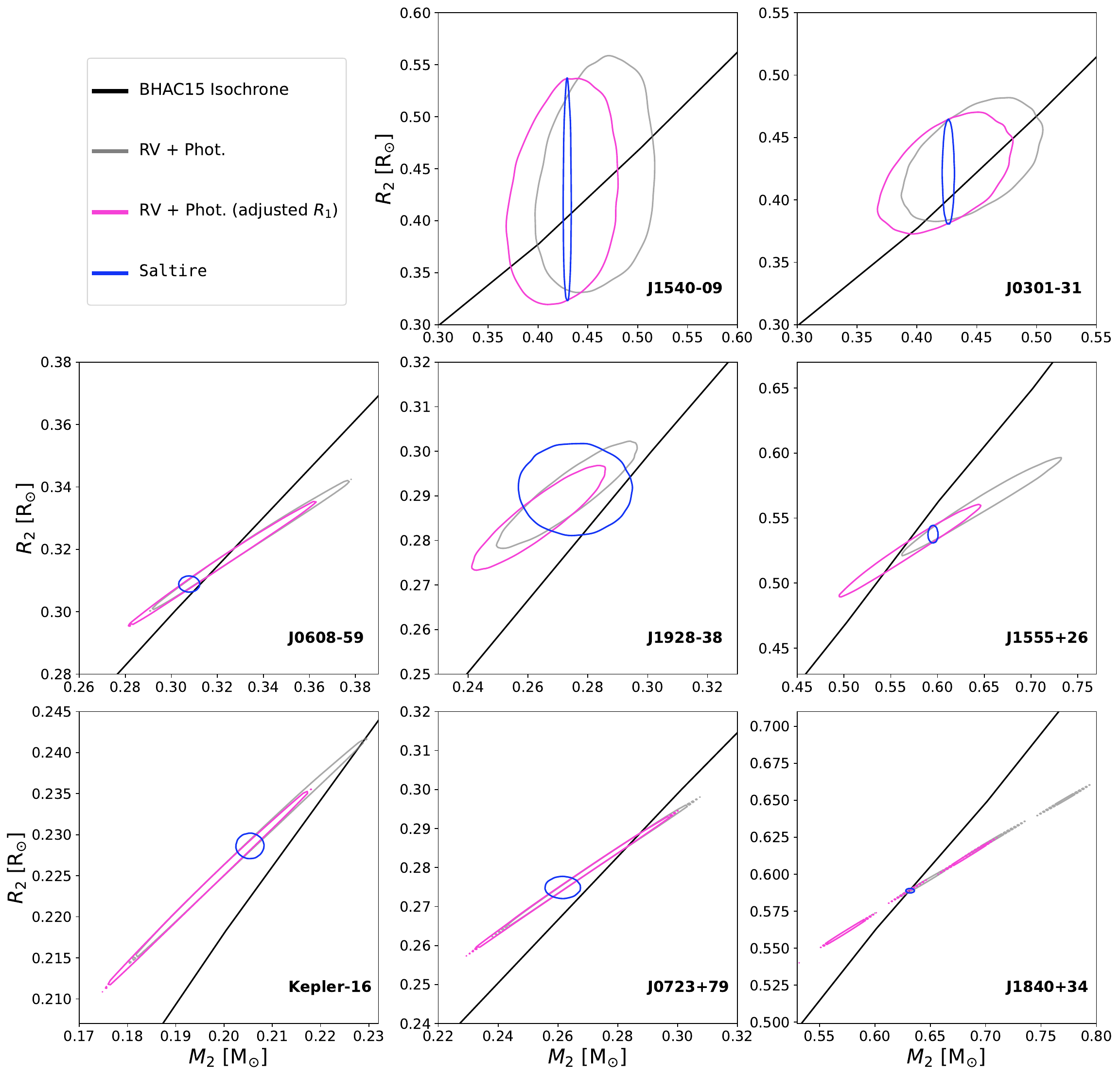}
    \caption{Mass--radius diagrams showing $1$-$\sigma$ contours for the eight systems compared in this work. The theoretical BHAC15 5\,Gyr isochrone is plotted in black \citep{Baraffe2015}. Blue ellipses show the \texttt{Saltire} results, with radii calculated by combining total mass $M$ with RV + Photometry posteriors. The original RV + Photometry results are shown as grey contours, and the pink ellipses show the results after correcting the MIST primary radius (Section~\ref{sec:correction}). 
    }
    \label{fig:MRdiagram}
\end{figure*}

To assess our results in terms of radius inflation, we plot the $1\sigma$ mass--radius contours for all eight systems\rev{}{in Fig.~\ref{fig:MRdiagram}}. The figure is split into sub-panels so that one can identify the position of each result with respect to the BHAC15 5\,Gyr isochrone, the difference caused by correcting $R_1$ in the RV + Photometry method, and the overall shape of each posterior distribution. As mentioned above, plotting a single $M$--$R$ diagram with marginalised values \citep[e.g.][]{Swayne2024-eblmXI,davis-2024,eblm-northern-sample}, does not allow one to see any correlation between the parameters. 

In Figure~\ref{fig:MRdiagram}, SB2 results from \texttt{Saltire} are shown in blue, and SB1 RV + Photometry results are in grey (original results), or pink (`corrected' input primary radii). With respect to the BHAC15 5\,Gyr isochrone, the movement caused by applying the radius correction is subtle, due to the shape of the posteriors. The ellipses typically remain at a fairly similar distance from the isochrone, and are entirely consistent between methodologies. 
The knowledge of an SB2 mass, or the fact that the input $R_1$ are likely to be overestimated for the high-$q$ systems, does not significantly alter the amount of radius inflation observed. 

Nevertheless, small deviations from the isochrone prediction start to matter now that we are pushing towards a considerably improved precision level. To present a finalised picture of radius inflation at the population level, we require the $R_1$ correction to be optimised (rather than using the first-order correction presented here), and an increased sample size. Both of these issues will be addressed in future works.

\section{Comparison to benchmark stars}\label{sec:benchmarks}

\citet{Maxted2025} recently published updated empirical relations linking measurable stellar properties to fundamental parameters like mass and radius. The author uses these relations to create a sample of stellar density `benchmark stars', which can be used to test the precision of stellar parameters for other solar-type stars.

Our small sample contains two of the EBLM binaries used as benchmark stars in \citet{Maxted2025}: J1928-38 and J0608-59\footnote{Though J0608-59 is included in DEBCat as TOI-1338 \citep{Southworth-debcat}, it was excluded from the calibration sample used to create the updated empirical relation \citep{Maxted2025}.}. We employ the same empirical relations (Equation~\ref{eqn:emp-rel}), though use differing atmospheric parameters, and find 
results that are consistent with those presented in \citet{Maxted2025}. For the primary stars in both systems, our use of the empirical relation results is also in good agreement with the \texttt{Saltire} SB2 mass measurement -- see Table~\ref{tab:empirical_rel_differences}. The empirical relations themselves are validated by \citet{Maxted2025} and our SB2 result.

\begin{table}
    \caption{
    A comparison of the primary masses for two ``benchmark'' systems, calculated using the \citet{Maxted2025} empirical relations, but with differing sets of input stellar atmospheric parameters, from the cited publications. The \texttt{Saltire} (SB2) mass measurement is included for ease of comparison. 
    }
    \label{tab:empirical_rel_differences}
\begin{adjustbox}{width=0.99\columnwidth}
\begin{tabular}{lcccccccccc}
\hline
\multirow{2}{*}{Star} & \multirow{2}{*}{\texttt{Saltire}} & \multicolumn{2}{c}{\citet{Maxted2025} Empirical Relation} &   \\
&  & \texttt{PAWS} (§\,\ref{sec:spectra}) & Other source\tnote{1}*  \\
\hline
J1928-38\,A & $1.027\pm 0.117$ & $1.144\pm 0.074$  & $1.050\pm 0.053$  \\
J0608-59\,A & $1.098\pm 0.017$ & $0.985\pm 0.066$ & $1.08\pm 0.06 $    \\
\hline
\end{tabular}
\end{adjustbox}
\begin{tablenotes}
  \item[1] *\citet{Maxted2025} use atmospheric parameters from \citet{Swayne2024-eblmXI} for J1928-38, and from \citet{fitzpatrick2025} for J0608-59. 
  \end{tablenotes}
\end{table}

For J1928-38, \citet{Maxted2025} report $T_{\rm eff}=5687\pm62$, ${\rm [Fe/H]}=-0.01\pm0.04$, and $\log{g} = 4.147\pm0.010$ \citep[derived in][by the \textit{CHEOPS} target characterisation working group using spectral fitting]{Swayne2024-eblmXI}. The parameters we use were calculated with \texttt{PAWS} to be $T_{\rm eff}=5738\pm153$, ${\rm [Fe/H]}=0.18\pm0.08$, and $\log{g} = 4.00\pm0.16$. There is a $\sim2\sigma$ difference between the metallicities, yet temperatures and surface gravities are consistent between these two sets of results\footnote{Note that we include $\log{g}$ here purely for comparison purposes, as it is not used in the particular empirical relation employed by \citet{Maxted2025}}. 

For J0608-59, \citet{Maxted2025} tabulate $T_{\rm eff}=6031\pm56$, ${\rm [Fe/H]}=0.01\pm0.05$, and $\log{g} = 4.240\pm0.010$ \citep{Maxted2024}. The atmospheric parameters employed in the present work are $T_{\rm eff}=5814\pm110$, ${\rm [Fe/H]}=-0.27\pm0.14$, and $\log{g} = 4.03\pm0.2$ \citep{freckelton-2024-bebopv}. We notice a statistically significant ($>1$-$\sigma$) difference between the values used in these two studies, though the resulting masses output from the empirical relations remain in 1-$\sigma$ agreement with each other (cf. Table~\ref{tab:empirical_rel_differences}).

\rev{With accurate $T_{\rm eff}$ and ${\rm [Fe/H]}$ estimates, one finds}{The main differences between the empirical relation primary mass estimates using our atmospheric parameters and those from \citet{Maxted2025} is caused by the differences in metallicity. Although the masses are technically in $1\sigma$ agreement, increasing the precision on metallicity will further improve the consistency. Regardless of these differences, using parameters accurate to within the quoted uncertainties results in} good agreement between dynamically and empirically derived $M_{1}$. 

These binaries are excellent benchmark systems where we can estimate effective temperature directly based on the stellar radii, parallax, and bolometric flux \citep[e.g.][and Section~\ref{sec:teff}]{Maxted2015,Maxted2022,Miller2020}, and use these to calibrate the atmospheric parameters derived through spectral analysis \citep{Maxted2025}. Extending the sample of well-characterised EBLMs allows for empirical relations to be created covering a broader metallicity, temperature, and surface gravity parameter space. Absolute dynamical mass measurements help to validate the results for benchmark systems, as we have shown here.


\section{Populating DEBCat}\label{sec:debcat}
By transforming EBLM targets into double-lined binaries, we are able to precisely and accurately determine the masses for both the M~dwarf \emph{and} the primary star in the system. The eight EBLM systems analysed in this way here can contribute to our catalogues of well-studied eclipsing binaries, and be used as benchmark stars where their physical properties inform relationships between parameters \citep[e.g.][]{Maxted2025}. To compare with known systems, we look to the `detached eclipsing binary catalogue' \citep[DEBCat;][]{Southworth-debcat}, and assess how our newly SB2 systems can expand upon the parameter space coverage. 

\begin{figure}
    \centering
    \includegraphics[width=0.99\columnwidth]{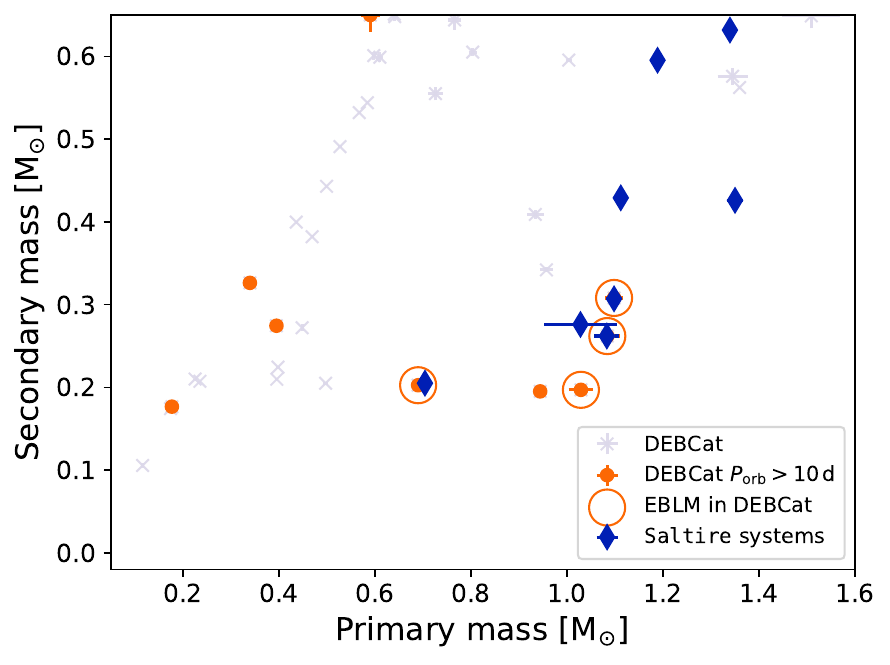}
    \caption{Masses of DEBCat systems, where binaries with period below 10\,d have been greyed-out/represented with cross markers. Three EBLM binaries compared in this work are already in DEBCat (J0608-59, J0723+79, Kepler-16), as is EBLM J0113+31, all indicated by the circled points. With \texttt{Saltire} SB2 masses overlaid in larger dark blue diamond markers, one can see that some previously included systems are not exactly centred, and have changed slightly. The additional systems aid in expanding coverage of parameter space for low-mass stars. \textit{N.B}: all points include error bars, yet most are smaller than the marker.}
    \label{fig:debcat+saltire}
\end{figure}

The low-mass region of DEBCat parameter space is shown in Figure~\ref{fig:debcat+saltire}. From a catalogue version downloaded on 2026 June 16, we have highlighted systems with orbital period more than 10~d -- to remove stars that are likely to be completely circularised, and therefore most strongly interacting. The period cut-off ($P_{\rm cut}$) for existence of eccentric orbits has no definitive value, and depends of the mass ratio of the binary in question. The `canonical' circularisation period is often quoted to be $\approx8$\,d \citep{Meibom2005,Maxted-2023-eblm}. For the EBLM sample, \citet{triaud-eblm-iv} calculated either $P_{\rm cut}=8.9$\,d, or $P_{\rm cut}=7.0$\,d when removing systems with primary stars more massive than $1.3 \,\rm{M}_{\odot}$ or with signs of a tertiary companion. We have chosen to use 10\,d as an approximate, though perhaps slightly conservative, cut-off to select binaries with components that \emph{should} be mostly representative of single stars.

In Figure~\ref{fig:debcat+saltire}, it is immediately apparent that there is a remarkable scarcity of systems with $P>10$\,d. Three of these nine systems included in the plot (orange circle markers) have already been characterised with \texttt{Saltire} as SB2s (J0608-59/BEBOP-1/TOI-1338, J0723-79/BEBOP-3, and Kepler-16, see Section~\ref{sec:sample}). EBLM systems transformed to SB2s, either in this work or in previous publications, are able to $\sim$double the number of sufficiently-detached secondary stars in this region of parameter space with accurately measured parameters, which can be used as benchmarks to assess other methodologies or derive empirical relationships. 

Employing the HRCCS \& \texttt{Saltire} methodology on an expanded sample, EBLM systems could dramatically improve our knowledge of the fundamental parameters of late-type stars. However, only so many systems are amendable to characterisation with SOPHIE and HARPS, and the lowest-mass companions remain elusive as they are inherently fainter, further reducing the contrast ratio between the two components. To transform more systems to SB2, we could employ spectrographs on larger telescopes, increasing the amount of light from the secondary that ultimately enters the fibre. \citet{sebastian-2024-bebop1} demonstrated the gain in significance for the HRCCS detection of J0608-59\,B, from $3$-$\sigma$ with HARPS to $11$-$\sigma$ with ESPRESSO at the VLT. Alternatively, one could instead observe in the near-infrared (NIR), where the contrast ratio between the stars becomes more favourable. By observing these systems using NIR echelle spectrographs such as NIRPS, CRIRES+, and SPIRou, far more of our EBLM sample will soon become accessible to HRCCS \citep{Maxted2022,2025CoSka..55c.412S}. Either approach provides a route towards measuring dynamical masses for a large sample of well-studied EBLM targets.

\section{Fundamental effective temperatures}\label{sec:teff}

Precise and accurate effective temperature measurements are vital for the testing of stellar atmosphere models and the following calibration of empirical relationships involving this quantity. There are few stars where it is possible to measure absolute temperatures, yet most of the population relies on indirect methods that suffer from systematic uncertainty, limiting the precision \citep{tayar-2022}. A fundamental approach, based on the Stefan-Boltzmann law, can be applied if both angular diameter and absolute flux are known. Measurement of angular diameters is typically only possible for nearby stars through interferometry, though can also be inferred for the components of eclipsing binaries with known radii. By combining this with the parallax from \textit{Gaia}, one can calculate accurate and model-independent effective temperatures \citep{Miller2020}. With our double-lined detections and photometric modelling, we have the absolute mass and radius measurements required to perform this analysis. We wish to estimate \teffone~and \tefftwo, the effective temperatures of the two stars in our EBLM systems, whilst accounting for known sources of uncertainty. The methodology and code, \textsc{teb}, are extensively described in works such as \citet{Miller2020} and \citet{Maxted-BEBOP-3}, so we only give a brief outline here.

Estimates of the spectral energy distributions (SEDs) for the stars in the binary are used to compute the Bayesian likelihood that photometric magnitudes, colours, and flux ratios would be observed, given a model with parameters $\mathbf{\Theta} = $ (\teffone, \tefftwo, ...). The SED function $f_{\lambda,i}$ for star $i$ is constructed by multiplying the SED computed from a grid of model stellar atmospheres with a normalisation factor and a smoothing function, $\Delta_i$. This smoothing accounts for any inaccuracies in the model SEDs. Thus, $f_{\lambda,i}$ contains realistic absorption features, yet is flexible enough that the overall shape is determined by the observed magnitudes and colours. Polynomial coefficients $c = (c_1,c_2,...)$ describing the smoothing function are also included in $\mathbf{\Theta}$. 

To compute the flux for each star in the system, $\mathcal{F}_{\lambda,i}$, we follow equation 2 of \citet{Maxted-BEBOP-3}:
\begin{equation}
    \mathcal{F}_{\lambda,i} = \frac{1}{4}~\sigma_{\rm SB}~\theta_{i}~T_{\rm eff, \mathit{i}}^{4}~f_{\lambda,i}~A_{\lambda},
\end{equation}
\noindent where the previously discussed parameters and functions are combined with the Stefan-Boltzmann constant ($\sigma_{\rm SB}$), the angular diameter ($\theta_i$), and the estimate for interstellar reddening ($A_{\lambda}$). With these fluxes, one can calculate the likelihood of the observed magnitudes, after correction for differing instrumental zero points via the \textsc{CALSPEC} absolute flux calibration \citep{Bohlin2014}. The ``distortion functions'' $\Delta_i$ need to be constrained due to the fact that we usually only have one or two flux ratio measurements for our EBLM systems. To achieve realistic SEDs, we assume that the colours of the two stars in the system are similar to other stars at the same $T_{\rm eff}$. Through use of colour -- $T_{\rm eff}$ empirical relationships, we infer the flux estimates at other wavelengths, and these are employed as flux ratio priors in the evaluation of the posterior probability.

\subsection{Reddening estimates}\label{sec:reddening}
To compute $\mathcal{F}_{\lambda,i}$ we first require an estimate of the reddening towards each target. We use the recent calibration from \citet{Maxted-EWs}, to relate the equivalent width (EW) of interstellar \ion{Na}{I} doublet lines to E(B$-$V). To measure EW(\ion{Na}{I} D1) and EW(\ion{Na}{I} D2), we follow the methodology described in \citet{Maxted2024,Maxted-BEBOP-3}. We first select all spectra for a given target with airmass $<1.4$, and shift these into the rest frame of the primary star. We then take the median at each wavelength to obtain a high SNR spectrum of the primary star in each system. After shifting each spectrum back to the barycentric rest frame, we divide through by the high-SNR master primary spectrum, and take the median of the residual spectra. This creates a high SNR spectrum of interstellar features. A schematic exemplifying this process is shown in Fig.~\ref{fig:sodium_example}

We measure the equivalent widths of the \ion{Na}{I} lines with numerical integration, and estimate uncertainty on the EW via the following equation:
\begin{equation}
    \Delta {\rm EW} \simeq1.5\frac{\sqrt{\mathit{FWHM}\times p}}{\mathit{SNR}},
\end{equation}
\noindent where the FWHM is the full width at half maximum of a Gaussian fit to the spectral line, $p$ is the pixel size, and SNR is the signal-to-noise ratio of the data in an average pixel \citep[e.g.][]{Cayrel1988}. For systems with EW $<10$\,m\AA, we follow \citet{Maxted-BEBOP-3} in assigning E(B$-$V) = $0 \pm 0.014$, based on the scatter observed for stars with little or no interstellar absorption \citep{Maxted-EWs}, though restrict that E(B$-$V) $\geq0$. The reddening estimates used as input to \textsc{teb} are listed in \rev{Table B1}{Table~\ref{tab:EWs}}.

\begin{table}
    \centering
    \caption{Estimates of interstellar reddening for stars in our sample with newly-estimated fundamental effective temperatures. These values of E(B$-$V) are derived from the EW of interstellar neutral sodium doublet lines. Note, J1540+09 is not included here, nor in Section~\ref{sec:teff}, because it lacks a secondary eclipse in \textit{TESS} photometry and hence has no flux ratio measurement.}
    \begin{tabular}{lc}
    \hline
    System  & E(B$-$V) \\
    \hline
    J0310-31   & $0.021 \pm 0.014$ \\
    J1555+26   & $0.011 \pm 0.014$ \\
    J1840+34   & $0.024 \pm 0.014$ \\
    J1928-38   & $0.059 \pm 0.019$ \\
    Kepler-16  & $0 \pm 0.014$\tnote{*}* \\
    \hline
    \end{tabular}

    \begin{tablenotes}
          \item[*] * No detectable \ion{Na}{I} D absorption with EW $>10$\,m\AA~\citep[cf.][]{Maxted-BEBOP-3}.
    \end{tablenotes}
    \label{tab:EWs}
\end{table}

\subsection{Results}\label{subsec:teff-results}
The results from our fundamental effective temperature analysis with \textsc{teb} are presented in Table~\ref{tab:teffs}. For each system we repeat the absolute masses from \texttt{Saltire} in Section~\ref{sec:results}, and include the absolute radii from Section~\ref{sec:validity-of-sb1}. These are needed for the angular diameter and $\log{g}$ inputs to \textsc{teb}. We report mean and standard error of the posterior probability distribution for \teffone~and \tefftwo,~sampled using the MCMC framework \textsc{emcee} \citep{emcee}. Each effective temperature has both random (rnd) and systematic (sys) error estimates: systematics in $T_{\rm eff}$, $\log{g}$, and apparent flux are calculated by \textsc{teb} to account for the uncertainty in the calibration of the CALSPEC flux scale. The number of coefficients ($N_{\rm c}$) in the distortion polynomial $\Delta_i$ is shown in the penultimate column. The $N_{\rm c}$ choice is made by optimising the Bayesian information criterion (BIC), as done in \citet{Maxted-BEBOP-3}. The BIC values are listed in the final column of the table. 

The results from our \textsc{teb} analysis are also shown in Figures~\ref{fig:teb_Kep16}--\ref{fig:teb_1555}. Each plot shows the best fit SED of the target system, including $1\sigma$ uncertainty bounds. Fluxes are plotted as points with errors, and the predicted fluxes based on flux-ratio priors are plotted as open circles. The corner plots for effective temperatures and reddening are also shown. 

As discussed by \citet{Maxted-BEBOP-3}, the results for secondary stars in EBLM systems with flux ratio of $\lesssim1$~per cent in the \textit{TESS} band, and only a single flux ratio measurement, are very dependent on the flux ratio priors calculated from the colour--temperature relations. As such, our measurements of \tefftwo~may be subject to additional systematic error. To properly quantify this systematic, one would need to measure the eclipse depth at NIR wavelengths, close to the peak of the M dwarf's SED. This source of systematic deviation is therefore difficult to quantify with the existing data and observations.

\begin{table*}
    \caption{Absolute masses and radii of EBLM systems used as input to our temperature analysis, and the resultant fundamental effective temperatures. The output parameter values are calculated using the mean and standard error of the posterior distribution. 
    The column $N_{\rm c}$ reports the number of distortion coefficients employed in the \textsc{teb} run with the optimal BIC, also shown in the final column. Systems are ordered by increasing \teffone.
    }
    \label{tab:teffs}
\begin{adjustbox}{width=0.99\textwidth}
\begin{tabular}{lcccccccccc}
\hline
Target & $M_{1}$ [M$_{\odot}$] & $M_{2}$ [M$_{\odot}$] & $R_{1}$ [R$_{\odot}$] & $R_{2}$ [R$_{\odot}$] & \teffone ($\pm$ rnd, $\pm$ sys) [K] & \tefftwo ($\pm$ rnd, $\pm$ sys) [K] & $N_{\rm c}$ & BIC \\
\hline
Kepler-16 & $0.704\pm 0.011$ & $0.205\pm 0.002$ & $0.653\pm0.003$ & $0.229\pm0.001$ & $4242\pm20\pm6$ & $3138\pm13\pm5$ & $3$ & $-57.817$ \\
J1928-38 & $1.027\pm 0.117$ & $0.276\pm 0.019$ & $1.508\pm0.032$ & $0.291\pm0.007$ & $5540\pm84\pm8$ & $3032\pm30\pm4 $ & $6$ & $-53.170$ \\
J0310-31 & $1.350\pm 0.017$ & $0.426\pm 0.003$ & $1.555\pm0.009$ & $0.422\pm0.028$ & $6233\pm47\pm9 $ & $3255\pm48\pm5$ & $6$ &  $-56.945$ \\
J1840+34 & $1.339\pm 0.009$ & $0.632\pm 0.003$ & $1.368\pm0.003$ & $0.589\pm0.001$ & $6372\pm50\pm10$ & $3847\pm15\pm6$ & $4$ & $-84.643$  \\
J1555+26 & $1.189\pm 0.012$ & $0.595\pm 0.004$ & $1.890\pm 0.006$ & $0.538\pm 0.004$ & $6645 \pm 60 \pm 10$ & $4423 \pm 49 \pm 6$  & 4 & $-76.847$ \\
\hline
\end{tabular}
\end{adjustbox}
\end{table*}

\subsection{Agreement with spectroscopy}
\begin{figure}
    \centering
    \includegraphics[width=0.99\columnwidth]{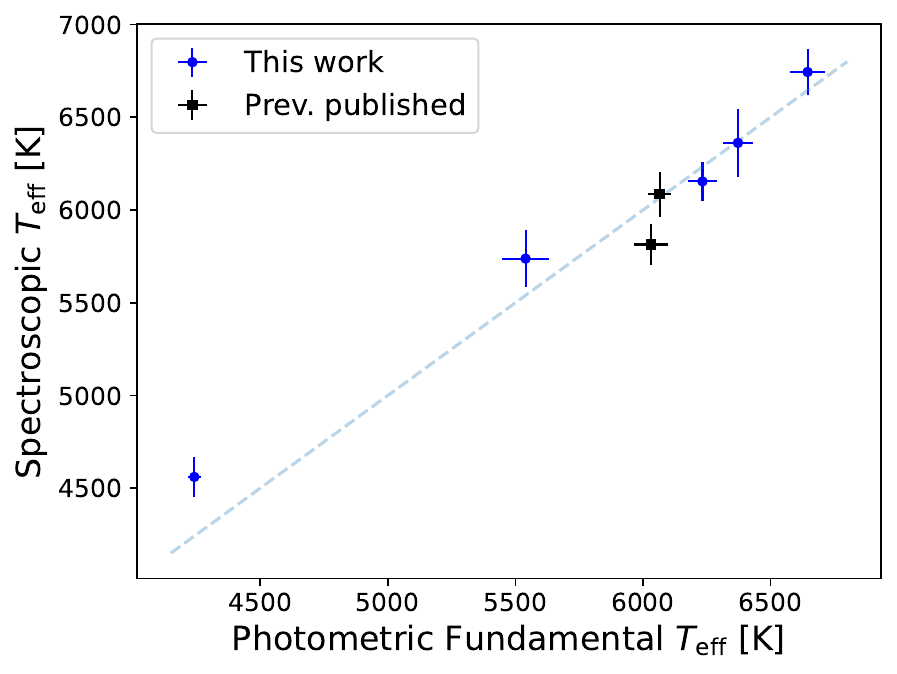}
    \caption{A comparison between the fundamental effective temperatures for 7 out of 8 stars in our sample, and $T_{\rm eff}$ determined via spectroscopic methods \citep[see Section~\ref{sec:spectra}, and][]{freckelton-2024-bebopv}. The dashed line shows 1:1 agreement. Blue circles are those where we have calculated the fundamental $T_{\rm eff}$ in this work, and black squares are those previously published \citep{Maxted2022,Maxted-BEBOP-3}}
    \label{fig:teff_comparison}
\end{figure}

The primary motivation behind calculating fundamental effective temperatures is the benchmarking of stellar atmosphere models \citep{Miller2020}. With our homogeneously analysed sample of EBLMs, we can begin to comment on the comparison between the effective temperatures found from SED-based fitting and from spectral analysis \citep[Section~\ref{sec:spectra} and][]{freckelton-2024-bebopv}. Our sample of seven systems with both measurements means we remain limited by small number statistics, yet take a first pass at assessing the agreement between the methods. In Figure~\ref{fig:teff_comparison}, we plot the spectroscopic $T_{\rm eff}$ against the \textsc{teb} \teffone  from this work, and from \citet{Maxted2022} for J0608-59 and \citet{Maxted-BEBOP-3} for J0723+79. Within $1\sigma$, the points are broadly consistent with the 1:1 agreement dashed line (RMS = 171\,K). 

Depending on the chosen analysis method, spectroscopic effective temperatures of FGK stars routinely exhibit differences of $200$--$300$\,K \citep{PASTEL,Jofre2019}. Recently, \citet{freckelton-2025-gr8stars} compared two sources of $T_{\rm eff}$, derived from spectral observations of a sample of bright single stars, and found an RMS difference of $101$\,K. This is fully consistent with the typical limiting systematic uncertainty on effective temperature \citep{tayar-2022}. 

\citet{freckelton-2025-gr8stars} also assessed the consistency between spectroscopic effective temperatures and those derived from photometric measurements via SED fitting, estimates that are more analogous to the fundamental $T_{\rm eff}$ derived in Section~\ref{subsec:teff-results}. Though some studies have found substantial differences between photometric and spectroscopic estimates \citep[e.g.][]{Groenewegen2020}, a Gaussian fit to the residuals $T_{\rm spec}-T_{\rm phot}$ yielded $\sigma=98$\,K. In light of the established scatter within spectroscopic methods themselves, and also between spectroscopy and photometry, the level of agreement we observe in Fig.~\ref{fig:teff_comparison} is satisfactory.

\section{Conclusions}\label{sec:conclusion}

We present dynamical mass measurements for a sample of EBLM binaries, where five systems have been newly transformed from single- to double-lined through the use of high resolution cross-correlation spectroscopy. By removing the spectral signature of the primary star and cross-correlating in the companion's rest frame, we have successfully identified the lines from the secondary, precisely measured the semi-amplitude with the \texttt{Saltire} model \citep{sebastian_2024_saltire}, and thus calculated model-independent masses for both stars.

Increasing the sample of EBLM targets analysed as a double-lined (SB2) binary, we confirm that this method provides a reliable avenue to estimate precise 
stellar masses, that we assume are also entirely accurate due to the evolutionary model independence.
We find consistency with previous works \citep{sebastian_2024_saltire,sebastian-2024-bebop1,sebastian_2025-kepler16,Baycroft-2025-bebop3} relating to the estimation of systematic uncertainty on extracted CCF map parameters.

We have investigated other methodologies used to estimate the masses of system constituents, from those typically employed for single-lined (SB1) eclipsing binary stars \citep{davis-2024}, to those used in exoplanetary searches -- where the primary mass estimate has to come from isochrone analysis or empirical relations linking masses and observables. We have performed a comparison between these methods, and the ``ground truth'' obtained by considering the systems as SB2s. We find good agreement between methods within quoted uncertainties, particularly for the SB1 and SB2 methods, though find that SB1 masses are systematically larger for targets with more massive secondary companions. We identify that this increase comes from an over-estimated primary radius assumed in the SB1 method, which in turn arises from treating the system as a single star when fitting a MIST isochrone to atmospheric parameters and photometric magnitudes. Our sample is naturally biased towards systems with less contrast between the two stars, hence show this behaviour more strongly than the EBLM sample in general. 

Despite the tendency to over-estimate the masses for the secondaries in higher mass-ratio systems, we find that the SB1 method provides good consistency with SB2 results in terms of quantifying radius inflation. Due to the correlated nature of the SB1 posteriors, placing a lower constraint on $M_2$ also results in a smaller radius estimate, moving the position of the star in mass-radius space in such a way that almost follows the isochrone model. For the fully-convective secondary stars that are of great interest to the EBLM project, these effects are minimised, and the SB1 method is valid for radius inflation studies. 

The masses derived for the secondary stars in this work are among the most accurate and precise possible measurements of fundamental parameters in the M~dwarf domain. We have begun an effort to increase the number of late-type stars characterised in this manner, by populating parts of parameter space that are often inaccessible to SB2 studies and are thus missing from resources like DEBCat \citep{Southworth-debcat}. We have used the absolute physical parameters calculated herein to estimate fundamental effective temperatures for the stars in these binaries, making these systems the ideal benchmarks for comparison to stellar atmospheric models and empirical relations in a number of ways. A large sample of stars below the fully-convective transition ($\sim0.35$\,M$_{\odot}$) is of particular importance to assess radius inflation \citep{eblm-northern-sample}, though early M~dwarfs can also help to calibrate relations spanning the tail end of the main sequence.

\section*{Acknowledgements}

The authors thank the referee, Dr. Samuel Gill, for their helpful comments which have improved the quality of the manuscript.\\

Based in part on observations collected at Observatoire de Haute-Provence (CNRS, UAR Pytheas), France, with the SOPHIE instrument on the 1.93m telescope. We thank the staff of the Observatoire de Haute-Provence for their support at 1.93m telescope. Based on observations collected at the European Southern Observatory under ESO programmes 099.C-0138(A), 1106.C-0912(A), 1106.C-0597(A), 1101.C-0721(A), 0113.C-0295(A). This research has made use of the services of the ESO Science Archive Facility. \\

We gratefully acknowledge all SOPHIE and HARPS instrument time-share observers, and the organisers of those programmes, for their aid in obtaining the spectra used herein. \\

ATS and AHMJT were supported by the ERC/UKRI Frontier Research Guarantee programme (EP/Z000327/1/CandY).
PM was supported by the UK Science and Technology Facilities Council (STFC) research grant numbers ST/R000638/1, ST/Y002563/1 and UKRI1193, UK Space Agency (UKSA) grant number UKRI966, and a travel grant from the STFC Panel for the Allocation of Telescope Time (PATT).
AVF acknowledges the support of the IOP through the Bell Burnell Graduate Scholarship Fund.
This work was supported by the "Programme National de Planétologie" (PNP) of CNRS/INSU co-funded by CNES. This work was supported by the "Action Thématique Physique Stellaire” of CNRS/INSU PN ASTRO co-funded by CEA and CNES.\\

This paper includes data collected with the \textit{TESS} mission, obtained from the MAST data archive at the Space Telescope Science Institute (STScI). Funding for the \textit{TESS} mission is provided by the NASA Explorer Program. STScI is operated by the Association of Universities for Research in Astronomy, Inc., under NASA contract NAS 5–26555.\\

This work has made use of data from the European Space Agency (ESA) mission
{\it Gaia} (\url{https://www.cosmos.esa.int/gaia}), processed by the {\it Gaia}
Data Processing and Analysis Consortium (DPAC,
\url{https://www.cosmos.esa.int/web/gaia/dpac/consortium}). Funding for the DPAC
has been provided by national institutions, in particular the institutions
participating in the {\it Gaia} Multilateral Agreement. \\

This research has made use of the Astrophysics Data System, funded by NASA under Cooperative Agreement 80NSSC21M0056. 
This research has made use of the SIMBAD database, CDS, Strasbourg Astronomical Observatory, France.
Radial velocity data were analysed with the \textsc{kima} package, freely accessible \href{https://github.com/kima-org/kima}{online} and described in \citet{kima-joss}.
Stellar atmospheric parameters were determined using \texttt{PAWS}, accessible openly from \url{https://github.com/alixviolet/PAWS}. \\

\textit{Software:} \textsc{allesfitter}, \textsc{astropy}, \textsc{ellc}, \textsc{isochrones}, \textsc{kima}, \textsc{numpy}, \textsc{matplotlib}, \textsc{pandas}, \textsc{pycheops}, \textsc{saltire}, \textsc{scipy}, \textsc{seaborn}, \textsc{teb}.

\section*{Data Availability}
 

All \textit{TESS} data used in this publication is available to download via the MAST portal (\url{https://mast.stsci.edu}). Spectroscopic data are available on the SOPHIE (\url{http://atlas.obs-hp.fr/sophie/}) and HARPS (\url{https://archive.eso.org}) archives.



\bibliographystyle{mnras}
\bibliography{refs} 

@ARTICLE{prsa-2016-iau,
       author = {{Pr{\v{s}}a}, Andrej and {Harmanec}, Petr and {Torres}, Guillermo and {Mamajek}, Eric and {Asplund}, Martin and {Capitaine}, Nicole and {Christensen-Dalsgaard}, J{\o}rgen and {Depagne}, {\'E}ric and {Haberreiter}, Margit and {Hekker}, Saskia and {Hilton}, James and {Kopp}, Greg and {Kostov}, Veselin and {Kurtz}, Donald W. and {Laskar}, Jacques and {Mason}, Brian D. and {Milone}, Eugene F. and {Montgomery}, Michele and {Richards}, Mercedes and {Schmutz}, Werner and {Schou}, Jesper and {Stewart}, Susan G.},
        title = "{Nominal Values for Selected Solar and Planetary Quantities: IAU 2015 Resolution B3}",
      journal = {\aj},
         year = 2016,
        month = aug,
       volume = {152},
       number = {2},
          eid = {41},
        pages = {41},
          doi = {10.3847/0004-6256/152/2/41},
archivePrefix = {arXiv},
       eprint = {1605.09788},
 primaryClass = {astro-ph.SR},
       adsurl = {https://ui.adsabs.harvard.edu/abs/2016AJ....152...41P}
}

@ARTICLE{Thomsen25,
       author = {{Thomsen}, J.~S. and {Miglio}, A. and {Brogaard}, K. and {Montalb{\'a}n}, J. and {Tailo}, M. and {van Rossem}, W.~E. and {Casali}, G. and {Jones}, D. and {Arentoft}, T. and {Casagrande}, L. and {Sebastian}, D. and {Buldgen}, G. and {Triaud}, A.~H.~M.~J. and {Matteuzzi}, M. and {Stokholm}, A. and {Lund}, M.~N. and {Mosser}, B. and {Maxted}, P.~F.~L. and {Southworth}, J. and {Gadeberg}, J.~T. and {Koivisto}, N. and {Gray}, Z. and {Pinter}, V. and {Matilainen}, K. and {Djupvik}, A.~A. and {Jessen-Hansen}, J. and {Grundahl}, F. and {Slumstrup}, D. and {Frandsen}, S.},
        title = "{Advancing the accuracy in age determinations of old-disk stars using an oscillating red giant in an eclipsing binary}",
      journal = {\aap},
         year = 2025,
        month = jul,
       volume = {699},
          eid = {A152},
        pages = {A152},
          doi = {10.1051/0004-6361/202453347},
archivePrefix = {arXiv},
       eprint = {2504.17853},
 primaryClass = {astro-ph.SR},
       adsurl = {https://ui.adsabs.harvard.edu/abs/2025A&A...699A.152T}
}

@ARTICLE{triaud24,
       author = {{Triaud}, Amaury H.~M.~J. and {de Wit}, Julien and {Klein}, Frieder and {Turbet}, Martin and {Rackham}, Benjamin V. and {Niraula}, Prajwal and {Glidden}, Ana and {Jagoutz}, Oliver E. and {Pe{\v{c}}}, Matej and {Petkowski}, Janusz J. and {Seager}, Sara and {Selsis}, Franck},
        title = "{Atmospheric carbon depletion as a tracer of water oceans and biomass on temperate terrestrial exoplanets}",
      journal = {Nature Astronomy},
         year = 2024,
        month = jan,
       volume = {8},
       number = {1},
        pages = {17-29},
          doi = {10.1038/s41550-023-02157-9},
archivePrefix = {arXiv},
       eprint = {2310.14987},
 primaryClass = {astro-ph.EP},
       adsurl = {https://ui.adsabs.harvard.edu/abs/2024NatAs...8...17T}
}

@ARTICLE{sebastian-2024-bebop1,
       author = {{Sebastian}, D. and {Triaud}, A.~H.~M.~J. and {Brogi}, M. and {Baycroft}, Thomas A. and {Standing}, Matthew R. and {Maxted}, Pierre F.~L. and {Martin}, David V. and {Sairam}, Lalitha and {Nielsen}, Martin B.},
        title = "{The EBLM project - XIII. The absolute dynamical masses of the circumbinary planet host TOI-1338/BEBOP-1, and applications to the study of exoplanet atmospheres}",
      journal = {\mnras},
         year = 2024,
        month = may,
       volume = {530},
       number = {3},
        pages = {2572-2589},
          doi = {10.1093/mnras/stae459},
archivePrefix = {arXiv},
       eprint = {2402.06449},
 primaryClass = {astro-ph.EP},
       adsurl = {https://ui.adsabs.harvard.edu/abs/2024MNRAS.530.2572S}
}

@ARTICLE{sebastian_2024_saltire,
       author = {{Sebastian}, D. and {Triaud}, A.~H.~M.~J. and {Brogi}, M.},
        title = "{Saltire: a model to measure dynamical masses for high-contrast binaries and exoplanets with high-resolution spectroscopy}",
      journal = {\mnras},
         year = 2024,
        month = feb,
       volume = {527},
       number = {4},
        pages = {10921-10936},
          doi = {10.1093/mnras/stad3765},
archivePrefix = {arXiv},
       eprint = {2312.01924},
 primaryClass = {astro-ph.EP},
       adsurl = {https://ui.adsabs.harvard.edu/abs/2024MNRAS.52710921S}
}

@ARTICLE{sebastian_2025-kepler16,
       author = {{Sebastian}, D. and {Boisse}, I. and {Santerne}, A. and {Triaud}, A.~H.~M.~J. and {Baycroft}, T.~A. and {Davis}, Y.~T. and {Deleuil}, M. and {Grouffal}, S. and {H{\'e}brard}, G. and {Heidari}, N. and {Martin}, D.~V. and {Maxted}, P.~F.~L. and {Nelson}, R.~P. and {Lalitha}, S. and {Scott}, M.~G. and {Scutt}, O.~J. and {Standing}, M.},
        title = "{EBLM XV - revised dynamical masses for the circumbinary planet host Kepler-16 AB, using the SOPHIE spectrograph}",
      journal = {\mnras},
         year = 2025,
        month = jul,
       volume = {540},
       number = {4},
        pages = {2914-2922},
          doi = {10.1093/mnras/staf863},
archivePrefix = {arXiv},
       eprint = {2505.19718},
 primaryClass = {astro-ph.EP},
       adsurl = {https://ui.adsabs.harvard.edu/abs/2025MNRAS.540.2914S}
}

@ARTICLE{davis-2024,
       author = {{Davis}, Yasmin T. and {Triaud}, Amaury H.~M.~J. and {Freckelton}, Alix V. and {Mortier}, Annelies and {Sebastian}, Daniel and {Brahm}, Rafael and {Baycroft}, Thomas and {Dransfield}, Georgina and {Duck}, Alison and {Henning}, Thomas and {Hobson}, Melissa J. and {Jord{\'a}n}, Andr{\'e}s and {Kunovac}, Vedad and {Martin}, David V. and {Maxted}, Pierre F.~L. and {Sairam}, Lalitha and {Standing}, Matthew R. and {Swayne}, Matthew I. and {Trifonov}, Trifon and {Udry}, St{\'e}phane},
        title = "{The EBLM Project XII. An eccentric, long-period eclipsing binary with a companion near the hydrogen-burning limit}",
      journal = {\mnras},
         year = 2024,
        month = may,
       volume = {530},
       number = {3},
        pages = {2565-2571},
          doi = {10.1093/mnras/stae842},
archivePrefix = {arXiv},
       eprint = {2312.09156},
 primaryClass = {astro-ph.SR},
       adsurl = {https://ui.adsabs.harvard.edu/abs/2024MNRAS.530.2565D}
}

@ARTICLE{freckelton-2024-bebopv,
       author = {{Freckelton}, Alix V. and {Sebastian}, Daniel and {Mortier}, Annelies and {Triaud}, Amaury H.~M.~J. and {Maxted}, Pierre F.~L. and {Acu{\~n}a}, Lorena and {Armstrong}, David J. and {Battley}, Matthew P. and {Baycroft}, Thomas A. and {Boisse}, Isabelle and {Bourrier}, Vincent and {Carmona}, Andres and {Coleman}, Gavin A.~L. and {Cameron}, Andrew Collier and {Cort{\'e}s-Zuleta}, P{\'\i}a and {Delfosse}, Xavier and {Dransfield}, Georgina and {Duck}, Alison and {Forveille}, Thierry and {French}, Jenni R. and {Hara}, Nathan and {Heidari}, Neda and {Hellier}, Coel and {Kunovac}, Vedad and {Martin}, David V. and {Martioli}, Eder and {McCormac}, James J. and {Nelson}, Richard P. and {Sairam}, Lalitha and {Sousa}, S{\'e}rgio G. and {Standing}, Matthew R. and {Willett}, Emma},
        title = "{BEBOP V. Homogeneous stellar analysis of potential circumbinary planet hosts}",
      journal = {\mnras},
         year = 2024,
        month = jul,
       volume = {531},
       number = {4},
        pages = {4085-4098},
          doi = {10.1093/mnras/stae1405},
archivePrefix = {arXiv},
       eprint = {2406.03094},
 primaryClass = {astro-ph.SR},
       adsurl = {https://ui.adsabs.harvard.edu/abs/2024MNRAS.531.4085F}
}

@ARTICLE{baycroft-2023-binaries,
       author = {{Baycroft}, Thomas A. and {Triaud}, Amaury H.~M.~J. and {Faria}, Jo{\~a}o and {Correia}, Alexandre C.~M. and {Standing}, Matthew R.},
        title = "{Improving circumbinary planet detections by fitting their binary's apsidal precession}",
      journal = {\mnras},
         year = 2023,
        month = may,
       volume = {521},
       number = {2},
        pages = {1871-1879},
          doi = {10.1093/mnras/stad607},
archivePrefix = {arXiv},
       eprint = {2301.11372},
 primaryClass = {astro-ph.EP},
       adsurl = {https://ui.adsabs.harvard.edu/abs/2023MNRAS.521.1871B}
}

@ARTICLE{standing-2022-bebopii,
       author = {{Standing}, Matthew R. and {Triaud}, Amaury H.~M.~J. and {Faria}, Jo{\~a}o P. and {Martin}, David V. and {Boisse}, Isabelle and {Correia}, Alexandre C.~M. and {Deleuil}, Magali and {Dransfield}, Georgina and {Gillon}, Micha{\"e}l and {H{\'e}brard}, Guillaume and {Hellier}, Coel and {Kunovac}, Vedad and {Maxted}, Pierre F.~L. and {Mardling}, Rosemary and {Santerne}, Alexandre and {Sairam}, Lalitha and {Udry}, St{\'e}phane},
        title = "{BEBOP II: sensitivity to sub-Saturn circumbinary planets using radial-velocities}",
      journal = {\mnras},
         year = 2022,
        month = apr,
       volume = {511},
       number = {3},
        pages = {3571-3583},
          doi = {10.1093/mnras/stac113},
archivePrefix = {arXiv},
       eprint = {2112.05652},
 primaryClass = {astro-ph.EP},
       adsurl = {https://ui.adsabs.harvard.edu/abs/2022MNRAS.511.3571S}
}

@ARTICLE{triaud-2013-eblm-i,
       author = {{Triaud}, A.~H.~M.~J. and {Hebb}, L. and {Anderson}, D.~R. and {Cargile}, P. and {Collier Cameron}, A. and {Doyle}, A.~P. and {Faedi}, F. and {Gillon}, M. and {Gomez Maqueo Chew}, Y. and {Hellier}, C. and {Jehin}, E. and {Maxted}, P. and {Naef}, D. and {Pepe}, F. and {Pollacco}, D. and {Queloz}, D. and {S{\'e}gransan}, D. and {Smalley}, B. and {Stassun}, K. and {Udry}, S. and {West}, R.~G.},
        title = "{The EBLM project. I. Physical and orbital parameters, including spin-orbit angles, of two low-mass eclipsing binaries on opposite sides of the brown dwarf limit}",
      journal = {\aap},
         year = 2013,
        month = jan,
       volume = {549},
          eid = {A18},
        pages = {A18},
          doi = {10.1051/0004-6361/201219643},
archivePrefix = {arXiv},
       eprint = {1208.4940},
 primaryClass = {astro-ph.SR},
       adsurl = {https://ui.adsabs.harvard.edu/abs/2013A&A...549A..18T}
}

@software{allesfitter-2019-ascl,
       author = {{G{\"u}nther}, Maximilian N. and {Daylan}, Tansu},
        title = "{allesfitter: Flexible star and exoplanet inference from photometry and radial velocity}",
 howpublished = {Astrophysics Source Code Library, record ascl:1903.003},
         year = 2019,
        month = mar,
          eid = {ascl:1903.003},
       adsurl = {https://ui.adsabs.harvard.edu/abs/2019ascl.soft03003G}
}

@ARTICLE{allesfitter-2021-apjs,
       author = {{G{\"u}nther}, Maximilian N. and {Daylan}, Tansu},
        title = "{Allesfitter: Flexible Star and Exoplanet Inference from Photometry and Radial Velocity}",
      journal = {\apjs},
         year = 2021,
        month = may,
       volume = {254},
       number = {1},
          eid = {13},
        pages = {13},
          doi = {10.3847/1538-4365/abe70e},
archivePrefix = {arXiv},
       eprint = {2003.14371},
 primaryClass = {astro-ph.EP},
       adsurl = {https://ui.adsabs.harvard.edu/abs/2021ApJS..254...13G}
}

@ARTICLE{stassun-2019-revisedTIC,
       author = {{Stassun}, Keivan G. and {Oelkers}, Ryan J. and {Paegert}, Martin and {Torres}, Guillermo and {Pepper}, Joshua and {De Lee}, Nathan and {Collins}, Kevin and {Latham}, David W. and {Muirhead}, Philip S. and {Chittidi}, Jay and {Rojas-Ayala}, B{\'a}rbara and {Fleming}, Scott W. and {Rose}, Mark E. and {Tenenbaum}, Peter and {Ting}, Eric B. and {Kane}, Stephen R. and {Barclay}, Thomas and {Bean}, Jacob L. and {Brassuer}, C.~E. and {Charbonneau}, David and {Ge}, Jian and {Lissauer}, Jack J. and {Mann}, Andrew W. and {McLean}, Brian and {Mullally}, Susan and {Narita}, Norio and {Plavchan}, Peter and {Ricker}, George R. and {Sasselov}, Dimitar and {Seager}, S. and {Sharma}, Sanjib and {Shiao}, Bernie and {Sozzetti}, Alessandro and {Stello}, Dennis and {Vanderspek}, Roland and {Wallace}, Geoff and {Winn}, Joshua N.},
        title = "{The Revised TESS Input Catalog and Candidate Target List}",
      journal = {\aj},
         year = 2019,
        month = oct,
       volume = {158},
       number = {4},
          eid = {138},
        pages = {138},
          doi = {10.3847/1538-3881/ab3467},
archivePrefix = {arXiv},
       eprint = {1905.10694},
 primaryClass = {astro-ph.SR},
       adsurl = {https://ui.adsabs.harvard.edu/abs/2019AJ....158..138S}
}

@ARTICLE{duchene-kraus-2013,
       author = {{Duch{\^e}ne}, Gaspard and {Kraus}, Adam},
        title = "{Stellar Multiplicity}",
      journal = {\araa},
         year = 2013,
        month = aug,
       volume = {51},
       number = {1},
        pages = {269-310},
          doi = {10.1146/annurev-astro-081710-102602},
archivePrefix = {arXiv},
       eprint = {1303.3028},
 primaryClass = {astro-ph.SR},
       adsurl = {https://ui.adsabs.harvard.edu/abs/2013ARA&A..51..269D}
}

@ARTICLE{tayar-2022,
       author = {{Tayar}, Jamie and {Claytor}, Zachary R. and {Huber}, Daniel and {van Saders}, Jennifer},
        title = "{A Guide to Realistic Uncertainties on the Fundamental Properties of Solar-type Exoplanet Host Stars}",
      journal = {\apj},
         year = 2022,
        month = mar,
       volume = {927},
       number = {1},
          eid = {31},
        pages = {31},
          doi = {10.3847/1538-4357/ac4bbc},
archivePrefix = {arXiv},
       eprint = {2012.07957},
 primaryClass = {astro-ph.EP},
       adsurl = {https://ui.adsabs.harvard.edu/abs/2022ApJ...927...31T}
}

@ARTICLE{freckelton-2025-gr8stars,
       author = {{Freckelton}, Alix Violet and {Mortier}, Annelies and {Bedell}, Megan and {Morrell}, Sam and {Naylor}, Tim and {Buchhave}, Lars A. and {Davies}, Guy R. and {Gonz{\'a}lez Hern{\'a}ndez}, J.~I. and {Klein}, Baptiste and {de Mooij}, Ernst J.~W. and {Passegger}, Vera Maria and {Quirrenbach}, Andreas and {Roy}, Arpita and {Santos}, Nuno C. and {Sousa}, S{\'e}rgio G. and {Su{\'a}rez Mascare{\~n}o}, A. and {Tsantaki}, Maria and {Zhao}, Lily L.},
        title = "{gr8stars ─ I. A homogeneous spectroscopic study of bright FGKM dwarfs and a public library of their high-resolution spectra}",
      journal = {\mnras},
         year = 2025,
        month = jun,
       volume = {540},
       number = {2},
        pages = {1786-1799},
          doi = {10.1093/mnras/staf825},
archivePrefix = {arXiv},
       eprint = {2505.12945},
 primaryClass = {astro-ph.SR},
       adsurl = {https://ui.adsabs.harvard.edu/abs/2025MNRAS.540.1786F}
}

@ARTICLE{sousa-2011,
       author = {{Sousa}, S.~G. and {Santos}, N.~C. and {Israelian}, G. and {Lovis}, C. and {Mayor}, M. and {Silva}, P.~B. and {Udry}, S.},
        title = "{Spectroscopic characterization of a sample of metal-poor solar-type stars from the HARPS planet search program. Precise spectroscopic parameters and mass estimation}",
      journal = {\aap},
         year = 2011,
        month = feb,
       volume = {526},
          eid = {A99},
        pages = {A99},
          doi = {10.1051/0004-6361/201015646},
archivePrefix = {arXiv},
       eprint = {1012.1528},
 primaryClass = {astro-ph.SR},
       adsurl = {https://ui.adsabs.harvard.edu/abs/2011A&A...526A..99S}
}

@conference{roy-vetterli-2007,
    author = {{Roy}, O. and {Vetterli}, M.},
    booktitle ={15th European Signal Processing Conference},
    title ={The effective rank: a measure of effective dimensionality},
    year = 2007,
    publisher={IEEE},
    pages = {606-610.},
}

@ARTICLE{martin-2019-bebopi,
       author = {{Martin}, David V. and {Triaud}, Amaury H.~M.~J. and {Udry}, St{\'e}phane and {Marmier}, Maxime and {Maxted}, Pierre F.~L. and {Collier Cameron}, Andrew and {Hellier}, Coel and {Pepe}, Francesco and {Pollacco}, Don and {S{\'e}gransan}, Damien and {West}, Richard},
        title = "{The BEBOP radial-velocity survey for circumbinary planets. I. Eight years of CORALIE observations of 47 single-line eclipsing binaries and abundance constraints on the masses of circumbinary planets}",
      journal = {\aap},
         year = 2019,
        month = apr,
       volume = {624},
          eid = {A68},
        pages = {A68},
          doi = {10.1051/0004-6361/201833669},
archivePrefix = {arXiv},
       eprint = {1901.01627},
 primaryClass = {astro-ph.EP},
       adsurl = {https://ui.adsabs.harvard.edu/abs/2019A&A...624A..68M}
}

@ARTICLE{triaud-2022-bebopiii,
       author = {{Triaud}, Amaury H.~M.~J. and {Standing}, Matthew R. and {Heidari}, Neda and {Martin}, David V. and {Boisse}, Isabelle and {Santerne}, Alexandre and {Correia}, Alexandre C.~M. and {Acu{\~n}a}, Lorena and {Battley}, Matthew and {Bonfils}, Xavier and {Carmona}, Andr{\'e}s and {Collier Cameron}, Andrew and {Cort{\'e}s-Zuleta}, P{\'\i}a and {Dransfield}, Georgina and {Dalal}, Shweta and {Deleuil}, Magali and {Delfosse}, Xavier and {Faria}, Jo{\~a}o and {Forveille}, Thierry and {Hara}, Nathan C. and {H{\'e}brard}, Guillaume and {Hoyer}, Sergio and {Kiefer}, Flavien and {Kunovac}, Vedad and {Maxted}, Pierre F.~L. and {Martioli}, Eder and {Miller}, Nicola J. and {Nelson}, Richard P. and {Poveda}, Mathilde and {Rein}, Hanno and {Sairam}, Lalitha and {Udry}, St{\'e}phane and {Willett}, Emma},
        title = "{BEBOP III. Observations and an independent mass measurement of Kepler-16 (AB) b - the first circumbinary planet detected with radial velocities}",
      journal = {\mnras},
         year = 2022,
        month = apr,
       volume = {511},
       number = {3},
        pages = {3561-3570},
          doi = {10.1093/mnras/stab3712},
archivePrefix = {arXiv},
       eprint = {2112.06584},
 primaryClass = {astro-ph.EP},
       adsurl = {https://ui.adsabs.harvard.edu/abs/2022MNRAS.511.3561T}
}

@ARTICLE{standing-2023-bebop1,
       author = {{Standing}, Matthew R. and {Sairam}, Lalitha and {Martin}, David V. and {Triaud}, Amaury H.~M.~J. and {Correia}, Alexandre C.~M. and {Coleman}, Gavin A.~L. and {Baycroft}, Thomas A. and {Kunovac}, Vedad and {Boisse}, Isabelle and {Cameron}, Andrew Collier and {Dransfield}, Georgina and {Faria}, Jo{\~a}o P. and {Gillon}, Micha{\"e}l and {Hara}, Nathan C. and {Hellier}, Coel and {Howard}, Jonathan and {Lane}, Ellie and {Mardling}, Rosemary and {Maxted}, Pierre F.~L. and {Miller}, Nicola J. and {Nelson}, Richard P. and {Orosz}, Jerome A. and {Pepe}, Franscesco and {Santerne}, Alexandre and {Sebastian}, Daniel and {Udry}, St{\'e}phane and {Welsh}, William F.},
        title = "{Radial-velocity discovery of a second planet in the TOI-1338/BEBOP-1 circumbinary system}",
      journal = {Nature Astronomy},
         year = 2023,
        month = jun,
       volume = {7},
        pages = {702-714},
          doi = {10.1038/s41550-023-01948-4},
archivePrefix = {arXiv},
       eprint = {2301.10794},
 primaryClass = {astro-ph.EP},
       adsurl = {https://ui.adsabs.harvard.edu/abs/2023NatAs...7..702S}
}

@INPROCEEDINGS{perruchot-sophie,
       author = {{Perruchot}, S. and {Kohler}, D. and {Bouchy}, F. and {Richaud}, Y. and {Richaud}, P. and {Moreaux}, G. and {Merzougui}, M. and {Sottile}, R. and {Hill}, L. and {Knispel}, G. and {Regal}, X. and {Meunier}, J. -P. and {Ilovaisky}, S. and {Le Coroller}, H. and {Gillet}, D. and {Schmitt}, J. and {Pepe}, F. and {Fleury}, M. and {Sosnowska}, D. and {Vors}, P. and {M{\'e}gevand}, D. and {Blanc}, P.~E. and {Carol}, C. and {Point}, A. and {Laloge}, A. and {Brunel}, J. -C.},
        title = "{The SOPHIE spectrograph: design and technical key-points for high throughput and high stability}",
    booktitle = {Ground-based and Airborne Instrumentation for Astronomy II},
         year = 2008,
       editor = {{McLean}, Ian S. and {Casali}, Mark M.},
       series = {Society of Photo-Optical Instrumentation Engineers (SPIE) Conference Series},
       volume = {7014},
        month = jul,
          eid = {70140J},
        pages = {70140J},
          doi = {10.1117/12.787379},
       adsurl = {https://ui.adsabs.harvard.edu/abs/2008SPIE.7014E..0JP}
}

@ARTICLE{mayor-harps,
       author = {{Mayor}, M. and {Pepe}, F. and {Queloz}, D. and {Bouchy}, F. and {Rupprecht}, G. and {Lo Curto}, G. and {Avila}, G. and {Benz}, W. and {Bertaux}, J. -L. and {Bonfils}, X. and {Dall}, Th. and {Dekker}, H. and {Delabre}, B. and {Eckert}, W. and {Fleury}, M. and {Gilliotte}, A. and {Gojak}, D. and {Guzman}, J.~C. and {Kohler}, D. and {Lizon}, J. -L. and {Longinotti}, A. and {Lovis}, C. and {Megevand}, D. and {Pasquini}, L. and {Reyes}, J. and {Sivan}, J. -P. and {Sosnowska}, D. and {Soto}, R. and {Udry}, S. and {van Kesteren}, A. and {Weber}, L. and {Weilenmann}, U.},
        title = "{Setting New Standards with HARPS}",
      journal = {The Messenger},
         year = 2003,
        month = dec,
       volume = {114},
        pages = {20-24},
       adsurl = {https://ui.adsabs.harvard.edu/abs/2003Msngr.114...20M}
}

@ARTICLE{Baycroft-2025-bebop3,
       author = {{Baycroft}, Thomas A. and {Santerne}, Alexandre and {Triaud}, Amaury H.~M.~J. and {Heidari}, Neda and {Sebastian}, Daniel and {Davis}, Yasmin T. and {Correia}, Alexandre C.~M. and {Sairam}, Lalitha and {Freckelton}, Alix V. and {Adamson}, Aleyna and {Boisse}, Isabelle and {Coleman}, Gavin A.~L. and {Dransfield}, Georgina and {Faria}, Jo{\~a}o and {Grouffal}, Salom{\'e} and {Hara}, Nathan and {H{\'e}brard}, Guillaume and {Kunovac}, Vedad and {Martin}, David V. and {Maxted}, Pierre F.~L. and {Nelson}, Richard P. and {Scott}, Madison G. and {Scutt}, Owen J. and {Standing}, Matthew R.},
        title = "{BEBOP VII. SOPHIE discovery of BEBOP-3b, a circumbinary giant planet on an eccentric orbit}",
      journal = {\mnras},
         year = 2025,
        month = aug,
       volume = {541},
       number = {3},
        pages = {2801-2814},
          doi = {10.1093/mnras/staf1184},
archivePrefix = {arXiv},
       eprint = {2506.14615},
 primaryClass = {astro-ph.EP},
       adsurl = {https://ui.adsabs.harvard.edu/abs/2025MNRAS.541.2801B}
}

@ARTICLE{Maxted2025,
       author = {{Maxted}, P.~F.~L.},
        title = "{Benchmark stars for mean stellar density and surface gravity estimates of solar-type stars}",
      journal = {\mnras},
         year = 2025,
        month = jun,
       volume = {539},
       number = {4},
        pages = {3741-3757},
          doi = {10.1093/mnras/staf735},
archivePrefix = {arXiv},
       eprint = {2505.01154},
 primaryClass = {astro-ph.SR},
       adsurl = {https://ui.adsabs.harvard.edu/abs/2025MNRAS.539.3741M}
}

@ARTICLE{Bourrier2018,
       author = {{Bourrier}, Vincent and {Lovis}, Christophe and {Beust}, Herv{\'e} and {Ehrenreich}, David and {Henry}, Gregory W. and {Astudillo-Defru}, Nicola and {Allart}, Romain and {Bonfils}, Xavier and {S{\'e}gransan}, Damien and {Delfosse}, Xavier and {Cegla}, Heather M. and {Wyttenbach}, Aur{\'e}lien and {Heng}, Kevin and {Lavie}, Baptiste and {Pepe}, Francesco},
        title = "{Orbital misalignment of the Neptune-mass exoplanet GJ 436b with the spin of its cool star}",
      journal = {\nat},
         year = 2018,
        month = jan,
       volume = {553},
       number = {7689},
        pages = {477-480},
          doi = {10.1038/nature24677},
archivePrefix = {arXiv},
       eprint = {1712.06638},
 primaryClass = {astro-ph.EP},
       adsurl = {https://ui.adsabs.harvard.edu/abs/2018Natur.553..477B}
}

@MISC{kima-ascl,
       author = {{Faria}, J.~P. and {Santos}, N.~C. and {Figueira}, P. and {Brewer}, B.~J.},
        title = "{kima: Exoplanet detection in RVs with DNest4 and GPs}",
 howpublished = {Astrophysics Source Code Library, record ascl:2302.014},
         year = 2023,
        month = feb,
          eid = {ascl:2302.014},
        pages = {ascl:2302.014},
archivePrefix = {ascl},
       eprint = {2302.014},
       adsurl = {https://ui.adsabs.harvard.edu/abs/2023ascl.soft02014F}
}

@ARTICLE{kima-joss,
       author = {{Faria}, J.~P. and {Santos}, N.~C. and {Figueira}, P. and {Brewer}, B.~J.},
        title = "{kima: Exoplanet detection in radial velocities}",
      journal = {The Journal of Open Source Software},
         year = 2018,
        month = jun,
       volume = {3},
       number = {26},
        pages = {487},
          doi = {10.21105/joss.00487},
archivePrefix = {arXiv},
       eprint = {1806.08305},
 primaryClass = {astro-ph.IM},
       adsurl = {https://ui.adsabs.harvard.edu/abs/2018JOSS....3..487F}
}

@ARTICLE{Moya2018-empiricalrels,
       author = {{Moya}, Andy and {Zuccarino}, Federico and {Chaplin}, William J. and {Davies}, Guy R.},
        title = "{Empirical Relations for the Accurate Estimation of Stellar Masses and Radii}",
      journal = {\apjs},
         year = 2018,
        month = aug,
       volume = {237},
       number = {2},
          eid = {21},
        pages = {21},
          doi = {10.3847/1538-4365/aacdae},
archivePrefix = {arXiv},
       eprint = {1806.06574},
 primaryClass = {astro-ph.SR},
       adsurl = {https://ui.adsabs.harvard.edu/abs/2018ApJS..237...21M}
}

@ARTICLE{Torres2010,
       author = {{Torres}, G. and {Andersen}, J. and {Gim{\'e}nez}, A.},
        title = "{Accurate masses and radii of normal stars: modern results and applications}",
      journal = {\aapr},
         year = 2010,
        month = feb,
       volume = {18},
       number = {1-2},
        pages = {67-126},
          doi = {10.1007/s00159-009-0025-1},
archivePrefix = {arXiv},
       eprint = {0908.2624},
 primaryClass = {astro-ph.SR},
       adsurl = {https://ui.adsabs.harvard.edu/abs/2010A&ARv..18...67T}
}

@ARTICLE{Stevenson2023-BDD,
       author = {{Stevenson}, Adam T. and {Haswell}, Carole A. and {Barnes}, John R. and {Barstow}, Joanna K.},
        title = "{Combing the brown dwarf desert with Gaia DR3}",
      journal = {\mnras},
         year = 2023,
        month = dec,
       volume = {526},
       number = {4},
        pages = {5155-5171},
          doi = {10.1093/mnras/stad3041},
archivePrefix = {arXiv},
       eprint = {2310.02695},
 primaryClass = {astro-ph.SR},
       adsurl = {https://ui.adsabs.harvard.edu/abs/2023MNRAS.526.5155S}
}

@ARTICLE{Enoch2010,
       author = {{Enoch}, B. and {Collier Cameron}, A. and {Parley}, N.~R. and {Hebb}, L.},
        title = "{An improved method for estimating the masses of stars with transiting planets}",
      journal = {\aap},
         year = 2010,
        month = jun,
       volume = {516},
          eid = {A33},
        pages = {A33},
          doi = {10.1051/0004-6361/201014326},
archivePrefix = {arXiv},
       eprint = {1004.1991},
 primaryClass = {astro-ph.EP},
       adsurl = {https://ui.adsabs.harvard.edu/abs/2010A&A...516A..33E}
}

@INPROCEEDINGS{Southworth-debcat,
       author = {{Southworth}, J.},
        title = "{DEBCat: A Catalog of Detached Eclipsing Binary Stars}",
    booktitle = {Living Together: Planets, Host Stars and Binaries},
         year = 2015,
       editor = {{Rucinski}, S.~M. and {Torres}, G. and {Zejda}, M.},
       series = {Astronomical Society of the Pacific Conference Series},
       volume = {496},
        month = jul,
        pages = {164},
          doi = {10.48550/arXiv.1411.1219},
archivePrefix = {arXiv},
       eprint = {1411.1219},
 primaryClass = {astro-ph.SR},
       adsurl = {https://ui.adsabs.harvard.edu/abs/2015ASPC..496..164S}
}

@ARTICLE{Rodel2025,
       author = {{Rodel}, Toby and {Watson}, Christopher A. and {Ulmer-Moll}, Sol{\`e}ne and {Gill}, Samuel and {Maxted}, Pierre F.~L. and {Casewell}, Sarah L. and {Brahm}, Rafael and {Wilson}, Thomas G. and {Costes}, Jean C. and {Eschen}, Yoshi Nike Emilia and {Doyle}, Lauren and {Freckelton}, Alix V. and {Alves}, Douglas R. and {Apergis}, Ioannis and {Bayliss}, Daniel and {Bouchy}, Francois and {Burleigh}, Matthew R. and {Dumusque}, Xavier and {Eberhardt}, Jan and {Fern{\'a}ndez Fern{\'a}ndez}, Jorge and {Gillen}, Edward and {Goad}, Michael R. and {Hawthorn}, Faith and {Helled}, Ravit and {Henning}, Thomas and {Hobbs}, Katlyn L. and {Jenkins}, James S. and {Jord{\'a}n}, Andr{\'e}s and {Kendall}, Alicia and {Lendl}, Monika and {McCormac}, James and {de Mooij}, Ernst J.~W. and {O'Brien}, Sean M. and {Saha}, Suman and {Pinto}, Marcelo Tala and {Trifonov}, Trifon and {Udry}, St{\'e}phane and {Wheatley}, Peter J.},
        title = "{NGTS-EB-7, an eccentric, long-period, low-mass eclipsing binary}",
      journal = {\mnras},
         year = 2025,
        month = feb,
       volume = {537},
       number = {1},
        pages = {35-55},
          doi = {10.1093/mnras/stae2799},
archivePrefix = {arXiv},
       eprint = {2501.04523},
 primaryClass = {astro-ph.SR},
       adsurl = {https://ui.adsabs.harvard.edu/abs/2025MNRAS.537...35R}
}

@software{morton2015,
       author = {{Morton}, Timothy D.},
        title = "{isochrones: Stellar model grid package}",
 howpublished = {Astrophysics Source Code Library, record ascl:1503.010},
         year = 2015,
        month = mar,
          eid = {ascl:1503.010},
       adsurl = {https://ui.adsabs.harvard.edu/abs/2015ascl.soft03010M}
}

@ARTICLE{dotter2016,
       author = {{Dotter}, Aaron},
        title = "{MESA Isochrones and Stellar Tracks (MIST) 0: Methods for the Construction of Stellar Isochrones}",
      journal = {\apjs},
         year = 2016,
        month = jan,
       volume = {222},
       number = {1},
          eid = {8},
        pages = {8},
          doi = {10.3847/0067-0049/222/1/8},
archivePrefix = {arXiv},
       eprint = {1601.05144},
 primaryClass = {astro-ph.SR},
       adsurl = {https://ui.adsabs.harvard.edu/abs/2016ApJS..222....8D}
}

@ARTICLE{multinest,
       author = {{Feroz}, F. and {Hobson}, M.~P. and {Bridges}, M.},
        title = "{MULTINEST: an efficient and robust Bayesian inference tool for cosmology and particle physics}",
      journal = {\mnras},
         year = 2009,
        month = oct,
       volume = {398},
       number = {4},
        pages = {1601-1614},
          doi = {10.1111/j.1365-2966.2009.14548.x},
archivePrefix = {arXiv},
       eprint = {0809.3437},
 primaryClass = {astro-ph},
       adsurl = {https://ui.adsabs.harvard.edu/abs/2009MNRAS.398.1601F}
}

@ARTICLE{Groenewegen2021,
       author = {{Groenewegen}, M.~A.~T.},
        title = "{The parallax zero-point offset from Gaia EDR3 data}",
      journal = {\aap},
         year = 2021,
        month = oct,
       volume = {654},
          eid = {A20},
        pages = {A20},
          doi = {10.1051/0004-6361/202140862},
archivePrefix = {arXiv},
       eprint = {2106.08128},
 primaryClass = {astro-ph.GA},
       adsurl = {https://ui.adsabs.harvard.edu/abs/2021A&A...654A..20G}
}

@ARTICLE{Maxted-2023-eblm,
       author = {{Maxted}, Pierre F.~L. and {Triaud}, Amaury H.~M.~J. and {Martin}, David V.},
        title = "{The EBLM Project{\textemdash}From False Positives to Benchmark Stars and Circumbinary Exoplanets}",
      journal = {Universe},
         year = 2023,
        month = nov,
       volume = {9},
       number = {12},
          eid = {498},
        pages = {498},
          doi = {10.3390/universe9120498},
archivePrefix = {arXiv},
       eprint = {2311.16677},
 primaryClass = {astro-ph.SR},
       adsurl = {https://ui.adsabs.harvard.edu/abs/2023Univ....9..498M}
}

@ARTICLE{triaud-eblm-iv,
       author = {{Triaud}, Amaury H.~M.~J. and {Martin}, David V. and {S{\'e}gransan}, Damien and {Smalley}, Barry and {Maxted}, Pierre F.~L. and {Anderson}, David R. and {Bouchy}, Fran{\c{c}}ois and {Collier Cameron}, Andrew and {Faedi}, Francesca and {G{\'o}mez Maqueo Chew}, Yilen and {Hebb}, Leslie and {Hellier}, Coel and {Marmier}, Maxime and {Pepe}, Francesco and {Pollacco}, Don and {Queloz}, Didier and {Udry}, St{\'e}phane and {West}, Richard},
        title = "{The EBLM Project. IV. Spectroscopic orbits of over 100 eclipsing M dwarfs masquerading as transiting hot Jupiters}",
      journal = {\aap},
         year = 2017,
        month = dec,
       volume = {608},
          eid = {A129},
        pages = {A129},
          doi = {10.1051/0004-6361/201730993},
archivePrefix = {arXiv},
       eprint = {1707.07521},
 primaryClass = {astro-ph.SR},
       adsurl = {https://ui.adsabs.harvard.edu/abs/2017A&A...608A.129T}
}

@ARTICLE{Meibom2005,
       author = {{Meibom}, S{\o}ren and {Mathieu}, Robert D.},
        title = "{A Robust Measure of Tidal Circularization in Coeval Binary Populations: The Solar-Type Spectroscopic Binary Population in the Open Cluster M35}",
      journal = {\apj},
         year = 2005,
        month = feb,
       volume = {620},
       number = {2},
        pages = {970-983},
          doi = {10.1086/427082},
archivePrefix = {arXiv},
       eprint = {astro-ph/0412147},
 primaryClass = {astro-ph},
       adsurl = {https://ui.adsabs.harvard.edu/abs/2005ApJ...620..970M}
}

@ARTICLE{Trappist-1,
       author = {{Gillon}, Micha{\"e}l and {Triaud}, Amaury H.~M.~J. and {Demory}, Brice-Olivier and {Jehin}, Emmanu{\"e}l and {Agol}, Eric and {Deck}, Katherine M. and {Lederer}, Susan M. and {de Wit}, Julien and {Burdanov}, Artem and {Ingalls}, James G. and {Bolmont}, Emeline and {Leconte}, Jeremy and {Raymond}, Sean N. and {Selsis}, Franck and {Turbet}, Martin and {Barkaoui}, Khalid and {Burgasser}, Adam and {Burleigh}, Matthew R. and {Carey}, Sean J. and {Chaushev}, Aleksander and {Copperwheat}, Chris M. and {Delrez}, Laetitia and {Fernandes}, Catarina S. and {Holdsworth}, Daniel L. and {Kotze}, Enrico J. and {Van Grootel}, Val{\'e}rie and {Almleaky}, Yaseen and {Benkhaldoun}, Zouhair and {Magain}, Pierre and {Queloz}, Didier},
        title = "{Seven temperate terrestrial planets around the nearby ultracool dwarf star TRAPPIST-1}",
      journal = {\nat},
         year = 2017,
        month = feb,
       volume = {542},
       number = {7642},
        pages = {456-460},
          doi = {10.1038/nature21360},
archivePrefix = {arXiv},
       eprint = {1703.01424},
 primaryClass = {astro-ph.EP},
       adsurl = {https://ui.adsabs.harvard.edu/abs/2017Natur.542..456G}
}

@ARTICLE{Proxima-b,
       author = {{Anglada-Escud{\'e}}, Guillem and {Amado}, Pedro J. and {Barnes}, John and {Berdi{\~n}as}, Zaira M. and {Butler}, R. Paul and {Coleman}, Gavin A.~L. and {de La Cueva}, Ignacio and {Dreizler}, Stefan and {Endl}, Michael and {Giesers}, Benjamin and {Jeffers}, Sandra V. and {Jenkins}, James S. and {Jones}, Hugh R.~A. and {Kiraga}, Marcin and {K{\"u}rster}, Martin and {L{\'o}pez-Gonz{\'a}lez}, Mar{\'\i}a J. and {Marvin}, Christopher J. and {Morales}, Nicol{\'a}s and {Morin}, Julien and {Nelson}, Richard P. and {Ortiz}, Jos{\'e} L. and {Ofir}, Aviv and {Paardekooper}, Sijme-Jan and {Reiners}, Ansgar and {Rodr{\'\i}guez}, Eloy and {Rodr{\'\i}guez-L{\'o}pez}, Cristina and {Sarmiento}, Luis F. and {Strachan}, John P. and {Tsapras}, Yiannis and {Tuomi}, Mikko and {Zechmeister}, Mathias},
        title = "{A terrestrial planet candidate in a temperate orbit around Proxima Centauri}",
      journal = {\nat},
         year = 2016,
        month = aug,
       volume = {536},
       number = {7617},
        pages = {437-440},
          doi = {10.1038/nature19106},
archivePrefix = {arXiv},
       eprint = {1609.03449},
 primaryClass = {astro-ph.EP},
       adsurl = {https://ui.adsabs.harvard.edu/abs/2016Natur.536..437A}
}

@ARTICLE{Turner2025,
       author = {{Turner}, Daisy A. and {Eschen}, Yoshi Nike Emilia and {Murgas}, Felipe and {Mortier}, Annelies and {Wilson}, Thomas G. and {Fern{\'a}ndez Fern{\'a}ndez}, Jorge and {Gromek}, Nicole and {Morello}, Giuseppe and {Tabernero}, Hugo M. and {Egger}, Jo Ann and {Vissapragada}, Shreyas and {Caballero}, Jos{\'e} A. and {Dreizler}, Stefan and {Freckelton}, Alix Violet and {Hatzes}, Artie P. and {Lakeland}, Ben Scott and {Nagel}, Evangelos and {Naponiello}, Luca and {Vanaverbeke}, Siegfried and {Venner}, Alexander and {Zapatero Osorio}, Mar{\'\i}a Rosa and {Amado}, Pedro J. and {B{\'e}jar}, V{\'\i}ctor J.~S. and {Bonomo}, Aldo Stefano and {Buchhave}, Lars A. and {Collier Cameron}, Andrew and {Carleo}, Ilaria and {Chaturvedi}, Priyanka and {Cloutier}, Ryan and {Damasso}, Mario and {Daspute}, Mangesh and {Dholakia}, Shishir and {Dufoer}, Sjoerd and {Dumusque}, Xavier and {Martinez Fiorenzano}, Aldo Fabricio and {Ghedina}, Adriano and {Harutyunyan}, Avet and {Herrero}, Enrique and {John}, Ancy Anna and {Lillo-Box}, Jorge and {Lodieu}, Nicolas and {L{\'o}pez-Morales}, Mercedes and {Malavolta}, Luca and {Mancini}, Luigi and {Mantovan}, Giacomo and {Montes}, David and {Morales}, Juan Carlos and {Nicholson}, Belinda and {Orell-Miquel}, Jaume and {Palethorpe}, Larissa and {Palle}, Enric and {Quirrenbach}, Andreas and {Reffert}, Sabine and {Reiners}, Ansgar and {Ribas}, Ignasi and {Rice}, Ken and {Silva}, Andr{\'e} M. and {Sozzetti}, Alessandro and {Stalport}, Manu and {Tal-Or}, Lev and {Trifonov}, Trifon and {Udry}, St{\'e}phane and {Zechmeister}, Mathias},
        title = "{The mass of the exo-Venus Gliese 12 b, as revealed by HARPS-N, ESPRESSO, and CARMENES}",
      journal = {\mnras},
         year = 2026,
        month = jan,
       volume = {545},
       number = {1},
          eid = {staf1703},
        pages = {staf1703},
          doi = {10.1093/mnras/staf1703},
archivePrefix = {arXiv},
       eprint = {2506.20564},
 primaryClass = {astro-ph.EP},
       adsurl = {https://ui.adsabs.harvard.edu/abs/2026MNRAS.545f1703T}
}

@ARTICLE{Brady2025,
       author = {{Brady}, Madison and {Bean}, Jacob L. and {Basant}, Ritvik and {Brown}, Nina and {Das}, Tanya and {Nixon}, Matthew C. and {Luque}, Rafael and {Piaulet-Ghorayeb}, Caroline and {Radica}, Michael and {Seifahrt}, Andreas and {St{\"u}rmer}, Julian and {Zhao}, Lily},
        title = "{An Earthlike Density for the Temperate Earth-sized Planet GJ 12 b}",
      journal = {\aj},
         year = 2025,
        month = dec,
       volume = {170},
       number = {6},
          eid = {335},
        pages = {335},
          doi = {10.3847/1538-3881/ae0f98},
archivePrefix = {arXiv},
       eprint = {2506.20561},
 primaryClass = {astro-ph.EP},
       adsurl = {https://ui.adsabs.harvard.edu/abs/2025AJ....170..335B}
}

@ARTICLE{SM2023-GJ1002,
       author = {{Su{\'a}rez Mascare{\~n}o}, A. and {Gonz{\'a}lez-{\'A}lvarez}, E. and {Zapatero Osorio}, M.~R. and {Lillo-Box}, J. and {Faria}, J.~P. and {Passegger}, V.~M. and {Gonz{\'a}lez Hern{\'a}ndez}, J.~I. and {Figueira}, P. and {Sozzetti}, A. and {Rebolo}, R. and {Pepe}, F. and {Santos}, N.~C. and {Cristiani}, S. and {Lovis}, C. and {Silva}, A.~M. and {Ribas}, I. and {Amado}, P.~J. and {Caballero}, J.~A. and {Quirrenbach}, A. and {Reiners}, A. and {Zechmeister}, M. and {Adibekyan}, V. and {Alibert}, Y. and {B{\'e}jar}, V.~J.~S. and {Benatti}, S. and {D'Odorico}, V. and {Damasso}, M. and {Delisle}, J. -B. and {Di Marcantonio}, P. and {Dreizler}, S. and {Ehrenreich}, D. and {Hatzes}, A.~P. and {Hara}, N.~C. and {Henning}, Th. and {Kaminski}, A. and {L{\'o}pez-Gonz{\'a}lez}, M.~J. and {Martins}, C.~J.~A.~P. and {Micela}, G. and {Montes}, D. and {Pall{\'e}}, E. and {Pedraz}, S. and {Rodr{\'\i}guez}, E. and {Rodr{\'\i}guez-L{\'o}pez}, C. and {Tal-Or}, L. and {Sousa}, S. and {Udry}, S.},
        title = "{Two temperate Earth-mass planets orbiting the nearby star GJ 1002}",
      journal = {\aap},
         year = 2023,
        month = feb,
       volume = {670},
          eid = {A5},
        pages = {A5},
          doi = {10.1051/0004-6361/202244991},
archivePrefix = {arXiv},
       eprint = {2212.07332},
 primaryClass = {astro-ph.EP},
       adsurl = {https://ui.adsabs.harvard.edu/abs/2023A&A...670A...5S}
}

@ARTICLE{Swayne2024-eblmXI,
       author = {{Swayne}, M.~I. and {Maxted}, P.~F.~L. and {Triaud}, A.~H.~M.~J. and {Sousa}, S.~G. and {Deline}, A. and {Ehrenreich}, D. and {Hoyer}, S. and {Olofsson}, G. and {Boisse}, I. and {Duck}, A. and {Gill}, S. and {Martin}, D. and {McCormac}, J. and {Persson}, C.~M. and {Santerne}, A. and {Sebastian}, D. and {Standing}, M.~R. and {Acu{\~n}a}, L. and {Alibert}, Y. and {Alonso}, R. and {Anglada}, G. and {B{\'a}rczy}, T. and {Barrado Navascues}, D. and {Barros}, S.~C.~C. and {Baumjohann}, W. and {Baycroft}, T.~A. and {Beck}, M. and {Beck}, T. and {Benz}, W. and {Billot}, N. and {Bonfils}, X. and {Borsato}, L. and {Bourrier}, V. and {Brandeker}, A. and {Broeg}, C. and {Carmona}, A. and {Charnoz}, S. and {Collier Cameron}, A. and {Cort{\'e}s-Zuleta}, P. and {Csizmadia}, Sz and {Cubillos}, P.~E. and {Davies}, M.~B. and {Deleuil}, M. and {Delfosse}, X. and {Delrez}, L. and {Demangeon}, O.~D.~S. and {Demory}, B. -O. and {Dransfield}, G. and {Erikson}, A. and {Fortier}, A. and {Forveille}, T. and {Fossati}, L. and {Fridlund}, M. and {Gandolfi}, D. and {Gillon}, M. and {G{\"u}del}, M. and {G{\"u}nther}, M.~N. and {Hara}, N. and {H{\'e}brard}, G. and {Heidari}, N. and {Hellier}, C. and {Helling}, Ch and {Isaak}, K.~G. and {Kerschbaum}, F. and {Kiefer}, F. and {Kiss}, L.~L. and {Kunovac}, V. and {Lalitha}, S. and {Lam}, K.~W.~F. and {Laskar}, J. and {Lecavelier des Etangs}, A. and {Lendl}, M. and {Magrin}, D. and {Marafatto}, L. and {Martioli}, E. and {Miller}, N.~J. and {Mordasini}, C. and {Moutou}, C. and {Nascimbeni}, V. and {Ottensamer}, R. and {Pagano}, I. and {Pall{\'e}}, E. and {Peter}, G. and {Piazza}, D. and {Piotto}, G. and {Pollacco}, D. and {Queloz}, D. and {Ragazzoni}, R. and {Rando}, N. and {Rauer}, H. and {Ribas}, I. and {Santos}, N.~C. and {Scandariato}, G. and {S{\'e}gransan}, D. and {Simon}, A.~E. and {Smith}, A.~M.~S. and {Southworth}, R. and {Stalport}, M. and {Szab{\'o}}, Gy M. and {Thomas}, N. and {Udry}, S. and {Ulmer}, B. and {Van Grootel}, V. and {Venturini}, J. and {Walton}, N.~A. and {Willett}, E. and {Wilson}, T.~G.},
        title = "{The EBLM Project- XI. Mass, radius, and effective temperature measurements for 23 M-dwarf companions to solar-type stars observed with CHEOPS}",
      journal = {\mnras},
         year = 2024,
        month = mar,
       volume = {528},
       number = {4},
        pages = {5703-5722},
          doi = {10.1093/mnras/stad3866},
archivePrefix = {arXiv},
       eprint = {2312.11339},
 primaryClass = {astro-ph.SR},
       adsurl = {https://ui.adsabs.harvard.edu/abs/2024MNRAS.528.5703S}
}

@ARTICLE{Banerjee2024,
       author = {{Banerjee}, Agnibha and {Barstow}, Joanna K. and {Gressier}, Am{\'e}lie and {Espinoza}, N{\'e}stor and {Sing}, David K. and {Allen}, Natalie H. and {Birkmann}, Stephan M. and {Challener}, Ryan C. and {Crouzet}, Nicolas and {Haswell}, Carole A. and {Lewis}, Nikole K. and {Lewis}, Stephen R. and {Yang}, Jingxuan},
        title = "{Atmospheric Retrievals Suggest the Presence of a Secondary Atmosphere and Possible Sulfur Species on L98-59 d from JWST Nirspec G395H Transmission Spectroscopy}",
      journal = {\apjl},
         year = 2024,
        month = nov,
       volume = {975},
       number = {1},
          eid = {L11},
        pages = {L11},
          doi = {10.3847/2041-8213/ad73d0},
archivePrefix = {arXiv},
       eprint = {2408.15707},
 primaryClass = {astro-ph.EP},
       adsurl = {https://ui.adsabs.harvard.edu/abs/2024ApJ...975L..11B}
}

@ARTICLE{JWST-trappist-community-initiative,
       author = {{TRAPPIST-1 JWST Community Initiative} and {de Wit}, Julien and {Doyon}, Ren{\'e} and {Rackham}, Benjamin V. and {Lim}, Olivia and {Ducrot}, Elsa and {Kreidberg}, Laura and {Benneke}, Bj{\"o}rn and {Ribas}, Ignasi and {Berardo}, David and {Niraula}, Prajwal and {Iyer}, Aishwarya and {Shapiro}, Alexander and {Kostogryz}, Nadiia and {Witzke}, Veronika and {Gillon}, Micha{\"e}l and {Agol}, Eric and {Meadows}, Victoria and {Burgasser}, Adam J. and {Owen}, James E. and {Fortney}, Jonathan J. and {Selsis}, Franck and {Bello-Arufe}, Aaron and {de Beurs}, Zo{\"e} and {Bolmont}, Emeline and {Cowan}, Nicolas and {Dong}, Chuanfei and {Drake}, Jeremy J. and {Garcia}, Lionel and {Greene}, Thomas and {Haworth}, Thomas and {Hu}, Renyu and {Kane}, Stephen R. and {Kervella}, Pierre and {Koll}, Daniel and {Krissansen-Totton}, Joshua and {Lagage}, Pierre-Olivier and {Lichtenberg}, Tim and {Lustig-Yaeger}, Jacob and {Lingam}, Manasvi and {Turbet}, Martin and {Seager}, Sara and {Barkaoui}, Khalid and {Bell}, Taylor J. and {Burdanov}, Artem and {Cadieux}, Charles and {Charnay}, Benjamin and {Cloutier}, Ryan and {Cook}, Neil J. and {Correia}, Alexandre C.~M. and {Dang}, Lisa and {Daylan}, Tansu and {Delrez}, Laetitia and {Edwards}, Billy and {Fauchez}, Thomas J. and {Flagg}, Laura and {Fraschetti}, Federico and {Haqq-Misra}, Jacob and {Huang}, Ziyu and {Iro}, Nicolas and {Jayawardhana}, Ray and {Jehin}, Emmanuel and {Jin}, Meng and {Kite}, Edwin and {Kitzmann}, Daniel and {Kral}, Quentin and {Lafreni{\`e}re}, David and {Libert}, Anne-Sophie and {Liu}, Beibei and {Mohanty}, Subhanjoy and {Morris}, Brett M. and {Murray}, Catriona A. and {Piaulet}, Caroline and {Pozuelos}, Francisco J. and {Radica}, Michael and {Ranjan}, Sukrit and {Rathcke}, Alexander and {Roy}, Pierre-Alexis and {Schwieterman}, Edward W. and {Turner}, Jake D. and {Triaud}, Amaury and {Way}, Michael J.},
        title = "{A roadmap for the atmospheric characterization of terrestrial exoplanets with JWST}",
      journal = {Nature Astronomy},
         year = 2024,
        month = jul,
       volume = {8},
        pages = {810-818},
          doi = {10.1038/s41550-024-02298-5},
archivePrefix = {arXiv},
       eprint = {2310.15895},
 primaryClass = {astro-ph.EP},
       adsurl = {https://ui.adsabs.harvard.edu/abs/2024NatAs...8..810T}
}

@ARTICLE{Taylor2025-GJ357b,
       author = {{Taylor}, Jake and {Radica}, Michael and {Chatterjee}, Richard D. and {Hammond}, Mark and {Meier}, Tobias and {Aigrain}, Suzanne and {MacDonald}, Ryan J. and {Albert}, Loic and {Benneke}, Bj{\"o}rn and {Coulombe}, Louis-Philippe and {Cowan}, Nicolas B. and {Dang}, Lisa and {Doyon}, Ren{\'e} and {Flagg}, Laura and {Johnstone}, Doug and {Kaltenegger}, Lisa and {Lafreni{\`e}re}, David and {Pelletier}, Stefan and {Piaulet-Ghorayeb}, Caroline and {Rowe}, Jason F. and {Roy}, Pierre-Alexis},
        title = "{JWST NIRISS transmission spectroscopy of the super-Earth GJ 357b, a favourable target for atmospheric retention}",
      journal = {\mnras},
         year = 2025,
        month = jul,
       volume = {540},
       number = {4},
        pages = {3677-3692},
          doi = {10.1093/mnras/staf894},
archivePrefix = {arXiv},
       eprint = {2505.24462},
 primaryClass = {astro-ph.EP},
       adsurl = {https://ui.adsabs.harvard.edu/abs/2025MNRAS.540.3677T}
}

@ARTICLE{Lustig-Yaeger2023,
       author = {{Lustig-Yaeger}, Jacob and {Fu}, Guangwei and {May}, E.~M. and {Ceballos}, Kevin N. Ortiz and {Moran}, Sarah E. and {Peacock}, Sarah and {Stevenson}, Kevin B. and {Kirk}, James and {L{\'o}pez-Morales}, Mercedes and {MacDonald}, Ryan J. and {Mayorga}, L.~C. and {Sing}, David K. and {Sotzen}, Kristin S. and {Valenti}, Jeff A. and {Redai}, J{\'e}a I. Adams and {Alam}, Munazza K. and {Batalha}, Natasha E. and {Bennett}, Katherine A. and {Gonzalez-Quiles}, Junellie and {Kruse}, Ethan and {Lothringer}, Joshua D. and {Rustamkulov}, Zafar and {Wakeford}, Hannah R.},
        title = "{A JWST transmission spectrum of the nearby Earth-sized exoplanet LHS 475 b}",
      journal = {Nature Astronomy},
         year = 2023,
        month = nov,
       volume = {7},
        pages = {1317-1328},
          doi = {10.1038/s41550-023-02064-z},
archivePrefix = {arXiv},
       eprint = {2301.04191},
 primaryClass = {astro-ph.EP},
       adsurl = {https://ui.adsabs.harvard.edu/abs/2023NatAs...7.1317L}
}

@ARTICLE{Kreidberg-Stevenson-2025,
       author = {{Kreidberg}, Laura and {Stevenson}, Kevin B.},
        title = "{A first look at rocky exoplanets with JWST}",
      journal = {Proceedings of the National Academy of Science},
         year = 2025,
        month = sep,
       volume = {122},
       number = {39},
          eid = {e2416190122},
        pages = {e2416190122},
          doi = {10.1073/pnas.2416190122},
archivePrefix = {arXiv},
       eprint = {2507.00933},
 primaryClass = {astro-ph.EP},
       adsurl = {https://ui.adsabs.harvard.edu/abs/2025PNAS..12216190K}
}

@ARTICLE{Andersen1991,
       author = {{Andersen}, J.},
        title = "{Accurate masses and radii of normal stars}",
      journal = {\aapr},
         year = 1991,
        month = jan,
       volume = {3},
       number = {2},
        pages = {91-126},
          doi = {10.1007/BF00873538},
       adsurl = {https://ui.adsabs.harvard.edu/abs/1991A&ARv...3...91A}
}

@ARTICLE{GaiaEDR3,
       author = {{Gaia Collaboration} and {Brown}, A.~G.~A. and {Vallenari}, A. and {Prusti}, T. and {de Bruijne}, J.~H.~J. and {Babusiaux}, C. and {Biermann}, M. and {Creevey}, O.~L. and {Evans}, D.~W. and {Eyer}, L. and {Hutton}, A. and {Jansen}, F. and {Jordi}, C. and {Klioner}, S.~A. and {Lammers}, U. and {Lindegren}, L. and {Luri}, X. and {Mignard}, F. and {Panem}, C. and {Pourbaix}, D. and {Randich}, S. and {Sartoretti}, P. and {Soubiran}, C. and {Walton}, N.~A. and {Arenou}, F. and {Bailer-Jones}, C.~A.~L. and {Bastian}, U. and {Cropper}, M. and {Drimmel}, R. and {Katz}, D. and {Lattanzi}, M.~G. and {van Leeuwen}, F. and {Bakker}, J. and {Cacciari}, C. and {Casta{\~n}eda}, J. and {De Angeli}, F. and {Ducourant}, C. and {Fabricius}, C. and {Fouesneau}, M. and {Fr{\'e}mat}, Y. and {Guerra}, R. and {Guerrier}, A. and {Guiraud}, J. and {Jean-Antoine Piccolo}, A. and {Masana}, E. and {Messineo}, R. and {Mowlavi}, N. and {Nicolas}, C. and {Nienartowicz}, K. and {Pailler}, F. and {Panuzzo}, P. and {Riclet}, F. and {Roux}, W. and {Seabroke}, G.~M. and {Sordo}, R. and {Tanga}, P. and {Th{\'e}venin}, F. and {Gracia-Abril}, G. and {Portell}, J. and {Teyssier}, D. and {Altmann}, M. and {Andrae}, R. and {Bellas-Velidis}, I. and {Benson}, K. and {Berthier}, J. and {Blomme}, R. and {Brugaletta}, E. and {Burgess}, P.~W. and {Busso}, G. and {Carry}, B. and {Cellino}, A. and {Cheek}, N. and {Clementini}, G. and {Damerdji}, Y. and {Davidson}, M. and {Delchambre}, L. and {Dell'Oro}, A. and {Fern{\'a}ndez-Hern{\'a}ndez}, J. and {Galluccio}, L. and {Garc{\'\i}a-Lario}, P. and {Garcia-Reinaldos}, M. and {Gonz{\'a}lez-N{\'u}{\~n}ez}, J. and {Gosset}, E. and {Haigron}, R. and {Halbwachs}, J. -L. and {Hambly}, N.~C. and {Harrison}, D.~L. and {Hatzidimitriou}, D. and {Heiter}, U. and {Hern{\'a}ndez}, J. and {Hestroffer}, D. and {Hodgkin}, S.~T. and {Holl}, B. and {Jan{\ss}en}, K. and {Jevardat de Fombelle}, G. and {Jordan}, S. and {Krone-Martins}, A. and {Lanzafame}, A.~C. and {L{\"o}ffler}, W. and {Lorca}, A. and {Manteiga}, M. and {Marchal}, O. and {Marrese}, P.~M. and {Moitinho}, A. and {Mora}, A. and {Muinonen}, K. and {Osborne}, P. and {Pancino}, E. and {Pauwels}, T. and {Petit}, J. -M. and {Recio-Blanco}, A. and {Richards}, P.~J. and {Riello}, M. and {Rimoldini}, L. and {Robin}, A.~C. and {Roegiers}, T. and {Rybizki}, J. and {Sarro}, L.~M. and {Siopis}, C. and {Smith}, M. and {Sozzetti}, A. and {Ulla}, A. and {Utrilla}, E. and {van Leeuwen}, M. and {van Reeven}, W. and {Abbas}, U. and {Abreu Aramburu}, A. and {Accart}, S. and {Aerts}, C. and {Aguado}, J.~J. and {Ajaj}, M. and {Altavilla}, G. and {{\'A}lvarez}, M.~A. and {{\'A}lvarez Cid-Fuentes}, J. and {Alves}, J. and {Anderson}, R.~I. and {Anglada Varela}, E. and {Antoja}, T. and {Audard}, M. and {Baines}, D. and {Baker}, S.~G. and {Balaguer-N{\'u}{\~n}ez}, L. and {Balbinot}, E. and {Balog}, Z. and {Barache}, C. and {Barbato}, D. and {Barros}, M. and {Barstow}, M.~A. and {Bartolom{\'e}}, S. and {Bassilana}, J. -L. and {Bauchet}, N. and {Baudesson-Stella}, A. and {Becciani}, U. and {Bellazzini}, M. and {Bernet}, M. and {Bertone}, S. and {Bianchi}, L. and {Blanco-Cuaresma}, S. and {Boch}, T. and {Bombrun}, A. and {Bossini}, D. and {Bouquillon}, S. and {Bragaglia}, A. and {Bramante}, L. and {Breedt}, E. and {Bressan}, A. and {Brouillet}, N. and {Bucciarelli}, B. and {Burlacu}, A. and {Busonero}, D. and {Butkevich}, A.~G. and {Buzzi}, R. and {Caffau}, E. and {Cancelliere}, R. and {C{\'a}novas}, H. and {Cantat-Gaudin}, T. and {Carballo}, R. and {Carlucci}, T. and {Carnerero}, M.~I. and {Carrasco}, J.~M. and {Casamiquela}, L. and {Castellani}, M. and {Castro-Ginard}, A. and {Castro Sampol}, P. and {Chaoul}, L. and {Charlot}, P. and {Chemin}, L. and {Chiavassa}, A. and {Cioni}, M. -R.~L. and {Comoretto}, G. and {Cooper}, W.~J. and {Cornez}, T. and {Cowell}, S. and {Crifo}, F. and {Crosta}, M. and {Crowley}, C. and {Dafonte}, C. and {Dapergolas}, A. and {David}, M. and {David}, P.},
        title = "{Gaia Early Data Release 3. Summary of the contents and survey properties}",
      journal = {\aap},
         year = 2021,
        month = may,
       volume = {649},
          eid = {A1},
        pages = {A1},
          doi = {10.1051/0004-6361/202039657},
archivePrefix = {arXiv},
       eprint = {2012.01533},
 primaryClass = {astro-ph.GA},
       adsurl = {https://ui.adsabs.harvard.edu/abs/2021A&A...649A...1G}
}

@ARTICLE{Lindegren2018,
       author = {{Lindegren}, L. and {Hern{\'a}ndez}, J. and {Bombrun}, A. and {Klioner}, S. and {Bastian}, U. and {Ramos-Lerate}, M. and {de Torres}, A. and {Steidelm{\"u}ller}, H. and {Stephenson}, C. and {Hobbs}, D. and {Lammers}, U. and {Biermann}, M. and {Geyer}, R. and {Hilger}, T. and {Michalik}, D. and {Stampa}, U. and {McMillan}, P.~J. and {Casta{\~n}eda}, J. and {Clotet}, M. and {Comoretto}, G. and {Davidson}, M. and {Fabricius}, C. and {Gracia}, G. and {Hambly}, N.~C. and {Hutton}, A. and {Mora}, A. and {Portell}, J. and {van Leeuwen}, F. and {Abbas}, U. and {Abreu}, A. and {Altmann}, M. and {Andrei}, A. and {Anglada}, E. and {Balaguer-N{\'u}{\~n}ez}, L. and {Barache}, C. and {Becciani}, U. and {Bertone}, S. and {Bianchi}, L. and {Bouquillon}, S. and {Bourda}, G. and {Br{\"u}semeister}, T. and {Bucciarelli}, B. and {Busonero}, D. and {Buzzi}, R. and {Cancelliere}, R. and {Carlucci}, T. and {Charlot}, P. and {Cheek}, N. and {Crosta}, M. and {Crowley}, C. and {de Bruijne}, J. and {de Felice}, F. and {Drimmel}, R. and {Esquej}, P. and {Fienga}, A. and {Fraile}, E. and {Gai}, M. and {Garralda}, N. and {Gonz{\'a}lez-Vidal}, J.~J. and {Guerra}, R. and {Hauser}, M. and {Hofmann}, W. and {Holl}, B. and {Jordan}, S. and {Lattanzi}, M.~G. and {Lenhardt}, H. and {Liao}, S. and {Licata}, E. and {Lister}, T. and {L{\"o}ffler}, W. and {Marchant}, J. and {Martin-Fleitas}, J. -M. and {Messineo}, R. and {Mignard}, F. and {Morbidelli}, R. and {Poggio}, E. and {Riva}, A. and {Rowell}, N. and {Salguero}, E. and {Sarasso}, M. and {Sciacca}, E. and {Siddiqui}, H. and {Smart}, R.~L. and {Spagna}, A. and {Steele}, I. and {Taris}, F. and {Torra}, J. and {van Elteren}, A. and {van Reeven}, W. and {Vecchiato}, A.},
        title = "{Gaia Data Release 2. The astrometric solution}",
      journal = {\aap},
         year = 2018,
        month = aug,
       volume = {616},
          eid = {A2},
        pages = {A2},
          doi = {10.1051/0004-6361/201832727},
archivePrefix = {arXiv},
       eprint = {1804.09366},
 primaryClass = {astro-ph.IM},
       adsurl = {https://ui.adsabs.harvard.edu/abs/2018A&A...616A...2L}
}

@ARTICLE{Lindegren2021,
       author = {{Lindegren}, L. and {Bastian}, U. and {Biermann}, M. and {Bombrun}, A. and {de Torres}, A. and {Gerlach}, E. and {Geyer}, R. and {Hern{\'a}ndez}, J. and {Hilger}, T. and {Hobbs}, D. and {Klioner}, S.~A. and {Lammers}, U. and {McMillan}, P.~J. and {Ramos-Lerate}, M. and {Steidelm{\"u}ller}, H. and {Stephenson}, C.~A. and {van Leeuwen}, F.},
        title = "{Gaia Early Data Release 3. Parallax bias versus magnitude, colour, and position}",
      journal = {\aap},
         year = 2021,
        month = may,
       volume = {649},
          eid = {A4},
        pages = {A4},
          doi = {10.1051/0004-6361/202039653},
archivePrefix = {arXiv},
       eprint = {2012.01742},
 primaryClass = {astro-ph.IM},
       adsurl = {https://ui.adsabs.harvard.edu/abs/2021A&A...649A...4L}
}

@ARTICLE{May2023,
       author = {{May}, E.~M. and {MacDonald}, Ryan J. and {Bennett}, Katherine A. and {Moran}, Sarah E. and {Wakeford}, Hannah R. and {Peacock}, Sarah and {Lustig-Yaeger}, Jacob and {Highland}, Alicia N. and {Stevenson}, Kevin B. and {Sing}, David K. and {Mayorga}, L.~C. and {Batalha}, Natasha E. and {Kirk}, James and {L{\'o}pez-Morales}, Mercedes and {Valenti}, Jeff A. and {Alam}, Munazza K. and {Alderson}, Lili and {Fu}, Guangwei and {Gonzalez-Quiles}, Junellie and {Lothringer}, Joshua D. and {Rustamkulov}, Zafar and {Sotzen}, Kristin S.},
        title = "{Double Trouble: Two Transits of the Super-Earth GJ 1132 b Observed with JWST NIRSpec G395H}",
      journal = {\apjl},
         year = 2023,
        month = dec,
       volume = {959},
       number = {1},
          eid = {L9},
        pages = {L9},
          doi = {10.3847/2041-8213/ad054f},
archivePrefix = {arXiv},
       eprint = {2310.10711},
 primaryClass = {astro-ph.EP},
       adsurl = {https://ui.adsabs.harvard.edu/abs/2023ApJ...959L...9M}
}

@ARTICLE{IAU2015b3,
       author = {{Mamajek}, E.~E. and {Prsa}, A. and {Torres}, G. and {Harmanec}, P. and {Asplund}, M. and {Bennett}, P.~D. and {Capitaine}, N. and {Christensen-Dalsgaard}, J. and {Depagne}, E. and {Folkner}, W.~M. and {Haberreiter}, M. and {Hekker}, S. and {Hilton}, J.~L. and {Kostov}, V. and {Kurtz}, D.~W. and {Laskar}, J. and {Mason}, B.~D. and {Milone}, E.~F. and {Montgomery}, M.~M. and {Richards}, M.~T. and {Schou}, J. and {Stewart}, S.~G.},
        title = "{IAU 2015 Resolution B3 on Recommended Nominal Conversion Constants for Selected Solar and Planetary Properties}",
      journal = {arXiv e-prints},
         year = 2015,
        month = oct,
          eid = {arXiv:1510.07674},
        pages = {arXiv:1510.07674},
          doi = {10.48550/arXiv.1510.07674},
archivePrefix = {arXiv},
       eprint = {1510.07674},
 primaryClass = {astro-ph.SR},
       adsurl = {https://ui.adsabs.harvard.edu/abs/2015arXiv151007674M}
}

@ARTICLE{CODATA2018,
       author = {{Tiesinga}, Eite and {Mohr}, Peter J. and {Newell}, David B. and {Taylor}, Barry N.},
        title = "{CODATA recommended values of the fundamental physical constants: 2018}",
      journal = {Reviews of Modern Physics},
         year = 2021,
        month = apr,
       volume = {93},
       number = {2},
          eid = {025010},
        pages = {025010},
          doi = {10.1103/RevModPhys.93.025010},
       adsurl = {https://ui.adsabs.harvard.edu/abs/2021RvMP...93b5010T}
}

@ARTICLE{Baraffe2015,
       author = {{Baraffe}, Isabelle and {Homeier}, Derek and {Allard}, France and {Chabrier}, Gilles},
        title = "{New evolutionary models for pre-main sequence and main sequence low-mass stars down to the hydrogen-burning limit}",
      journal = {\aap},
         year = 2015,
        month = may,
       volume = {577},
          eid = {A42},
        pages = {A42},
          doi = {10.1051/0004-6361/201425481},
archivePrefix = {arXiv},
       eprint = {1503.04107},
 primaryClass = {astro-ph.SR},
       adsurl = {https://ui.adsabs.harvard.edu/abs/2015A&A...577A..42B}
}

@ARTICLE{Pass2023,
       author = {{Pass}, Emily K. and {Charbonneau}, David},
        title = "{G 68-34: A Double-lined M-dwarf Eclipsing Binary in a Hierarchical Triple System}",
      journal = {\apj},
         year = 2023,
        month = jun,
       volume = {949},
       number = {2},
          eid = {37},
        pages = {37},
          doi = {10.3847/1538-4357/accadf},
archivePrefix = {arXiv},
       eprint = {2304.02466},
 primaryClass = {astro-ph.SR},
       adsurl = {https://ui.adsabs.harvard.edu/abs/2023ApJ...949...37P}
}

@ARTICLE{Dittmann2017,
       author = {{Dittmann}, Jason A. and {Irwin}, Jonathan M. and {Charbonneau}, David and {Bonfils}, Xavier and {Astudillo-Defru}, Nicola and {Haywood}, Rapha{\"e}lle D. and {Berta-Thompson}, Zachory K. and {Newton}, Elisabeth R. and {Rodriguez}, Joseph E. and {Winters}, Jennifer G. and {Tan}, Thiam-Guan and {Almenara}, Jose-Manuel and {Bouchy}, Fran{\c{c}}ois and {Delfosse}, Xavier and {Forveille}, Thierry and {Lovis}, Christophe and {Murgas}, Felipe and {Pepe}, Francesco and {Santos}, Nuno C. and {Udry}, Stephane and {W{\"u}nsche}, Ana{\"e}l and {Esquerdo}, Gilbert A. and {Latham}, David W. and {Dressing}, Courtney D.},
        title = "{A temperate rocky super-Earth transiting a nearby cool star}",
      journal = {\nat},
         year = 2017,
        month = apr,
       volume = {544},
       number = {7650},
        pages = {333-336},
          doi = {10.1038/nature22055},
archivePrefix = {arXiv},
       eprint = {1704.05556},
 primaryClass = {astro-ph.EP},
       adsurl = {https://ui.adsabs.harvard.edu/abs/2017Natur.544..333D}
}

@ARTICLE{Scott2026,
       author = {{Scott}, Madison G. and {Dransfield}, Georgina and {Timmermans}, Mathilde and {Triaud}, Amaury H.~M.~J. and {Rackham}, Benjamin V. and {Barkaoui}, Khalid and {Burgasser}, Adam J. and {Collins}, Karen A. and {Gillon}, Micha{\"e}l and {Howell}, Steve B. and {Levine}, Alan M. and {Pozuelos}, Francisco J. and {Stassun}, Keivan G. and {Ziegler}, Carl and {Chew}, Yilen Gomez Maqueo and {Clark}, Catherine A. and {Davis}, Yasmin and {Davoudi}, Fatemeh and {Daylan}, Tansu and {Demory}, Brice-Olivier and {Feliz}, Dax and {Fukui}, Akihiko and {G{\"u}nther}, Maximilian N. and {Jehin}, Emmanu{\"e}l and {Lienhard}, Florian and {Mann}, Andrew W. and {Mu{\~n}oz}, Cl{\`a}udia Jan{\'o} and {Narita}, Norio and {Pedersen}, Peter P. and {Schwarz}, Richard P. and {Shporer}, Avi and {Soubkiou}, Abderahmane and {Z{\'u}{\~n}iga-Fern{\'a}ndez}, Sebasti{\'a}n},
        title = "{Two temperate Earth- and Neptune-sized planets orbiting fully convective M dwarfs}",
      journal = {\mnras},
         year = 2026,
        month = mar,
       volume = {547},
       number = {1},
          eid = {stag070},
        pages = {stag070},
          doi = {10.1093/mnras/stag070},
archivePrefix = {arXiv},
       eprint = {2601.05799},
 primaryClass = {astro-ph.EP},
       adsurl = {https://ui.adsabs.harvard.edu/abs/2026MNRAS.547ag070S}
}

@misc{eblm-northern-sample,
   author = {{Davis}, Yasmin T. and {Triaud}, Amaury H. M. J. and {Other}, A. N and {Other}, A. N and {Other}, A. N and {Other}, A. N and {Other}, A. N and {Other}, A. N and {Other}, A. N},
   year = {subm}
}

@ARTICLE{Schweitzer2019,
       author = {{Schweitzer}, A. and {Passegger}, V.~M. and {Cifuentes}, C. and {B{\'e}jar}, V.~J.~S. and {Cort{\'e}s-Contreras}, M. and {Caballero}, J.~A. and {del Burgo}, C. and {Czesla}, S. and {K{\"u}rster}, M. and {Montes}, D. and {Zapatero Osorio}, M.~R. and {Ribas}, I. and {Reiners}, A. and {Quirrenbach}, A. and {Amado}, P.~J. and {Aceituno}, J. and {Anglada-Escud{\'e}}, G. and {Bauer}, F.~F. and {Dreizler}, S. and {Jeffers}, S.~V. and {Guenther}, E.~W. and {Henning}, T. and {Kaminski}, A. and {Lafarga}, M. and {Marfil}, E. and {Morales}, J.~C. and {Schmitt}, J.~H.~M.~M. and {Seifert}, W. and {Solano}, E. and {Tabernero}, H.~M. and {Zechmeister}, M.},
        title = "{The CARMENES search for exoplanets around M dwarfs. Different roads to radii and masses of the target stars}",
      journal = {\aap},
         year = 2019,
        month = may,
       volume = {625},
          eid = {A68},
        pages = {A68},
          doi = {10.1051/0004-6361/201834965},
archivePrefix = {arXiv},
       eprint = {1904.03231},
 primaryClass = {astro-ph.SR},
       adsurl = {https://ui.adsabs.harvard.edu/abs/2019A&A...625A..68S}
}

@ARTICLE{Kroupa2002,
       author = {{Kroupa}, Pavel},
        title = "{The Initial Mass Function of Stars: Evidence for Uniformity in Variable Systems}",
      journal = {Science},
         year = 2002,
        month = jan,
       volume = {295},
       number = {5552},
        pages = {82-91},
          doi = {10.1126/science.1067524},
archivePrefix = {arXiv},
       eprint = {astro-ph/0201098},
 primaryClass = {astro-ph},
       adsurl = {https://ui.adsabs.harvard.edu/abs/2002Sci...295...82K}
}

@ARTICLE{Chabrier2003,
       author = {{Chabrier}, Gilles},
        title = "{Galactic Stellar and Substellar Initial Mass Function}",
      journal = {\pasp},
         year = 2003,
        month = jul,
       volume = {115},
       number = {809},
        pages = {763-795},
          doi = {10.1086/376392},
archivePrefix = {arXiv},
       eprint = {astro-ph/0304382},
 primaryClass = {astro-ph},
       adsurl = {https://ui.adsabs.harvard.edu/abs/2003PASP..115..763C}
}

@ARTICLE{Winters2019,
       author = {{Winters}, Jennifer G. and {Henry}, Todd J. and {Jao}, Wei-Chun and {Subasavage}, John P. and {Chatelain}, Joseph P. and {Slatten}, Ken and {Riedel}, Adric R. and {Silverstein}, Michele L. and {Payne}, Matthew J.},
        title = "{The Solar Neighborhood. XLV. The Stellar Multiplicity Rate of M Dwarfs Within 25 pc}",
      journal = {\aj},
         year = 2019,
        month = jun,
       volume = {157},
       number = {6},
          eid = {216},
        pages = {216},
          doi = {10.3847/1538-3881/ab05dc},
archivePrefix = {arXiv},
       eprint = {1901.06364},
 primaryClass = {astro-ph.SR},
       adsurl = {https://ui.adsabs.harvard.edu/abs/2019AJ....157..216W}
}

@ARTICLE{Salpeter1955,
       author = {{Salpeter}, Edwin E.},
        title = "{The Luminosity Function and Stellar Evolution.}",
      journal = {\apj},
         year = 1955,
        month = jan,
       volume = {121},
        pages = {161},
          doi = {10.1086/145971},
       adsurl = {https://ui.adsabs.harvard.edu/abs/1955ApJ...121..161S}
}

@ARTICLE{BrogiLine2019,
       author = {{Brogi}, Matteo and {Line}, Michael R.},
        title = "{Retrieving Temperatures and Abundances of Exoplanet Atmospheres with High-resolution Cross-correlation Spectroscopy}",
      journal = {\aj},
         year = 2019,
        month = mar,
       volume = {157},
       number = {3},
          eid = {114},
        pages = {114},
          doi = {10.3847/1538-3881/aaffd3},
archivePrefix = {arXiv},
       eprint = {1811.01681},
 primaryClass = {astro-ph.EP},
       adsurl = {https://ui.adsabs.harvard.edu/abs/2019AJ....157..114B}
}

@ARTICLE{Lafarga2023,
       author = {{Lafarga}, M. and {Brogi}, M. and {Gandhi}, S. and {Cegla}, H.~M. and {Seidel}, J.~V. and {Doyle}, L. and {Allart}, R. and {Buchschacher}, N. and {Lendl}, M. and {Lovis}, C. and {Sosnowska}, D.},
        title = "{The hot Neptune WASP-166 b with ESPRESSO - III. A blue-shifted tentative water signal constrains the presence of clouds}",
      journal = {\mnras},
         year = 2023,
        month = may,
       volume = {521},
       number = {1},
        pages = {1233-1252},
          doi = {10.1093/mnras/stad480},
archivePrefix = {arXiv},
       eprint = {2302.04794},
 primaryClass = {astro-ph.EP},
       adsurl = {https://ui.adsabs.harvard.edu/abs/2023MNRAS.521.1233L}
}

@ARTICLE{Snellen2025,
       author = {{Snellen}, Ignas A.~G.},
        title = "{Exoplanet Atmospheres at High Spectral Resolution}",
      journal = {\araa},
         year = 2025,
        month = aug,
       volume = {63},
       number = {1},
        pages = {83-125},
          doi = {10.1146/annurev-astro-052622-031342},
archivePrefix = {arXiv},
       eprint = {2505.08926},
 primaryClass = {astro-ph.EP},
       adsurl = {https://ui.adsabs.harvard.edu/abs/2025ARA&A..63...83S}
}

@ARTICLE{Snellen2008,
       author = {{Snellen}, I.~A.~G. and {Albrecht}, S. and {de Mooij}, E.~J.~W. and {Le Poole}, R.~S.},
        title = "{Ground-based detection of sodium in the transmission spectrum of exoplanet HD 209458b}",
      journal = {\aap},
         year = 2008,
        month = aug,
       volume = {487},
       number = {1},
        pages = {357-362},
          doi = {10.1051/0004-6361:200809762},
archivePrefix = {arXiv},
       eprint = {0805.0789},
 primaryClass = {astro-ph},
       adsurl = {https://ui.adsabs.harvard.edu/abs/2008A&A...487..357S}
}

@ARTICLE{Snellen2010,
       author = {{Snellen}, Ignas A.~G. and {de Kok}, Remco J. and {de Mooij}, Ernst J.~W. and {Albrecht}, Simon},
        title = "{The orbital motion, absolute mass and high-altitude winds of exoplanet HD209458b}",
      journal = {\nat},
         year = 2010,
        month = jun,
       volume = {465},
       number = {7301},
        pages = {1049-1051},
          doi = {10.1038/nature09111},
archivePrefix = {arXiv},
       eprint = {1006.4364},
 primaryClass = {astro-ph.EP},
       adsurl = {https://ui.adsabs.harvard.edu/abs/2010Natur.465.1049S}
}

@ARTICLE{Peterson2023,
       author = {{Peterson}, Merrin S. and {Benneke}, Bj{\"o}rn and {Collins}, Karen and {Piaulet}, Caroline and {Crossfield}, Ian J.~M. and {Ali-Dib}, Mohamad and {Christiansen}, Jessie L. and {Gagn{\'e}}, Jonathan and {Faherty}, Jackie and {Kite}, Edwin and {Dressing}, Courtney and {Charbonneau}, David and {Murgas}, Felipe and {Cointepas}, Marion and {Almenara}, Jose Manuel and {Bonfils}, Xavier and {Kane}, Stephen and {Werner}, Michael W. and {Gorjian}, Varoujan and {Roy}, Pierre-Alexis and {Shporer}, Avi and {Pozuelos}, Francisco J. and {Socia}, Quentin Jay and {Cloutier}, Ryan and {Dietrich}, Jeremy and {Irwin}, Jonathan and {Weiss}, Lauren and {Waalkes}, William and {Berta-Thomson}, Zach and {Evans}, Thomas and {Apai}, Daniel and {Parviainen}, Hannu and {Pall{\'e}}, Enric and {Narita}, Norio and {Howard}, Andrew W. and {Dragomir}, Diana and {Barkaoui}, Khalid and {Gillon}, Micha{\"e}l and {Jehin}, Emmanuel and {Ducrot}, Elsa and {Benkhaldoun}, Zouhair and {Fukui}, Akihiko and {Mori}, Mayuko and {Nishiumi}, Taku and {Kawauchi}, Kiyoe and {Ricker}, George and {Latham}, David W. and {Winn}, Joshua N. and {Seager}, Sara and {Isaacson}, Howard and {Bixel}, Alex and {Gibbs}, Aidan and {Jenkins}, Jon M. and {Smith}, Jeffrey C. and {Chavez}, Jose Perez and {Rackham}, Benjamin V. and {Henning}, Thomas and {Gabor}, Paul and {Chen}, Wen-Ping and {Espinoza}, Nestor and {Jensen}, Eric L.~N. and {Collins}, Kevin I. and {Schwarz}, Richard P. and {Conti}, Dennis M. and {Wang}, Gavin and {Kielkopf}, John F. and {Mao}, Shude and {Horne}, Keith and {Sefako}, Ramotholo and {Quinn}, Samuel N. and {Moldovan}, Dan and {Fausnaugh}, Michael and {F{\.z}{\.z}r{\'e}sz}, G{\'a}bor and {Barclay}, Thomas},
        title = "{A temperate Earth-sized planet with tidal heating transiting an M6 star}",
      journal = {\nat},
         year = 2023,
        month = may,
       volume = {617},
       number = {7962},
        pages = {701-705},
          doi = {10.1038/s41586-023-05934-8},
       adsurl = {https://ui.adsabs.harvard.edu/abs/2023Natur.617..701P}
}

@ARTICLE{Bonfils2013,
       author = {{Bonfils}, X. and {Delfosse}, X. and {Udry}, S. and {Forveille}, T. and {Mayor}, M. and {Perrier}, C. and {Bouchy}, F. and {Gillon}, M. and {Lovis}, C. and {Pepe}, F. and {Queloz}, D. and {Santos}, N.~C. and {S{\'e}gransan}, D. and {Bertaux}, J. -L.},
        title = "{The HARPS search for southern extra-solar planets. XXXI. The M-dwarf sample}",
      journal = {\aap},
         year = 2013,
        month = jan,
       volume = {549},
          eid = {A109},
        pages = {A109},
          doi = {10.1051/0004-6361/201014704},
archivePrefix = {arXiv},
       eprint = {1111.5019},
 primaryClass = {astro-ph.EP},
       adsurl = {https://ui.adsabs.harvard.edu/abs/2013A&A...549A.109B}
}

@article{Skilling2006,
author = {John Skilling},
title = {{Nested sampling for general Bayesian computation}},
volume = {1},
journal = {Bayesian Analysis},
number = {4},
publisher = {International Society for Bayesian Analysis},
pages = {833 -- 859},
year = {2006},
doi = {10.1214/06-BA127},
URL = {https://doi.org/10.1214/06-BA127}
}

@ARTICLE{mortier2020,
       author = {{Mortier}, A. and {Zapatero Osorio}, M.~R. and {Malavolta}, L. and {Alibert}, Y. and {Rice}, K. and {Lillo-Box}, J. and {Vanderburg}, A. and {Oshagh}, M. and {Buchhave}, L. and {Adibekyan}, V. and {Delgado Mena}, E. and {Lopez-Morales}, M. and {Charbonneau}, D. and {Sousa}, S.~G. and {Lovis}, C. and {Affer}, L. and {Allende Prieto}, C. and {Barros}, S.~C.~C. and {Benatti}, S. and {Bonomo}, A.~S. and {Boschin}, W. and {Bouchy}, F. and {Cabral}, A. and {Collier Cameron}, A. and {Cosentino}, R. and {Cristiani}, S. and {Demangeon}, O.~D.~S. and {Di Marcantonio}, P. and {D'Odorico}, V. and {Dumusque}, X. and {Ehrenreich}, D. and {Figueira}, P. and {Fiorenzano}, A. and {Ghedina}, A. and {Gonz{\'a}lez Hern{\'a}ndez}, J.~I. and {Haldemann}, J. and {Harutyunyan}, A. and {Haywood}, R.~D. and {Latham}, D.~W. and {Lavie}, B. and {Lo Curto}, G. and {Maldonado}, J. and {Manescau}, A. and {Martins}, C.~J.~A.~P. and {Mayor}, M. and {M{\'e}gevand}, D. and {Mehner}, A. and {Micela}, G. and {Molaro}, P. and {Molinari}, E. and {Nunes}, N.~J. and {Pepe}, F.~A. and {Palle}, E. and {Phillips}, D. and {Piotto}, G. and {Pinamonti}, M. and {Poretti}, E. and {Riva}, M. and {Rebolo}, R. and {Santos}, N.~C. and {Sasselov}, D. and {Sozzetti}, A. and {Su{\'a}rez Mascare{\~n}o}, A. and {Udry}, S. and {West}, R.~G. and {Watson}, C.~A. and {Wilson}, T.~G.},
        title = "{K2-111: an old system with two planets in near-resonance}",
      journal = {\mnras},
         year = 2020,
        month = dec,
       volume = {499},
       number = {4},
        pages = {5004-5021},
          doi = {10.1093/mnras/staa3144},
archivePrefix = {arXiv},
       eprint = {2010.01993},
 primaryClass = {astro-ph.EP},
       adsurl = {https://ui.adsabs.harvard.edu/abs/2020MNRAS.499.5004M}
}

@ARTICLE{skrutskie2006,
       author = {{Skrutskie}, M.~F. and {Cutri}, R.~M. and {Stiening}, R. and {Weinberg}, M.~D. and {Schneider}, S. and {Carpenter}, J.~M. and {Beichman}, C. and {Capps}, R. and {Chester}, T. and {Elias}, J. and {Huchra}, J. and {Liebert}, J. and {Lonsdale}, C. and {Monet}, D.~G. and {Price}, S. and {Seitzer}, P. and {Jarrett}, T. and {Kirkpatrick}, J.~D. and {Gizis}, J.~E. and {Howard}, E. and {Evans}, T. and {Fowler}, J. and {Fullmer}, L. and {Hurt}, R. and {Light}, R. and {Kopan}, E.~L. and {Marsh}, K.~A. and {McCallon}, H.~L. and {Tam}, R. and {Van Dyk}, S. and {Wheelock}, S.},
        title = "{The Two Micron All Sky Survey (2MASS)}",
      journal = {\aj},
         year = 2006,
        month = feb,
       volume = {131},
       number = {2},
        pages = {1163-1183},
          doi = {10.1086/498708},
       adsurl = {https://ui.adsabs.harvard.edu/abs/2006AJ....131.1163S}
}

@dataset{cutri2013,
       author = {{Cutri}, R.~M. and {Wright}, E.~L. and {Conrow}, T. and {Fowler}, J.~W. and {Eisenhardt}, P.~R.~M. and {Grillmair}, C. and {Kirkpatrick}, J.~D. and {Masci}, F. and {McCallon}, H.~L. and {Wheelock}, S.~L. and {Fajardo-Acosta}, S. and {Yan}, L. and {Benford}, D. and {Harbut}, M. and {Jarrett}, T. and {Lake}, S. and {Leisawitz}, D. and {Ressler}, M.~E. and {Stanford}, S.~A. and {Tsai}, C. -W. and {Liu}, F. and {Helou}, G. and {Mainzer}, A. and {Gettngs}, D. and {Gonzalez}, A. and {Hoffman}, D. and {Marsh}, K.~A. and {Padgett}, D. and {Skrutskie}, M.~F. and {Beck}, R. and {Papin}, M. and {Wittman}, M.},
        title = "{VizieR Online Data Catalog: AllWISE Data Release (Cutri+ 2013)}",
 howpublished = {VizieR On-line Data Catalog: II/328.  Originally published in: IPAC/Caltech (2013)},
         year = 2021,
        month = feb,
          eid = {II/328},
       adsurl = {https://ui.adsabs.harvard.edu/abs/2014yCat.2328....0C}
}

@ARTICLE{Hoxie1970,
       author = {{Hoxie}, Dwight T.},
        title = "{The Structure and Evolution of Stars of Very Low Mass}",
      journal = {\apj},
         year = 1970,
        month = sep,
       volume = {161},
        pages = {1083},
          doi = {10.1086/150609},
       adsurl = {https://ui.adsabs.harvard.edu/abs/1970ApJ...161.1083H}
}

@ARTICLE{Hoxie1973,
       author = {{Hoxie}, D.~T.},
        title = "{The low-mass main-sequence: the comparison between theory and observation.}",
      journal = {\aap},
         year = 1973,
        month = aug,
       volume = {26},
        pages = {437},
       adsurl = {https://ui.adsabs.harvard.edu/abs/1973A&A....26..437H}
}

@ARTICLE{Lacy1977,
       author = {{Lacy}, C.~H.},
        title = "{Radii of nearby stars: an application of the Barnes-Evans relation.}",
      journal = {\apjs},
         year = 1977,
        month = aug,
       volume = {34},
        pages = {479-492},
          doi = {10.1086/190459},
       adsurl = {https://ui.adsabs.harvard.edu/abs/1977ApJS...34..479L}
}

@ARTICLE{Popper1997,
       author = {{Popper}, Daniel M.},
        title = "{Orbits of detached main-sequence eclipsing binaries of types late F to K. II. UV leonis, UV piscium, and BH virginis.}",
      journal = {\aj},
         year = 1997,
        month = sep,
       volume = {114},
        pages = {1195},
          doi = {10.1086/118552},
       adsurl = {https://ui.adsabs.harvard.edu/abs/1997AJ....114.1195P}
}

@ARTICLE{Swayne2021,
       author = {{Swayne}, M.~I. and {Maxted}, P.~F.~L. and {Triaud}, A.~H.~M.~J. and {Sousa}, S.~G. and {Broeg}, C. and {Flor{\'e}n}, H. -G. and {Guterman}, P. and {Simon}, A.~E. and {Boisse}, I. and {Bonfanti}, A. and {Martin}, D. and {Santerne}, A. and {Salmon}, S. and {Standing}, M.~R. and {Van Grootel}, V. and {Wilson}, T.~G. and {Alibert}, Y. and {Alonso}, R. and {Anglada Escud{\'e}}, G. and {Asquier}, J. and {B{\'a}rczy}, T. and {Barrado}, D. and {Barros}, S.~C.~C. and {Battley}, M. and {Baumjohann}, W. and {Beck}, M. and {Beck}, T. and {Bekkelien}, A. and {Benz}, W. and {Billot}, N. and {Bonfils}, X. and {Brandeker}, A. and {Busch}, M. -D. and {Cabrera}, J. and {Charnoz}, S. and {Collier Cameron}, A. and {Csizmadia}, Sz and {Davies}, M.~B. and {Deleuil}, M. and {Deline}, A. and {Delrez}, L. and {Demangeon}, O.~D.~S. and {Demory}, B. -O. and {Dransfield}, G. and {Ehrenreich}, D. and {Erikson}, A. and {Fortier}, A. and {Fossati}, L. and {Fridlund}, M. and {Futyan}, D. and {Gandolfi}, D. and {Gillon}, M. and {Guedel}, M. and {H{\'e}brard}, G. and {Heidari}, N. and {Hellier}, C. and {Heng}, K. and {Hobson}, M. and {Hoyer}, S. and {Isaak}, K.~G. and {Kiss}, L. and {Kunovac Hod{\v{z}}i{\'c}}, V. and {Lalitha}, S. and {Laskar}, J. and {Lecavelier des Etangs}, A. and {Lendl}, M. and {Lovis}, C. and {Magrin}, D. and {Marafatto}, L. and {McCormac}, J. and {Miller}, N. and {Nascimbeni}, V. and {Olofsson}, G. and {Ottensamer}, R. and {Pagano}, I. and {Pall{\'e}}, E. and {Peter}, G. and {Piotto}, G. and {Pollacco}, D. and {Queloz}, D. and {Ragazzoni}, R. and {Rando}, N. and {Rauer}, H. and {Ribas}, I. and {Santos}, N.~C. and {Scandariato}, G. and {S{\'e}gransan}, D. and {Smith}, A.~M.~S. and {Steinberger}, M. and {Steller}, M. and {Szab{\'o}}, Gy M. and {Thomas}, N. and {Udry}, S. and {Walter}, I. and {Walton}, N.~A. and {Willett}, E.},
        title = "{The EBLM project - VIII. First results for M-dwarf mass, radius, and effective temperature measurements using CHEOPS light curves}",
      journal = {\mnras},
         year = 2021,
        month = sep,
       volume = {506},
       number = {1},
        pages = {306-322},
          doi = {10.1093/mnras/stab1687},
archivePrefix = {arXiv},
       eprint = {2106.07276},
 primaryClass = {astro-ph.SR},
       adsurl = {https://ui.adsabs.harvard.edu/abs/2021MNRAS.506..306S}
}

@INPROCEEDINGS{Clausen1999,
       author = {{Clausen}, J.~V. and {Baraffe}, I. and {Claret}, A. and {Vandenberg}, D.~A.},
        title = "{Do 0.7-1.1 M$_{sun}$ Eclipsing Binaries Pose a Problem for Current Stellar Evolutionary Models?}",
    booktitle = {Stellar Structure:  Theory and Test of Connective Energy Transport},
         year = 1999,
       editor = {{Gimenez}, Alvaro and {Guinan}, Edward F. and {Montesinos}, Benjamin},
       series = {Astronomical Society of the Pacific Conference Series},
       volume = {173},
        month = jan,
        pages = {265},
       adsurl = {https://ui.adsabs.harvard.edu/abs/1999ASPC..173..265C}
}

@ARTICLE{TorresRibas2002,
       author = {{Torres}, Guillermo and {Ribas}, Ignasi},
        title = "{Absolute Dimensions of the M-Type Eclipsing Binary YY Geminorum (Castor C): A Challenge to Evolutionary Models in the Lower Main Sequence}",
      journal = {\apj},
         year = 2002,
        month = mar,
       volume = {567},
       number = {2},
        pages = {1140-1165},
          doi = {10.1086/338587},
archivePrefix = {arXiv},
       eprint = {astro-ph/0111167},
 primaryClass = {astro-ph},
       adsurl = {https://ui.adsabs.harvard.edu/abs/2002ApJ...567.1140T}
}

@ARTICLE{2025CoSka..55c.412S,
       author = {{Sebastian}, D. and {Triaud}, A.~H.~M.~J. and {Maxted}, P.~F.~L. and {Brogi}, M.},
        title = "{Accurate dynamical masses from binaries with extreme brightness ratios}",
      journal = {Contributions of the Astronomical Observatory Skalnate Pleso},
         year = 2025,
        month = apr,
       volume = {55},
       number = {3},
        pages = {412-420},
          doi = {10.31577/caosp.2025.55.3.412},
       adsurl = {https://ui.adsabs.harvard.edu/abs/2025CoSka..55c.412S}
}

@ARTICLE{Maxted2022,
       author = {{Maxted}, P.~F.~L. and {Miller}, N.~J. and {Hoyer}, S. and {Adibekyan}, V. and {Sousa}, S.~G. and {Billot}, N. and {Fortier}, A. and {Simon}, A.~E. and {Collier Cameron}, A. and {Swayne}, M.~I. and {Gutermann}, P. and {Triaud}, A.~H.~M.~J. and {Southworth}, J. and {Alibert}, Y. and {Alonso}, R. and {Anglada}, G. and {B{\'a}rczy}, T. and {Barrado Navascues}, D. and {Barros}, S.~C.~C. and {Baumjohann}, W. and {Beck}, M. and {Beck}, T. and {Benz}, W. and {Bonfils}, X. and {Brandeker}, A. and {Broeg}, C. and {Buder}, M. and {Cabrera}, J. and {Charnoz}, S. and {Corral van Damme}, C. and {Csizmadia}, Sz and {Davies}, M.~B. and {Deleuil}, M. and {Delrez}, L. and {Demangeon}, O. and {Demory}, B. -O. and {Ehrenreich}, D. and {Erikson}, A. and {Fossati}, L. and {Fridlund}, M. and {Gandolfi}, D. and {Gillon}, M. and {G{\"u}del}, M. and {Heng}, K. and {Hern{\'a}ndez Leon}, J.~E. and {Isaak}, K.~G. and {Kiss}, L.~L. and {Laskar}, J. and {Lecavelier des Etangs}, A. and {Lendl}, M. and {Lovis}, C. and {Magrin}, D. and {Munari}, M. and {Nascimbeni}, V. and {Olofsson}, G. and {Ottensamer}, R. and {Pagano}, I. and {Pall{\'e}}, E. and {Peter}, G. and {Piotto}, G. and {Pollacco}, D. and {Queloz}, D. and {Ragazzoni}, R. and {Rando}, N. and {Rauer}, H. and {Ribas}, I. and {Santos}, N.~C. and {Scandariato}, G. and {S{\'e}gransan}, D. and {Smith}, A.~M.~S. and {Steinberger}, M. and {Steller}, M. and {Szab{\'o}}, Gy M. and {Thomas}, N. and {Udry}, S. and {Van Grootel}, V. and {Walton}, N.},
        title = "{Fundamental effective temperature measurements for eclipsing binary stars - III. SPIRou near-infrared spectroscopy and CHEOPS photometry of the benchmark G0V star EBLM J0113+31}",
      journal = {\mnras},
         year = 2022,
        month = jul,
       volume = {513},
       number = {4},
        pages = {6042-6057},
          doi = {10.1093/mnras/stac1270},
archivePrefix = {arXiv},
       eprint = {2205.01466},
 primaryClass = {astro-ph.SR},
       adsurl = {https://ui.adsabs.harvard.edu/abs/2022MNRAS.513.6042M}
}

@ARTICLE{ChabrierBaraffe1997,
       author = {{Chabrier}, Gilles and {Baraffe}, Isabelle},
        title = "{Structure and evolution of low-mass stars}",
      journal = {\aap},
         year = 1997,
        month = nov,
       volume = {327},
        pages = {1039-1053},
          doi = {10.48550/arXiv.astro-ph/9704118},
archivePrefix = {arXiv},
       eprint = {astro-ph/9704118},
 primaryClass = {astro-ph},
       adsurl = {https://ui.adsabs.harvard.edu/abs/1997A&A...327.1039C}
}

@ARTICLE{LD2003,
       author = {{Lindegren}, Lennart and {Dravins}, Dainis},
        title = "{The fundamental definition of ``radial velocity''}",
      journal = {\aap},
         year = 2003,
        month = apr,
       volume = {401},
        pages = {1185-1201},
          doi = {10.1051/0004-6361:20030181},
archivePrefix = {arXiv},
       eprint = {astro-ph/0302522},
 primaryClass = {astro-ph},
       adsurl = {https://ui.adsabs.harvard.edu/abs/2003A&A...401.1185L}
}

@ARTICLE{deKok2013,
       author = {{de Kok}, R.~J. and {Brogi}, M. and {Snellen}, I.~A.~G. and {Birkby}, J. and {Albrecht}, S. and {de Mooij}, E.~J.~W.},
        title = "{Detection of carbon monoxide in the high-resolution day-side spectrum of the exoplanet HD 189733b}",
      journal = {\aap},
         year = 2013,
        month = jun,
       volume = {554},
          eid = {A82},
        pages = {A82},
          doi = {10.1051/0004-6361/201321381},
archivePrefix = {arXiv},
       eprint = {1304.4014},
 primaryClass = {astro-ph.EP},
       adsurl = {https://ui.adsabs.harvard.edu/abs/2013A&A...554A..82D}
}

@ARTICLE{Maxted2015,
       author = {{Maxted}, P.~F.~L. and {Serenelli}, A.~M. and {Southworth}, J.},
        title = "{Bayesian mass and age estimates for transiting exoplanet host stars}",
      journal = {\aap},
         year = 2015,
        month = mar,
       volume = {575},
          eid = {A36},
        pages = {A36},
          doi = {10.1051/0004-6361/201425331},
archivePrefix = {arXiv},
       eprint = {1412.7891},
 primaryClass = {astro-ph.IM},
       adsurl = {https://ui.adsabs.harvard.edu/abs/2015A&A...575A..36M}
}

@ARTICLE{Maxted2024,
       author = {{Maxted}, P.~F.~L. and {Miller}, N.~J. and {Sebastian}, D. and {Triaud}, A.~H.~M.~J. and {Martin}, D.~V. and {Duck}, A.},
        title = "{Fundamental effective temperature measurements for eclipsing binary stars - V. The circumbinary planet system EBLM J0608-59}",
      journal = {\mnras},
         year = 2024,
        month = jul,
       volume = {531},
       number = {4},
        pages = {4577-4583},
          doi = {10.1093/mnras/stae1434},
archivePrefix = {arXiv},
       eprint = {2406.04204},
 primaryClass = {astro-ph.SR},
       adsurl = {https://ui.adsabs.harvard.edu/abs/2024MNRAS.531.4577M}
}

@ARTICLE{Miller2020,
       author = {{Miller}, N.~J. and {Maxted}, P.~F.~L. and {Smalley}, B.},
        title = "{Fundamental effective temperature measurements for eclipsing binary stars - I. Development of the method and application to AI Phoenicis}",
      journal = {\mnras},
         year = 2020,
        month = sep,
       volume = {497},
       number = {3},
        pages = {2899-2909},
          doi = {10.1093/mnras/staa2167},
archivePrefix = {arXiv},
       eprint = {2004.04568},
 primaryClass = {astro-ph.SR},
       adsurl = {https://ui.adsabs.harvard.edu/abs/2020MNRAS.497.2899M}
}

@ARTICLE{triaud-bebop4,
       author = {{Triaud}, Amaury H.~M.~J. and {Baycroft}, Thomas A. and {Heidari}, Neda and {Santerne}, Alexandre and {Adamson}, Aleyna and {Boisse}, Isabelle and {Coleman}, Gavin A.~L. and {Correia}, Alexandre C.~M. and {Davis}, Yasmin T. and {Deleuil}, Magali and {H{\'e}brard}, Guillaume and {Martin}, David V. and {Maxted}, Pierre F.~L. and {Nelson}, Richard P. and {Sairam}, Lalitha and {Sebastian}, Daniel and {Standing}, Matthew R. and {Stevenson}, Adam T. and {Stokholm}, Amalie and {Timmermans}, Mathilde and {Udry}, St{\'e}phane},
        title = "{BEBOP VIII. SOPHIE radial velocities reveal an eccentric, circumbinary brown dwarf}",
      journal = {\mnras},
         year = 2025,
        month = dec,
       volume = {544},
       number = {2},
        pages = {2180-2192},
          doi = {10.1093/mnras/staf1884},
archivePrefix = {arXiv},
       eprint = {2510.25329},
 primaryClass = {astro-ph.EP},
       adsurl = {https://ui.adsabs.harvard.edu/abs/2025MNRAS.544.2180T}
}

@INPROCEEDINGS{spoc,
       author = {{Jenkins}, Jon M. and {Twicken}, Joseph D. and {McCauliff}, Sean and {Campbell}, Jennifer and {Sanderfer}, Dwight and {Lung}, David and {Mansouri-Samani}, Masoud and {Girouard}, Forrest and {Tenenbaum}, Peter and {Klaus}, Todd and {Smith}, Jeffrey C. and {Caldwell}, Douglas A. and {Chacon}, A.~D. and {Henze}, Christopher and {Heiges}, Cory and {Latham}, David W. and {Morgan}, Edward and {Swade}, Daryl and {Rinehart}, Stephen and {Vanderspek}, Roland},
        title = "{The TESS science processing operations center}",
    booktitle = {Software and Cyberinfrastructure for Astronomy IV},
         year = 2016,
       editor = {{Chiozzi}, Gianluca and {Guzman}, Juan C.},
       series = {Society of Photo-Optical Instrumentation Engineers (SPIE) Conference Series},
       volume = {9913},
        month = aug,
          eid = {99133E},
        pages = {99133E},
          doi = {10.1117/12.2233418},
       adsurl = {https://ui.adsabs.harvard.edu/abs/2016SPIE.9913E..3EJ}
}

@ARTICLE{PontEyer04,
       author = {{Pont}, Fr{\'e}d{\'e}ric and {Eyer}, Laurent},
        title = "{Isochrone ages for field dwarfs: method and application to the age-metallicity relation}",
      journal = {\mnras},
         year = 2004,
        month = jun,
       volume = {351},
       number = {2},
        pages = {487-504},
          doi = {10.1111/j.1365-2966.2004.07780.x},
archivePrefix = {arXiv},
       eprint = {astro-ph/0401418},
 primaryClass = {astro-ph},
       adsurl = {https://ui.adsabs.harvard.edu/abs/2004MNRAS.351..487P}
}

@INPROCEEDINGS{PontEyer05,
       author = {{Pont}, F. and {Eyer}, L.},
        title = "{The Trouble with Isochrone Ages for Field Stars: A Cautionary Tale for Solar Neighbourhood Studies}",
    booktitle = {The Three-Dimensional Universe with Gaia},
         year = 2005,
       editor = {{Turon}, C. and {O'Flaherty}, K.~S. and {Perryman}, M.~A.~C.},
       series = {ESA Special Publication},
       volume = {576},
        month = jan,
        pages = {187},
       adsurl = {https://ui.adsabs.harvard.edu/abs/2005ESASP.576..187P}
}

@ARTICLE{Davies2016,
       author = {{Davies}, G.~R. and {Miglio}, A.},
        title = "{Asteroseismology of red giants: From analysing light curves to estimating ages}",
      journal = {Astronomische Nachrichten},
         year = 2016,
        month = sep,
       volume = {337},
       number = {8-9},
        pages = {774},
          doi = {10.1002/asna.201612371},
archivePrefix = {arXiv},
       eprint = {1601.02802},
 primaryClass = {astro-ph.SR},
       adsurl = {https://ui.adsabs.harvard.edu/abs/2016AN....337..774D}
}

@ARTICLE{sse-lectures,
       author = {{Christensen-Dalsgaard}, J{\o}rgen},
        title = "{Solar structure and evolution}",
      journal = {Living Reviews in Solar Physics},
         year = 2021,
        month = dec,
       volume = {18},
       number = {1},
          eid = {2},
        pages = {2},
          doi = {10.1007/s41116-020-00028-3},
archivePrefix = {arXiv},
       eprint = {2007.06488},
 primaryClass = {astro-ph.SR},
       adsurl = {https://ui.adsabs.harvard.edu/abs/2021LRSP...18....2C}
}

@ARTICLE{Wright2013,
       author = {{Wright}, Jason T. and {Roy}, Arpita and {Mahadevan}, Suvrath and {Wang}, Sharon X. and {Ford}, Eric B. and {Payne}, Matt and {Lee}, Brian L. and {Wang}, Ji and {Crepp}, Justin R. and {Gaudi}, B. Scott and {Eastman}, Jason and {Pepper}, Joshua and {Ge}, Jian and {Fleming}, Scott W. and {Ghezzi}, Luan and {Gonz{\'a}lez-Hern{\'a}ndez}, Jonay I. and {Cargile}, Phillip and {Stassun}, Keivan G. and {Wisniewski}, John and {Dutra-Ferreira}, Leticia and {Porto de Mello}, Gustavo F. and {Maia}, M{\'a}rcio A.~G. and {Nicolaci da Costa}, Luiz and {Ogando}, Ricardo L.~C. and {Santiago}, Basilio X. and {Schneider}, Donald P. and {Hearty}, Fred R.},
        title = "{MARVELS-1: A Face-on Double-lined Binary Star Masquerading as a Resonant Planetary System and Consideration of Rare False Positives in Radial Velocity Planet Searches}",
      journal = {\apj},
         year = 2013,
        month = jun,
       volume = {770},
       number = {2},
          eid = {119},
        pages = {119},
          doi = {10.1088/0004-637X/770/2/119},
archivePrefix = {arXiv},
       eprint = {1305.0280},
 primaryClass = {astro-ph.SR},
       adsurl = {https://ui.adsabs.harvard.edu/abs/2013ApJ...770..119W}
}

@ARTICLE{PojmanskiUdalski1997,
       author = {{Pojmanski}, G. and {Udalski}, A.},
        title = "{An Orbital Solution for RS CVn Binary Star XY UMa}",
      journal = {\actaa},
         year = 1997,
        month = oct,
       volume = {47},
        pages = {451-466},
       adsurl = {https://ui.adsabs.harvard.edu/abs/1997AcA....47..451P}
}

@ARTICLE{Ding2024,
       author = {{Ding}, Yuedan and {Zhang}, Shidi and {Han}, Henggeng and {Cui}, Wenyuan and {Wang}, Song and {Fang}, Min and {Gao}, Yawei},
        title = "{Double-edged Sword: The Influence of Tidal Interaction on Stellar Activity in Binaries}",
      journal = {\apj},
         year = 2024,
        month = dec,
       volume = {976},
       number = {2},
          eid = {243},
        pages = {243},
          doi = {10.3847/1538-4357/ad8eb9},
archivePrefix = {arXiv},
       eprint = {2410.15039},
 primaryClass = {astro-ph.SR},
       adsurl = {https://ui.adsabs.harvard.edu/abs/2024ApJ...976..243D}
}

@ARTICLE{Strassmeier2011,
       author = {{Strassmeier}, K.~G. and {Carroll}, T.~A. and {Weber}, M. and {Granzer}, T. and {Bartus}, J. and {Ol{\'a}h}, K. and {Rice}, J.~B.},
        title = "{Binary-induced magnetic activity?. Time-series echelle spectroscopy and photometry of HD 123351 = CZ CVn}",
      journal = {\aap},
         year = 2011,
        month = nov,
       volume = {535},
          eid = {A98},
        pages = {A98},
          doi = {10.1051/0004-6361/201117167},
archivePrefix = {arXiv},
       eprint = {1108.4673},
 primaryClass = {astro-ph.SR},
       adsurl = {https://ui.adsabs.harvard.edu/abs/2011A&A...535A..98S}
}

@ARTICLE{Queloz2009,
       author = {{Queloz}, D. and {Bouchy}, F. and {Moutou}, C. and {Hatzes}, A. and {H{\'e}brard}, G. and {Alonso}, R. and {Auvergne}, M. and {Baglin}, A. and {Barbieri}, M. and {Barge}, P. and {Benz}, W. and {Bord{\'e}}, P. and {Deeg}, H.~J. and {Deleuil}, M. and {Dvorak}, R. and {Erikson}, A. and {Ferraz Mello}, S. and {Fridlund}, M. and {Gandolfi}, D. and {Gillon}, M. and {Guenther}, E. and {Guillot}, T. and {Jorda}, L. and {Hartmann}, M. and {Lammer}, H. and {L{\'e}ger}, A. and {Llebaria}, A. and {Lovis}, C. and {Magain}, P. and {Mayor}, M. and {Mazeh}, T. and {Ollivier}, M. and {P{\"a}tzold}, M. and {Pepe}, F. and {Rauer}, H. and {Rouan}, D. and {Schneider}, J. and {Segransan}, D. and {Udry}, S. and {Wuchterl}, G.},
        title = "{The CoRoT-7 planetary system: two orbiting super-Earths}",
      journal = {\aap},
         year = 2009,
        month = oct,
       volume = {506},
       number = {1},
        pages = {303-319},
          doi = {10.1051/0004-6361/200913096},
       adsurl = {https://ui.adsabs.harvard.edu/abs/2009A&A...506..303Q}
}

@ARTICLE{CollierCameron2019,
       author = {{Collier Cameron}, A. and {Mortier}, A. and {Phillips}, D. and {Dumusque}, X. and {Haywood}, R.~D. and {Langellier}, N. and {Watson}, C.~A. and {Cegla}, H.~M. and {Costes}, J. and {Charbonneau}, D. and {Coffinet}, A. and {Latham}, D.~W. and {Lopez-Morales}, M. and {Malavolta}, L. and {Maldonado}, J. and {Micela}, G. and {Milbourne}, T. and {Molinari}, E. and {Saar}, S.~H. and {Thompson}, S. and {Buchschacher}, N. and {Cecconi}, M. and {Cosentino}, R. and {Ghedina}, A. and {Glenday}, A. and {Gonzalez}, M. and {Li}, C.-H. and {Lodi}, M. and {Lovis}, C. and {Pepe}, F. and {Poretti}, E. and {Rice}, K. and {Sasselov}, D. and {Sozzetti}, A. and {Szentgyorgyi}, A. and {Udry}, S. and {Walsworth}, R.},
        title = "{Three years of Sun-as-a-star radial-velocity observations on the approach to solar minimum}",
      journal = {\mnras},
         year = 2019,
        month = jul,
       volume = {487},
       number = {1},
        pages = {1082-1100},
          doi = {10.1093/mnras/stz1215},
archivePrefix = {arXiv},
       eprint = {1904.12186},
 primaryClass = {astro-ph.SR},
       adsurl = {https://ui.adsabs.harvard.edu/abs/2019MNRAS.487.1082C}
}

@ARTICLE{Barnes2024,
       author = {{Barnes}, J.~R. and {Jeffers}, S.~V. and {Haswell}, C.~A. and {Damasso}, M. and {Del Sordo}, F. and {Liebing}, F. and {Perger}, M. and {Anglada-Escud{\'e}}, G.},
        title = "{Identifying activity induced RV periodicities and correlations using central line moments}",
      journal = {\mnras},
         year = 2024,
        month = oct,
       volume = {534},
       number = {2},
        pages = {1257-1282},
          doi = {10.1093/mnras/stae2125},
archivePrefix = {arXiv},
       eprint = {2409.10306},
 primaryClass = {astro-ph.SR},
       adsurl = {https://ui.adsabs.harvard.edu/abs/2024MNRAS.534.1257B}
}

@ARTICLE{Queloz2001,
       author = {{Queloz}, D. and {Henry}, G.~W. and {Sivan}, J.~P. and {Baliunas}, S.~L. and {Beuzit}, J.~L. and {Donahue}, R.~A. and {Mayor}, M. and {Naef}, D. and {Perrier}, C. and {Udry}, S.},
        title = "{No planet for HD 166435}",
      journal = {\aap},
         year = 2001,
        month = nov,
       volume = {379},
        pages = {279-287},
          doi = {10.1051/0004-6361:20011308},
archivePrefix = {arXiv},
       eprint = {astro-ph/0109491},
 primaryClass = {astro-ph},
       adsurl = {https://ui.adsabs.harvard.edu/abs/2001A&A...379..279Q}
}

@ARTICLE{fitzpatrick2025,
       author = {{Fitzpatrick}, Jay and {Maxted}, Pierre F.~L. and {Freckelton}, Alix V. and {Triaud}, A.~H.~M.~J. and {Martin}, D.~V. and {Duck}, A.},
        title = "{The EBLM project - XIV. TESS light curves for eclipsing binaries with very low mass companions}",
      journal = {\mnras},
         year = 2025,
        month = mar,
       volume = {537},
       number = {4},
        pages = {3943-3953},
          doi = {10.1093/mnras/staf268},
archivePrefix = {arXiv},
       eprint = {2502.09267},
 primaryClass = {astro-ph.SR},
       adsurl = {https://ui.adsabs.harvard.edu/abs/2025MNRAS.537.3943F}
}

@ARTICLE{Costes2021,
       author = {{Costes}, Jean C. and {Watson}, Christopher A. and {de Mooij}, Ernst and {Saar}, Steven H. and {Dumusque}, Xavier and {Cameron}, Collier and {Phillips}, David F. and {G{\"u}nther}, Maximilian N. and {Jenkins}, James S. and {Mortier}, Annelies and {Thompson}, Andrew P.~G.},
        title = "{Long-term stellar activity variations and their effect on radial-velocity measurements}",
      journal = {\mnras},
         year = 2021,
        month = jul,
       volume = {505},
       number = {1},
        pages = {830-850},
          doi = {10.1093/mnras/stab1183},
archivePrefix = {arXiv},
       eprint = {2105.01915},
 primaryClass = {astro-ph.SR},
       adsurl = {https://ui.adsabs.harvard.edu/abs/2021MNRAS.505..830C}
}

@ARTICLE{Desort2007,
       author = {{Desort}, M. and {Lagrange}, A.-M. and {Galland}, F. and {Udry}, S. and {Mayor}, M.},
        title = "{Search for exoplanets with the radial-velocity technique: quantitative diagnostics of stellar activity}",
      journal = {\aap},
         year = 2007,
        month = oct,
       volume = {473},
       number = {3},
        pages = {983-993},
          doi = {10.1051/0004-6361:20078144},
archivePrefix = {arXiv},
       eprint = {0708.4338},
 primaryClass = {astro-ph},
       adsurl = {https://ui.adsabs.harvard.edu/abs/2007A&A...473..983D}
}

@ARTICLE{Haywood2014,
       author = {{Haywood}, R.~D. and {Collier Cameron}, A. and {Queloz}, D. and {Barros}, S.~C.~C. and {Deleuil}, M. and {Fares}, R. and {Gillon}, M. and {Lanza}, A.~F. and {Lovis}, C. and {Moutou}, C. and {Pepe}, F. and {Pollacco}, D. and {Santerne}, A. and {S{\'e}gransan}, D. and {Unruh}, Y.~C.},
        title = "{Planets and stellar activity: hide and seek in the CoRoT-7 system}",
      journal = {\mnras},
         year = 2014,
        month = sep,
       volume = {443},
       number = {3},
        pages = {2517-2531},
          doi = {10.1093/mnras/stu1320},
archivePrefix = {arXiv},
       eprint = {1407.1044},
 primaryClass = {astro-ph.EP},
       adsurl = {https://ui.adsabs.harvard.edu/abs/2014MNRAS.443.2517H}
}

@ARTICLE{Lafarga2021,
       author = {{Lafarga}, M. and {Ribas}, I. and {Reiners}, A. and {Quirrenbach}, A. and {Amado}, P.~J. and {Caballero}, J.~A. and {Azzaro}, M. and {B{\'e}jar}, V.~J.~S. and {Cort{\'e}s-Contreras}, M. and {Dreizler}, S. and {Hatzes}, A.~P. and {Henning}, Th. and {Jeffers}, S.~V. and {Kaminski}, A. and {K{\"u}rster}, M. and {Montes}, D. and {Morales}, J.~C. and {Oshagh}, M. and {Rodr{\'\i}guez-L{\'o}pez}, C. and {Sch{\"o}fer}, P. and {Schweitzer}, A. and {Zechmeister}, M.},
        title = "{The CARMENES search for exoplanets around M dwarfs. Mapping stellar activity indicators across the M dwarf domain}",
      journal = {\aap},
         year = 2021,
        month = aug,
       volume = {652},
          eid = {A28},
        pages = {A28},
          doi = {10.1051/0004-6361/202140605},
archivePrefix = {arXiv},
       eprint = {2105.13467},
 primaryClass = {astro-ph.SR},
       adsurl = {https://ui.adsabs.harvard.edu/abs/2021A&A...652A..28L}
}

@ARTICLE{Aigrain2023,
       author = {{Aigrain}, Suzanne and {Foreman-Mackey}, Daniel},
        title = "{Gaussian Process Regression for Astronomical Time Series}",
      journal = {\araa},
         year = 2023,
        month = aug,
       volume = {61},
        pages = {329-371},
          doi = {10.1146/annurev-astro-052920-103508},
archivePrefix = {arXiv},
       eprint = {2209.08940},
 primaryClass = {astro-ph.IM},
       adsurl = {https://ui.adsabs.harvard.edu/abs/2023ARA&A..61..329A}
}

@ARTICLE{Shkolnik2005,
       author = {{Shkolnik}, E. and {Walker}, G.~A.~H. and {Bohlender}, D.~A. and {Gu}, P.-G. and {K{\"u}rster}, M.},
        title = "{Hot Jupiters and Hot Spots: The Short- and Long-Term Chromospheric Activity on Stars with Giant Planets}",
      journal = {\apj},
         year = 2005,
        month = apr,
       volume = {622},
       number = {2},
        pages = {1075-1090},
          doi = {10.1086/428037},
archivePrefix = {arXiv},
       eprint = {astro-ph/0411655},
 primaryClass = {astro-ph},
       adsurl = {https://ui.adsabs.harvard.edu/abs/2005ApJ...622.1075S}
}

@ARTICLE{Dransfield2024,
       author = {{Dransfield}, Georgina and {Timmermans}, Mathilde and {Triaud}, Amaury H.~M.~J. and {D{\'e}vora-Pajares}, Mart{\'\i}n and {Aganze}, Christian and {Barkaoui}, Khalid and {Burgasser}, Adam J. and {Collins}, Karen A. and {Cointepas}, Marion and {Ducrot}, Elsa and {G{\"u}nther}, Maximilian N. and {Howell}, Steve B. and {Murray}, Catriona A. and {Niraula}, Prajwal and {Rackham}, Benjamin V. and {Sebastian}, Daniel and {Stassun}, Keivan G. and {Z{\'u}{\~n}iga-Fern{\'a}ndez}, Sebasti{\'a}n and {Almenara}, Jos{\'e} Manuel and {Bonfils}, Xavier and {Bouchy}, Fran{\c{c}}ois and {Burke}, Christopher J. and {Charbonneau}, David and {Christiansen}, Jessie L. and {Delrez}, Laetitia and {Gan}, Tianjun and {Garc{\'\i}a}, Lionel J. and {Gillon}, Micha{\"e}l and {G{\'o}mez Maqueo Chew}, Yilen and {Hesse}, Katharine M. and {Hooton}, Matthew J. and {Isopi}, Giovanni and {Jehin}, Emmanu{\"e}l and {Jenkins}, Jon M. and {Latham}, David W. and {Mallia}, Franco and {Murgas}, Felipe and {Pedersen}, Peter P. and {Pozuelos}, Francisco J. and {Queloz}, Didier and {Rodriguez}, David R. and {Schanche}, Nicole and {Seager}, Sara and {Srdoc}, Gregor and {Stockdale}, Chris and {Twicken}, Joseph D. and {Vanderspek}, Roland and {Wells}, Robert and {Winn}, Joshua N. and {de Wit}, Julien and {Zapparata}, Aldo},
        title = "{A 1.55 R$_{{\ensuremath{\oplus}}}$ habitable-zone planet hosted by TOI-715, an M4 star near the ecliptic South Pole}",
      journal = {\mnras},
         year = 2024,
        month = jan,
       volume = {527},
       number = {1},
        pages = {35-52},
          doi = {10.1093/mnras/stad1439},
archivePrefix = {arXiv},
       eprint = {2305.06206},
 primaryClass = {astro-ph.EP},
       adsurl = {https://ui.adsabs.harvard.edu/abs/2024MNRAS.527...35D}
}

@ARTICLE{Delrez2022-speculoos2,
       author = {{Delrez}, L. and {Murray}, C.~A. and {Pozuelos}, F.~J. and {Narita}, N. and {Ducrot}, E. and {Timmermans}, M. and {Watanabe}, N. and {Burgasser}, A.~J. and {Hirano}, T. and {Rackham}, B.~V. and {Stassun}, K.~G. and {Van Grootel}, V. and {Aganze}, C. and {Cointepas}, M. and {Howell}, S. and {Kaltenegger}, L. and {Niraula}, P. and {Sebastian}, D. and {Almenara}, J.~M. and {Barkaoui}, K. and {Baycroft}, T.~A. and {Bonfils}, X. and {Bouchy}, F. and {Burdanov}, A. and {Caldwell}, D.~A. and {Charbonneau}, D. and {Ciardi}, D.~R. and {Collins}, K.~A. and {Daylan}, T. and {Demory}, B.-O. and {de Wit}, J. and {Dransfield}, G. and {Fajardo-Acosta}, S.~B. and {Fausnaugh}, M. and {Fukui}, A. and {Furlan}, E. and {Garcia}, L.~J. and {Gnilka}, C.~L. and {G{\'o}mez Maqueo Chew}, Y. and {G{\'o}mez-Mu{\~n}oz}, M.~A. and {G{\"u}nther}, M.~N. and {Harakawa}, H. and {Heng}, K. and {Hooton}, M.~J. and {Hori}, Y. and {Ikoma}, M. and {Jehin}, E. and {Jenkins}, J.~M. and {Kagetani}, T. and {Kawauchi}, K. and {Kimura}, T. and {Kodama}, T. and {Kotani}, T. and {Krishnamurthy}, V. and {Kudo}, T. and {Kunovac}, V. and {Kusakabe}, N. and {Latham}, D.~W. and {Littlefield}, C. and {McCormac}, J. and {Melis}, C. and {Mori}, M. and {Murgas}, F. and {Palle}, E. and {Pedersen}, P.~P. and {Queloz}, D. and {Ricker}, G. and {Sabin}, L. and {Schanche}, N. and {Schroffenegger}, U. and {Seager}, S. and {Shiao}, B. and {Sohy}, S. and {Standing}, M.~R. and {Tamura}, M. and {Theissen}, C.~A. and {Thompson}, S.~J. and {Triaud}, A.~H.~M.~J. and {Vanderspek}, R. and {Vievard}, S. and {Wells}, R.~D. and {Winn}, J.~N. and {Zou}, Y. and {Z{\'u}{\~n}iga-Fern{\'a}ndez}, S. and {Gillon}, M.},
        title = "{Two temperate super-Earths transiting a nearby late-type M dwarf}",
      journal = {\aap},
         year = 2022,
        month = nov,
       volume = {667},
          eid = {A59},
        pages = {A59},
          doi = {10.1051/0004-6361/202244041},
archivePrefix = {arXiv},
       eprint = {2209.02831},
 primaryClass = {astro-ph.EP},
       adsurl = {https://ui.adsabs.harvard.edu/abs/2022A&A...667A..59D}
}

@ARTICLE{Ducrot2025,
       author = {{Ducrot}, Elsa and {Lagage}, Pierre-Olivier and {Min}, Michiel and {Gillon}, Micha{\"e}l and {Bell}, Taylor J. and {Tremblin}, Pascal and {Greene}, Thomas and {Dyrek}, Achr{\`e}ne and {Bouwman}, Jeroen and {Waters}, Rens and {G{\"u}del}, Manuel and {Henning}, Thomas and {Vandenbussche}, Bart and {Absil}, Olivier and {Barrado}, David and {Boccaletti}, Anthony and {Coulais}, Alain and {Decin}, Leen and {Edwards}, Billy and {Gastaud}, Ren{\'e} and {Glasse}, Alistair and {Kendrew}, Sarah and {Olofsson}, Goran and {Patapis}, Polychronis and {Pye}, John and {Rouan}, Daniel and {Whiteford}, Niall and {Argyriou}, Ioannis and {Cossou}, Christophe and {Glauser}, Adrian M. and {Krause}, Oliver and {Lahuis}, Fred and {Royer}, Pierre and {Scheithauer}, Silvia and {Colina}, Luis and {van Dishoeck}, Ewine F. and {Ostlin}, G{\"o}ran and {Ray}, Tom P. and {Wright}, Gillian},
        title = "{Combined analysis of the 12.8 and 15 {\ensuremath{\mu}}m JWST/MIRI eclipse observations of TRAPPIST-1 b}",
      journal = {Nature Astronomy},
         year = 2025,
        month = mar,
       volume = {9},
        pages = {358-369},
          doi = {10.1038/s41550-024-02428-z},
archivePrefix = {arXiv},
       eprint = {2412.11627},
 primaryClass = {astro-ph.EP},
       adsurl = {https://ui.adsabs.harvard.edu/abs/2025NatAs...9..358D}
}

@ARTICLE{Henry2006,
       author = {{Henry}, Todd J. and {Jao}, Wei-Chun and {Subasavage}, John P. and {Beaulieu}, Thomas D. and {Ianna}, Philip A. and {Costa}, Edgardo and {M{\'e}ndez}, Ren{\'e} A.},
        title = "{The Solar Neighborhood. XVII. Parallax Results from the CTIOPI 0.9 m Program: 20 New Members of the RECONS 10 Parsec Sample}",
      journal = {\aj},
         year = 2006,
        month = dec,
       volume = {132},
       number = {6},
        pages = {2360-2371},
          doi = {10.1086/508233},
archivePrefix = {arXiv},
       eprint = {astro-ph/0608230},
 primaryClass = {astro-ph},
       adsurl = {https://ui.adsabs.harvard.edu/abs/2006AJ....132.2360H}
}

@ARTICLE{Hurley2002,
       author = {{Hurley}, Jarrod R. and {Tout}, Christopher A. and {Pols}, Onno R.},
        title = "{Evolution of binary stars and the effect of tides on binary populations}",
      journal = {\mnras},
         year = 2002,
        month = feb,
       volume = {329},
       number = {4},
        pages = {897-928},
          doi = {10.1046/j.1365-8711.2002.05038.x},
archivePrefix = {arXiv},
       eprint = {astro-ph/0201220},
 primaryClass = {astro-ph},
       adsurl = {https://ui.adsabs.harvard.edu/abs/2002MNRAS.329..897H}
}

@ARTICLE{Skumanich1972,
       author = {{Skumanich}, A.},
        title = "{Time Scales for Ca II Emission Decay, Rotational Braking, and Lithium Depletion}",
      journal = {\apj},
         year = 1972,
        month = feb,
       volume = {171},
        pages = {565},
          doi = {10.1086/151310},
       adsurl = {https://ui.adsabs.harvard.edu/abs/1972ApJ...171..565S}
}

@ARTICLE{Brogi2012-tauBoo,
       author = {{Brogi}, Matteo and {Snellen}, Ignas A.~G. and {de Kok}, Remco J. and {Albrecht}, Simon and {Birkby}, Jayne and {de Mooij}, Ernst J.~W.},
        title = "{The signature of orbital motion from the dayside of the planet {\ensuremath{\tau}} Bo{\"o}tis b}",
      journal = {\nat},
         year = 2012,
        month = jun,
       volume = {486},
       number = {7404},
        pages = {502-504},
          doi = {10.1038/nature11161},
archivePrefix = {arXiv},
       eprint = {1206.6109},
 primaryClass = {astro-ph.EP},
       adsurl = {https://ui.adsabs.harvard.edu/abs/2012Natur.486..502B}
}

@ARTICLE{Birkby2017-51peg,
       author = {{Birkby}, J.~L. and {de Kok}, R.~J. and {Brogi}, M. and {Schwarz}, H. and {Snellen}, I.~A.~G.},
        title = "{Discovery of Water at High Spectral Resolution in the Atmosphere of 51 Peg b}",
      journal = {\aj},
         year = 2017,
        month = mar,
       volume = {153},
       number = {3},
          eid = {138},
        pages = {138},
          doi = {10.3847/1538-3881/aa5c87},
archivePrefix = {arXiv},
       eprint = {1701.07257},
 primaryClass = {astro-ph.EP},
       adsurl = {https://ui.adsabs.harvard.edu/abs/2017AJ....153..138B}
}

@Inbook{Topping1972,
author="Topping, J.",
title="Some Statistical Ideas",
bookTitle="Errors of Observation and their Treatment",
year="1972",
publisher="Springer Netherlands",
address="Dordrecht",
pages="29--71",
isbn="978-94-011-6928-8",
doi="10.1007/978-94-011-6928-8_2",
url="https://doi.org/10.1007/978-94-011-6928-8_2"
}

@book{UaaoS,
publisher = {Oxford University Press},
author = {Reichmann, W. J.},
booktitle = {Use and abuse of statistics /},
year = {1962},
title = {Use and abuse of statistics},
language = {eng},
}

@ARTICLE{dynesty,
       author = {{Speagle}, Joshua S.},
        title = "{DYNESTY: a dynamic nested sampling package for estimating Bayesian posteriors and evidences}",
      journal = {\mnras},
         year = 2020,
        month = apr,
       volume = {493},
       number = {3},
        pages = {3132-3158},
          doi = {10.1093/mnras/staa278},
archivePrefix = {arXiv},
       eprint = {1904.02180},
 primaryClass = {astro-ph.IM},
       adsurl = {https://ui.adsabs.harvard.edu/abs/2020MNRAS.493.3132S}
}

@ARTICLE{SME1996,
       author = {{Valenti}, J.~A. and {Piskunov}, N.},
        title = "{Spectroscopy made easy: A new tool for fitting observations with synthetic spectra.}",
      journal = {\aaps},
         year = 1996,
        month = sep,
       volume = {118},
        pages = {595-603},
       adsurl = {https://ui.adsabs.harvard.edu/abs/1996A&AS..118..595V}
}

@ARTICLE{Mortier2013,
       author = {{Mortier}, A. and {Santos}, N.~C. and {Sousa}, S.~G. and {Fernandes}, J.~M. and {Adibekyan}, V. Zh. and {Delgado Mena}, E. and {Montalto}, M. and {Israelian}, G.},
        title = "{New and updated stellar parameters for 90 transit hosts. The effect of the surface gravity}",
      journal = {\aap},
         year = 2013,
        month = oct,
       volume = {558},
          eid = {A106},
        pages = {A106},
          doi = {10.1051/0004-6361/201322240},
archivePrefix = {arXiv},
       eprint = {1309.1998},
 primaryClass = {astro-ph.EP},
       adsurl = {https://ui.adsabs.harvard.edu/abs/2013A&A...558A.106M}
}

@ARTICLE{SethiMartin2024,
       author = {{Sethi}, Ritika and {Martin}, David V.},
        title = "{Tight stellar binaries favour active longitudes at sub- and antistellar points}",
      journal = {\mnras},
         year = 2024,
        month = apr,
       volume = {529},
       number = {4},
        pages = {4442-4454},
          doi = {10.1093/mnras/stae717},
archivePrefix = {arXiv},
       eprint = {2403.03065},
 primaryClass = {astro-ph.SR},
       adsurl = {https://ui.adsabs.harvard.edu/abs/2024MNRAS.529.4442S}
}

@ARTICLE{Faria2016,
       author = {{Faria}, J.~P. and {Haywood}, R.~D. and {Brewer}, B.~J. and {Figueira}, P. and {Oshagh}, M. and {Santerne}, A. and {Santos}, N.~C.},
        title = "{Uncovering the planets and stellar activity of CoRoT-7 using only radial velocities}",
      journal = {\aap},
         year = 2016,
        month = apr,
       volume = {588},
          eid = {A31},
        pages = {A31},
          doi = {10.1051/0004-6361/201527899},
archivePrefix = {arXiv},
       eprint = {1601.07495},
 primaryClass = {astro-ph.EP},
       adsurl = {https://ui.adsabs.harvard.edu/abs/2016A&A...588A..31F}
}

@ARTICLE{Berger2020,
       author = {{Berger}, Travis A. and {Huber}, Daniel and {van Saders}, Jennifer L. and {Gaidos}, Eric and {Tayar}, Jamie and {Kraus}, Adam L.},
        title = "{The Gaia-Kepler Stellar Properties Catalog. I. Homogeneous Fundamental Properties for 186,301 Kepler Stars}",
      journal = {\aj},
         year = 2020,
        month = jun,
       volume = {159},
       number = {6},
          eid = {280},
        pages = {280},
          doi = {10.3847/1538-3881/159/6/280},
archivePrefix = {arXiv},
       eprint = {2001.07737},
 primaryClass = {astro-ph.SR},
       adsurl = {https://ui.adsabs.harvard.edu/abs/2020AJ....159..280B}
}

@ARTICLE{maxted-pycheops,
       author = {{Maxted}, P.~F.~L. and {Ehrenreich}, D. and {Wilson}, T.~G. and {Alibert}, Y. and {Cameron}, A. Collier and {Hoyer}, S. and {Sousa}, S.~G. and {Olofsson}, G. and {Bekkelien}, A. and {Deline}, A. and {Delrez}, L. and {Bonfanti}, A. and {Borsato}, L. and {Alonso}, R. and {Anglada Escud{\'e}}, G. and {Barrado}, D. and {Barros}, S.~C.~C. and {Baumjohann}, W. and {Beck}, M. and {Beck}, T. and {Benz}, W. and {Billot}, N. and {Biondi}, F. and {Bonfils}, X. and {Brandeker}, A. and {Broeg}, C. and {B{\'a}rczy}, T. and {Cabrera}, J. and {Charnoz}, S. and {Corral Van Damme}, C. and {Csizmadia}, Sz and {Davies}, M.~B. and {Deleuil}, M. and {Demangeon}, O.~D.~S. and {Demory}, B.-O. and {Erikson}, A. and {Flor{\'e}n}, H.~G. and {Fortier}, A. and {Fossati}, L. and {Fridlund}, M. and {Futyan}, D. and {Gandolfi}, D. and {Gillon}, M. and {Guedel}, M. and {Guterman}, P. and {Heng}, K. and {Isaak}, K.~G. and {Kiss}, L. and {Laskar}, J. and {Lecavelier des Etangs}, A. and {Lendl}, M. and {Lovis}, C. and {Magrin}, D. and {Nascimbeni}, V. and {Ottensamer}, R. and {Pagano}, I. and {Pall{\'e}}, E. and {Peter}, G. and {Piotto}, G. and {Pollacco}, D. and {Pozuelos}, F.~J. and {Queloz}, D. and {Ragazzoni}, R. and {Rando}, N. and {Rauer}, H. and {Reimers}, C. and {Ribas}, I. and {Salmon}, S. and {Santos}, N.~C. and {Scandariato}, G. and {Simon}, A.~E. and {Smith}, A.~M.~S. and {Steller}, M. and {Swayne}, M.~I. and {Szab{\'o}}, Gy M. and {S{\'e}gransan}, D. and {Thomas}, N. and {Udry}, S. and {Van Grootel}, V. and {Walton}, N.~A.},
        title = "{Analysis of Early Science observations with the CHaracterising ExOPlanets Satellite (CHEOPS) using PYCHEOPS}",
      journal = {\mnras},
         year = 2022,
        month = jul,
       volume = {514},
       number = {1},
        pages = {77-104},
          doi = {10.1093/mnras/stab3371},
archivePrefix = {arXiv},
       eprint = {2111.08828},
 primaryClass = {astro-ph.EP},
       adsurl = {https://ui.adsabs.harvard.edu/abs/2022MNRAS.514...77M}
}

@ARTICLE{ElBadry2018,
       author = {{El-Badry}, Kareem and {Rix}, Hans-Walter and {Ting}, Yuan-Sen and {Weisz}, Daniel R. and {Bergemann}, Maria and {Cargile}, Phillip and {Conroy}, Charlie and {Eilers}, Anna-Christina},
        title = "{Signatures of unresolved binaries in stellar spectra: implications for spectral fitting}",
      journal = {\mnras},
         year = 2018,
        month = feb,
       volume = {473},
       number = {4},
        pages = {5043-5049},
          doi = {10.1093/mnras/stx2758},
archivePrefix = {arXiv},
       eprint = {1709.03983},
 primaryClass = {astro-ph.SR},
       adsurl = {https://ui.adsabs.harvard.edu/abs/2018MNRAS.473.5043E}
}

@ARTICLE{Winters2026,
       author = {{Winters}, Jennifer G. and {Pass}, Emily K. and {Charbonneau}, David and {Irwin}, Jonathan and {Latham}, David W. and {Mink}, Jessica and {Esquerdo}, Gilbert A. and {Berlind}, Perry and {Calkins}, Michael L. and {Payne}, Matthew J.},
        title = "{Radial and Rotational Velocities of a Volume-Complete Sample of M Dwarfs with Masses 0.1-0.3 Msun within 15 parsecs}",
      journal = {arXiv e-prints},
         year = 2026,
        month = jan,
          eid = {arXiv:2602.00382},
        pages = {arXiv:2602.00382},
          doi = {10.48550/arXiv.2602.00382},
archivePrefix = {arXiv},
       eprint = {2602.00382},
 primaryClass = {astro-ph.SR},
       adsurl = {https://ui.adsabs.harvard.edu/abs/2026arXiv260200382W}
}

@ARTICLE{espresso,
       author = {{Pepe}, F. and {Cristiani}, S. and {Rebolo}, R. and {Santos}, N.~C. and {Dekker}, H. and {Cabral}, A. and {Di Marcantonio}, P. and {Figueira}, P. and {Lo Curto}, G. and {Lovis}, C. and {Mayor}, M. and {M{\'e}gevand}, D. and {Molaro}, P. and {Riva}, M. and {Zapatero Osorio}, M.~R. and {Amate}, M. and {Manescau}, A. and {Pasquini}, L. and {Zerbi}, F.~M. and {Adibekyan}, V. and {Abreu}, M. and {Affolter}, M. and {Alibert}, Y. and {Aliverti}, M. and {Allart}, R. and {Allende Prieto}, C. and {{\'A}lvarez}, D. and {Alves}, D. and {Avila}, G. and {Baldini}, V. and {Bandy}, T. and {Barros}, S.~C.~C. and {Benz}, W. and {Bianco}, A. and {Borsa}, F. and {Bourrier}, V. and {Bouchy}, F. and {Broeg}, C. and {Calderone}, G. and {Cirami}, R. and {Coelho}, J. and {Conconi}, P. and {Coretti}, I. and {Cumani}, C. and {Cupani}, G. and {D'Odorico}, V. and {Damasso}, M. and {Deiries}, S. and {Delabre}, B. and {Demangeon}, O.~D.~S. and {Dumusque}, X. and {Ehrenreich}, D. and {Faria}, J.~P. and {Fragoso}, A. and {Genolet}, L. and {Genoni}, M. and {G{\'e}nova Santos}, R. and {Gonz{\'a}lez Hern{\'a}ndez}, J.~I. and {Hughes}, I. and {Iwert}, O. and {Kerber}, F. and {Knudstrup}, J. and {Landoni}, M. and {Lavie}, B. and {Lillo-Box}, J. and {Lizon}, J.-L. and {Maire}, C. and {Martins}, C.~J.~A.~P. and {Mehner}, A. and {Micela}, G. and {Modigliani}, A. and {Monteiro}, M.~A. and {Monteiro}, M.~J.~P.~F.~G. and {Moschetti}, M. and {Murphy}, M.~T. and {Nunes}, N. and {Oggioni}, L. and {Oliveira}, A. and {Oshagh}, M. and {Pall{\'e}}, E. and {Pariani}, G. and {Poretti}, E. and {Rasilla}, J.~L. and {Rebord{\~a}o}, J. and {Redaelli}, E.~M. and {Santana Tschudi}, S. and {Santin}, P. and {Santos}, P. and {S{\'e}gransan}, D. and {Schmidt}, T.~M. and {Segovia}, A. and {Sosnowska}, D. and {Sozzetti}, A. and {Sousa}, S.~G. and {Span{\`o}}, P. and {Su{\'a}rez Mascare{\~n}o}, A. and {Tabernero}, H. and {Tenegi}, F. and {Udry}, S. and {Zanutta}, A.},
        title = "{ESPRESSO at VLT. On-sky performance and first results}",
      journal = {\aap},
         year = 2021,
        month = jan,
       volume = {645},
          eid = {A96},
        pages = {A96},
          doi = {10.1051/0004-6361/202038306},
archivePrefix = {arXiv},
       eprint = {2010.00316},
 primaryClass = {astro-ph.IM},
       adsurl = {https://ui.adsabs.harvard.edu/abs/2021A&A...645A..96P}
}

@ARTICLE{DOLBY,
       author = {{Sairam}, Lalitha and {Baycroft}, Thomas A. and {Boisse}, Isabelle and {Heidari}, Neda and {Santerne}, Alexandre and {Triaud}, Amaury H.~M.~J. and {Coleman}, Gavin A.~L. and {Davis}, Yasmin T. and {Deleuil}, Magali and {H{\'e}brard}, Guillaume and {Martin}, David V. and {Maxted}, Pierre F.~L. and {Nelson}, Richard P. and {Sebastian}, Daniel and {Scutt}, Owen J. and {Standing}, Matthew R.},
        title = "{BEBOP VI. Enabling the detection of circumbinary planets orbiting double-lined binaries with the DOLBY method of radial-velocity extraction}",
      journal = {\mnras},
         year = 2024,
        month = nov,
       volume = {534},
       number = {4},
        pages = {3999-4017},
          doi = {10.1093/mnras/stae2317},
archivePrefix = {arXiv},
       eprint = {2410.02573},
 primaryClass = {astro-ph.EP},
       adsurl = {https://ui.adsabs.harvard.edu/abs/2024MNRAS.534.3999S}
}

@INPROCEEDINGS{Offner2023,
       author = {{Offner}, S.~S.~R. and {Moe}, M. and {Kratter}, K.~M. and {Sadavoy}, S.~I. and {Jensen}, E.~L.~N. and {Tobin}, J.~J.},
        title = "{The Origin and Evolution of Multiple Star Systems}",
    booktitle = {Protostars and Planets VII},
         year = 2023,
       editor = {{Inutsuka}, S. and {Aikawa}, Y. and {Muto}, T. and {Tomida}, K. and {Tamura}, M.},
       series = {Astronomical Society of the Pacific Conference Series},
       volume = {534},
        month = jul,
        pages = {275},
          doi = {10.48550/arXiv.2203.10066},
archivePrefix = {arXiv},
       eprint = {2203.10066},
 primaryClass = {astro-ph.SR},
       adsurl = {https://ui.adsabs.harvard.edu/abs/2023ASPC..534..275O}
}

@ARTICLE{GaiaDR3,
       author = {{Gaia Collaboration} and {Vallenari}, A. and {Brown}, A.~G.~A. and {Prusti}, T. and {de Bruijne}, J.~H.~J. and {Arenou}, F. and {Babusiaux}, C. and {Biermann}, M. and {Creevey}, O.~L. and {Ducourant}, C. and {Evans}, D.~W. and {Eyer}, L. and {Guerra}, R. and {Hutton}, A. and {Jordi}, C. and {Klioner}, S.~A. and {Lammers}, U.~L. and {Lindegren}, L. and {Luri}, X. and {Mignard}, F. and {Panem}, C. and {Pourbaix}, D. and {Randich}, S. and {Sartoretti}, P. and {Soubiran}, C. and {Tanga}, P. and {Walton}, N.~A. and {Bailer-Jones}, C.~A.~L. and {Bastian}, U. and {Drimmel}, R. and {Jansen}, F. and {Katz}, D. and {Lattanzi}, M.~G. and {van Leeuwen}, F. and {Bakker}, J. and {Cacciari}, C. and {Casta{\~n}eda}, J. and {De Angeli}, F. and {Fabricius}, C. and {Fouesneau}, M. and {Fr{\'e}mat}, Y. and {Galluccio}, L. and {Guerrier}, A. and {Heiter}, U. and {Masana}, E. and {Messineo}, R. and {Mowlavi}, N. and {Nicolas}, C. and {Nienartowicz}, K. and {Pailler}, F. and {Panuzzo}, P. and {Riclet}, F. and {Roux}, W. and {Seabroke}, G.~M. and {Sordo}, R. and {Th{\'e}venin}, F. and {Gracia-Abril}, G. and {Portell}, J. and {Teyssier}, D. and {Altmann}, M. and {Andrae}, R. and {Audard}, M. and {Bellas-Velidis}, I. and {Benson}, K. and {Berthier}, J. and {Blomme}, R. and {Burgess}, P.~W. and {Busonero}, D. and {Busso}, G. and {C{\'a}novas}, H. and {Carry}, B. and {Cellino}, A. and {Cheek}, N. and {Clementini}, G. and {Damerdji}, Y. and {Davidson}, M. and {de Teodoro}, P. and {Nu{\~n}ez Campos}, M. and {Delchambre}, L. and {Dell'Oro}, A. and {Esquej}, P. and {Fern{\'a}ndez-Hern{\'a}ndez}, J. and {Fraile}, E. and {Garabato}, D. and {Garc{\'\i}a-Lario}, P. and {Gosset}, E. and {Haigron}, R. and {Halbwachs}, J.-L. and {Hambly}, N.~C. and {Harrison}, D.~L. and {Hern{\'a}ndez}, J. and {Hestroffer}, D. and {Hodgkin}, S.~T. and {Holl}, B. and {Jan{\ss}en}, K. and {Jevardat de Fombelle}, G. and {Jordan}, S. and {Krone-Martins}, A. and {Lanzafame}, A.~C. and {L{\"o}ffler}, W. and {Marchal}, O. and {Marrese}, P.~M. and {Moitinho}, A. and {Muinonen}, K. and {Osborne}, P. and {Pancino}, E. and {Pauwels}, T. and {Recio-Blanco}, A. and {Reyl{\'e}}, C. and {Riello}, M. and {Rimoldini}, L. and {Roegiers}, T. and {Rybizki}, J. and {Sarro}, L.~M. and {Siopis}, C. and {Smith}, M. and {Sozzetti}, A. and {Utrilla}, E. and {van Leeuwen}, M. and {Abbas}, U. and {{\'A}brah{\'a}m}, P. and {Abreu Aramburu}, A. and {Aerts}, C. and {Aguado}, J.~J. and {Ajaj}, M. and {Aldea-Montero}, F. and {Altavilla}, G. and {{\'A}lvarez}, M.~A. and {Alves}, J. and {Anders}, F. and {Anderson}, R.~I. and {Anglada Varela}, E. and {Antoja}, T. and {Baines}, D. and {Baker}, S.~G. and {Balaguer-N{\'u}{\~n}ez}, L. and {Balbinot}, E. and {Balog}, Z. and {Barache}, C. and {Barbato}, D. and {Barros}, M. and {Barstow}, M.~A. and {Bartolom{\'e}}, S. and {Bassilana}, J.-L. and {Bauchet}, N. and {Becciani}, U. and {Bellazzini}, M. and {Berihuete}, A. and {Bernet}, M. and {Bertone}, S. and {Bianchi}, L. and {Binnenfeld}, A. and {Blanco-Cuaresma}, S. and {Blazere}, A. and {Boch}, T. and {Bombrun}, A. and {Bossini}, D. and {Bouquillon}, S. and {Bragaglia}, A. and {Bramante}, L. and {Breedt}, E. and {Bressan}, A. and {Brouillet}, N. and {Brugaletta}, E. and {Bucciarelli}, B. and {Burlacu}, A. and {Butkevich}, A.~G. and {Buzzi}, R. and {Caffau}, E. and {Cancelliere}, R. and {Cantat-Gaudin}, T. and {Carballo}, R. and {Carlucci}, T. and {Carnerero}, M.~I. and {Carrasco}, J.~M. and {Casamiquela}, L. and {Castellani}, M. and {Castro-Ginard}, A. and {Chaoul}, L. and {Charlot}, P. and {Chemin}, L. and {Chiaramida}, V. and {Chiavassa}, A. and {Chornay}, N. and {Comoretto}, G. and {Contursi}, G. and {Cooper}, W.~J. and {Cornez}, T. and {Cowell}, S. and {Crifo}, F. and {Cropper}, M. and {Crosta}, M. and {Crowley}, C. and {Dafonte}, C. and {Dapergolas}, A. and {David}, M. and {David}, P. and {de Laverny}, P. and {De Luise}, F. and {De March}, R.},
        title = "{Gaia Data Release 3. Summary of the content and survey properties}",
      journal = {\aap},
         year = 2023,
        month = jun,
       volume = {674},
          eid = {A1},
        pages = {A1},
          doi = {10.1051/0004-6361/202243940},
archivePrefix = {arXiv},
       eprint = {2208.00211},
 primaryClass = {astro-ph.GA},
       adsurl = {https://ui.adsabs.harvard.edu/abs/2023A&A...674A...1G}
}

@ARTICLE{Maxted-EWs,
       author = {{Maxted}, Pierre F.~L.},
        title = "{Equivalent Width of the Interstellar Na I D$_{1}$ and D$_{2}$ Absorption Lines for Stars with E(B {\ensuremath{-}} V) < 0.15}",
      journal = {Research Notes of the American Astronomical Society},
         year = 2025,
        month = jun,
       volume = {9},
       number = {6},
          eid = {146},
        pages = {146},
          doi = {10.3847/2515-5172/ade39b},
       adsurl = {https://ui.adsabs.harvard.edu/abs/2025RNAAS...9..146M}
}

@INPROCEEDINGS{Cayrel1988,
       author = {{Cayrel}, R.},
        title = "{Data Analysis}",
    booktitle = {The Impact of Very High S/N Spectroscopy on Stellar Physics},
         year = 1988,
       editor = {{Cayrel de Strobel}, G. and {Spite}, Monique},
       series = {IAU Symposium},
       volume = {132},
        month = jan,
        pages = {345},
       adsurl = {https://ui.adsabs.harvard.edu/abs/1988IAUS..132..345C}
}

@ARTICLE{Maxted-BEBOP-3,
       author = {{Maxted}, P.~F.~L. and {Miller}, N.~J. and {Baycroft}, T.~A. and {Sebastian}, D. and {Triaud}, A.~H.~M.~J. and {Martin}, D.~V.},
        title = "{Fundamental effective temperature measurements for eclipsing binary stars ─ VI. Improved methodology and application to the circumbinary planet host star BEBOP-3}",
      journal = {\mnras},
         year = 2025,
        month = dec,
       volume = {544},
       number = {4},
        pages = {4611-4620},
          doi = {10.1093/mnras/staf1856},
archivePrefix = {arXiv},
       eprint = {2510.23218},
 primaryClass = {astro-ph.SR},
       adsurl = {https://ui.adsabs.harvard.edu/abs/2025MNRAS.544.4611M}
}

@ARTICLE{Bohlin2014,
       author = {{Bohlin}, Ralph C. and {Gordon}, Karl D. and {Tremblay}, P.-E.},
        title = "{Techniques and Review of Absolute Flux Calibration from the Ultraviolet to the Mid-Infrared}",
      journal = {\pasp},
         year = 2014,
        month = aug,
       volume = {126},
       number = {942},
        pages = {711},
          doi = {10.1086/677655},
archivePrefix = {arXiv},
       eprint = {1406.1707},
 primaryClass = {astro-ph.IM},
       adsurl = {https://ui.adsabs.harvard.edu/abs/2014PASP..126..711B}
}

@ARTICLE{emcee,
       author = {{Foreman-Mackey}, Daniel and {Hogg}, David W. and {Lang}, Dustin and {Goodman}, Jonathan},
        title = "{emcee: The MCMC Hammer}",
      journal = {\pasp},
         year = 2013,
        month = mar,
       volume = {125},
       number = {925},
        pages = {306},
          doi = {10.1086/670067},
archivePrefix = {arXiv},
       eprint = {1202.3665},
 primaryClass = {astro-ph.IM},
       adsurl = {https://ui.adsabs.harvard.edu/abs/2013PASP..125..306F}
}

@ARTICLE{Jofre2019,
       author = {{Jofr{\'e}}, Paula and {Heiter}, Ulrike and {Soubiran}, Caroline},
        title = "{Accuracy and Precision of Industrial Stellar Abundances}",
      journal = {\araa},
         year = 2019,
        month = aug,
       volume = {57},
        pages = {571-616},
          doi = {10.1146/annurev-astro-091918-104509},
archivePrefix = {arXiv},
       eprint = {1811.08041},
 primaryClass = {astro-ph.SR},
       adsurl = {https://ui.adsabs.harvard.edu/abs/2019ARA&A..57..571J}
}

@ARTICLE{PASTEL,
       author = {{Soubiran}, Caroline and {Le Campion}, Jean-Fran{\c{c}}ois and {Brouillet}, Nathalie and {Chemin}, Laurent},
        title = "{The PASTEL catalogue: 2016 version}",
      journal = {\aap},
         year = 2016,
        month = jun,
       volume = {591},
          eid = {A118},
        pages = {A118},
          doi = {10.1051/0004-6361/201628497},
archivePrefix = {arXiv},
       eprint = {1605.07384},
 primaryClass = {astro-ph.SR},
       adsurl = {https://ui.adsabs.harvard.edu/abs/2016A&A...591A.118S}
}

@ARTICLE{Groenewegen2020,
       author = {{Groenewegen}, M.~A.~T.},
        title = "{Analysing the spectral energy distributions of Galactic classical Cepheids}",
      journal = {\aap},
         year = 2020,
        month = mar,
       volume = {635},
          eid = {A33},
        pages = {A33},
          doi = {10.1051/0004-6361/201937060},
archivePrefix = {arXiv},
       eprint = {2002.02186},
 primaryClass = {astro-ph.SR},
       adsurl = {https://ui.adsabs.harvard.edu/abs/2020A&A...635A..33G}
}

@ARTICLE{Martin2024-CMDra,
       author = {{Martin}, David V. and {Sethi}, Ritika and {Armitage}, Tayt and {Gilbert}, Gregory J. and {Rodr{\'\i}guez Mart{\'\i}nez}, Romy and {Gilbert}, Emily A.},
        title = "{The benchmark M dwarf eclipsing binary CM Draconis with TESS: spots, flares, and ultra-precise parameters}",
      journal = {\mnras},
         year = 2024,
        month = feb,
       volume = {528},
       number = {1},
        pages = {963-975},
          doi = {10.1093/mnras/stae015},
archivePrefix = {arXiv},
       eprint = {2301.10858},
 primaryClass = {astro-ph.SR},
       adsurl = {https://ui.adsabs.harvard.edu/abs/2024MNRAS.528..963M}
}

@ARTICLE{maxted-ellc,
       author = {{Maxted}, P.~F.~L.},
        title = "{ellc: A fast, flexible light curve model for detached eclipsing binary stars and transiting exoplanets}",
      journal = {\aap},
         year = 2016,
        month = jun,
       volume = {591},
          eid = {A111},
        pages = {A111},
          doi = {10.1051/0004-6361/201628579},
archivePrefix = {arXiv},
       eprint = {1603.08484},
 primaryClass = {astro-ph.IM},
       adsurl = {https://ui.adsabs.harvard.edu/abs/2016A&A...591A.111M}
}

@ARTICLE{Bouchy2009,
       author = {{Bouchy}, F. and {H{\'e}brard}, G. and {Udry}, S. and {Delfosse}, X. and {Boisse}, I. and {Desort}, M. and {Bonfils}, X. and {Eggenberger}, A. and {Ehrenreich}, D. and {Forveille}, T. and {Lagrange}, A.~M. and {Le Coroller}, H. and {Lovis}, C. and {Moutou}, C. and {Pepe}, F. and {Perrier}, C. and {Pont}, F. and {Queloz}, D. and {Santos}, N.~C. and {S{\'e}gransan}, D. and {Vidal-Madjar}, A.},
        title = "{The SOPHIE search for northern extrasolar planets . I. A companion around HD 16760 with mass close to the planet/brown-dwarf transition}",
      journal = {\aap},
         year = 2009,
        month = oct,
       volume = {505},
       number = {2},
        pages = {853-858},
          doi = {10.1051/0004-6361/200912427},
       adsurl = {https://ui.adsabs.harvard.edu/abs/2009A&A...505..853B}
}

@ARTICLE{Pepe2004,
       author = {{Pepe}, F. and {Mayor}, M. and {Queloz}, D. and {Benz}, W. and {Bonfils}, X. and {Bouchy}, F. and {Lo Curto}, G. and {Lovis}, C. and {M{\'e}gevand}, D. and {Moutou}, C. and et al.},
        title = "{The HARPS search for southern extra-solar planets. I. HD 330075 b: A new ``hot Jupiter''}",
      journal = {\aap},
         year = 2004,
        month = aug,
       volume = {423},
        pages = {385-389},
          doi = {10.1051/0004-6361:20040389},
archivePrefix = {arXiv},
       eprint = {astro-ph/0405252},
 primaryClass = {astro-ph},
       adsurl = {https://ui.adsabs.harvard.edu/abs/2004A&A...423..385P}
}




\appendix



\section{Additional Figures}

\begin{figure*}
    \centering
    \includegraphics[width=0.99\linewidth]{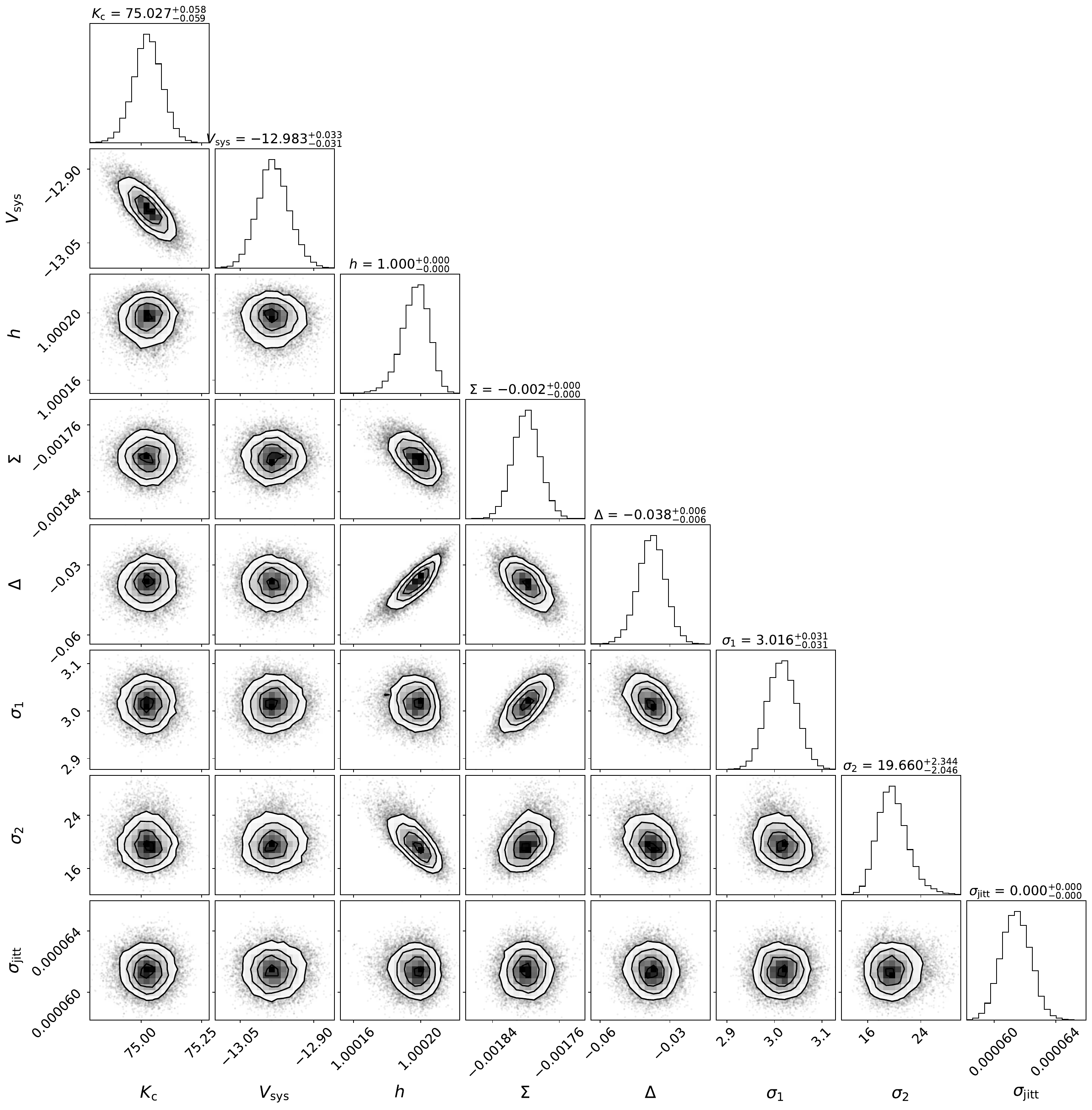}
    \caption{Corner plot showing posterior distributions for the \texttt{Saltire} MCMC fit of J1840+34. This corner plot is included as an example to demonstrate the symmetric nature of the distributions. Quoted uncertainties in titles are precision errors only.}
    \label{fig:example_corner}
\end{figure*}

\begin{figure*}
    \centering
    \includegraphics[width=0.99\linewidth]{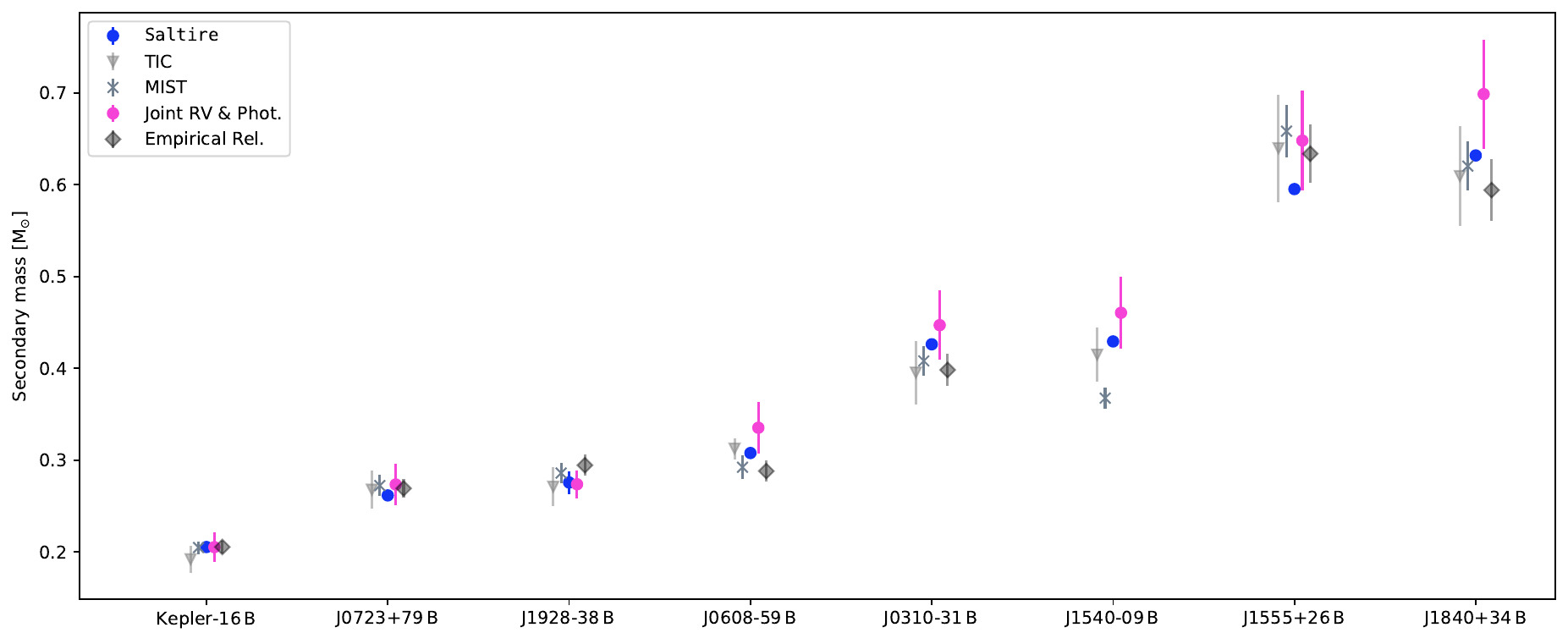}
    \includegraphics[width=0.99\linewidth]{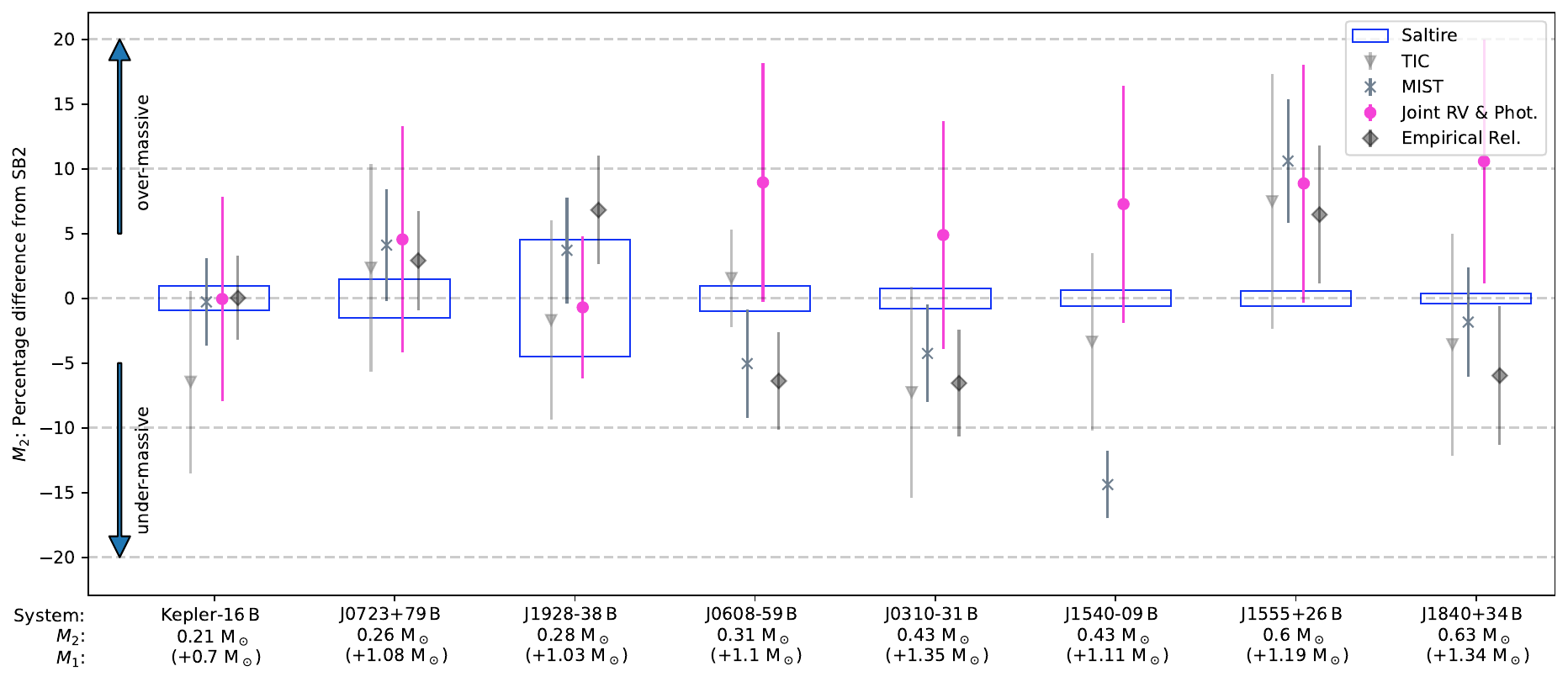}
    \caption{Figures comparing secondary masses from different methods. \textbf{Top:} secondary masses of the eight systems included in our sample (Table~\ref{tab:all-masses}). Secondary masses are estimated directly in the RV+photometry and \texttt{Saltire} methods, whereas the others are calculated from primary mass estimates in combination with inclination and the primary's RV semi-amplitude. Marker colours and styles represent each method. \textbf{Bottom:} Masses plotted as percentage deviation from the \texttt{Saltire} SB2 results (blue boxes, including uncertainty). Targets along the x-axes are ordered by primary mass, denoted below the name in the lower panel. A systematic increase is seen for the RV+Photometry (SB1) method. The 0~per~cent difference is denoted with a horizontal dashed line. Results from other sources are scattered around the plots.}
    \label{fig:secondaries}    
\end{figure*}

\begin{figure*}
    \centering
    \includegraphics[width=0.99\linewidth]{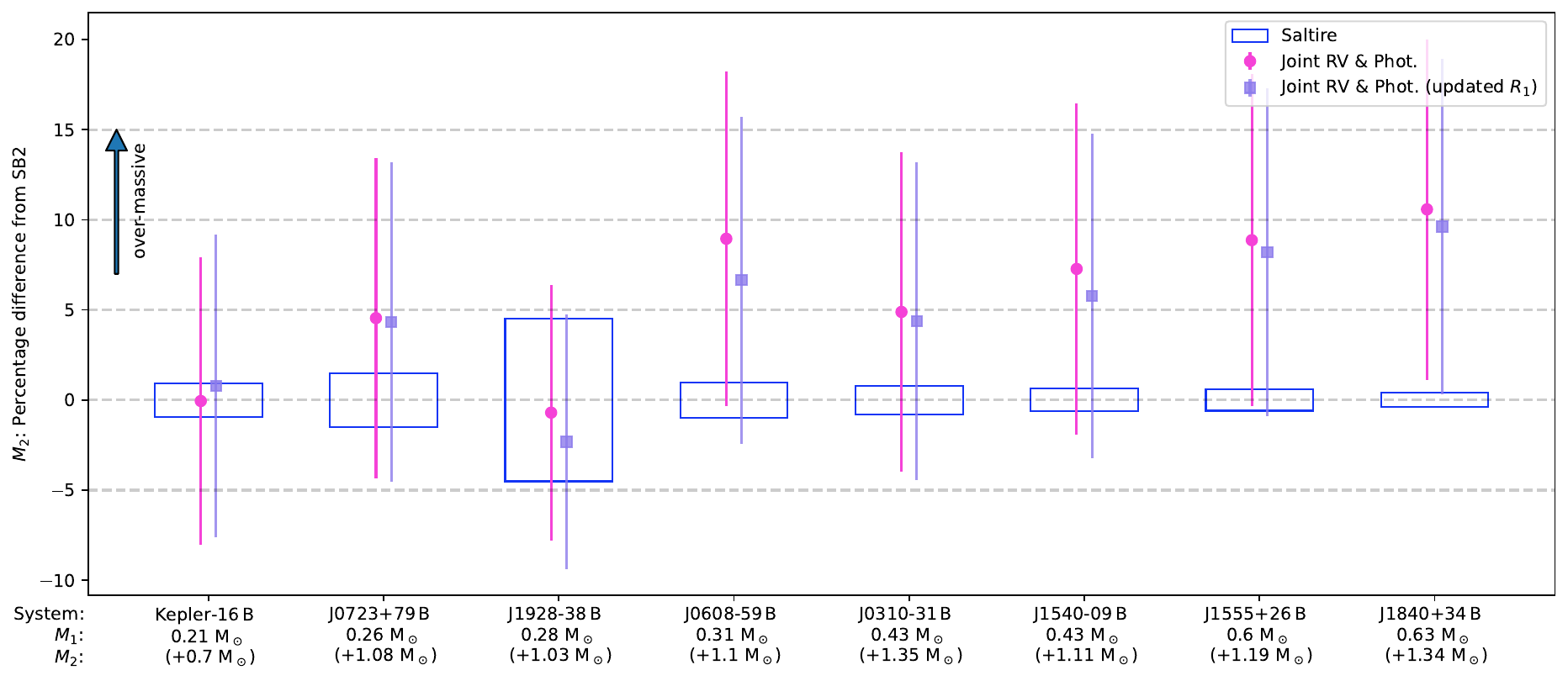}
    \caption{A comparison between SB1 results when using either original or parallax-zero-point-corrected primary radii. The values, secondary masses, are plotted as percentage deviation from the \texttt{Saltire} SB2 results. The updated SB1 masses are in a different colour, and indicated with square markers. For most systems, the difference is almost negligible, and we see that this does not account for the identified systematic bias in results from this method. J1928-38 shows a larger modification (albeit fully within $1\sigma$ uncertainties) that brings the quoted median mass more into alignment with the SB2 measurement. The updated radii have reduced the mass for J1840+34, now enabling formal agreement (within $1\sigma$) between SB1 and SB2 methods for all eight targets considered herein. Targets along the x-axes are ordered by \textbf{secondary} mass.
    }
    \label{fig:deviation-from-saltire-newR1-secondary}    
\end{figure*}

\begin{figure*}
    \centering
    \includegraphics[width=0.99\linewidth]{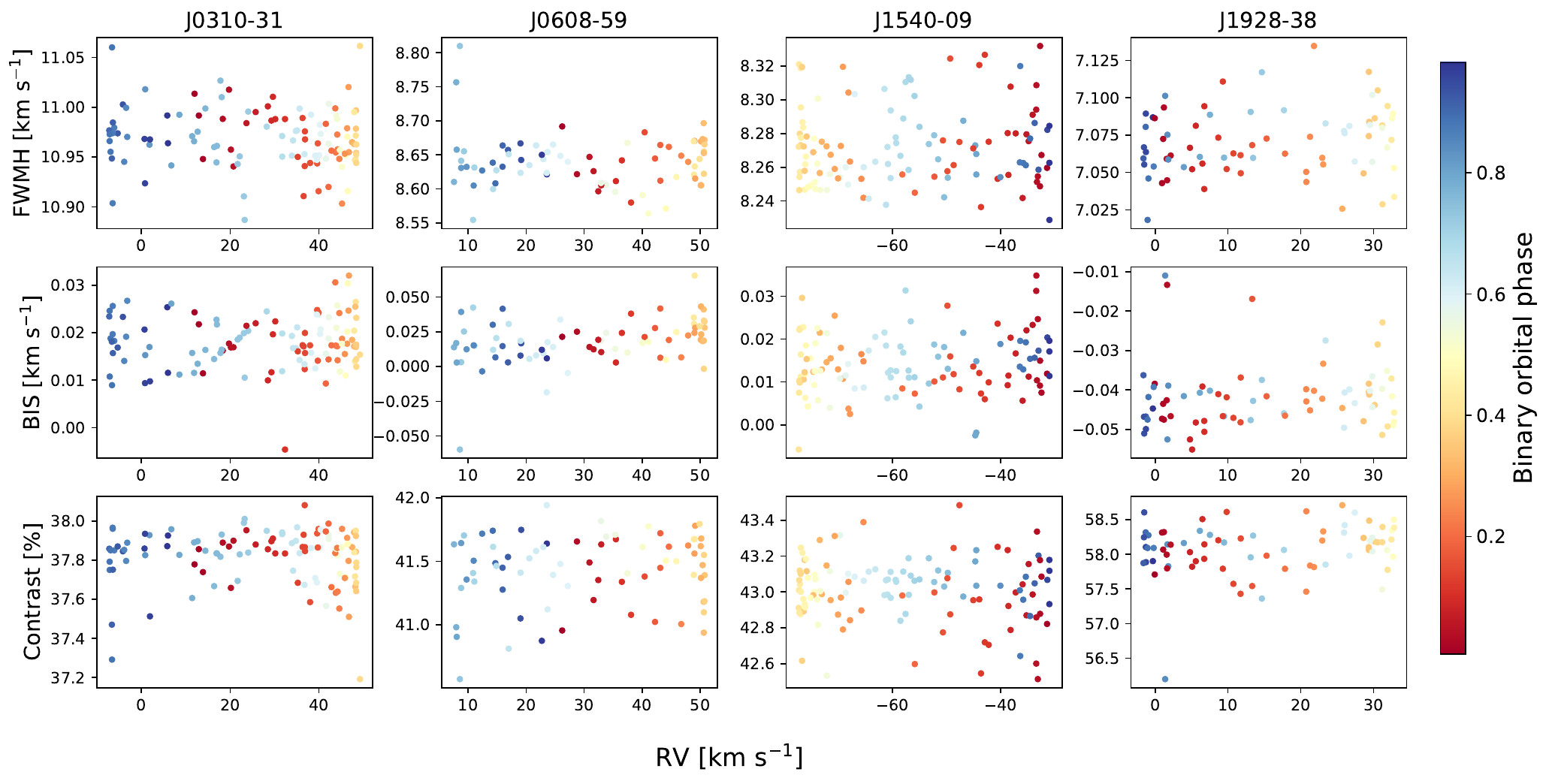}
    \caption{Plots showing correlation between measured RVs and commonly used CCF-based activity indicators. \textbf{Columns} of the plot grid relate to a particular system/primary star, and \textbf{rows} of the grid to each activity indicator. Points are coloured based on binary orbital phase [0,1]. The four systems considered herein with existing HARPS observations are included in this figure.}
    \label{fig:harps_correlations}    
\end{figure*}

\begin{figure*}
    \centering
    \includegraphics[width=0.99\linewidth]{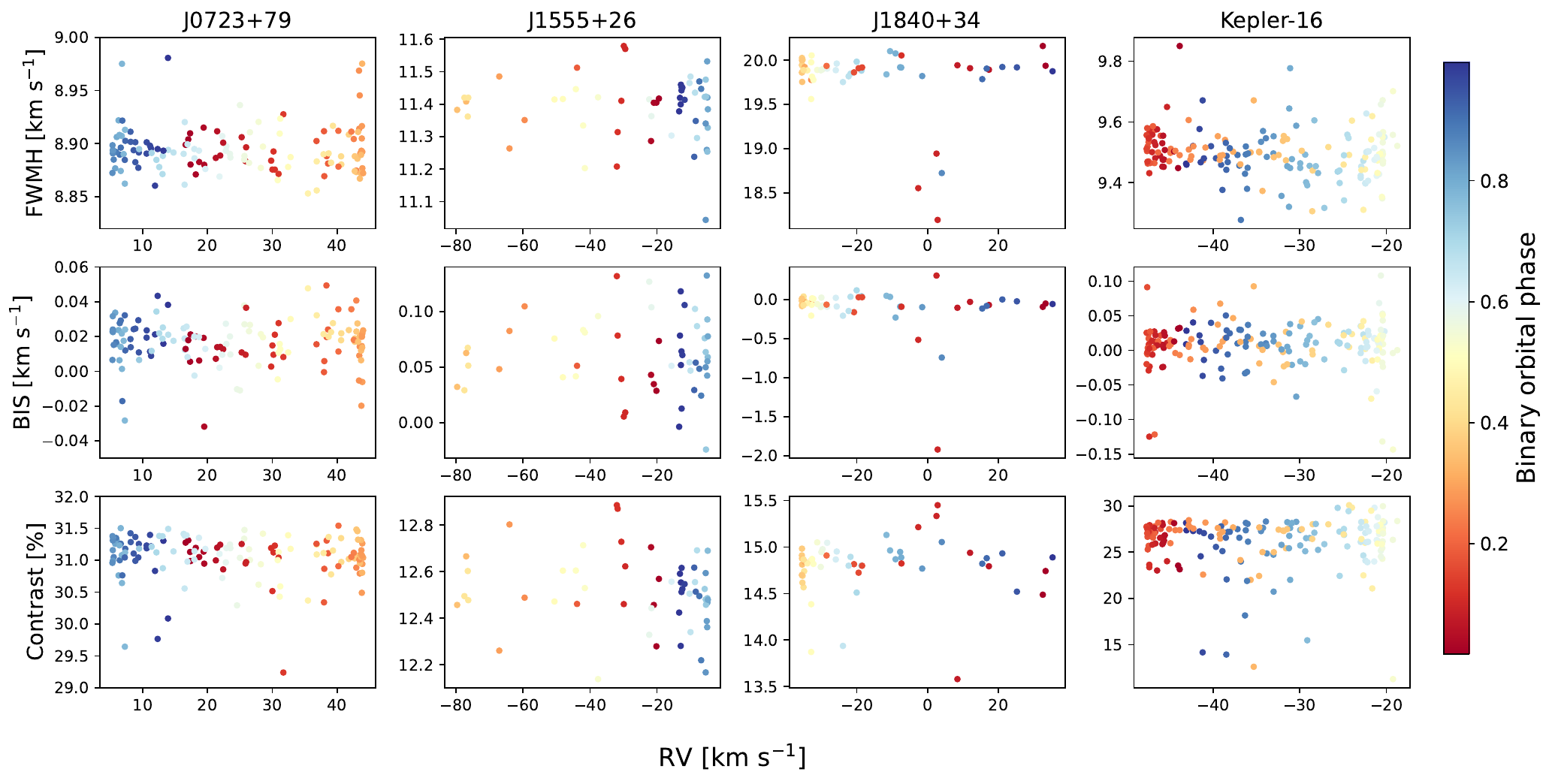}
    \caption{Identical to Figure~\ref{fig:harps_correlations}, now for northern/SOPHIE targets.}
    \label{fig:sophie_correlations}    
\end{figure*}

\begin{figure*}
    \centering
    \includegraphics[width=0.75\linewidth]{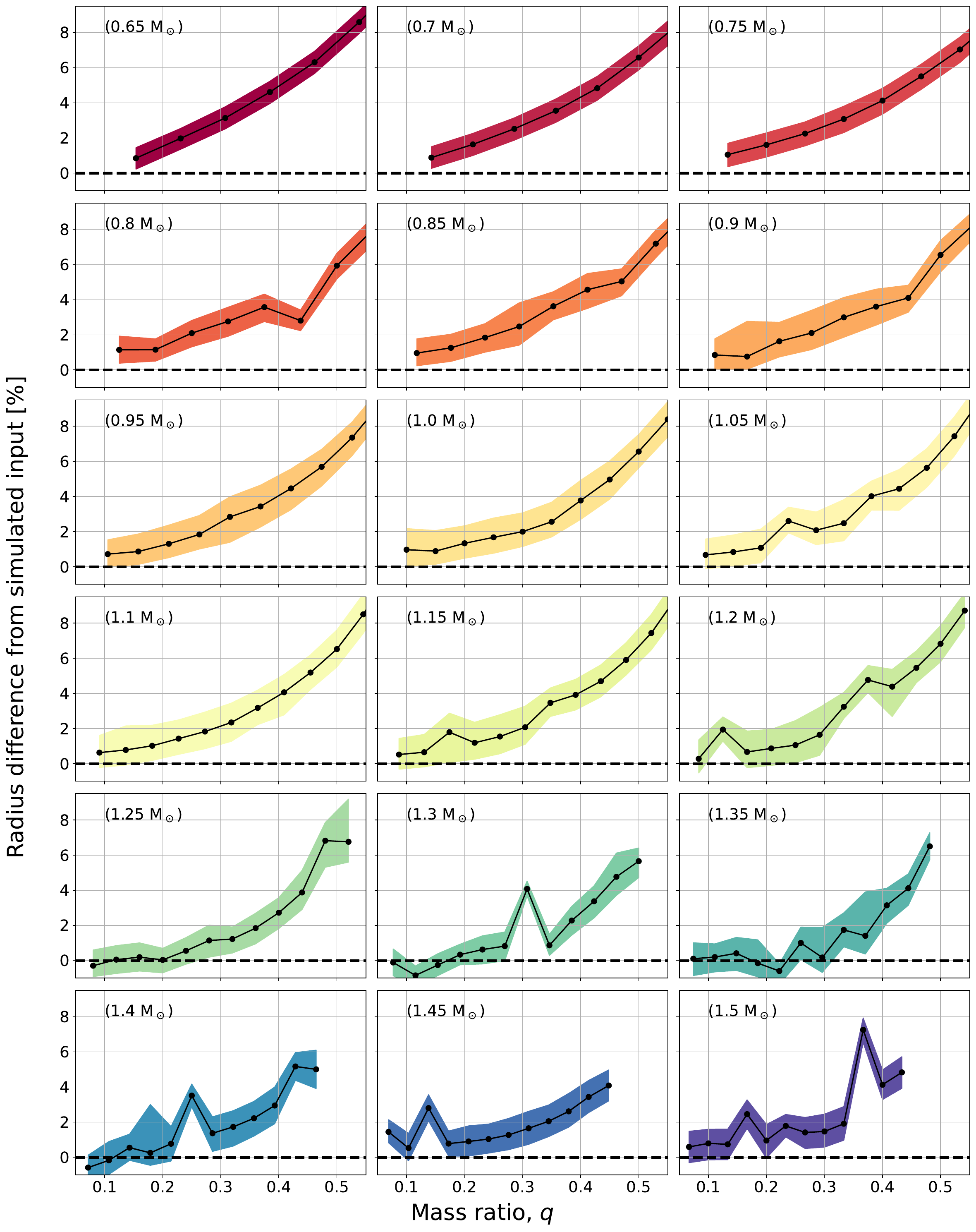}
    \caption{Each panel shows the radius difference between the recovered \texttt{SingleStarModel} isochrone for our simulated binary systems, and the ground truth primary radius from the MIST track used to generate the system parameters. The amount of radius over-estimation increases for larger mass ratios. The over-estimation is similar no matter the mass of the simulated primary, though is slightly stronger for the least massive primaries -- a consequence of cooler stars being closer in temperature to their M dwarf companions. Panels are separated by primary input mass, and the shaded regions aim to show the $1\sigma$ errors on the radii difference.} 
    \label{fig:radius_isochrone_18panel}
\end{figure*}

\begin{figure*}
    \centering
    \includegraphics[width=0.99\linewidth]{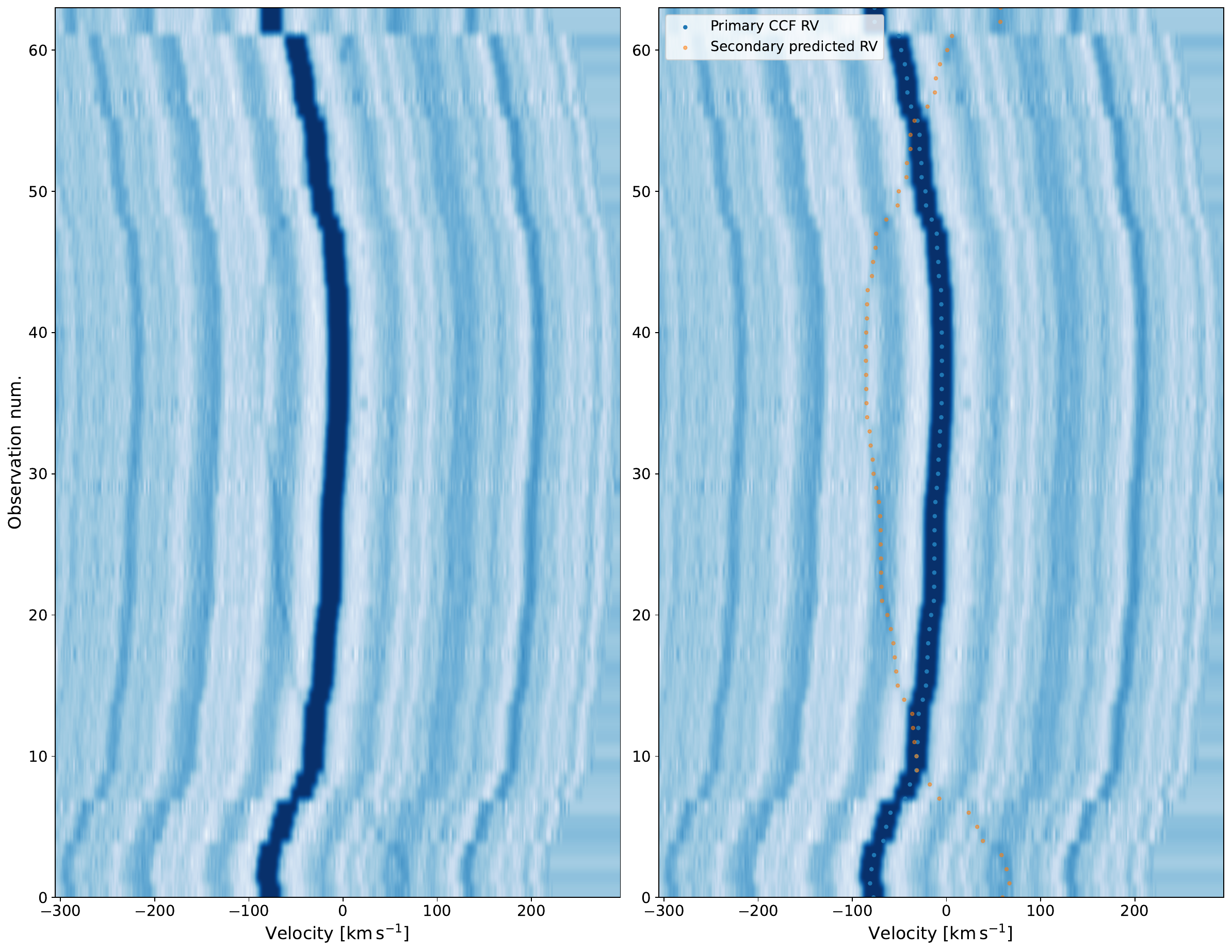}
    \caption{A visual representation of the SOPHIE CCFs (G2 mask) for J1555+26. By ordering the CCFs in orbital phase, the trace of the secondary becomes apparent, despite being indistinguishable from the noise in each single observation. The CCFs have been scaled by an arbitrary value to aid in visibility of the secondary component, where darker regions correspond to deeper troughs of the CCF profile. \textbf{Right panel:} the measured CCF velocity of the primary is overlaid, as is the predicted RV of the secondary at each epoch, calculated via $\varv_{2} = \varv_1\,M_1/M_2$ and inserting \texttt{Saltire} masses from Table~\ref{tab:all-masses}. The predicted velocities visually follow the secondary's trace.}
    \label{fig:ccftrails}
\end{figure*}

\begin{figure}
    \centering
    \includegraphics[width=0.99\columnwidth]{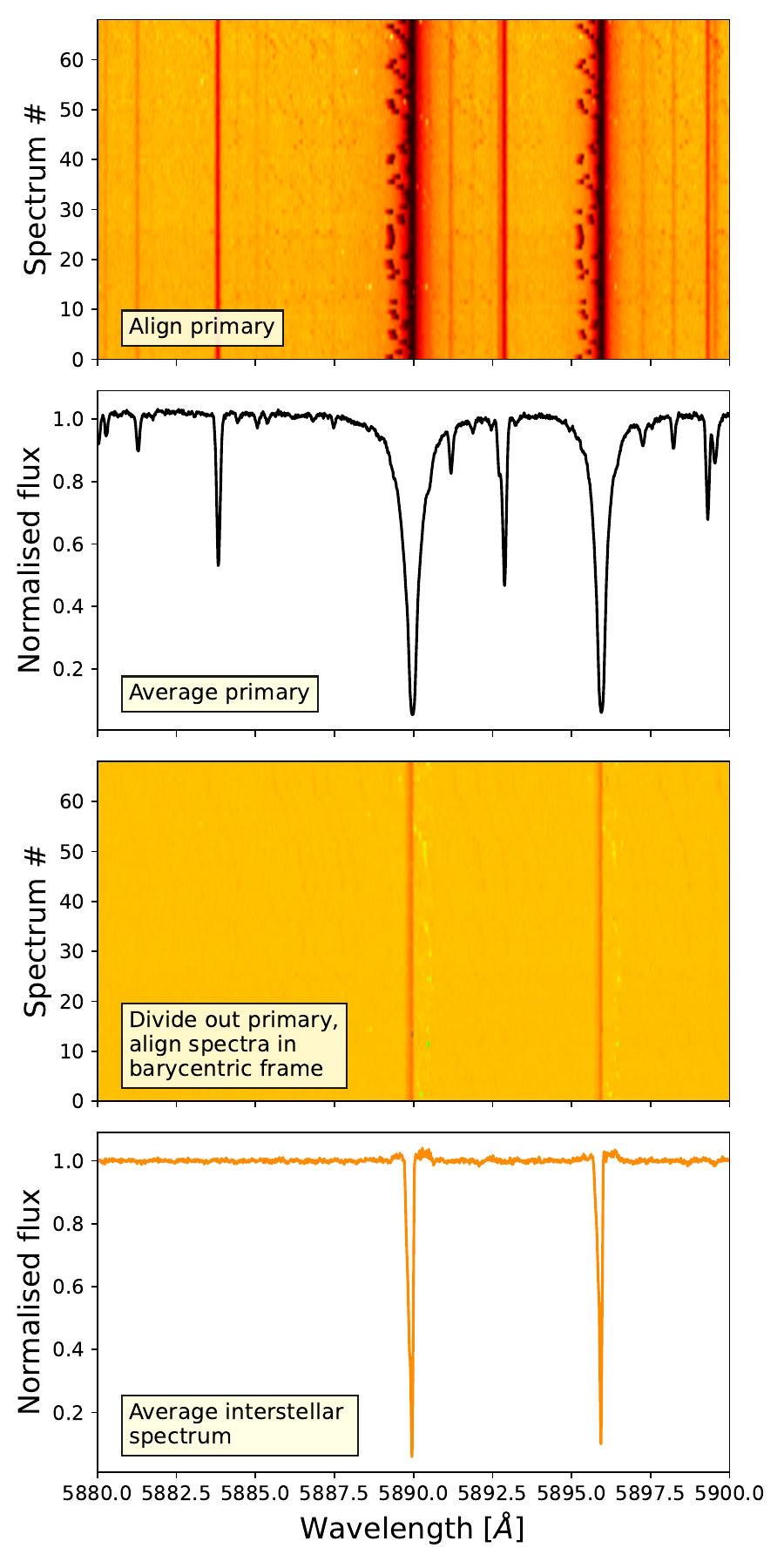}\\
    \caption{Figures illustrating the process of creating a master average primary spectrum, and removing this to identify interstellar sodium lines used to estimate reddening towards the targets, discussed in Section~\ref{sec:reddening}. \textbf{First panel:} Each normalised spectrum is plotted as a row of the image. Dark streaks show the primary's spectrum after alignment. \textbf{Second panel:} The averaged primary spectrum, showing high SNR features. \textbf{Third panel:} the average primary spectrum is divided out of each spectrum, and then returned to the barycentric frame. Orange streaks indicate features that are stationary with respect to the solar system barycentre, i.e. interstellar lines. \textbf{Fourth panel:} Averaging the residual spectrum creates a high SNR interstellar spectrum, where we can measure the EW of the lines to estimate extinction towards each target.}
    \label{fig:sodium_example}
\end{figure}

\begin{figure}
    \centering
    \includegraphics[width=0.99\columnwidth]{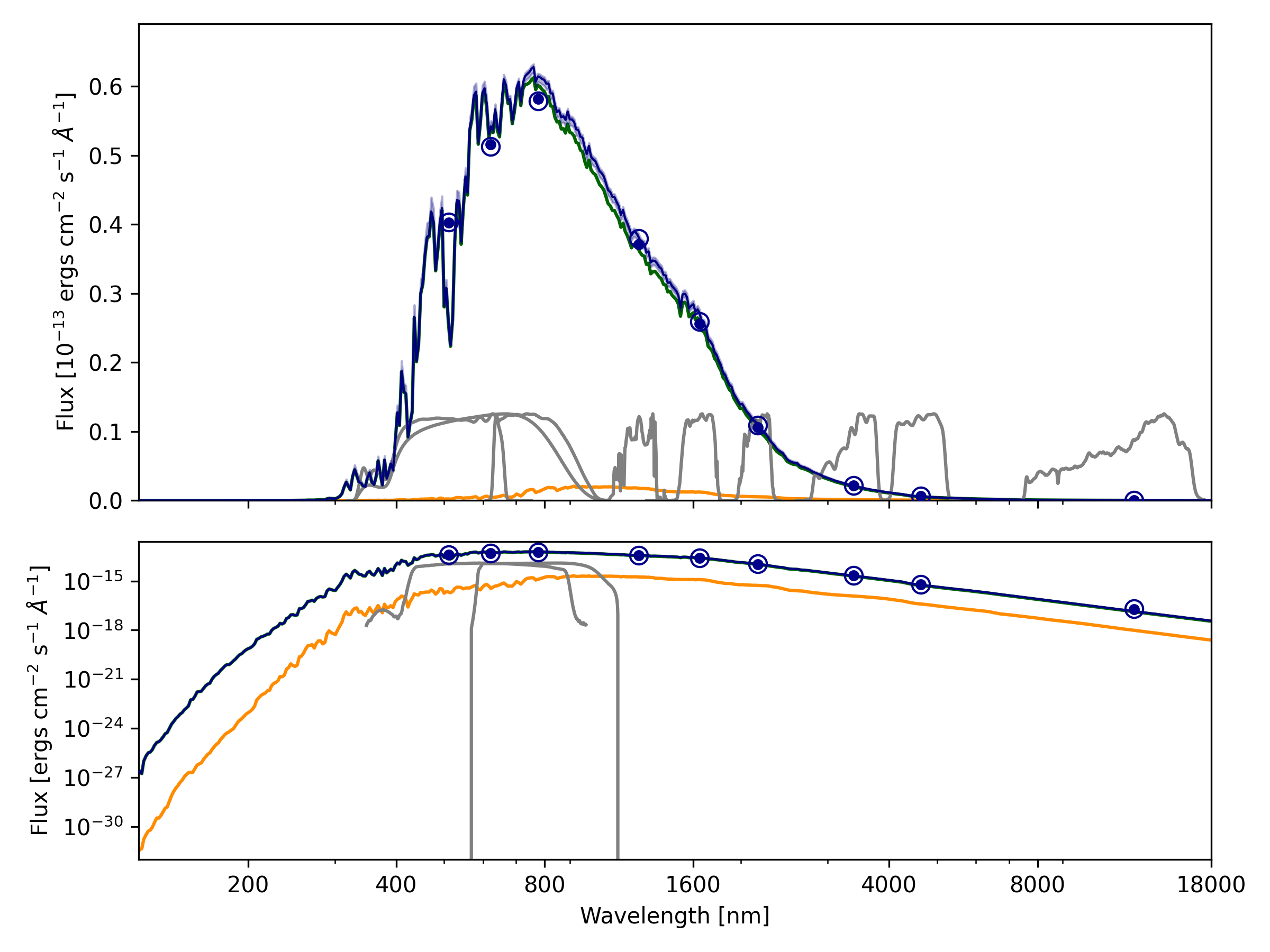}\\
    \includegraphics[width=0.99\columnwidth]{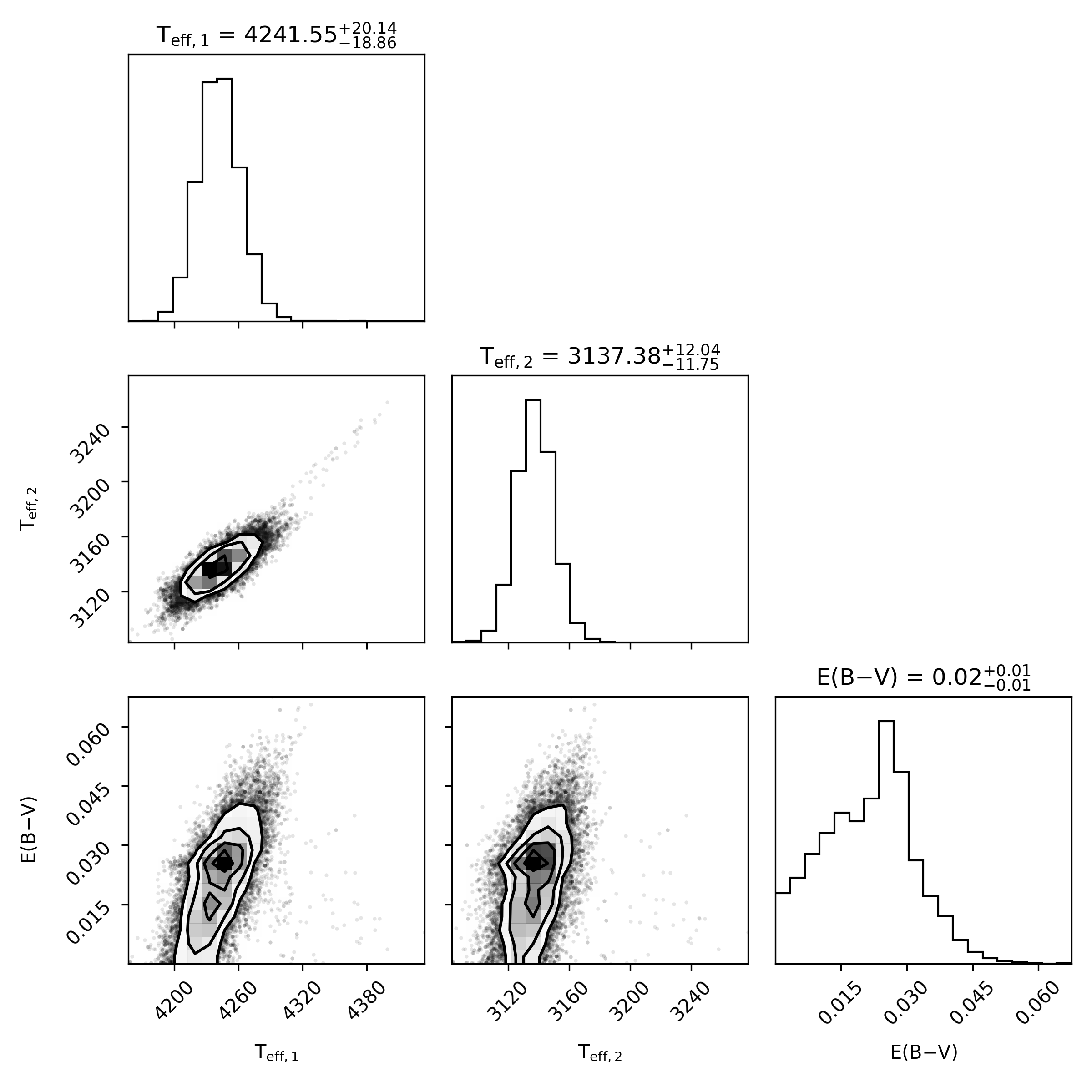}
    \caption{\textbf{Top:} \textit{Upper panel}: The SED of Kepler-16. The best-fit SED is plotted as a line and the mean SED $\pm 1-\sigma$ is plotted as a filled region. The observed fluxes are plotted as points with error bars and predicted fluxes for the best-fit SED integrated over the response functions shown are plotted with open circles.  The SEDs of the two stars are also plotted. \textit{Lower panel}: Same as the upper panel but with fluxes plotted on a logarithmic scale. Filters used to measure flux ratios are also plotted here. \textbf{Bottom:} Corner plots showing \textsc{teb} posterior distributions of \teffone, \tefftwo, and interstellar reddening, for Kepler-16.}
    \label{fig:teb_Kep16}
\end{figure}

\begin{figure}
    \centering
    \includegraphics[width=0.99\columnwidth]{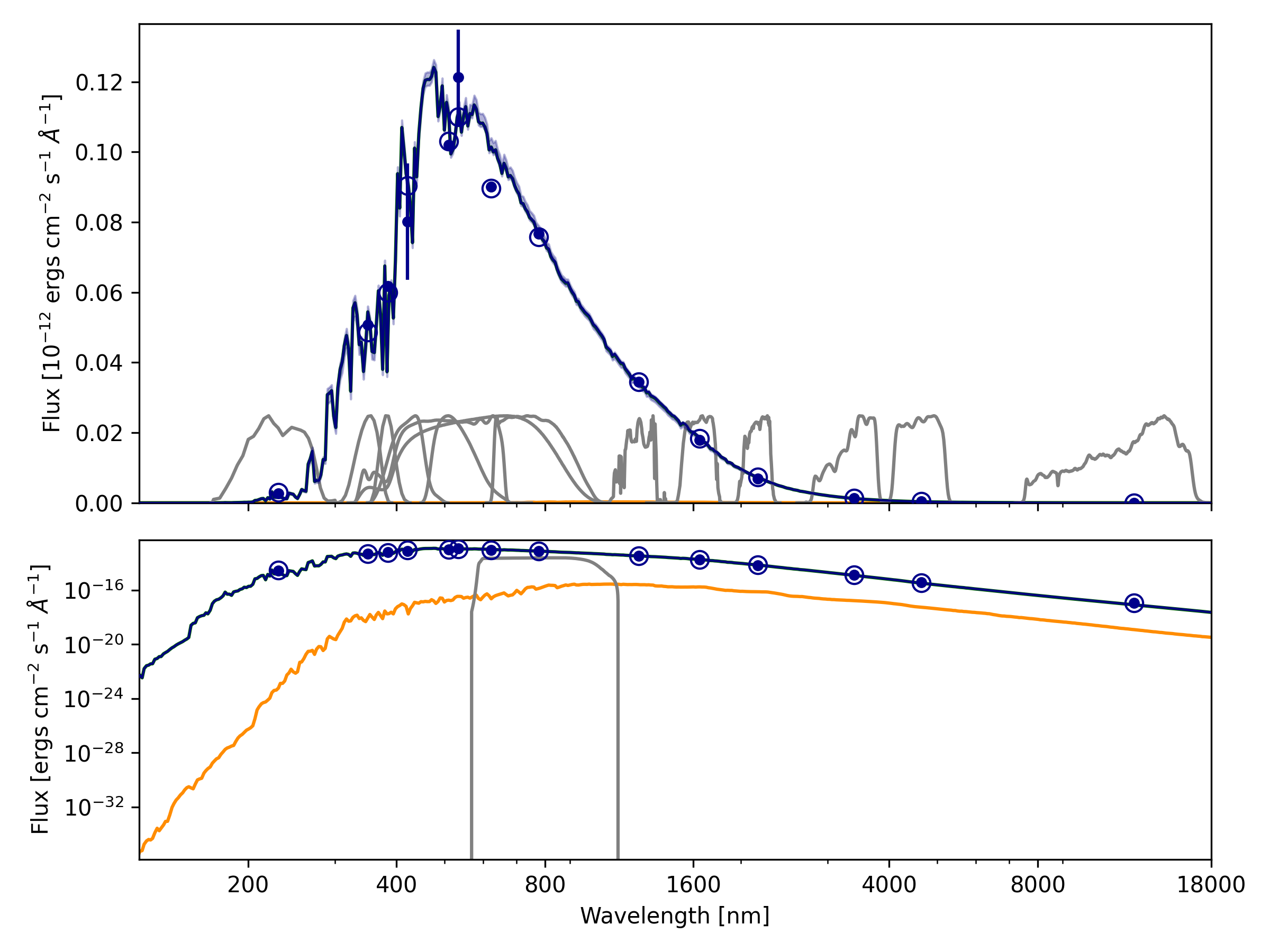}\\
    \includegraphics[width=0.99\columnwidth]{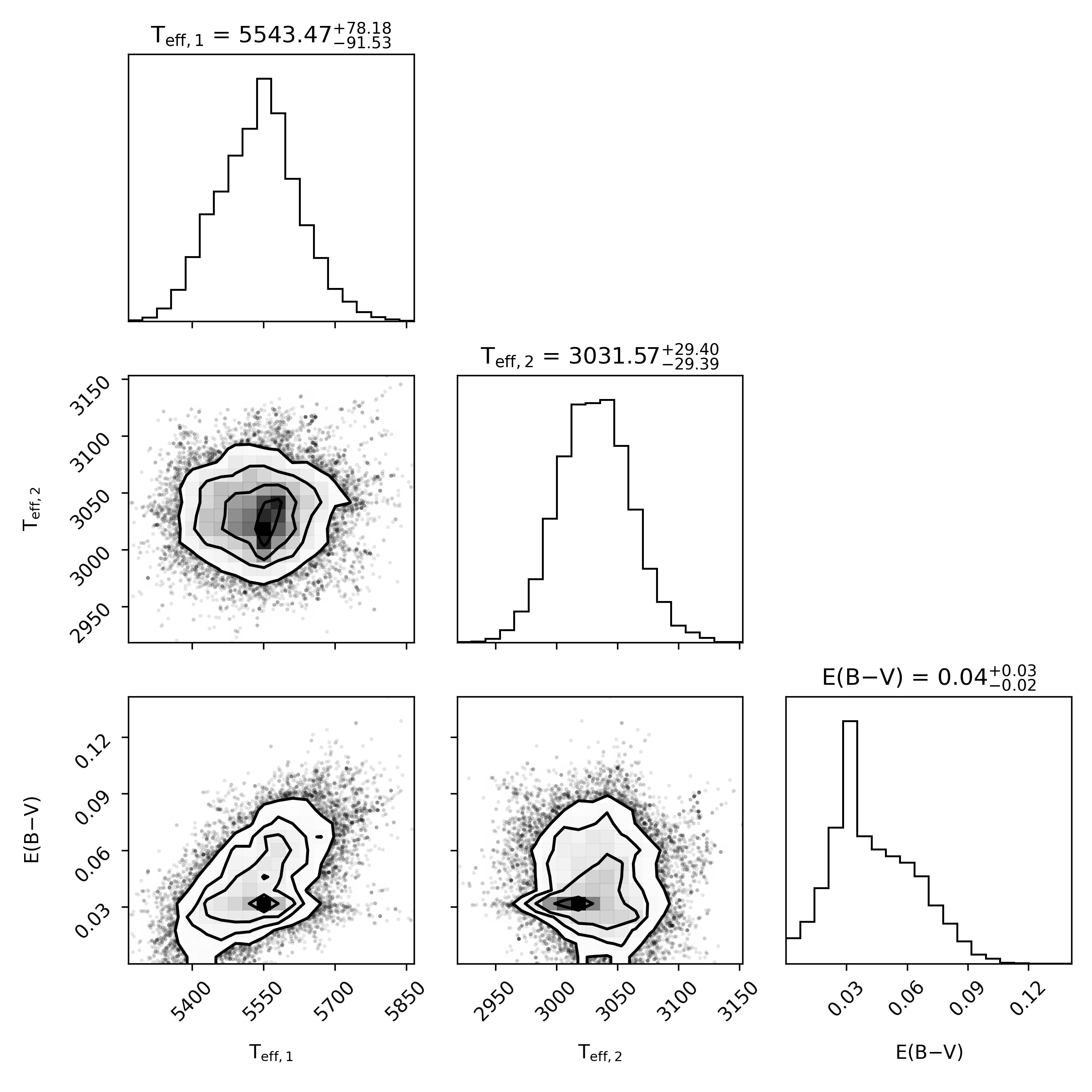}
    \caption{Same as Figure~\ref{fig:teb_Kep16}, but for J1928-38}
    \label{fig:teb_1928}
\end{figure}

\begin{figure}
    \centering
    \includegraphics[width=0.99\columnwidth]{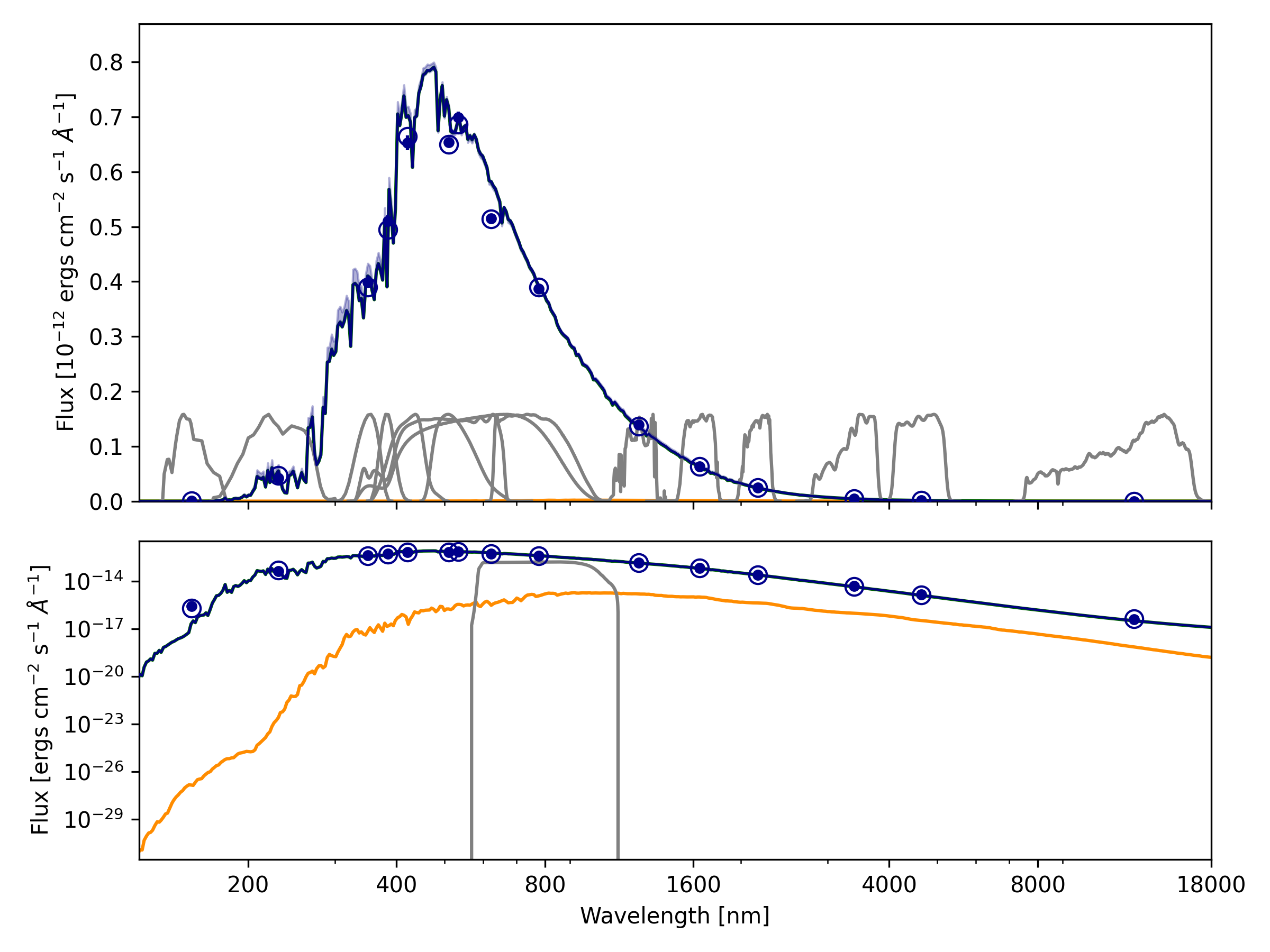}\\
    \includegraphics[width=0.99\columnwidth]{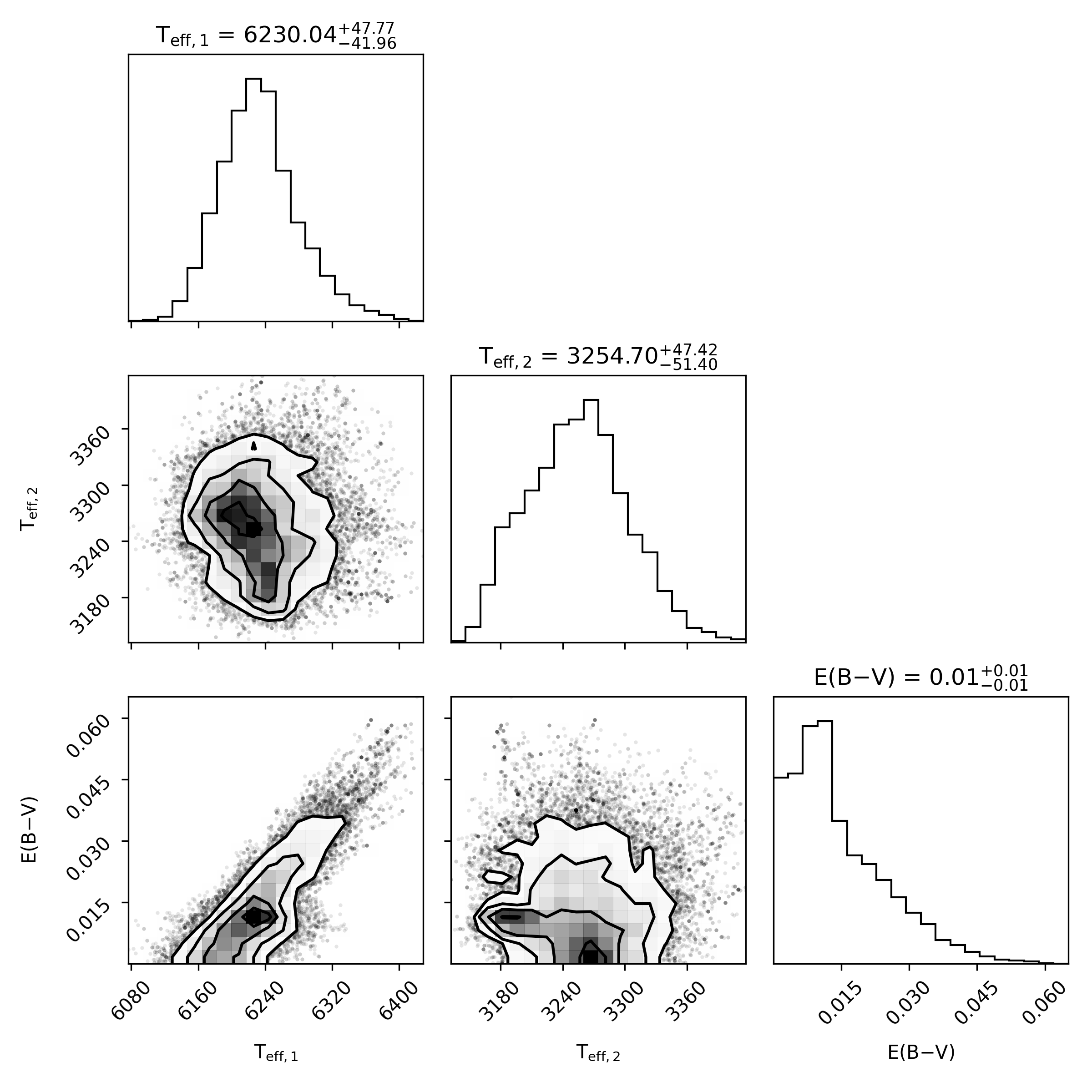}
    \caption{Same as Figure~\ref{fig:teb_Kep16}, but for J0310-31}
    \label{fig:teb_0310}
\end{figure}

\begin{figure}
    \centering
    \includegraphics[width=0.99\columnwidth]{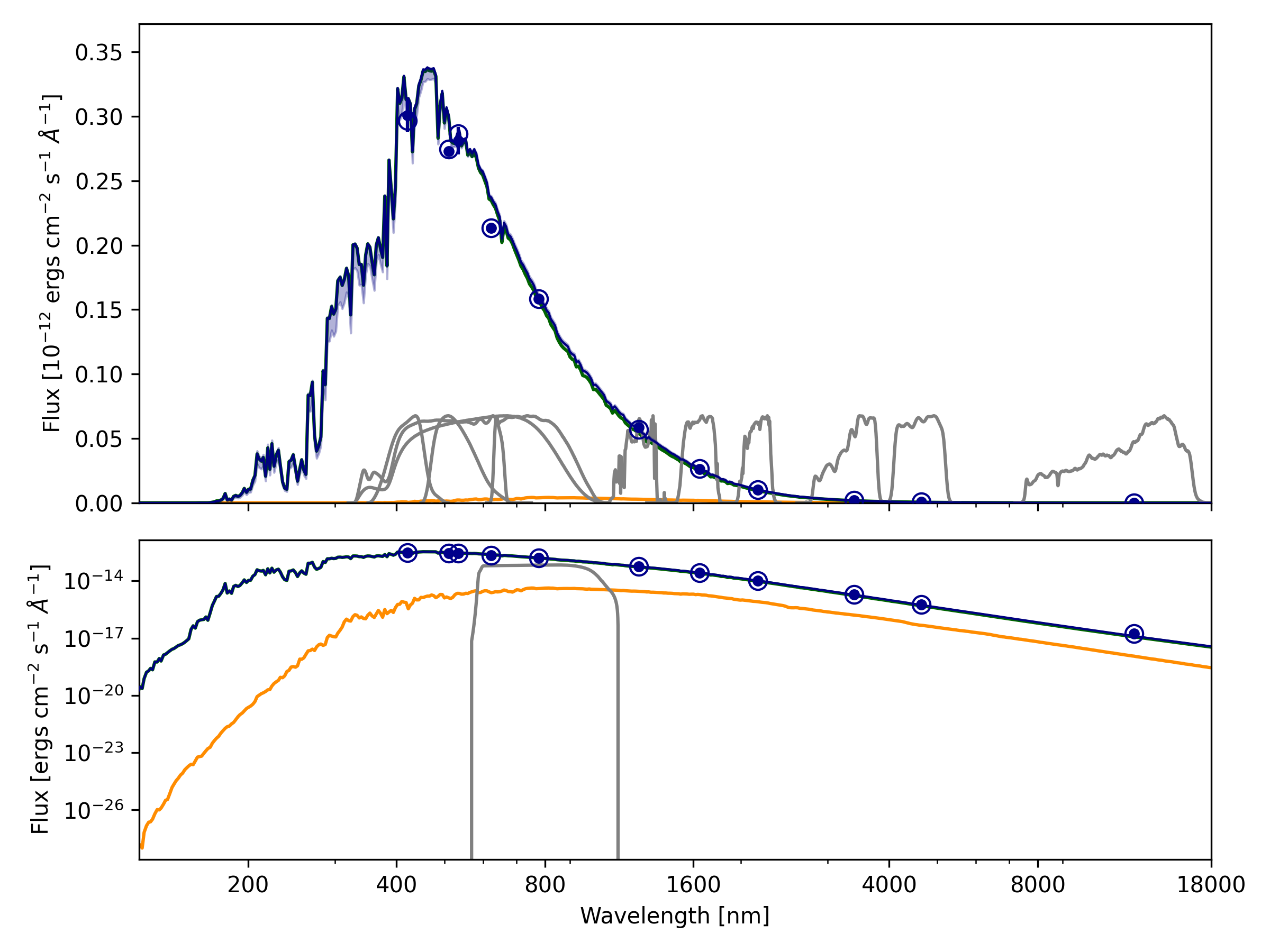}\\
    \includegraphics[width=0.99\columnwidth]{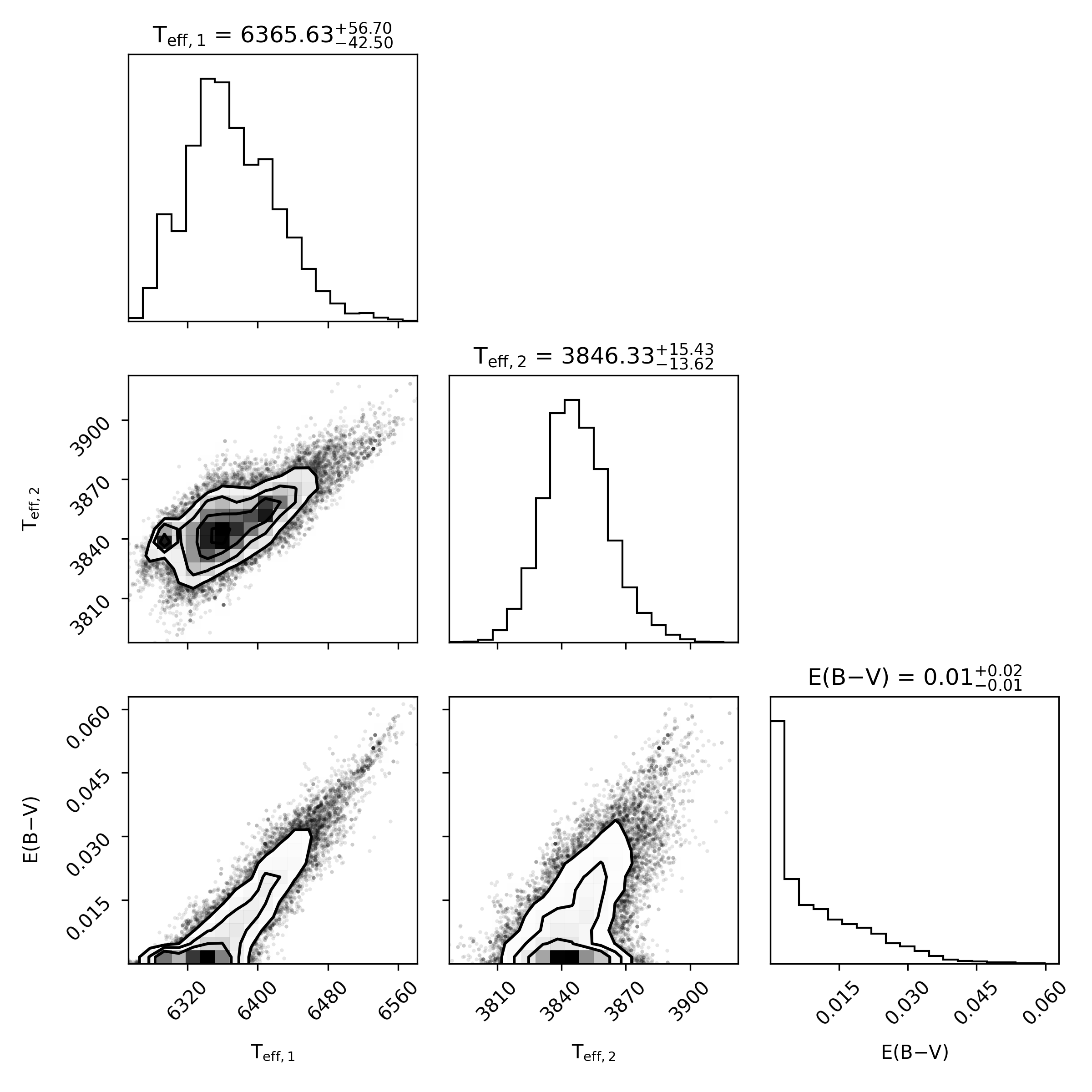}
    \caption{Same as Figure~\ref{fig:teb_Kep16}, but for J1840+34}
    \label{fig:teb_1840}
\end{figure}

\begin{figure}
    \centering
    \includegraphics[width=0.99\columnwidth]{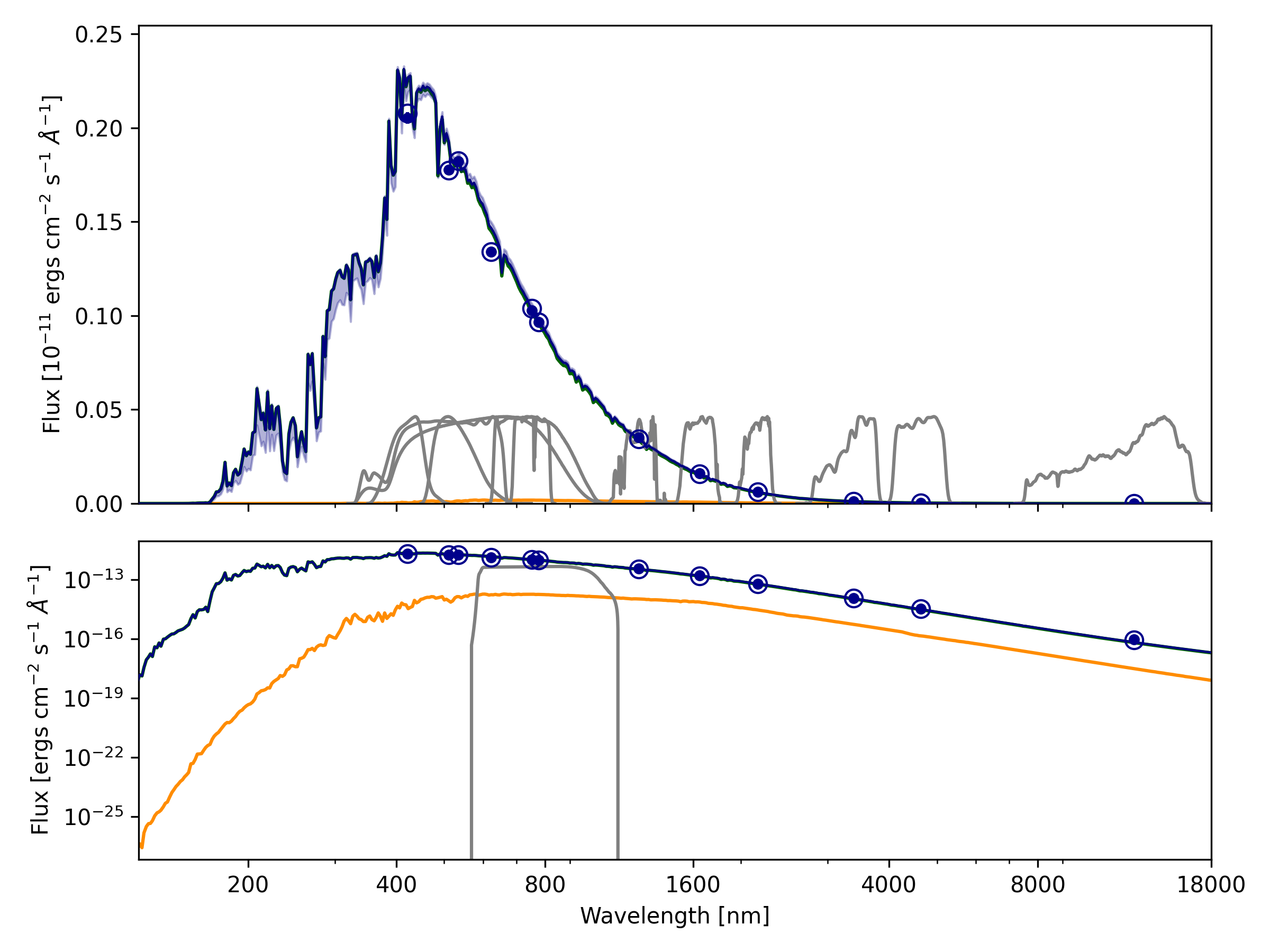}\\
    \includegraphics[width=0.99\columnwidth]{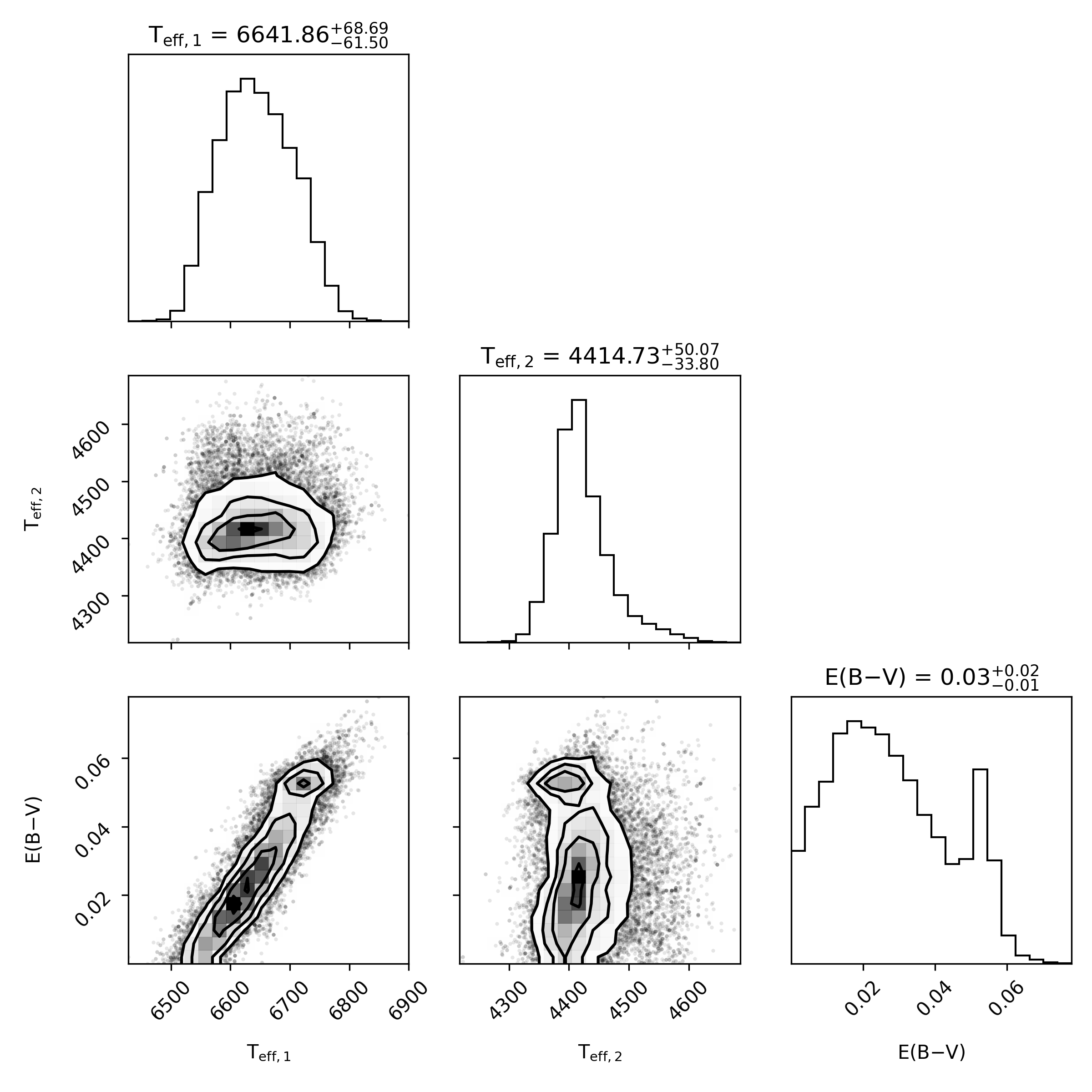}
    \caption{Same as Figure~\ref{fig:teb_Kep16}, but for J1555+26}
    \label{fig:teb_1555}
\end{figure}



\bsp	
\label{lastpage}
\end{document}